\documentclass[
 reprint, prd, superscriptaddress,
 amsmath,amssymb,
nofootinbib
]{revtex4-2}

\usepackage{tikz}
\usepackage{tikz-feynman}
\tikzfeynmanset{compat=1.0.0}
\usetikzlibrary{shapes,arrows,positioning,automata,backgrounds,calc,er,patterns}
\tikzfeynmanset{
doublefermion/.style={/tikz/double,
/tikz/decoration={name=none},
/tikz/postaction={/tikzfeynman/with arrow=0.5,}
}
}

\usepackage{amsmath,amssymb,amsfonts,latexsym,mathrsfs}
\usepackage{epsfig, graphicx, verbatim, xspace, multirow, mathtools}
\usepackage[utf8]{inputenc}
\usepackage{subfiles} 
\usepackage{cancel,xcolor,multirow}
\usepackage[normal]{caption}
\usepackage{subcaption}
\usepackage{float}   
\usepackage{placeins} 
\usepackage{slashed}
\usepackage{pifont}
\usepackage{color}
\usepackage{bm}
\usepackage{array}
\usepackage{enumitem}
\usepackage[normalem]{ulem}
\usepackage{booktabs}
\usepackage{makecell}
\usepackage{dcolumn}

\usepackage[colorlinks=true,urlcolor=purple,anchorcolor=blue,citecolor=blue,filecolor=blue,linkcolor=red,menucolor=blue]{hyperref}

\hypersetup{
  colorlinks=true,
  urlcolor=purple,
  anchorcolor=blue,
  citecolor=blue,
  filecolor=blue,
  linkcolor=blue,  
  menucolor=blue
}

\usepackage[percent]{overpic}

\newif\ifLinkDebug
\LinkDebugfalse 

\newcommand{\ExpDest}[1]{sec:cnst:#1}

\newcommand{\ClickBox}[2]{\makebox[#1][l]{\rule{0pt}{#2}}}

\newcommand{\ClickRect}[2]{%
  \ifLinkDebug
    \smash{\rlap{%
      {\setlength{\fboxsep}{0pt}%
       \setlength{\fboxrule}{0.6pt}%
       \color{magenta}\fbox{\ClickBox{#1}{#2}}}%
    }}%
  \fi
  \ClickBox{#1}{#2}%
}

\newcommand{\PlaceClick}[6][lb]{%
  \put(#2,#3){%
    \makebox(0,0)[#1]{%
      \hyperref[\ExpDest{#4}]{\ClickRect{#5}{#6}}%
    }%
  }%
}

\newcommand{\tanInv}{\text{tan}^{-1}\,}

\newcommand{\ve}{\varepsilon}
\newcommand{\vp}{\varphi}
\newcommand{\gbl}{g_{B-L}}

\newcommand{\snT}{\textbf{T}}
\newcommand{\snL}{\textbf{L}}
\newcommand{\snLT}{\textbf{L,T}}
\newcommand{\tCM}{\text{CM}}
\newcommand{\reP}{\text{Re} \, \Pi}
\newcommand{\imP}{\text{Im} \, \Pi}
\newcommand{\rfar}{R_\text{far}}
\newcommand{\km}{\text{km}}

\newcommand{\gev}{\text{GeV}}
\newcommand{\mev}{\text{MeV}}
\newcommand{\kev}{\text{keV}}
\newcommand{\ev}{\text{eV}}

\newcommand{\neff}{N_{\text{eff}}}
\newcommand{\dneff}{\Delta N_{\text{eff}}}

\newcommand{\ee}{e^{+}e^{-}}
\newcommand{\mumu}{\mu^{+}\mu^{-}}
\newcommand{\epm}{e^{\pm}}
\newcommand{\lra}{\leftrightarrow}

\newcommand{\cevns}{\text{CE}\nu\text{NS}}

\begin{document}

\begin{flushright}
    CERN-TH-2025-242
\end{flushright}

\title{\Large\textbf{Hiding a Light Vector Boson from 
Terrestrial Experiments: \\[0.5em]
A Chargephobic Dark Photon}}

\author{Haidar Esseili}
\affiliation{Department of Physics and Institute for Fundamental Science, 
University of Oregon, Eugene, Oregon 97403, USA}

\author{Graham D. Kribs}
\affiliation{Department of Physics and Institute for Fundamental Science, 
University of Oregon, Eugene, Oregon 97403, USA}
\affiliation{Theoretical Physics Department, CERN, 1211 Geneva 23, Switzerland}

\date{\today}

\begin{abstract}
We calculate the terrestrial, astrophysical and cosmological constraints
on a light vector boson that couples to an arbitrary combination 
of the electromagnetic and $B-L$ currents of the Standard Model.
The dark photon and a vector boson coupling to $B-L$ are special cases of our 
generalized flavor-universal anomaly-free vector boson, 
requiring just one additional parameter (the 
``dark mixing angle'' corresponding to the linear combination of 
the electromagnetic and $B-L$ currents) beyond that of the 
overall coupling strength and the vector boson mass, 
where we focus on the range $1$~MeV to $60$~GeV\@.
We perform a detailed investigation of a unique combination 
where the vector boson couplings to electrically charged leptons and protons are highly suppressed:  the 
``chargephobic dark photon''.
A chargephobic vector boson
is very weakly constrained
by current terrestrial experiments including beam dumps and collider experiments,
since they rely on couplings to electrons and protons.
Instead, neutrino scattering experiments (such as COHERENT),
astrophysical sources (supernova emission), and cosmology ($\Delta N_{\rm eff}$) provide the strongest
constraints due to the nonzero couplings of the chargephobic 
vector boson to neutrinos and neutrons. 
Indeed, we find that
supernova 
emission and $\Delta N_{\rm eff}$ 
provide constraints 
throughout the space of 
dark mixing angles,
demonstrating their importance to 
provide model-independent constraints.
For nearly all of the parameter space, a chargephobic vector boson 
is the most weakly constrained 
anomaly-free vector boson that couples
to flavor-independent or flavor-dependent combinations of standard model currents.  Finally, we highlight the importance of future experiments,
including SHiP, that \emph{are} able to probe new regions of the chargephobic parameter space due to the significantly improved detector capabilities.
\end{abstract}

\maketitle

\newpage
\section{Introduction}

Light vector bosons are ubiquitous in dark sectors, with their utility to connect the SM with the dark sector.  An essential ingredient to achieve a relic abundance of light dark matter consistent with cosmology requires a light mediator, and a light vector boson has long been considered a prime candidate for this interaction.  
The diversity of uses and implications of light vector boson mediators have motivated a vast range of experimental searches.  Extensive reviews 
(see, e.g., 
\cite{Ilten:2018crw,Bauer:2018onh,Fabbrichesi:2020wbt,Caputo:2021eaa,Caputo:2025avc}) have surveyed the terrestrial, astrophysical, and cosmological constraints on light vector boson couplings and masses.

The simplest model of a light vector boson mediator is the dark photon
$A'$.  The low energy effective theory of a massive dark photon is simply a vector boson with St\"uckelberg mass and kinetic mixing with electromagnetism\footnote{Above the electroweak scale, kinetic mixing with $U(1)_Y$.}, $\epsilon F_{A',\mu\nu} F^{\mu\nu}_{\rm em}$.  The kinetic mixing can be rotated away resulting in the dark photon
coupled to the electromagnetic current, $\epsilon e j_{\rm em}^\mu A'_{\mu}$.  Electromagnetic current conservation ensures the longitudinal mode of the massive vector boson does not couple to the SM (at zero temperature).  
Other anomaly-free currents in the SM
include $B-L$, the only other  flavor-universal conserved current, when three right-handed neutrinos are added to the model
(early discussions include
\cite{Davidson:1978pm, Marshak:1979fm,Mohapatra:1980qe, Fayet:1980ad, Fayet:1980rr, Wetterich:1981bx, Fayet:1986rh, Fayet:1989mq, Fayet:1990wx} and a review of massive $Z'$s \cite{Langacker:2008yv}),
while a light vector boson coupled to $B-L$ has been considered
in detail in \cite{Harnik:2012ni,Heeck:2014zfa,Bauer:2018onh,Ilten:2018crw}).  Other anomaly-free flavor-dependent combinations of $B$, $L_e$, $L_\mu$, $L_\tau$ have also been considered 
(there is a vast literature,
including \cite{Foot:1990mn,He:1991qd,
Foot:1994vd,Ma:1997nq,Ma:1998dp,Appelquist:2002mw, Boehm:2003hm, Fayet:2004bw, Fayet:2006sp, Fayet:2007ua, Salvioni:2009jp, Fayet:2016nyc, Fayet:2020bmb}
and more recently
\cite{Celis:2015ara,Bauer:2018egk,Altmannshofer:2018xyo,Heeck:2018nzc,Escudero:2019gzq,Han:2019zkz,Coloma:2020gfv,Bauer:2020itv, Ghosh:2024cxi}).

In this paper we consider a single, massive vector boson $X^\mu$, 
that couples to a \emph{linear combination} of the electromagnetic 
current and the $B-L$ current of the SM\@.  
The notion that a new vector boson could couple to an arbitrary linear 
combination of these currents is not new, 
and was emphasized long ago 
\cite{Fayet:1980rr,Fayet:1986rh,Fayet:1989mq,Fayet:1990wx},
in the context of $Z'$s detectable at LHC
\cite{Salvioni:2009mt}, as well as for light vector bosons
in \cite{Heeck:2014zfa}.
Indeed, Refs.~\cite{Salvioni:2009mt,Heeck:2014zfa} emphasized 
that it is expected
that a vector boson coupling purely to $B-L$ at a
high scale will automatically develop couplings to the 
hypercharge current (becoming a rescaled electromagnetic current below electroweak symmetry breaking) through renormalization group evolution.  The size of these
induced couplings depends critically on the field content and the extent of running.   
We will be agnostic towards the UV theory, and instead focus on the phenomenology within the low energy effective theory with no UV priors on the coupling strength, mixing angle, and vector boson mass.

A vector boson with a St\"uckelberg mass is a vector boson without gauge invariance \cite{Kribs:2022gri}.
Consequently, the vector boson can be directly coupled to an arbitrary vector current, i.e., $j_\mu X^\mu$, within or beyond the SM\@.  
That said, not all vector currents are the same, and in particular, nonconserved currents (such as baryon number) have significant additional constraints that arise from new processes enabled by anomalous couplings \cite{Dror:2017ehi,Dror:2017nsg}
(see also \cite{Kribs:2022gri}).  For this reason, in this paper we focus on coupling to anomaly-free conserved currents, and choose to restrict to (linear combinations of) flavor-independent currents of the SM.

Readers who are already familiar with the constraints on a dark photon and a $B-L$ vector boson will recognize
that there is a \emph{large} overlap of the terrestrial, astrophysical, and cosmological constraints on these two distinct ``pure'' vector mediators.  The overlap arises because just as a dark photon couples to charged leptons and protons, a $B-L$ vector boson also couples to leptons and hadrons.  The main distinction is that there are additional constraints that arise on $B-L$ due to its couplings to neutrinos (as well as couplings to neutrons).  Yet, much of the constraints that rely on couplings to electrons and/or protons broadly carry over from the case of a dark photon to the case of a $B-L$ vector boson.  

In this paper, we show that there is an anomaly-free light vector boson 
with \emph{qualitatively} different flavor-independent couplings:  a ``chargephobic'' vector boson.
A chargephobic vector boson couples to a specific linear combination of the electromagnetic current and the $B-L$ current, $- j_{\rm em}^\mu + j_{B-L}^\mu$, where the couplings to charged leptons and singly positively charged baryons
vanish.\footnote{Note that a vector boson with ``chargephobic'' couplings is \emph{neither} ``leptophobic'' (since there are couplings to neutrinos) nor ``protophobic'' \cite{Ilten:2018crw} (since protophobic vector bosons typically assumes there are couplings to charged leptons, while chargephobic does not couple to charged leptons).
Yet another flavor-dependent current that is distinct from what we consider is
$B-2 L_\mu - L_\tau + r_{\mu\tau} (L_\mu - L_\tau)$, 
where $r_{\mu\tau}$ is arbitrary.  This has been called 
``electrophobic'' \cite{Heeck:2018nzc}, since the vector boson does not couple to electrons, but does couple to muons and/or taus.}
In this case, the vast majority of constraints from terrestrial experiments \emph{disappear}.  This is because most terrestrial experiments rely on the coupling of the vector boson to electrons or muons and/or the coupling to protons.  At first glance, this has the potential to wipe clean the constraints in the coupling versus mass parameter space for a light vector boson mediator.
However, as we will see, several constraints remain:
terrestrial experiments that measure neutrino scattering off nuclei restrict much of the parameter space at larger couplings, while 
constraints on supernova cooling and constraints on modifications to early universe cosmology from $\Delta N_{\rm eff}$ restrict smaller couplings for smaller vector boson masses.\footnote{Ref.~\cite{Heeck:2014zfa} 
presaged several of these
constraints, but did not carry out a 
detailed investigation.}

The particular linear combination of currents that leads to chargephobic couplings is a tuned choice of parameters in the low energy effective theory.  
There is no technical obstacle to achieving this tuning from a UV theory.  Indeed, for larger couplings, Ref.~\cite{Salvioni:2009mt} 
suggested that $g_Y \sim - g_{B-L}$ could be an attractor of the renormalization group evolution
that is close (but not identical)
to achieving chargephobic couplings. 
Our interest will generally be much smaller couplings, and so we do not anticipate RG fixed points with SM couplings.  
We briefly discuss how
linear combinations of electromagnetism and $B-L$ could arise by renormalization group evolution from a high scale in Appendix \ref{generatingDarkMixingAppendix}.

What is more relevant for phenomenology is that there is 
renormalization group evolution 
of the dark mixing angle, 
and hence the couplings to
SM fields.   
As we show, this effect is smaller than one might think, leaving a substantial range of energies where chargephobic couplings are present; see Appendix \ref{RGEAppendix}.  Nevertheless, we take into account this coupling evolution in our analysis, and show the constraints from experiments that operate at larger energies (such as LHCb and CMS searches for vector boson resonances to $e^+e^-$ or $\mu^+\mu^-$), where the small regenerated coupling to charged leptons and protons is relevant.

Finally, in the course of calculating the constraints on a vector boson with chargephobic couplings, we also revisit the constraints on dark photons, $B-L$, and other linear combinations (that are not chargephobic).  
While we have employed recasting for a handful of constraints such as collider searches, we have also performed independent 
calculations of the major constraints that apply to a chargephobic gauge boson and other mixing angles $\vp$.
This includes a full out-of-equilibrium evolution of early universe dynamics of $X$ including finite temperature effects to determine the cosmological implications.  For supernova, we have carefully considered the production processes that arise as a function of $\vp$, and used the Garching simulations \cite{Garching} to model the detailed proto-neutron star properties that are essential to determine the cooling constraint.  This has not only allowed us to determine the constraint on the chargephobic case, but it has also allowed us to obtain an accurate cooling constraint on a $B-L$ vector boson (clearing up some tension in past literature). 
Finally, we have performed a thorough analysis of the constraints from neutrino scattering off nucleons, using the data releases from the experiments, to determine the constraints that arise for $\vp \not= 0$.  These constraints are the most powerful when couplings to electrons and protons are suppressed.

\section{Model}

\subsection{Vector Bosons Coupling to EM and $B-L$ Currents}
\label{sec:vectorbosonsEMBminusL}

We now provide an overview of the extension to the SM that we consider in the paper, specifying the parameters and couplings relevant to our analysis.
Consider a vector boson $X^\mu$ with Proca Lagrangian
\begin{eqnarray}
\mathcal{L}_X &=&{} -\frac{1}{4} F_{X,\mu\nu} F_X^{\mu\nu} + \frac{1}{2} m_X^2 X_\mu X^\mu \, ,
\end{eqnarray}
where we are interested broadly
in the mass range 
$1 \; {\rm MeV} \lesssim  m_X \lesssim 60 \; {\rm GeV}$.
For the purposes of this paper, the vector boson mass could arise from a spontaneously broken theory with suitable Higgs sector, where we assume the Higgs states are heavy enough to be neglected in our analysis, or it could simply be a St\"uckelberg mass. 
We will adopt the latter point of view, i.e., the vector field has a 
St\"uckelberg mass, 
and hence $X^\mu$ has no gauge invariance \cite{Kribs:2022gri}. 
This immediately implies that we are free to write interactions of $X^\mu$ to arbitrary currents 
of the SM\@.  In this paper we further restrict the 
interactions of $X^\mu$:
\begin{itemize}
    \item Couplings to the SM are flavor-independent;
    \item Couplings only to 
    conserved currents\footnote{More precisely, only linear interactions of $X^\mu$ with SM conserved currents. We do not include a Higgs portal interaction $H^\dagger H X^\mu X_\mu$.} of the SM; and,
    \item Dark sector
    interactions, if present,
    do not affect the couplings to the SM nor the decay width of $X^\mu$.
\end{itemize}
While there are rather stealthy 
anomaly-free combinations of baryon number and one or more
specific flavors of lepton number
(for example, $B - 3 L_\tau$), these interactions
have strong constraints from non-standard 
neutrino interactions (NSI), for example 
\cite{Heeck:2018nzc,Altmannshofer:2018xyo,Han:2019zkz,Coloma:2020gfv,Bauer:2020itv}. 
Couplings to non-conserved currents have
also been considered in \cite{Dror:2017ehi,Dror:2017nsg,Dror:2018wfl,Dror:2020fbh,Michaels:2020fzj,Kribs:2022gri}, where non-trivial phenomenology arises because 
longitudinal mode of $X^\mu$ also can
couple to the SM\@.  This can result in amplitudes involving $X^\mu$ that diverge
at large $\sqrt{s}/m_X$, leading to a 
host of other strong experimental 
constraints.
Finally, interactions between a light vector boson and the dark sector 
have been considered in a variety of 
contexts.  For our purposes, if 
$X^\mu$ couples to other dark sector fields, we assume these interactions do not affect the couplings to the SM, nor do they change the decay width of $X^\mu$.

The interactions of $X^\mu$ to the SM that we consider are thus\footnote{At high energies, mixing with the hypercharge current induces both the electromagnetic coupling and a suppressed coupling to the 
$Z$ current, see Appendix~\ref{generatingDarkMixingAppendix} for details.}
\begin{eqnarray}
\label{interactionLagrangian}
\mathcal{L}_{\text{int}} &=& \left( 
\ve e j^{\text{EM}}_\mu  + \gbl j^{B-L}_{\mu} 
\right) X^\mu \, ,
\end{eqnarray}
where
\begin{eqnarray}
j_\mu^{\rm EM} &=& \sum_f Q_f \bar{f} \gamma_\mu f \\
j_\mu^{B-L} &=& \sum_f q_f \bar{f} \gamma_\mu f 
\end{eqnarray}
and $Q_f = (-1,0,2/3,-1/3)$ are the usual electric charges of the SM fermions $(e,\nu,u,d)$ while 
\begin{equation}
q_f = (B - L)_f =
\left\{ \begin{array}{lcl} +1/3 &\quad& u, \; d \\
-1 &\quad& e, \; \nu_L, \; \nu_R \, ,
\end{array} \right.
\end{equation}
are also the usual $B-L$ charges of the SM fermions.
We have normalized the relative strength of $X^\mu$
couplings to the $B-L$ current with $\gbl$
and the EM current with $\ve e$ 
which is familiar from dark photon literature.
We have included right-handed neutrinos to ensure $B-L$ is anomaly-free, however we take the right-handed neutrinos to be heavy, above at least the weak scale, so that they do not directly affect the phenomenology of $X^\mu$.  A detailed discussion of the role of right-handed neutrinos, and the effects of taking them heavy can be found in our previous work \cite{Esseili:2023ldf}.

\begin{figure*}
\centering
\includegraphics[width=0.9\textwidth]{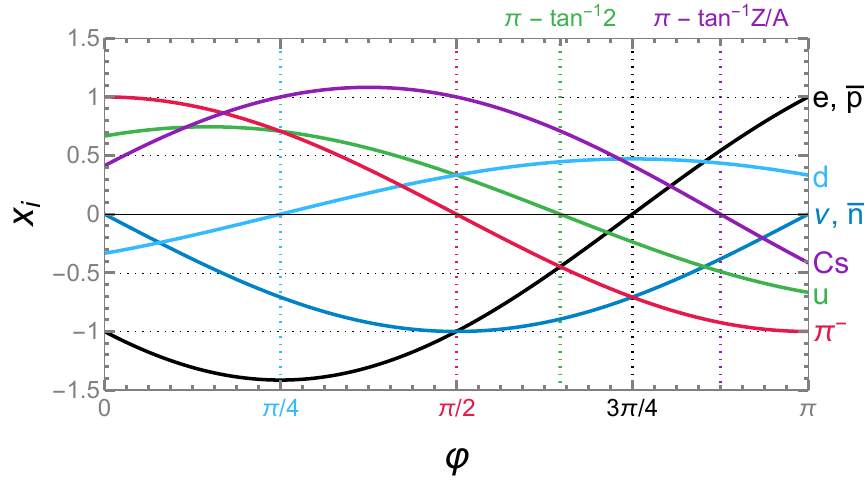}
\caption{ Charges $x_i$ defined in Eq.~ (\ref{electricChargeUX}). The proton, neutron, and $\pi^{+}$ charges are equal and opposite to the electron, neutrino, and $\pi^{-}$ respectively. We also show the scaled Cesium charge $x_{\rm Cs}/A_{\rm Cs}$ which is relevant for $\cevns$ experiments. The vertical dotted lines correspond to specific combinations of the electromagnetic and $B-L$ currents that are of interest, as shown in Fig.~\ref{angleParamSpaceFig}.}
\label{chargesFig}	
\end{figure*}

We can rewrite the interactions of $X^\mu$ as
\begin{eqnarray}
\mathcal{L}_X &=& g_X j_\mu^X X^\mu \\
j_\mu^X  &=& \cos\varphi \, j_\mu^{\rm EM} + \sin\varphi \, j_\mu^{B-L} \; = \; \sum_f x_f \, \bar{f} \gamma_\mu f \, ,
\end{eqnarray}
where we define the overall coupling strength 
and dark mixing angle as
\begin{eqnarray}
g_X & \equiv & \sqrt{(\ve e)^2+\gbl^2} \\
\vp & \equiv & \text{tan}^{-1} \, \frac{\gbl}{\ve e} \, ,
\label{mixingAngleDef}
\end{eqnarray}
such that the ``charges'' 
(or couplings) for an arbitrary fermion to the $j_\mu^X$ current are
\begin{eqnarray}
x_f &=& Q_f \cos\varphi + q_f \sin\varphi
\end{eqnarray}
and for the SM fermions,
\begin{align}
x_{e}    &= -\cos{\vp}-\sin{\vp}, \quad 
&& x_{\nu} = -\sin{\vp}, \quad \nonumber\\
x_u     &= \frac{2\,\cos{\vp}}{3}
        + \frac{\sin{\vp}}{3}, \quad  
&&x_{d}   = -\frac{\cos{\vp}}{3} + \frac{\sin{\vp}}{3} \, ,
\label{electricChargeUX}
\end{align}
that we illustrate in Fig.~\ref{chargesFig}. 
Note that under 
$X^\mu \rightarrow - X^\mu$, 
$\vp \rightarrow \vp + \pi$,
and so without loss of generality
we can restrict $0 \le \vp < \pi$.

Mesons obviously do not carry $B-L$ number, and so their couplings to $X^\mu$ are determined just by their couplings to the electromagnetic current.  For example, the pions
\begin{eqnarray}
x_{\pi^\pm} &=& \pm \cos\varphi \\
x_{\pi^0} &=& 0 \, .
\end{eqnarray}
Similarly, neutral baryons obviously do not 
carry electric charge,
with their coupling to $X^\mu$ being the same
as their coupling to a $B-L$ vector boson.
Positive electrically charged baryons
have a $\varphi$-dependent coupling to 
$X^\mu$, which is easy to see is identical
to the coupling of an 
electrically 
charged anti-lepton.  For example, for the proton, 
$x_{p} = -x_e$ (and $x_{\bar{p}} = x_e$). 
Similarly, the neutron and neutrino have equal and opposite 
couplings, $x_n = - x_\nu$
(and $x_{\bar{n}} = x_\nu$).
The couplings to baryons with different electrical charges 
differ from one another,
for example for the $\Delta$ multiplet:
\begin{eqnarray}
x_{\Delta^{++}} &=& 2 \cos\varphi + \sin\varphi \\
x_{\Delta^{+}} \;=\; x_{p} &=&  \cos\varphi + \sin\varphi \\
x_{\Delta^{0}} \;=\; x_{n} &=&   \sin\varphi \\
x_{\Delta^-} &=& -\cos\varphi + \sin\varphi \, .
\end{eqnarray}

\section{Overview of Results}

\subsection{Dark Photon and $B-L$ Results}

In general, small deviations in $\vp$ result in small 
modifications to observables, and thus unsurprisingly small modifications to the constraints for a given $\vp$.  This general result
does \emph{not} hold when the charge of a particle or set of particles is close to or exactly zero, 
i.e., $x_i = 0$.  In these special cases, 
unique and nontrivial phenomenology arises. 
The best known example is the dark photon,
corresponding to $\vp=0$, and thus 
$x_\nu = 0$: 
the dark photon has no coupling to neutrinos. 
The next best known example is the $B-L$ vector boson,
corresponding to $\vp=\pi/2$.
In this case, the couplings to mesons vanish,
while the proton and neutron couplings are identical.

Anticipating the many various constraints 
arising from terrestrial experiments,
astrophysics, and cosmology, we first 
show the bounds we obtain on the well-known cases of the dark photon and $B-L$ in Fig.~\ref{DPandBmLCurrentConstraintPlots}. Besides some minor differences in constraint shapes, the main difference between the dark photon and the 
$B-L$ vector boson cases is the additional set of constraints 
from neutrino-dependent observables 
that arise with $B-L$.  Our results are broadly consistent with 
prior equivalent results that can be found in
\cite{Ilten:2018crw,Bauer:2018onh,Caputo:2025avc}. 
The primary purpose of showing our
calculated constraints
is to demonstrate that our analysis pipeline
for general $\vp$ broadly reproduces 
existing results in the specific 
$\vp =0$ (dark photon) and $\vp = \pi/2$ 
($B-L$ vector boson) limits.
In later sections, we highlight a few 
specific distinctions between our results
and those in past literature.

In order to orient the reader to which results correspond to which mixing angle, we provide a mixing angle ``compass'' defined in Fig.~\ref{angleParamSpaceFig}.  This shows the cases of interest in the paper (dark photon, $B-L$, chargephobic) as well as several additional angles with their own unique properties.

\subsection{Chargephobic Dark Photon Results}

Our main purpose is to explore 
the chargephobic case, $\vp = 3\pi/4$.
The tree-level couplings to quarks, leptons, the proton, neutron, and charged pions are:
\begin{eqnarray}
\left.
\begin{array}{rcl}
x_u &=& -\frac{1}{3 \sqrt{2}} \\
x_d &=& \frac{\sqrt{2}}{3} 
\end{array}
\right\}
& \longrightarrow & 
\left\{
\begin{array}{rcl}
 x_p &=& 0 \\
 x_n &=& \frac{1}{\sqrt{2}} \\
 x_{\pi^{\pm}} &=& \mp \frac{1}{\sqrt{2}}
\end{array} 
\right.\nonumber  \\
x_e &=& 0 \\
x_\nu &=& -{\small \frac{1}{\sqrt{2}}} \nonumber
\end{eqnarray}
A chargephobic $X^\mu$
couples to quarks, and therefore
baryons, in a very specific way: 
$X^\mu$ is ``phobic'' only to 
singly positively charged baryons (and singly negatively charged anti-baryons).
Moreover, while $X^\mu$ couples to charged mesons,
there is no
coupling to neutral mesons.
The couplings to neutrons and
neutrinos implies important astrophysical 
and cosmological consequences.
For astrophysics, the 
constraint from excessive
cooling of SN1987A by $X^\mu$
emission 
remains strongly constrained 
in a way broadly comparable to the dark photon 
and $B-L$.  This occurs dominantly because of
$X^\mu$ bremsstrahlung and other emission processes
from neutrons in the supernova.
Differences are present, however, since once $X^\mu$ is emitted from a supernova, it decays nearly $100\%$ of the time into neutrinos, whereas both a dark photon and $B-L$ vector boson have a large branching fraction to charged leptons.
Similarly, for cosmology, $X^\mu$ can be 
produced from its interactions with neutrinos, 
modifying $\Delta N_{\rm eff}$, again broadly
comparable to $B-L$ (where production of $X$ can occur from all leptons)
or the dark photon (production of $X$ restricted to just charged 
leptons).

Our results for the chargephobic case are shown in
Fig.~\ref{CPConstraintPlots}.
The majority of constraints from beam dump experiments, fixed target experiments, colliders, precision constraints, neutrino-electron scattering experiments, and searches to invisible decays are either \emph{absent} or \emph{suppressed}.
What remains are constraints from coherent neutrino nuclear scattering experiments, 
a small region
related to hadron beam fixed target experiments that are specifically sensitive to neutrinos (DONuT and FASER$\nu$),
early universe modification to $\Delta \neff$, and excessive cooling of protoneutron star after supernova (SN1987A), 
and 
all of which are resilient to a vanishing charged lepton coupling. 

The astute reader will observe that Fig.~\ref{CPConstraintPlots} still contains weakened ``stalactite'' bounds in green from collider experiments,  despite the fact that these constraints rely on couplings to muons. 
The re-appearance of these bounds
arises because the mixing angle $\vp$
has a loop-induced energy dependence 
due to the renormalization group
evolution of the underlying 
couplings.  In 
Fig.~\ref{CPConstraintPlots}, we made the particular choice $\vp(\mu = m_e) = 3\pi/4$,
and evolved the angle using 
the renormalization of the 
couplings detailed in
App.~\ref{RGEAppendix}.
While the electromagnetic coupling
evolves with renormalization scale
above $\mu = m_e$ as usual, the dark mixing angle $\vp$ does not actually start to 
deviate from its low energy value until $\mu \gtrsim m_\pi$; see Appendix \ref{RGEAppendix}. 
This explains why the collider experiment
bounds appear only above about
$200$~MeV, and get slightly stronger with increasing mass up to $m_X = 60$~GeV at the upper right-edge of the plot.

Fig.~\ref{CPConstraintPlots}
is the main result of our paper, 
with the subsequent sections of the
paper detailing how the individual
terrestrial, astrophysical,
and cosmological constraints
are obtained.  
What should be clear from
Fig.~\ref{CPConstraintPlots}
is that current terrestrial experiments are highly insensitive to a chargephobic vector boson.  

\subsection{Other Dark Mixing Angles}

\begin{figure*}
\centering
\includegraphics[width=0.58\textwidth]{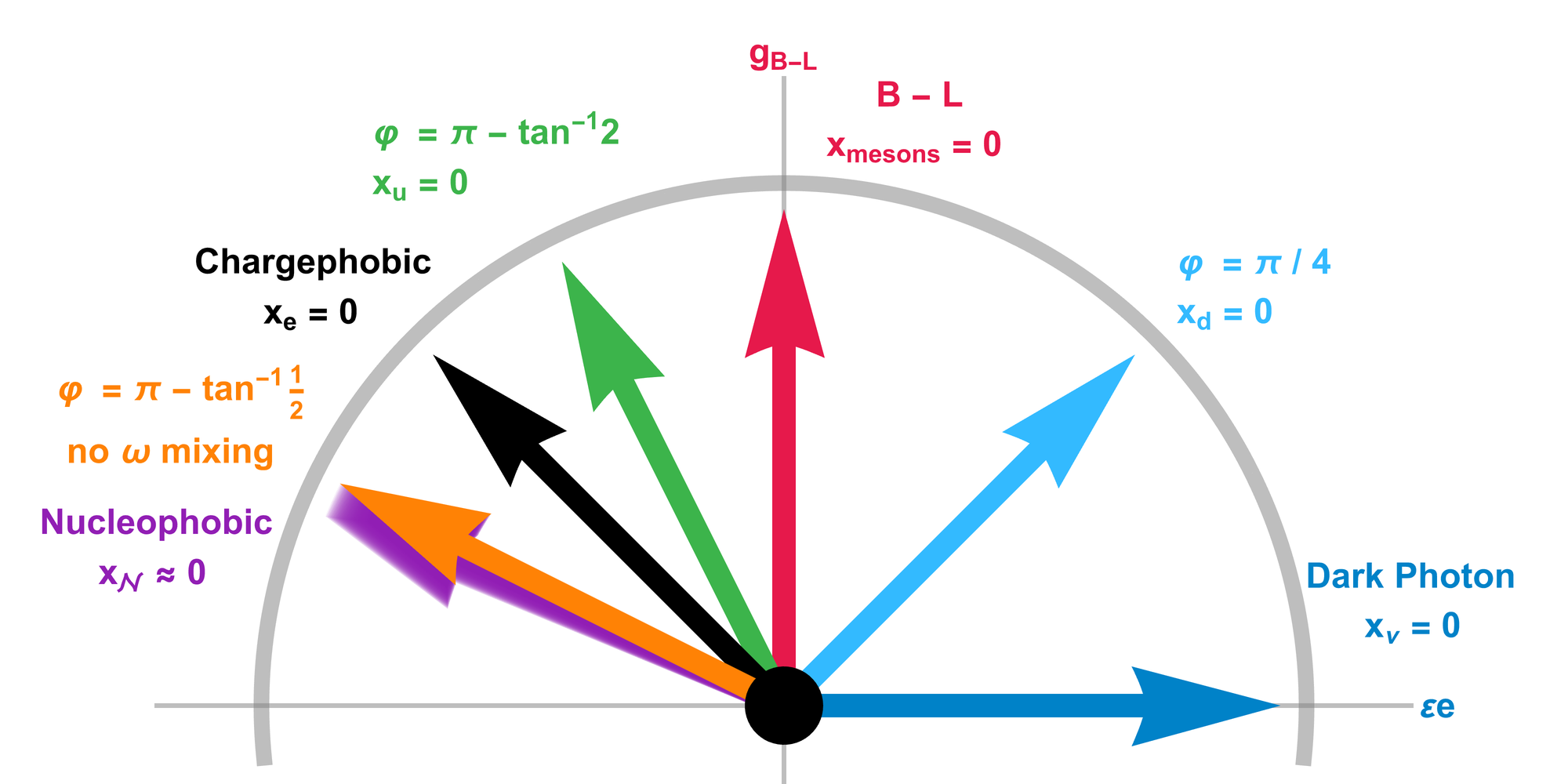}
\caption{The dark mixing angle ``compass''.  Each arrow corresponds to a dark mixing angle $\vp$ that has a  vanishing tree-level coupling to at least one SM state.  The compass is used in subsequent plots to easily illustrate which dark mixing angles are being considered.}
\label{angleParamSpaceFig}
\end{figure*}

There are several other dark mixing angles (beyond $\vp = 0, \pi/2$) that have their own interesting properties. 
In particular, consider
$\vp =\pi-\text{tan}^{-1}(Z_{\mathcal{N}}/A_{\mathcal{N}})$, where 
the coupling to an entire nucleus with charge number $Z_{\mathcal{N}}$ and mass number $A_{\mathcal{N}}$ can vanish. 
(The vanishing of the coupling assumes that $X^\mu$ couples
\emph{coherently} to the nucleus, i.e., 
the momentum transfer is small compared with inverse size of the nucleus.)
We will call this angle \textit{nucleophobic}, and consider the implications in the context of the $\cevns$ analysis. 
(A related idea, ``xenophobic'', tunes the proton and neutron couplings so that the coherent response of xenon nuclei is strongly suppressed for the purposes of suppressing
dark matter direct detection 
signals in \cite{Feng:2013vod}.)
Of particular interest to our analysis are the nuclei of cesium, iodine, and germanium. For example, $\vp=\pi-\text{tan}^{-1}(55/133)$ results in a model with vanishing tree level coupling to Cesium; here $x_p/x_n\approx -1.42$ which leads to $x_p Z_{\rm Cs} +x_n N_{\rm Cs}=0$. The angle has the added consequence of suppressing the coupling to other intermediate mass stable nuclei which share a similar number ratio of proton to neutrons; for example, $N_{\rm Cs}/Z_{\rm Cs}\approx N_{\rm I}/Z_{\rm I}\approx 1.4$. This is evident in the valley of stability relation as derived from the semi-empirical mass formula $A/2Z \approx 1+0.007A^{2/3}$. The suppression is still in effect but less severe for lighter nuclei such as germanium and argon where $N/Z\approx 1.2$. Explicitly, the nuclear coupling to $X$ are $\{x_{\rm Cs},\;x_{\rm I},\; x_{\rm Ge},\; x_{\rm Ar}\}_{B-L}= \{133,\; 127,\; 72,\; 40\}$ for $B-L$ and are suppressed to $\{x_{\rm Cs},\;x_{\rm I},\; x_{\rm Ge},\; x_{\rm Ar}\}_{\cancel{Cs}}= \{0,\; -0.4,\; -2,\; -1.3\}$ in the nucleophobic case with vanishing Cesium coupling.\footnote{Note that $\vp =\pi-\text{tan}^{-1} \, Z/A$ is quite close to $\vp =\pi-\text{tan}^{-1} \, 1/2$, which is the angle where $\omega$ mixing vanishes in Fig.~\ref{angleParamSpaceFig}. In the former case, $x_p=-x_n (A-Z)/Z$ and the coupling to the nucleus vanishes. As for the latter case $x_p=-x_n$ and $X$ effectively couples to the difference $A-2 Z$ rather than the full nucleus $A$ as it would occur for $B-L$. The two angles converge for light nuclei with $A = 2 Z$.}

\begin{figure*}
\centering
\begin{overpic}[width=0.49\textwidth]{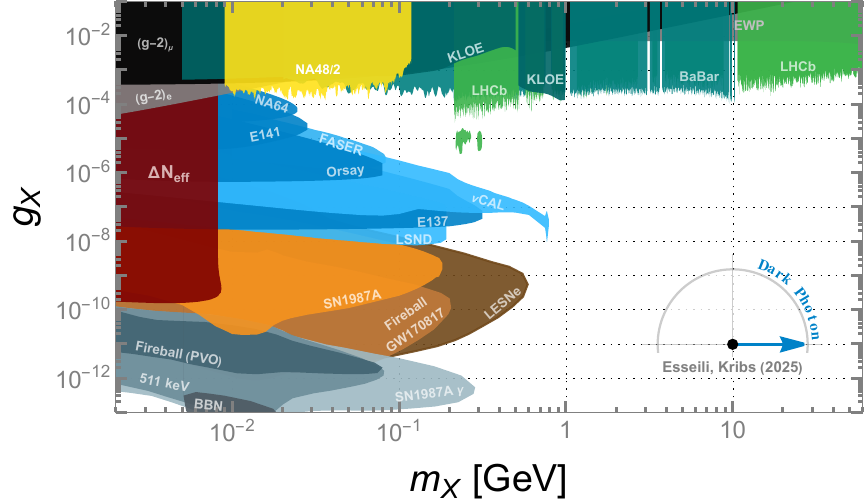}
  \input{cnstrntLabelLinks/dpCnstrntLinks}
\end{overpic}
\hfill
\begin{overpic}[width=0.49\textwidth]{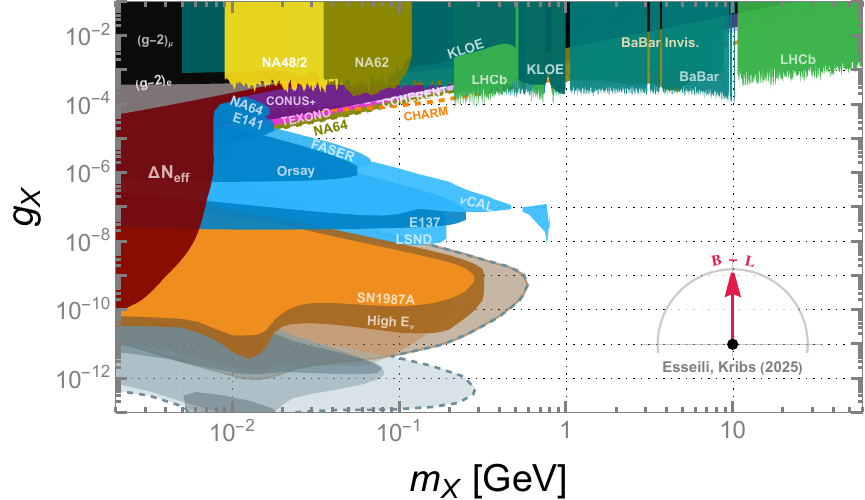}
  \input{cnstrntLabelLinks/bmlCnstrntLinks}
\end{overpic}
\caption{Terrestrial experiment constraints, astrophysical constraints, and cosmological constraints on the dark photon ($\vp=0$) and $B-L$ case ($\vp=\pi/2$), see text for full details. 
For $B-L$, there are additional constraints that arise due to non-zero couplings to neutrinos, in particular those labeled CHARM, COHERENT, CONUS, NA64, and TEXONO\@. 
For the dark photon case, the constraints labeled
BBN, LESNe, SN1987A $\gamma$, 511 keV line, Fireball PVO/GW170817
were taken directly from Ref.~\cite{Caputo:2025avc}.
Since the branching fraction to charged leptons in the $B-L$ case is comparable to the dark photon case, we anticipate the same processes that led to the bounds \cite{Caputo:2025avc} to also  apply to $B-L$.  We have taken the liberty to overlay these dark photon bounds
from \cite{Caputo:2025avc} 
as unlabeled light-shaded regions with dashed-lines in the $B-L$ case.
This is illustrative of where 
one might expect these constraints to appear, however, a detailed calculation for $B-L$ is beyond the scope of the paper.}
 \label{DPandBmLCurrentConstraintPlots}	
\end{figure*}
\begin{figure*}
\centering
\begin{overpic}[width=0.49\textwidth]{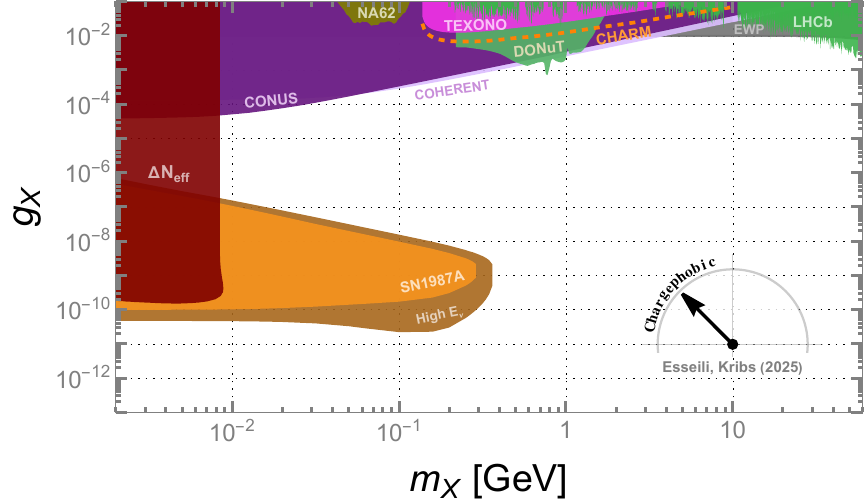}
  \input{cnstrntLabelLinks/chargephobicCnstrntLinks}
\end{overpic}
\hfill
\begin{overpic}[width=0.49\textwidth]{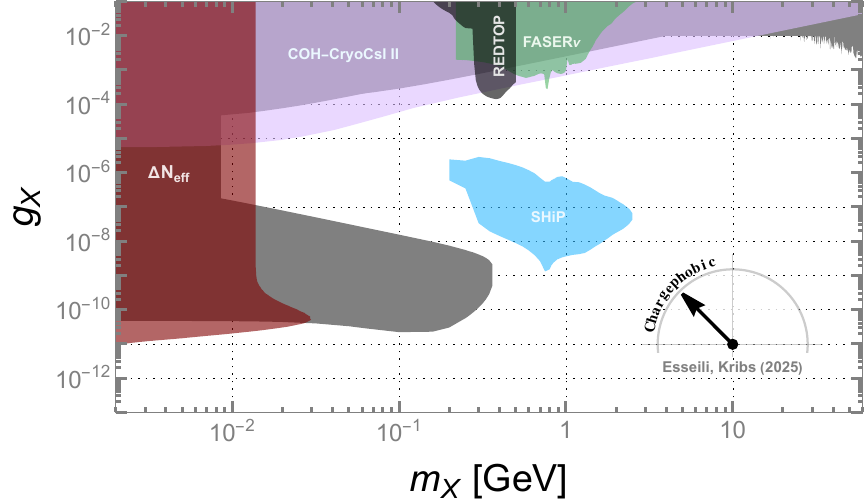}
  \input{cnstrntLabelLinks/chargephobicFutureCnstrntLinks}
\end{overpic}
\caption{ The central results of our paper: the current constraints (left) and anticipated future sensitivity (right) on a chargephobic dark photon 
($\vp=3\pi/4$).  The absence of the vast majority of terrestrial experiments (compare with Fig.~\ref{DPandBmLCurrentConstraintPlots}) arises due to the specific couplings that vanish for chargephobic dark photon: charged leptons and protons.   
See text for full details.}
 \label{CPConstraintPlots}	
\end{figure*}

\section{Production} \label{productionSubsection}

In constraining our parameter space, we encounter a variety of production mechanisms that depend nontrivially on $\{m_X, g_X, \vp\}$. These production mechanisms are well cataloged in the literature for some vector bosons,
such as the dark photon, denoted by $A'$. In this section, we generalize these production mechanisms to include dependence on the dark mixing angle, $\vp$, and compare the rates to their respective counterparts in dark photon case. Doing so will reveal the regions in the parameter space where we expect significant deviations from the dark photon constraints, to be distinguished from other regions where the change is expected to be modest.

The experiments/probes that will be used to set constraints on the generalized vector boson $X$ (with arbitrary $\vp$) can be broadly classified into five production mechanism categories: 
\begin{itemize}
    \item[i)] $X$ bremsstrahlung -- emission in nucleon-nucleon scattering, $NN \to NN X$, in the dense nuclear medium of a proto–neutron star, as well as in the production processes $eZ \to eZX$ and $pZ \to pZX$ in electron and proton beam dumps, with $Z$ the nucleus of the target material.
    \item[ii)]  Drell-Yan processes $f\bar{f} \to X$ with $f$ being a charged fermion -- an important production mode in the early universe evolution of $X$, and a key production mechanism in collider experiments.
    \item[iii)] Meson decay -- for example: $\pi^0 \to  X\gamma$ and $\eta \to  X\gamma$ in FASER \cite{FASER:2023tle, FASER:2018eoc} and $\phi \to X \eta$ in the KLOE \cite{KLOE-2:2011hhj}.
    \item[iv)] QCD vector meson $V=\omega, \rho, \phi$ mixing $V\to X$ using the vector meson dominance framework (VMD) \cite{Fujiwara:1984mp, Tulin:2014tya, Ilten:2018crw}. 
    \item[v)] Enhancement of electron-neutrino scattering $\nu e \longrightarrow \nu e$, such as TEXONO \cite{Bilmis:2015lja, Lindner:2018kjo},
    or coherent elastic neutrino-nucleus scattering (CE$\nu$NS) $\nu N \longrightarrow \nu N$ such as 
    COHERENT \cite{DeRomeri:2022twg}, both occurring through $X$-exchange. 
\end{itemize}

In order to most easily see the changes resulting from varying $\vp$, we consider 
the production ratio $\mathbb{P}_X$, defined by the cross section ratio $\sigma_X/\sigma_{A'}$, and the decay width ratio $\Gamma_X/\Gamma_{A'}$ for a given process. This ratio is summarized for various production mechanisms in Table~\ref{sigmaProdRatioRep} and Fig.~\ref{sigmaProdRatioRep}. Importantly, the production ratio vanishes for special values of $\vp$ as detailed in the previous section. 
Thus, when an experiment relies on one such production mechanism, it can lose 
sensitivity to $X$ for some values 
of $\varphi$.

\definecolor{omegaOrange}{RGB}{255, 130, 5}
\definecolor{rhoRed}{RGB}{229, 25, 74}
\definecolor{upGreen}{RGB}{61, 181, 74}
\definecolor{phiBlue}{RGB}{51, 186.15, 255}
\newcommand{\tblSp}{11pt}

\begin{figure*}
\centering
\begin{tabular}{@{}l c@{}} 
\toprule 
Process & $\mathbb{P}_X \equiv \sigma_X/\sigma_{A'}$ \\ 
\midrule
\vspace{\tblSp}
{\hspace{6pt}\color{omegaOrange}$\omega \longrightarrow X$} & {\color{omegaOrange}$(\cos\varphi + 2\sin\varphi)^2$} \\
\vspace{\tblSp}
\makecell[l]{$eZ \longrightarrow eZX$ \\[3pt] $pZ \longrightarrow pZX$} & $(\cos\varphi + \sin\varphi)^2$ \\ 
\vspace{\tblSp}
{\color{upGreen}$u\bar{u} \longrightarrow X$} & {\color{upGreen}$\dfrac{1}{4}(2\cos\varphi + \sin\varphi)^2$} \\
\vspace{\tblSp}
{\hspace{6pt}\color{rhoRed}$\rho \longrightarrow X$} & {\color{rhoRed}$\cos^2\varphi$} \\
\makecell[l]{ {\color{phiBlue}$d\bar{d} \longrightarrow X$} \\[3pt] \hspace{5pt}{\color{phiBlue}$\phi \longrightarrow X$}  } & {\color{phiBlue}$(\cos\varphi - \sin\varphi)^2$}\\
\bottomrule 
\end{tabular} 
  \quad  \quad
\raisebox{-0.52\height}{\includegraphics[width=0.6\textwidth]{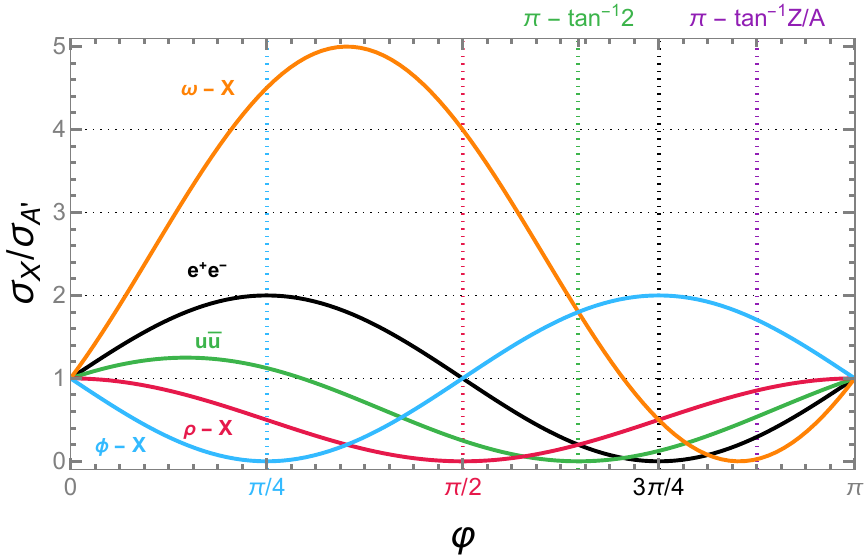}}
    \captionlistentry[table]{A table beside a figure}
    \caption{Tree level cross section ratio of $X$ to $A'$ production rate. All values in the table are multiplied by a factor of $g_X^2/(\varepsilon e)^2$; we set this extra term to one in the plot. The ratio also sets energies in both models equal when necessary. }
    \label{sigmaProdRatioRep}
 \end{figure*}

\begin{figure*}
  \centering
  \includegraphics[width=0.49\textwidth]{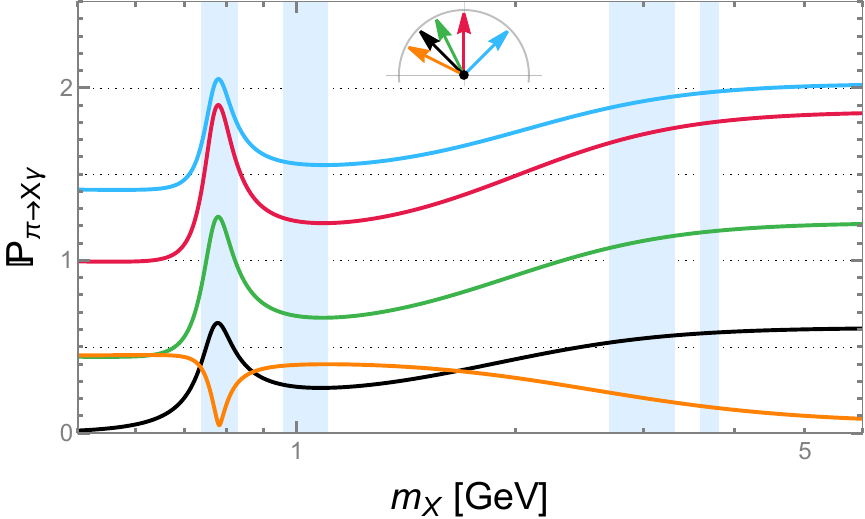}\hfill
  \includegraphics[width=0.49\textwidth]{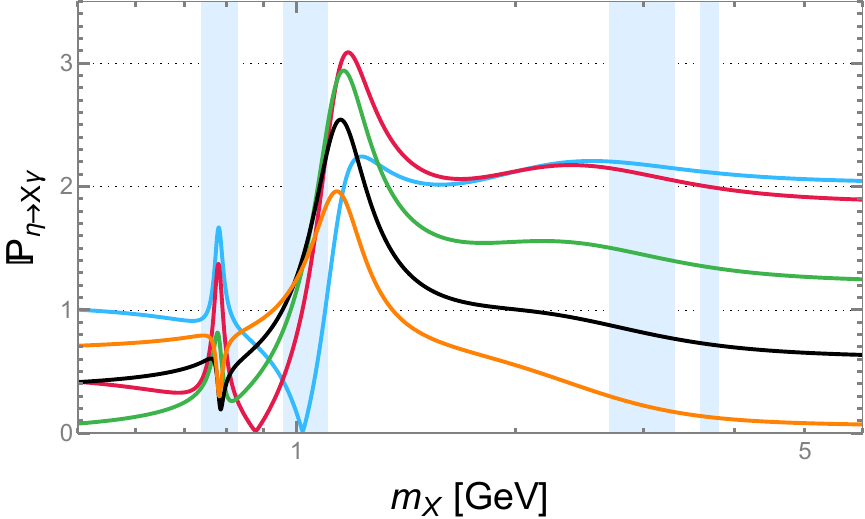}
\caption{ Decay widths for $\pi^0 \to X\gamma$ (left) and $\eta \to X\gamma$ (right), each normalized to the corresponding dark photon widths $\pi^0 \to A'\gamma$ and $\eta \to A'\gamma$ as given in Eqs.~(\ref{piToXgamma}) and (\ref{etaToXgamma}). The mixing angles are represented with colors indicated by the inset compass: $\vp=\pi/4$ (blue), $B-L$  (red), $\vp=\pi-\tanInv 2$ (green), chargephobic (black), and $\varphi=\pi - \tanInv(1/2)$ (orange).} 
 \label{mesonDecayWidthPlot}	
\end{figure*}

Next we discuss four relevant decay width ratios processes by adapting the VMD framework developed in \cite{Ilten:2018crw, Tulin:2014tya} to our case. These decay ratios are  
\begin{widetext}
\begin{eqnarray}
\mathbb{P}_{\omega \rightarrow \pi^0X}&=&\cos^2{\varphi} \\   
\mathbb{P}_{\phi \rightarrow X\eta}&=&(\cos{\varphi}-\sin{\varphi})^2 \\
\label{piToXgamma}
\mathbb{P}_{\pi^0 \rightarrow X\gamma}&=&\frac{\mid\text{BW}_{\rho}\cos{\varphi}+\text{BW}_{\omega}\, (\cos{\varphi}+2\sin{\varphi}) \mid^2}{\mid\text{BW}_{\rho}+\text{BW}_{\omega}\mid^2} \\ 
\label{etaToXgamma}
\mathbb{P}_{\eta \rightarrow X\gamma} &=& \frac{\mid9\,\text{BW}_{\rho}\cos{\varphi}- 2\,\text{BW}_{\phi}\, (\cos{\varphi}-\sin{\varphi}) +\text{BW}_{\omega}\, (\cos{\varphi}+2\sin{\varphi})\mid^2}{\mid9\,\text{BW}_{\rho}- 2\,\text{BW}_{\phi}+\text{BW}_{\omega}\mid^2} \\
\label{etaPrimeToXgamma}
\mathbb{P}_{\eta' \rightarrow X\gamma} &=& \frac{\mid9\,\text{BW}_{\rho}\cos{\varphi}+4\,\text{BW}_{\phi}\, (\cos{\varphi}-\sin{\varphi}) +\text{BW}_{\omega}\, (\cos{\varphi}+2\sin{\varphi})\mid^2}{\mid9\,\text{BW}_{\rho}+4\,\text{BW}_{\phi}+\text{BW}_{\omega}\mid^2} \, .
\end{eqnarray}
\end{widetext}
Here,
\begin{equation}
\label{breitwignerEqn}
\text{BW}_V(m_X) \; = \; \frac{m_V^2}{m_V^2+m_X^2-i m_X \Gamma_V(m_X)} \, ,
\end{equation}
is the Breit-Wigner expression for vector meson $V$ and $\Gamma_V(m_X)$ is the total decay width. The $\pi^0$ and $\eta$ decays ratios are shown in Fig.~\ref{mesonDecayWidthPlot} for some critical  values of $\varphi$. In Eqs.~(\ref{piToXgamma} --\ref{etaPrimeToXgamma}), the $\rho,\, \phi,$ and $\omega$ contributions to the decay width vanishes for $\vp=\pi/2, \pi/4,$ and $\pi-\text{tan}^{-1}(1/2)$
respectively.%
\footnote{For $\vp=\pi-\text{tan}^{-1}(1/2)$, the charges are given by $x_e=x_d=x_n=-x_\nu=-x_u=-x_p= 1/\sqrt{5}$ and pion charges $m_{\pi^\pm}=\mp 2/\sqrt{5}$.} 
Furthermore, $m_\rho \approx m_\omega$ leads to a partial cancellation between the two Breit-Wigner terms in the chargephobic limit at $\vp = 3\pi/4$. These limits correspond to dips in Fig.~\ref{mesonDecayWidthPlot} near the resonance of the relevant meson.

The production mechanisms discussed in this section have important implications to collider,  fixed target, and beam dump experiments and their associated constraints. Production in the early universe, proto-neutron star cooling, and neutrino scattering experiments require a more careful analysis and will be the subject of later sections.

\subsection{Branching Fraction} \label{BranchingSubsection}

We assume $X$ decays entirely to SM final states, and so the partial decay widths and branching ratios of $X$ are completely determined by $\{ g_X, m_X,\varphi\}$.
For fermions, the decay width is
 \begin{equation}
\label{partialDecayWidthToFermions}
\Gamma_{X\longrightarrow f\bar{f}} =\frac{\mathcal{C}_f\, g_X^2 x_f^2}{12\pi}m_X\left(1+2\frac{m_f^2}{m_X^2}\right) \sqrt{1-4\frac{m_f^2}{m_X^2}}
\end{equation}
where $x_f$ are the couplings in Eq.~(\ref{electricChargeUX}) and $\mathcal{C}_f$ is $1$ for charged leptons, $3$ for quarks, and $1/2$ for neutrinos. 
Decay into hadrons is considered separately in two regimes on either side of $m_X = 2$~GeV\@. The simpler of the two is $m_X > 2$~GeV where the decay is given by 
\begin{equation}
\label{partialDecayWidthToHadronsOver2}
\Gamma_{X\rightarrow \text{hadrons}} =\frac{\sum_q \Gamma_{X\longrightarrow q \bar{q}} }{\sum_q \Gamma_{\gamma^\ast \longrightarrow q \bar{q}}}\Gamma_{\gamma^\ast \longrightarrow \mu^{+}\mu^{-}} \mathcal{R}(m_X),
\end{equation}
with the sum ranging over all kinematically accessible quarks, and $\mathcal{R}(m_X)$ is the experimentally determined cross section ratio given by $\mathcal{R}(m_X)= \sigma_{e^{+}e^{-}\longrightarrow \text{hadrons}}/ \sigma_{e^{+}e^{-}\longrightarrow \mu^{+}\mu^{-}}$ \cite{10.1093/ptep/ptaa104, Ezhela:2003pp}. For $m_X \lesssim 2\;\text{GeV}$, QCD becomes non-perturbative and the decay into hadrons can no longer be estimated by summing the contribution of $q \bar{q}$ decays. In this range, we are also interested in decays into specific hadronic final states, i.e.,  $\pi^{+}\pi^{-}$, rather than the total sum of all hadronic decays. Accordingly, we employ the data driven analysis in \cite{Ilten:2018crw} based in the vector meson dominance (VMD) framework \cite{Tulin:2014tya}. In this approach, the data from the most important hadronic contributions $\mathcal{F}$ \cite{BaBar:2012bdw, BaBar:2004ytv, BaBar:2013jqz, BaBar:2007ceh, BaBar:2012sxt, BaBar:2017zmc, Achasov:2003ir,Achasov:2002ud} at low mass are fit to 
 \begin{equation}
\label{partialDecayWidthToHadronsUnder2NormFit}
\mathcal{R}_\mu^\mathcal{F}(m) \; = \; \frac{\sigma_{e^{+}e^{-}\longrightarrow \mathcal{F}}}{\sigma_{e^{+}e^{-}\longrightarrow \mu^{+}\mu^{-}}} \; = \; \frac{9}{\alpha_\text{EM}^2}|\mathcal{A}_\mathcal{F}|^2,
\end{equation}
where $\mathcal{A}_\mathcal{F}$ is the sum of a real function and functions of a Breit-Wigner form accounting for the contributions of the three vector mesons $\rho, \omega, \phi$. The fits are then used to decompose the decay into hadrons to $\rho\text{-like}\;\mathcal{R}_\mu^\mathcal{\rho} ,\;\omega\text{-like}\;\mathcal{R}_\mu^\mathcal{\omega}, \; \text{and}\;\phi\text{-like}\;\mathcal{R}_\mu^\mathcal{\phi}$ contributions giving
 \begin{equation}
\label{partialDecayWidthToHadronsUnder2}
\Gamma_{X\longrightarrow \text{hadrons}} \; = \; \frac{g_X^2 m_X}{12\pi}\left( \sum_V\mathcal{R}_X^V+\mathcal{R}_X^\mathcal{\omega-\phi} \right)
\end{equation}
with
\begin{eqnarray}
\mathcal{R}_X^\mathcal{\rho}(m)&=&\cos^2{\varphi}\,\mathcal{R}_\mu^\mathcal{\rho}(m)\nonumber\\
\mathcal{R}_X^\mathcal{\omega}(m)&=&(\cos{\varphi}+2\sin{\varphi})^2\,\mathcal{R}_\mu^\mathcal{\omega}(m)\nonumber\\
\mathcal{R}_X^\mathcal{\phi}(m)&=&(\cos{\varphi}-\sin{\varphi})^2\,\mathcal{R}_\mu^\mathcal{\phi}(m)\nonumber\\
\mathcal{R}_X^\mathcal{\omega-\phi}(m)&=&\left[1-3\cos{2\varphi}-2\sin{2\varphi}\right]\,\mathcal{R}_\mu^\mathcal{\omega-\phi}(m)
\label{mesonLikeBranching}
\end{eqnarray}
and $\mathcal{R}_X^\mathcal{\omega-\phi} $ is the only non-negligible interference term between $\omega\text{-like}$ and $\phi\text{-like}$ to the $\pi^{+}\pi^{-}\pi^{0}$ final hadronic decay state.

\begin{figure*}
\centering
\includegraphics[width=0.49\textwidth]{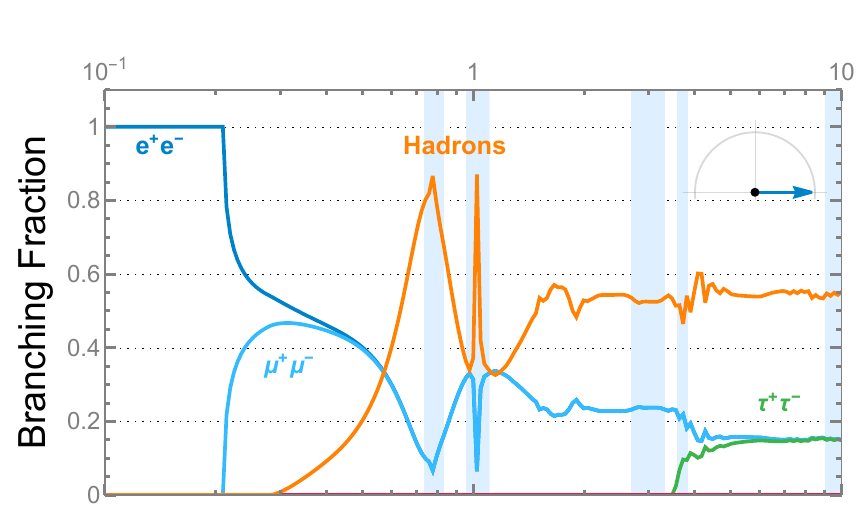}
\hfill
\includegraphics[width=0.49\textwidth]{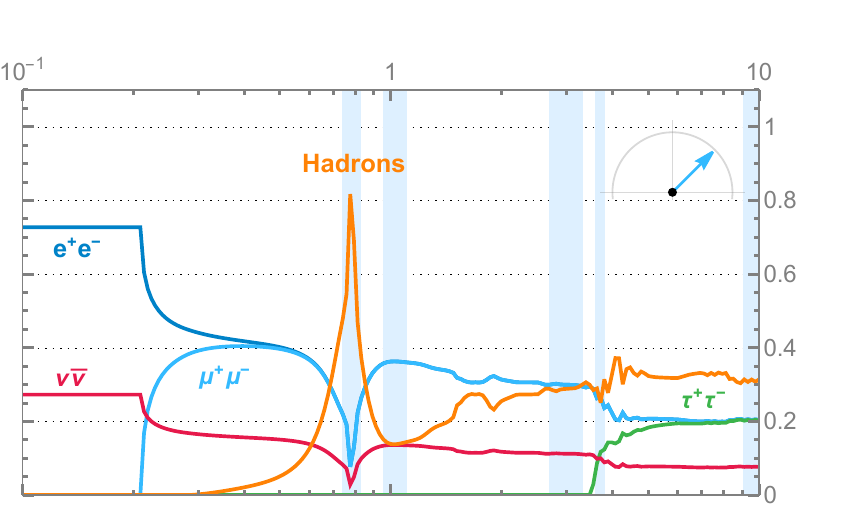}
\vfill
\includegraphics[width=0.49\textwidth]{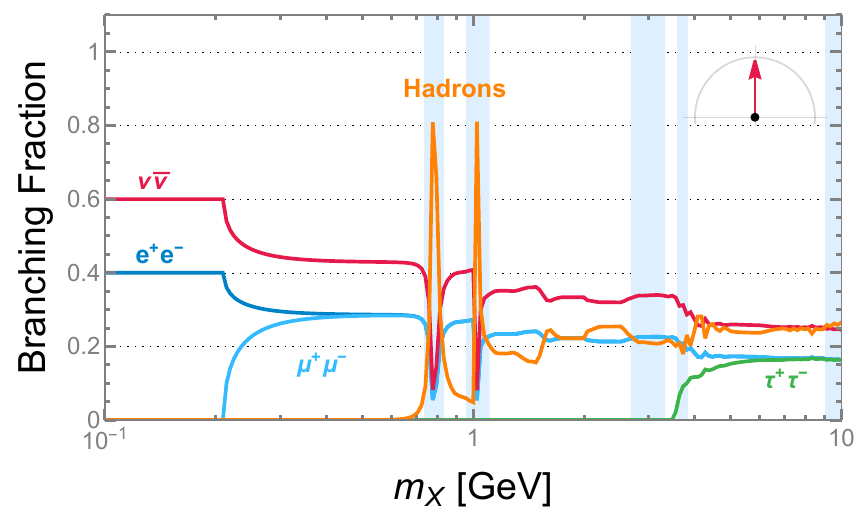}
\hfill
\includegraphics[width=0.49\textwidth]{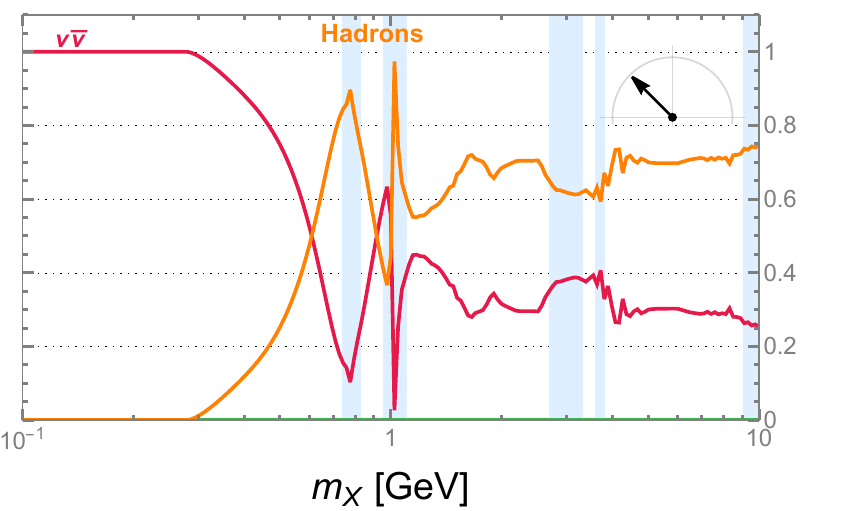}
 \caption{Branching fraction of $X$ with $\varphi= \{0, \frac{\pi}{4}, \frac{\pi}{2}, \frac{3\pi}{4}\}$ as indicated by the inset compass.
 The blue shaded regions correspond to meson resonances.} 
 \label{BranchingFractionFig}				
\end{figure*}
\begin{figure*}
\centering
\includegraphics[width=0.49\textwidth]{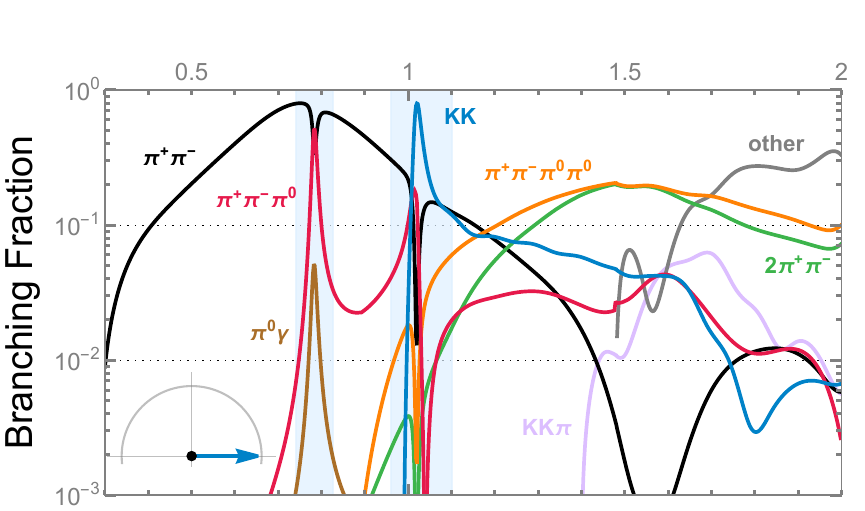}
\hfill
\includegraphics[width=0.49\textwidth]{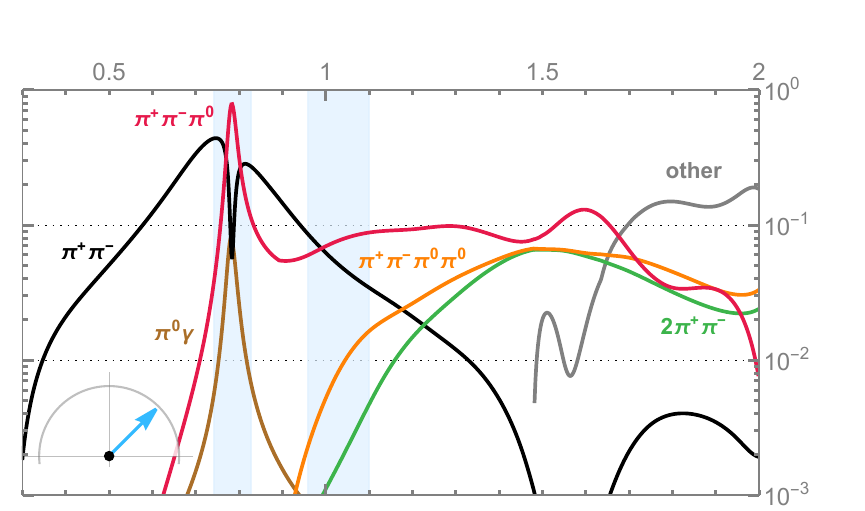}
\vfill
\includegraphics[width=0.49\textwidth]{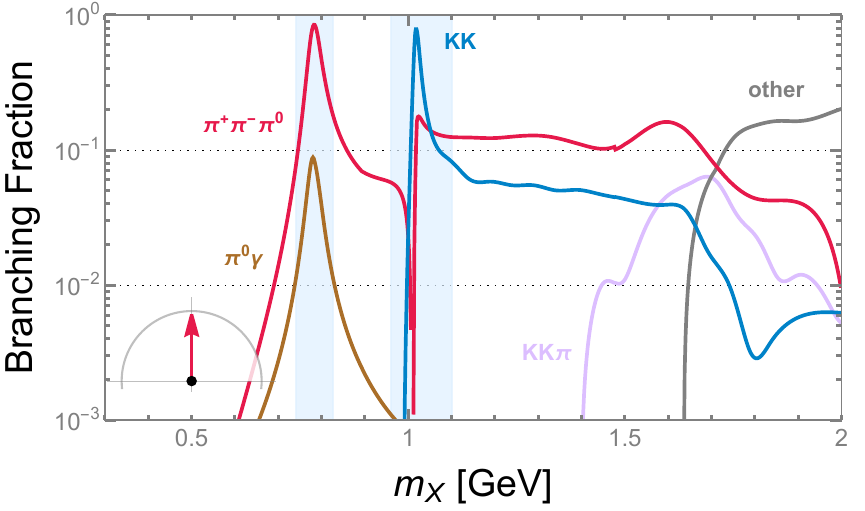}
\hfill
\includegraphics[width=0.49\textwidth]{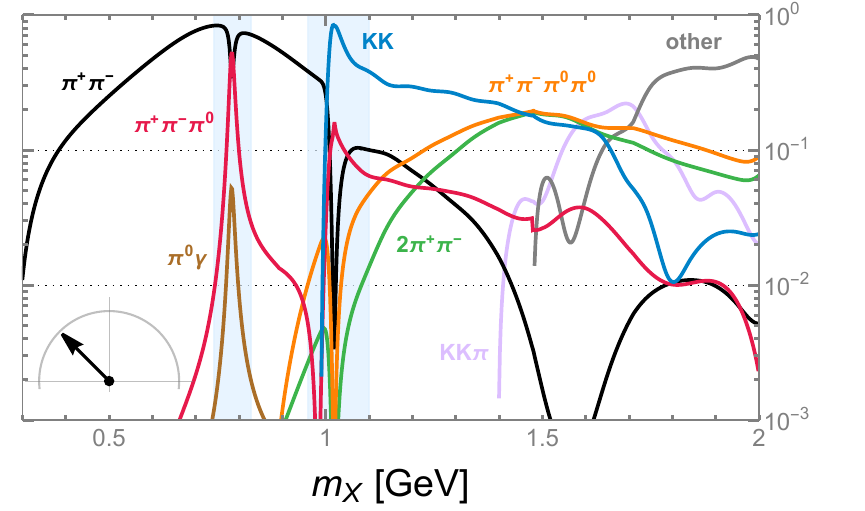}
 \caption{ Branching fraction of $X$ to individual hadronic 
 that compose the results in Fig. \ref{BranchingFractionFig} for $\varphi= \{0, \frac{\pi}{4}, \frac{\pi}{2}, \frac{3\pi}{4}\}$ as indicated by the inset compass.} 
 \label{VMDBranchingFractionFig}				
\end{figure*}

The branching ratios for four mixing angles $\varphi= \{0, \frac{\pi}{4}, \frac{\pi}{2}, \frac{3\pi}{4} \}$ are shown in Fig.~\ref{BranchingFractionFig}; of those, the branching ratio to individual hadronic modes for $\sqrt{s}<2\;\gev$ are shown in Fig.~\ref{VMDBranchingFractionFig}. The $\varphi=0$ and $\varphi = \pi/2$ correspond to the dark photon and $B-L$ cases respectively and have been discussed thoroughly in the literature. The $\varphi = \pi/4$ figure shows the transition between the two aforementioned models with a nonzero neutrino branching fraction, where the absence of $\phi - X$ mixing, as in Eq.~(\ref{mesonLikeBranching}), leads to the disappearance of the $\phi$ resonance and the subsequent $KK$ and $KK\pi$ decay modes.
The case $\vp=\pi/2$ ($B-L$) does not have $\rho-X$ mixing, leading to 
the disappearance of $\pi^{+}\pi^{-}$, $\pi^{+}\pi^{-}\pi^{0}\pi^{0}$, and $\pi^{+}\pi^{-}\pi^{+}\pi^{-}$ decay modes.  This distinction means that $X$ decay to hadronic modes for $B-L$ (and related flavor-dependent currents such as $B-3 L_\tau$) remain suppressed up to $m_X \simeq 700$~MeV, whereas for all of the mixing angles,  including the chargephobic case, there are hadronic decay channels ($X$ mixing $\rho$ decaying to $\pi^+\pi^-$) that start appearing once $m_X \gtrsim 300$~MeV\@.\footnote{For the case
$\vp =\pi-\text{tan}^{-1} \, 1/2$
(not shown), the absence of $\omega-X$ mixing 
leads to the disappearance of $\pi^{+}\pi^{-}\pi^{0}$ and $\pi^{0}\gamma$.}
Fig.~\ref{BranchingFractionFig} demonstrates
that the chargephobic case $\vp = 3\pi/4$ has the most striking differences in branching fraction with the other angles given the absence of a tree-level coupling to charged leptons.
The plots include the RGE evolution of $x_e$ as discussed in App.~\ref{RGEAppendix}, although the branching ratio to charged leptons is at most $\mathcal{O}(10^{-4})$. 

The blue shaded regions in Fig. \ref{BranchingFractionFig} correspond to the following meson resonances in mass ranges [MeV]: $\rho$ and $\omega$ at $(741, 827)$, $\phi$ at $(960, 1100)$, $J/\psi$ at $(2700, 3300)$, $\psi(2S)\; \&  \;\psi(3770)$ at $(3600, 3800) $, and $\Upsilon(1S)-\Upsilon(4S)$ at $(9100, 10600)$. These bands are obtained from \cite{CMS:2019buh} where prompt-like $A'$ searches are vetoed due to the large resonance contribution, and are meant only to illustrate an example of experimental veto window rather than a theoretical limitation. In our analysis, we have not explicitly included the $J/\psi,\psi(2S), \psi(3770),$ and $\Upsilon(1S)-\Upsilon(4S)$ resonances in hadronic decays, but their effects are consistently considered in the constraints.

\begin{figure*}
\includegraphics[width=0.49\textwidth]{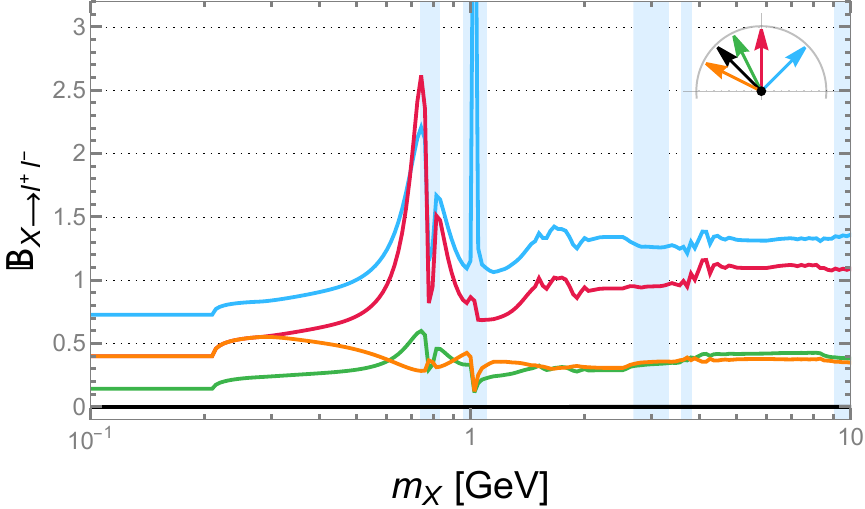}
\hfill
\includegraphics[width=0.49\textwidth]{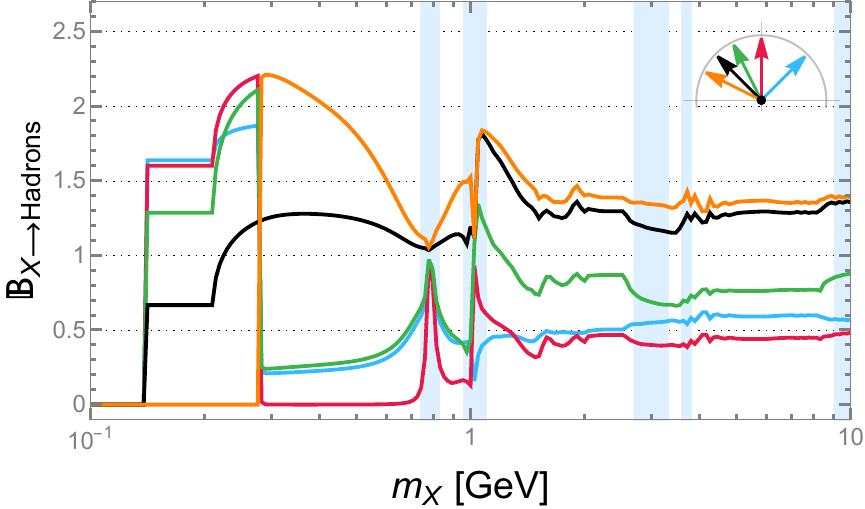}
 \caption{Ratio of the branching fractions of $X$ to those of the dark photon $A'$ into charged leptons (left) and hadrons (right). The mixing angles are represented with colors indicated by the inset compass: $\vp=\pi/4$ (blue), $B-L$  (red), $\vp=\pi-\tanInv 2$ (green), chargephobic (black), and $\varphi=\pi - \tanInv(1/2)$ (orange).
}
 \label{BranchingFractionRatioFig}
\end{figure*}

Experiments search for the decay products of $X$, or lack thereof, as a signal to set constraints on the $\{m_X,g_X\}$ parameter space. The branching fraction is a guide for experiments as to which SM particles can be suitably employed as a signal for a given mass $m_X$. 
The ratio of branching fraction of $X$ to $A'$ provides a quantitative guide to how the sensitivity of a given search to $X$ differs from the corresponding dark photon case. For some experiments, a constraint on $A'$ can be recast to a constraint on $X$ via $ \sigma_X \text{Br}_{X\longrightarrow \mathcal{F}} \text{Eff}(\tau_X)=\sigma_{A'} \text{Br}_{A' \longrightarrow \mathcal{F}} \text{Eff}(\tau_{A'} )$, where $\text{Br}$ is the branching ratio and $\text{Eff}(\tau_X)$ is the experimental efficiency with explicit dependence on $X$ lifetime. Equivalently, when applicable, constraints can be recast using
\begin{equation}
\mathbb{P}_X \, \mathbb{B}_{X\longrightarrow \mathcal{F}}\,  \mathbb{E}(\tau_X)=1,
\label{recastingEquation}
\end{equation}
where $\mathbb{P}_X$ is shown in Fig.~\ref{sigmaProdRatioRep} and Fig.~\ref{mesonDecayWidthPlot}, $\mathbb{B}_{X\longrightarrow \mathcal{F}}$ is the ratio of branching fractions and $\mathbb{E}(\tau_X)$ is the efficiency ratio. The ratio of branching fractions $\mathbb{B}_{X\rightarrow \mathbb{F}}$ for charged leptons and hadrons for $\varphi=\{ \frac{\pi}{4},\frac{\pi}{2}, \pi-\text{tan}^{-1} \, 2, \frac{3\pi}{4}, \pi-\text{tan}^{-1}(1/2) \} $ are shown in Fig.~\ref{BranchingFractionRatioFig}.

The recasting equation, Eq.~(\ref{recastingEquation}), is powerful and can be applied to a variety of constraints set by fixed target experiments, colliders, beam dumps, and searches to invisible final states. For these experiment categories, we use the methodology developed in \cite{Ilten:2018crw}%
 \footnote{The reference \cite{Ilten:2018crw} also provides a publicly available code \texttt{Darkcast} \url{https://gitlab.com/philten/darkcast} which we have recreated in \texttt{Mathematica} and used to recast constraints from some experiments as discussed in the text.}
to recast the constraints. Otherwise, we rederive the constraints set on beam dump experiments \footnote{We rederive constraints from electron beam dump experiments and the bremsstrahlung process in proton beam dump experiments even though those can be recast. Constraints set from meson decay in proton beam dumps are recast from dark photon constraints.}, neutrino experiments, SNe cooling, early universe $\Delta \neff$ modification and some precision experiments. In the following sections, we summarize the derivation of all aforementioned constraints.

\section{Constraints}

We now detail the derivation of the constraints in Figs.~\ref{DPandBmLCurrentConstraintPlots},\ref{CPConstraintPlots} that span a rich set of cosmological, astrophysical and terrestrial experiment observables. As we have seen, cosmology and astrophysics provide constraints throughout the range of dark mixing angles $\vp$.  
Consequently, we present these constraints first, 
since we have carried out independent calculations of the contribution of $X$ production and decay to $\Delta \neff$ during the early universe as well as our own $X$ emission from SN1987A\@.
Next, we derive constraints on neutrino experiments, since they provide some of the strongest constraints on the chargephobic case. We then consider beam dump experiments and colliders, and detail the effects of $X$ production and decay with the critically important $\vp$ dependence. Finally, we consider constraints from precision tests of the Standard Model, which provide leading bounds on the large $m_X$ in the chargephobic model.

\subsection{Cosmology: CMB \& BBN}
\label{neffSection}

\phantomsection
\label{sec:cnst:dneff}

The precision determination of the effective number of relativistic species in the early universe, $\neff$, from CMB power spectra \cite{Planck:2018nkj, Planck:2018vyg, AtacamaCosmologyTelescope:2025nti} and primordial element abundances \cite{Cyburt:2015mya, Yeh:2020mgl, Yeh:2022heq, Workman:2022ynf} is in good agreement with the SM prediction \cite{deSalas:2016ztq, Mangano:2005cc, Dolgov:2002wy, Dicus:1982bz, Hannestad:1995rs, Dodelson:1992km, Dolgov:1997mb, Esposito:2000hi, Mangano:2001iu, Birrell:2014uka, Grohs:2015tfy, Cielo:2023bqp, Akita:2020szl, Froustey:2020mcq, Bennett:2020zkv}. Hence, BSM physics that alters $\neff$  before recombination is strongly constrained \cite{Sarkar:1995dd, Iocco:2008va, Pospelov:2010hj, Blennow:2012de, Boehm:2012gr, Boehm:2013jpa, Brust:2013ova, Vogel:2013raa, Fradette:2014sza, Nollett:2014lwa, Buen-Abad:2015ova, Chacko:2015noa, Wilkinson:2016gsy, Huang:2017egl, Escudero:2018mvt, Escudero:2019gzq, Abazajian:2019oqj, Ibe:2019gpv, EscuderoAbenza:2020cmq, Coffey:2020oir, Luo:2020sho, Luo:2020fdt, Adshead:2022ovo, Eijima:2022dec, Sandner:2023ptm, Ghosh:2024cxi}. In our case, a well known effect is that an $X$ boson can ``freeze in'' \cite{Hall:2009bx}, become nonrelativistic, redshift as matter, and then decay back into SM states, modifying $\dneff$. This constraint was derived for a dark photon \cite{Ibe:2019gpv}, $U(1)_{\mu-\tau}$ \cite{Escudero:2019gzq}, and $U(1)_{B-L}$ \cite{Esseili:2023ldf,Li:2023puz}\footnote{The constraint in our previous work \cite{Esseili:2023ldf} is stronger than that in Ref. \cite{Li:2023puz} as the latter neglected including the coupling to charged leptons (electrons) and finite temperature effects in their ${B-L}$ constraint calculation. These effects play a crucial role in determining the constraint in a large portion of the parameter space.}.

In this paper,  we utilize the formalism and approximations outlined in \cite{Escudero:2018mvt,Escudero:2019gzq,EscuderoAbenza:2020cmq, Esseili:2023ldf} to derive constraints on $X$. Specifically, our previous work in \cite{Esseili:2023ldf} can be fully adapted to this case by modifying the couplings appropriately. For each given $\{m_X, g_X, \vp\}$, we solve the five Boltzmann equations 
\begin{widetext}
\begin{align} 
\label{fiveboltzmannGeneral}
\frac{dT_X}{dt}&= \mathbb{D}(T_X,\mu_X)\left[-3 H\left((\rho_X + P_X)\frac{\partial n_X}{\partial \mu_X} - n_X \frac{\partial \rho_X}{\partial \mu_X} \right)+\frac{\partial n_X}{\partial \mu_X}\frac{\delta \rho_X}{\delta t}-\frac{\partial \rho_X}{\partial \mu_X}\frac{\delta n_X}{\delta t} \right],\nonumber\\ 
\frac{dT_\nu}{dt}&= \mathbb{D}(T_\nu,\mu_\nu)\left[-3 H\left((\rho_\nu + P_\nu)\frac{\partial n_\nu}{\partial \mu_\nu} - n_\nu \frac{\partial \rho_\nu}{\partial \mu_\nu} \right)+\frac{\partial n_\nu}{\partial \mu_\nu}\frac{\delta \rho_\nu}{\delta t}-\frac{\partial \rho_\nu}{\partial \mu_\nu}\frac{\delta n_\nu}{\delta t} \right],\nonumber\\ 
\frac{dT_\gamma}{dt}&=\left(\frac{\partial \rho_\gamma}{\partial T_\gamma}+\frac{\partial \rho_e}{\partial T_\gamma}\right)^{-1}\left[ 4H\rho_\gamma +3H(\rho_e+P_e)+\frac{\delta\rho_e}{\delta t}\right],\\
\frac{d\mu_X}{dt}&= -\mathbb{D}(T_X,\mu_X)\left[-3 H\left((\rho_X + P_X)\frac{\partial n_X}{\partial T_X} - n_X \frac{\partial \rho_X}{\partial T_X} \right)+\frac{\partial n_X}{\partial T_X}\frac{\delta \rho_X}{\delta t}-\frac{\partial \rho_X}{\partial T_X}\frac{\delta n_X}{\delta t} \right],\nonumber\\
\frac{d\mu_\nu}{dt}&= -\mathbb{D}(T_\nu,\mu_\nu)\left[-3 H\left((\rho_\nu + P_\nu)\frac{\partial n_\nu}{\partial T_\nu} - n_\nu \frac{\partial \rho_\nu}{\partial T_\nu} \right)+\frac{\partial n_\nu}{\partial T_\nu}\frac{\delta \rho_\nu}{\delta t}-\frac{\partial \rho_\nu}{\partial T_\nu}\frac{\delta n_\nu}{\delta t} \right]\nonumber\\
\mathbb{D}(T, \mu)&=\left(\frac{\partial n}{\partial \mu}\frac{\partial \rho}{\partial T} - \frac{\partial n}{\partial T}\frac{\partial \rho}{\partial \mu} \right)^{-1}\nonumber,
\end{align}
\end{widetext}
numerically from $T\sim \mathcal{O}(100)\;\mev$ all the way down to recombination at $T\simeq 0.3 \;\ev$.

In writing these equations, we made the implicit assumption that fermions and bosons follow Fermi-Dirac (FD, positive) and Bose-Einstein (BE, negative), $f(E)=[e^{(E-\mu)/T}\pm 1]^{-1}$ distributions respectively. Here $n,\rho, P$ are the number, energy, and pressure densities for a particle $i$ with $d_i$ degrees of freedom obtained by integrating the particle distribution $f$ over $d_i \, d^3p/(2\pi)^3$, $d_i \, E \, d^3p/(2\pi)^3$, and $d_i \, \frac{p^2}{3E} \, d^3p/(2\pi)^3$ respectively. $\delta n/\delta t$ and $\delta \rho/\delta t$ are the number and energy transfer rates between particle species obtained by integrating collision terms over the same measures. Finally, $H=\sqrt{\rho_{\text{tot}}/(3 M_{\rm Pl}^2)}$ is the Hubble expansion rate and $M_{\rm Pl}=2.4 \times 10^{18}\;\gev$. The formulae for these thermodynamic quantities and their derivatives can be found in Appendix~A.6 of \cite{EscuderoAbenza:2020cmq}.

The $\delta n/\delta t$ and $\delta \rho/\delta t$ terms account for all relevant interactions driving the dynamics. At the temperature range of interest,  $e^{+}e^{-}\lra\bar{\nu}_\alpha\nu_\alpha, \; e^{\pm}\nu_\alpha \lra e^{\pm}\nu_\alpha$, and $e^{\pm}\bar{\nu}_\alpha \lra e^{\pm}\bar{\nu}_\alpha$ are the only SM interactions needed. These interactions, mediated by $W$ and $Z$ exchange, drive the neutrino and electron population to equilibrium for temperatures above neutrino decoupling $T\sim 2\;\mev$. If $X$ couples to neutrinos and electrons, these interactions can also be mediated by $X$. We  also include the interactions $X \lra \nu\bar{\nu}$, $X\lra \ee$,  $\gamma X\lra\ee$, and $e^{\pm} X\lra e^{\pm}\gamma$ when applicable. Interactions such as $\ee \lra \gamma X$,  $ \nu\bar{\nu}  \lra XX$, $X \epm \lra X \epm$, and $X\nu \lra X\nu$ are subdominant to the aforementioned interactions and can be neglected. The rates, along with a detailed discussion of the dynamics, can be found in \cite{Esseili:2023ldf}. These rates are trivially adapted to our case with the appropriate coupling substitutions.

Following the discussion in \cite{Esseili:2023ldf, Yeh:2022heq} and based on the datasets in \cite{Planck:2018nkj, Planck:2018vyg, Aver:2015iza, Pettini:2012ph,Cooke:2013cba,Riemer-Sorensen:2014aoa,Cooke:2016rky,Balashev:2015hoe,Riemer-Sorensen:2017pey,Zavarygin:2018ara,Cooke:2017cwo, Aver:2020fon,Valerdi:2019beb,Fernandez:2019hds,Kurichin:2021ppm,Hsyu:2020uqb}, constraints in Fig. \ref{CPConstraintPlots} are set on modifications to $\neff$ of $|\dneff| \geq 0.4$ and/or helium mass fraction $Y_p$ of $\Delta Y_p>0.008$; we also show future projected sensitivity at $|\dneff| \geq 0.05$. The first constraint is set from $X$ modifying the effective number of relativistic species in the early universe defined as
 \begin{equation}
\label{NeffEqn}
\neff \;=\; \frac{8}{7}
\left( \frac{11}{4}  \right)^{4/3}
\left( \frac{\rho_{\text{rad}}-\rho_\gamma}{\rho_\gamma} \right),
\end{equation}
which is the ratio of non-photon to photon radiation density. The state-of-the-art calculation in the SM gives
$\neff^{\text{SM}} = 3.045$-$3.046$, 
that takes into higher order corrections, non-thermal neutrino
distribution functions, neutrino oscillations,
etc.~\cite{Mangano:2005cc, deSalas:2016ztq}. Using the method outlined above without $X$, we obtain
the photon-to-neutrino temperature ratio, the neutrino chemical
potential, and $N_{\rm eff}$ in the SM
 \begin{equation}
\label{NeffinSM}
T_\gamma/T_\nu = 1.394 \, ,
\quad 
\mu_\nu/T_\nu = - 4.8\times 10^{-3} \, ,
\quad 
\neff^{\text{SM}} = 3.042 \, . 
\end{equation}
This demonstrates that our calculation is able to reproduce the
non-instantaneous decoupling of neutrinos in the SM to an accuracy
of order $\Delta N_{\rm eff} \sim 0.01$. The value of $\neff$ in Eq.~(\ref{NeffinSM}) will be modified when solving the Boltzmann equations in Eq.~(\ref{fiveboltzmannGeneral}) including $X$ and a constraint is set if $\dneff \geq 0.4$. The constraint from $\Delta Y_p$ can also be derived as in \cite{Esseili:2023ldf}; however, this constraint is weaker than that derived from $\dneff$ everywhere in our parameter space. 

\begin{figure}
\centering
\includegraphics[width=0.49\textwidth]{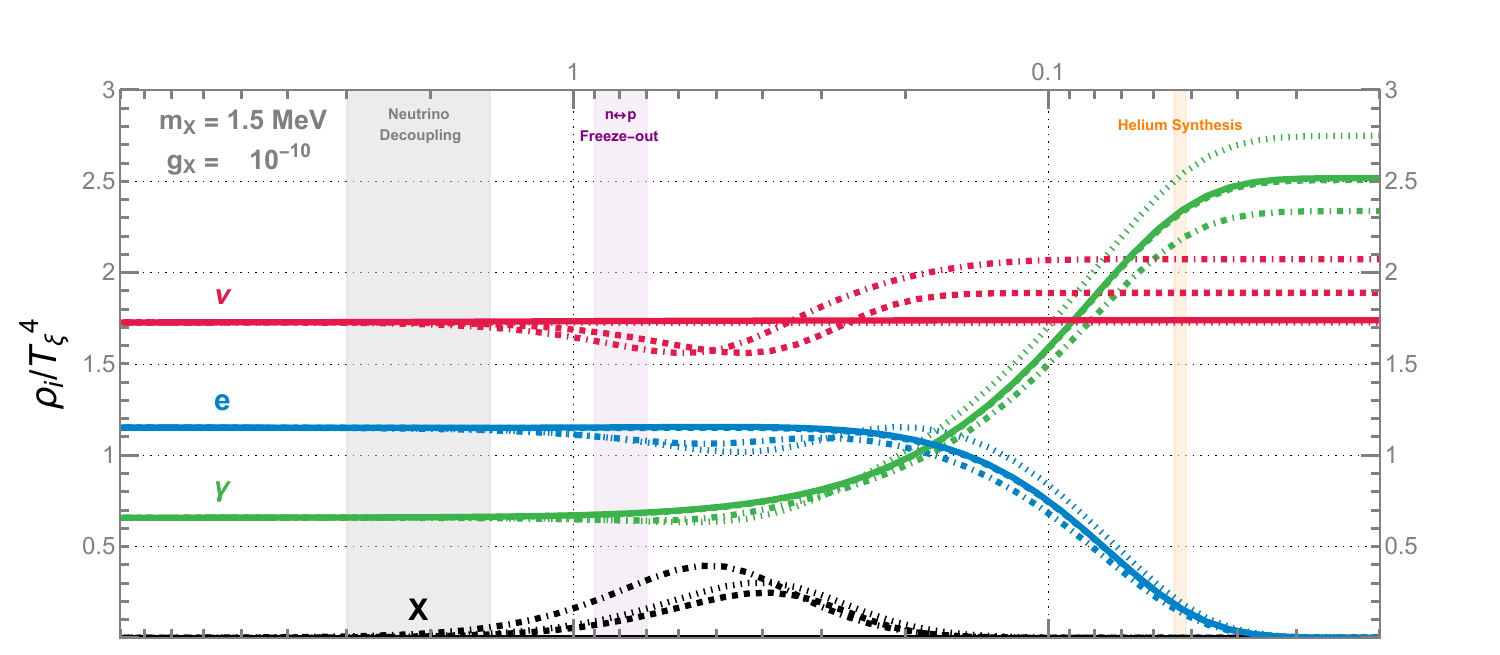}
\vspace{0.1cm}
\includegraphics[width=0.49\textwidth]{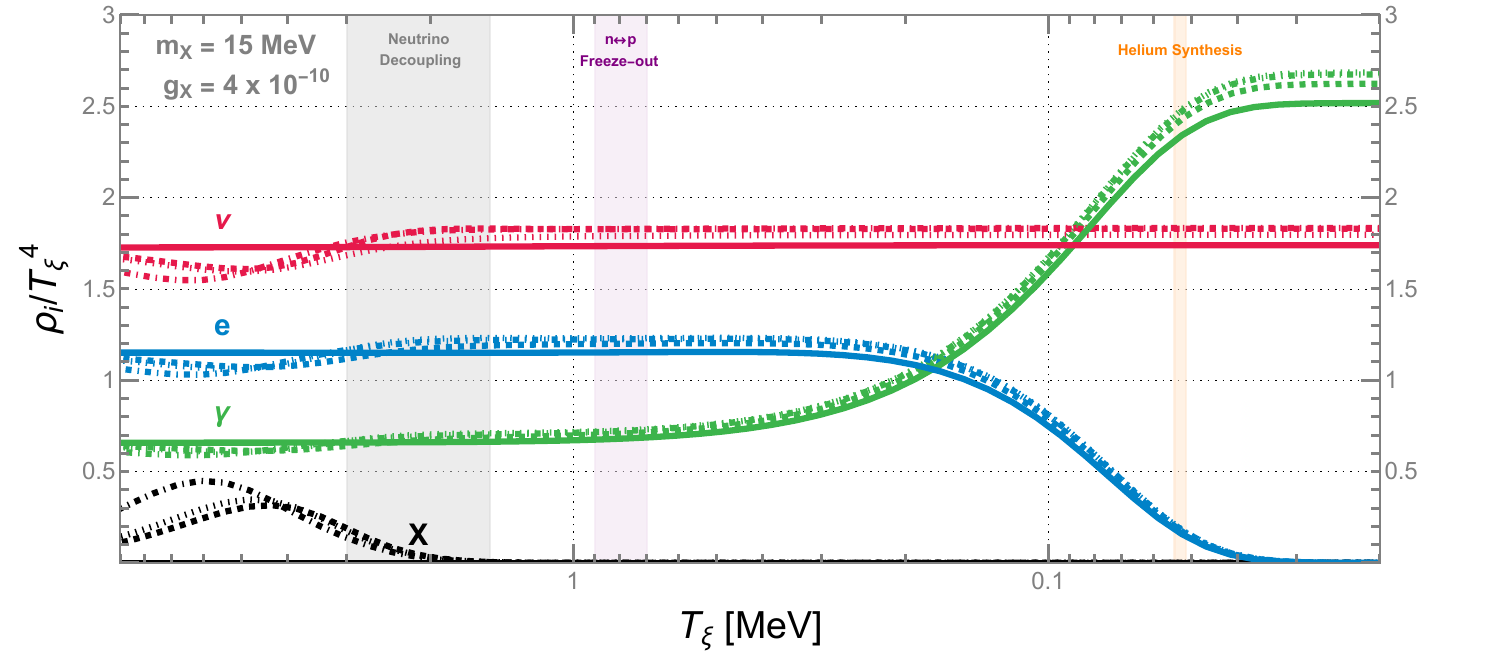}
 \caption{The normalized energy density evolution $\rho/T_\xi^4$ of the neutrino (red), electron (blue), photon (green), and $X$ (black) around the BBN era as a function of a reference temperature $T_\xi$ explained in the text. (Top) $m_X=1.5\;\mev$ with $g_X=10^{-10}$ and (bottom) $m_X=15\;\mev$ with $g_X=4 \times 10^{-10}$. 
 In both plots, (solid) lines are SM evolution with no $X$ included, (dotted) dark photon $(\vp=0)$, (dot-dashed) $B-L$ $(\vp=\pi/2)$ and (dashed) chargephobic case $(\vp=3\pi/4)$.} 
 \label{energyDensityEvolutionFig}		
\end{figure}

The normalized density evolution for two representative points in the parameter space is shown in Fig.~\ref{energyDensityEvolutionFig}. Here, $T_\xi$ is a reference temperature as in \cite{Esseili:2023ldf} that can be thought of as the evolution of a fictitious massless particle with a temperature that is initially set to the SM temperature but is actually decoupled from all other particles. Using $T_\xi$ makes it simple to compare different values of $\vp$ where $X$ couples differentially to neutrinos and electron photon bath.

When the mass of $X$ is of order a few $\mev$, the majority of the production and decay dynamics occur after neutrino decoupling as can be seen in the representative example $m_X= 1.5 \;\mev$ in Fig.~\ref{energyDensityEvolutionFig}. For this mass and in the chargephobic limit (dashed), $X$ is slowly populated at the expense of the $\nu$ population through $\nu \bar{\nu} \rightarrow X$. The rates are maximal around $T_\xi \sim m_X/3$ where the $X$ population becomes non-relativistic and starts decaying back into neutrinos. Importantly, the $X$ distribution evolves as matter $\rho_X \propto a^{-3}$ before dumping its entropy back into neutrinos, that themselves have been evolving as radiation $\rho_\nu \propto a^{-4}$. Hence, this back and forth exchange in energy heats up the neutrino plasma. The result is a neutrino energy density that is larger than it would have been in the SM as compared to the photon energy density. This effect \textit{increases} the value of $\neff$ in Eq.~(\ref{NeffEqn}) and gives $\dneff = 0.27$ for $m_X= 1.5 \;\mev$ and $g_X = 10^{-10}$. The small shift in the electron and photon densities here is due to the dynamics commencing slightly before neutrinos are fully decoupled. 

The dark photon limit (dotted) for $m_X= 1.5 \;\mev$ proceeds similarly with the dynamics now occurring between $X$ and the electron-photon bath instead of neutrinos. The result is a photon energy density that is larger than it would have been in the SM as compared to neutrino density and a \textit{decrease} in $\neff$; in this case $\dneff=-0.27$. In the $B-L$ case (dot-dashed), $X$ couples to $e$ and $\nu$ and has access to a larger energy density bath. This leads to a larger $X$ population produced at a slightly earlier time as compared to the previous two cases. The $X$ population then decays back into both $e$ and $\nu$. However, this decay occurs around the time that electrons also become non-relativistic and the dynamics become highly nontrivial as discussed in \cite{Esseili:2023ldf}. The upshot is a preferential heating of the neutrinos and an increase in $\neff$ giving $\dneff = 0.86$ for $m_X= 1.5 \;\mev$ and $g_X= 10^{-10}$.

For $m_X= 15\;\mev$ and $g_X=4 \times 10^{-10}$, the majority of $X$ production occurs before the neutrinos are fully decoupled. This means that regardless of the charge assignment, $X$ has access to the entire SM thermal bath. For example, in the chargephobic limit, $X$ is produced via $\nu \bar{\nu} \rightarrow X$ which effectively cools the neutrino population. However, in this case, processes such as $e^{+}e^{-}\lra\bar{\nu}_\alpha\nu_\alpha$, which have previously kept neutrinos and the electron-photon bath in equilibrium, are still active. These processes transfer energy from the electron-photon bath to neutrinos driving the system to a new equilibrium at a lower temperature. These two effects occur simultaneously and proceed until the neutrinos decouple from SM. After $X$ decays and heats up the neutrino plasma, some of this energy is transferred into the electron-photon bath through  $e^{+}e^{-}\lra\bar{\nu}_\alpha\nu_\alpha$ interactions which are still partially active. This leads to an attenuated increase in $\neff$ giving $\dneff=0.03$ for $m_X= 15 \;\mev$ and $g_X=4 \times 10^{-10}$. A similar effect is seen in dark photon case where $\dneff=-0.1$. Lastly, in the $B-L$ case, $X$ decays and heats up both neutrinos and electron-photon bath with comparable amounts leading to a weaker increase in $\neff$; in this case $\dneff=-0.02$. For larger values of $m_X$, the production and decay of $X$ terminates completely before neutrino decoupling. This gives the SM enough time to return to thermal equilibrium leading to no change in $\neff$. 

\begin{figure*}
\centering
\includegraphics[width=0.325\textwidth]{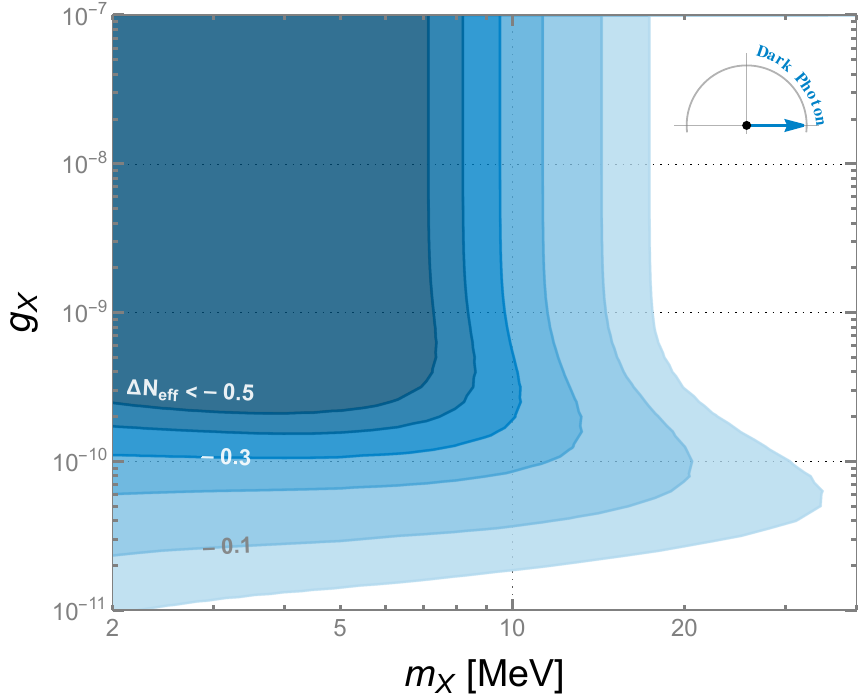}
\hfill
\includegraphics[width=0.325\textwidth]{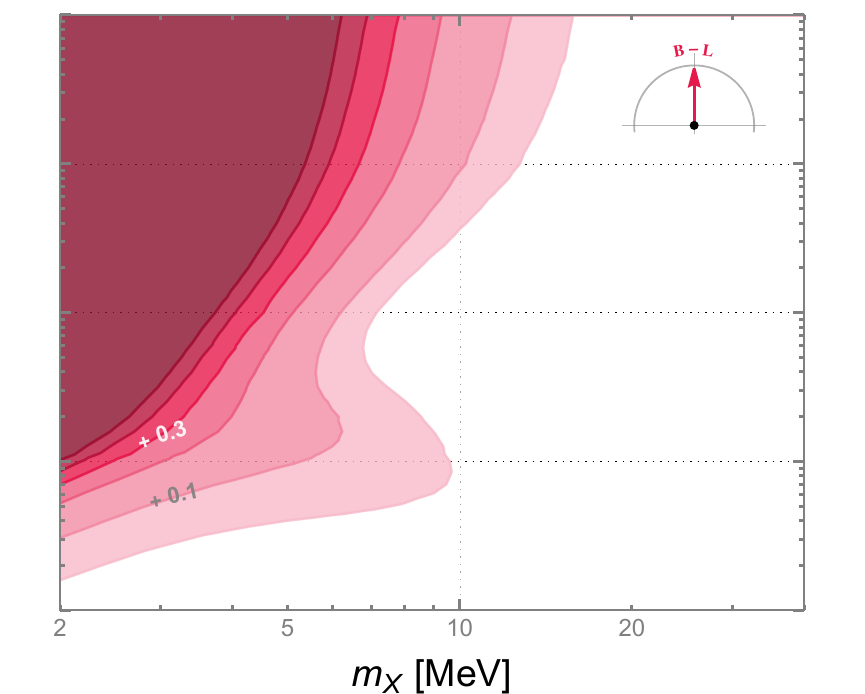}
\hfill
\includegraphics[width=0.325\textwidth]{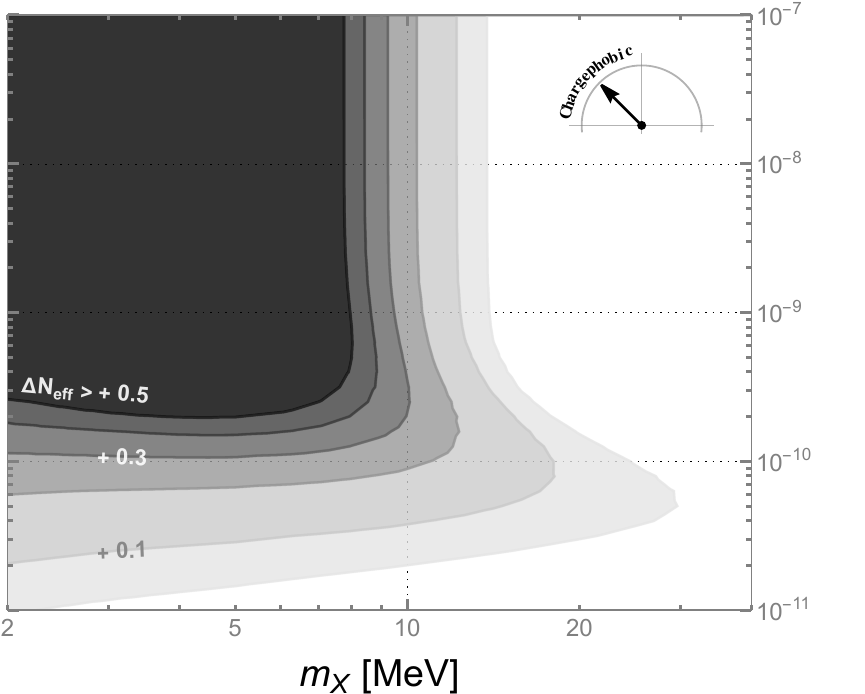}
 \caption{
 $\dneff$ contours for the  dark photon, $B-L$, and chargephobic cases as indicated by compass inset plot.  The color shading from lighter to darker corresponds to $|\dneff| = \{0.05, \;0.1,\;0.2,\;0.3,\;0.4,\;0.5\}$ and contours greater than $0.5$ are not shown. Note that the dark photon case drags $\neff$ below its SM value (negative $\dneff$) as opposed to $B-L$ and chargephobic cases with positive $\dneff$ values. For the three models, the constraint is set on $|\dneff=0.4|$, and the future CMB experimental sensitivity is taken to be $|\dneff|=0.05$.
 } 
 \label{neffContourFigures}			
\end{figure*}

Fig.~\ref{neffContourFigures} contains a summary of the $\dneff$ constraints on a dark photon, a $B-L$ vector boson, and on a chargephobic vector boson.  
Given the discussion above, 
we find that $\dneff$ receives 
positive contributions in the $B-L$ and chargephobic cases, and negative contributions in the dark photon case.  This last result is potentially interesting in the context of some CMB results that find a best fit to $\dneff < 0$ \cite{AtacamaCosmologyTelescope:2025nti}.

\subsection{Astrophysics}
\label{astrophysicsSec}

The explosion of the blue supergiant Sanduleak $-69\,202$ in the Large Magellanic Cloud and its subsequent detection is known as supernova 1987A (SN1987A). 
The supernova is estimated to have 
released $\approx 3\times10^{53}\, \text{erg}$ in the form of neutrinos. 
The Kamiokande II \cite{Kamiokande-II:1987idp} and IMB collaboration \cite{Bionta:1987qt}
detection of $\sim\,20$ neutrino events
within $\sim\,10$ seconds of the explosion
agrees with the predictions of the Standard Model for total energy, average neutrino energy, and burst duration based on simulation of neutrino emission from stellar core collapse for SN1987A \cite{Burrows:1986me, Burrows:1987zz,Fiorillo:2023frv}.  For reviews of supernova core collapse theory see \cite{Raffelt:1990yz,Raffelt:1996wa,Burrows:2018qjy, Janka:2006fh}.
This agreement allows constraints to be established on BSM models that provide an additional cooling mechanism to the proto-neutron star (PNS) after collapse \cite{Raffelt:1996wa},
as discussed in Sec.~\ref{pnsCoolingSec}.
In addition, there are  constraints on
the nonobservation of high energy neutrino events
during the same time window that low energy neutrinos were observed, that is relevant to models where such emission is possible
\cite{Fiorillo:2022cdq, Akita:2022etk, Syvolap:2023trc, Akita:2023iwq, Syvolap:2024hdh, Telalovic:2024cot, Blinov:2025aha},
as discussed in Sec.~\ref{highEnuSec}.
Finally, there are also several supernova based constraints arising from the production and decay to visible modes (especially $e^+e^-$) \cite{Pastorello:2003tc, Prantzos:2010wi, Spiro_2014, Sung:2019xie,  Caputo:2022mah, Diamond:2023cto, Fiorillo:2025yzf, Candon:2025ypl}; these bounds were most recently derived on a dark photon in \cite{Caputo:2025avc} and will be the discussed in Sec.~\ref{EMSSsec}.

\subsubsection{Protoneutron Star Cooling: SN1987A}
\label{pnsCoolingSec}

\phantomsection
\label{sec:cnst:snCooling}

For dark photons, the PNS cooling constraint was determined by Refs.~\cite{Chang:2016ntp, Hardy:2016kme, Mori:2025baz,  Caputo:2025avc}, which improved on previous constraints \cite{Bjorken:2009mm, Dent:2012mx, Kazanas:2014mca, Rrapaj:2015wgs} by including finite temperature and density effects. For a $B-L$ vector boson, calculations were carried out in Refs.~\cite{Croon:2020lrf, Shin:2021bvz}, however the resulting constraint differs between the two references in certain regions of parameter space by up to an order of magnitude in the values of $g_{B-L}$.   As discussed below, we rederive the constraints from PNS cooling in our parameter space. In the dark photon case, our results are in good agreement with the previous results \cite{Chang:2016ntp, Hardy:2016kme, Caputo:2025avc}.
In the $B-L$ case, our results resolve the discrepancy between the constraints in \cite{Croon:2020lrf} and \cite{Shin:2021bvz}, by including effects neglected in both analyses. Finally, we derive PNS cooling constraints in the chargephobic case; this is an important result since PNS cooling is one of the few ways the chargephobic vector boson couplings can be constrained. 

The PNS cooling constraint is traditionally set by demanding the instantaneous luminosity of the new particles $L_X(m_X,g_X)$ not exceed the ``Raffelt Criterion", $L_\nu= 3\times 10^{52} \text{erg}/\text{s}$ \cite{Raffelt:1996wa} . Specifically, this limit is applied around $t=1\,\text{s}$ post-bounce where the PNS core reaches peak density. The duration of SN1987A neutrino burst is reduced by half if the luminosity exceeds this limit which is inconsistent with observations. In order to obtain the constraints on 
$X$ emission, we have utilized
Refs.~\cite{Caputo:2021rux, Caputo:2022rca, Blinov:2025aha}, calculating the instantaneous luminosity 
\begin{widetext}
\begin{align}
\label{Luminosity}
L_X(m_X, g_X, R_\nu)&=\int_0^{R_\nu}dr 4\pi r^2 N(r)^2 \int_{m_X/N(r)}^\infty
 d\omega \frac{\omega^3 v}{2\pi^2} \Gamma_{\text{prod} }(r,\omega) \,O(r,\omega),\\\nonumber\\
 O(r,\omega)&=\frac{1}{2}\int_{-1}^{+1} d\cos \theta \exp \left[
-\int_{0}^{s_{\rm max}} ds \frac{\Gamma_{\rm abs} (\sqrt{r^2 +s^2 +2rs \cos\theta},\omega)}{v}
\right].
\end{align}
\end{widetext}
The luminosity accounts for $X$ production inside the neutrinosphere, $r<R_\nu$, that escapes the PNS, therefore constituting energy loss or cooling. The optical depth ensures that $X$ decay/absorption that occurs inside the PNS does not contribute to the luminosity (see also Refs.~\cite{Fiorillo:2023cas, Fiorillo:2023ytr}). This is done by taking the direction-averaged absorption of $X$ along its propagation path parameterized by $(s, \theta)$, the chord distance between the production and escape radii and emission angle respectively. In this context, the chord distance from $r$ to $R_{\rm max}$ is 
\begin{equation}
\label{sMaxChordDistance}
s_{\rm max}=-r\cos\theta +\sqrt{R_{\rm max}^2-r^2+r^2 \cos^2\theta}.
\end{equation}
In this work, we take $R_{\rm max}=R_\nu \sim 25 \, \km$ as advocated for in Ref. \cite{Lai:2024mse}. 
Finally, general relativistic effects of the PNS gravitational well are encoded in gravitational lapse factor, $N(r)$ \cite{Caputo:2022mah, Blinov:2025aha}. The $N(r)^2$ term accounts for the time dilation of the production rate and the redshifting of emitted $X$, whereas the $m_X/N(r)$ term accounts for necessary escape velocity.

\begin{figure*}
\includegraphics[width=0.49\textwidth]{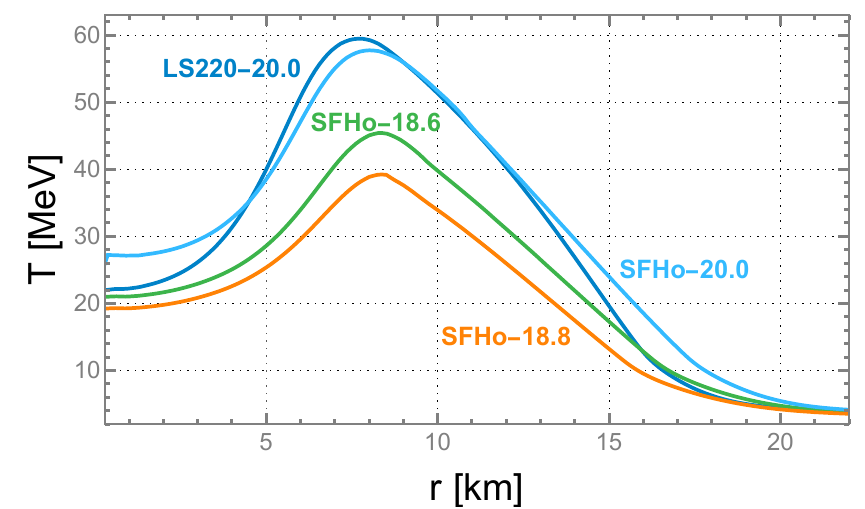}
\hfill
\includegraphics[width=0.49\textwidth]{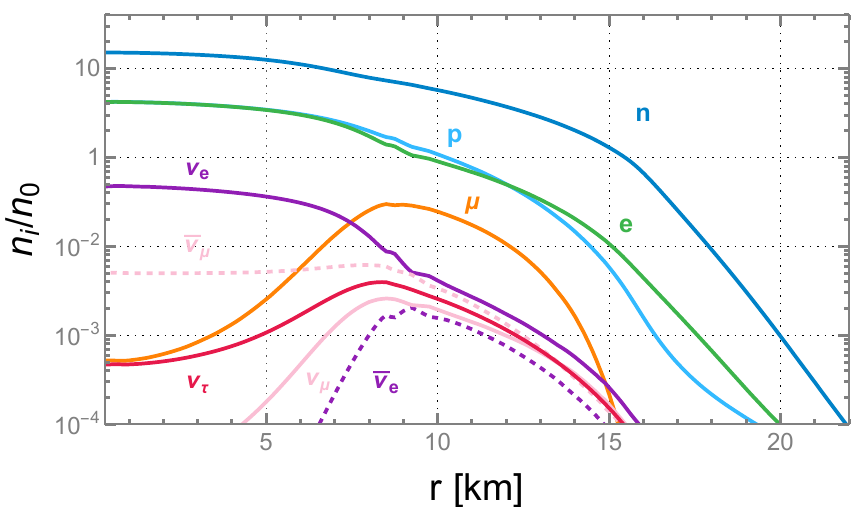}
\caption{(Left) temperature profile of the 4 models used, and (right) particle number density profile in the SFHo-$18.8$ model normalized to $n_0=0.181 \;\text{fm}^{-3}$ corresponding to nuclear density of $3\times 10^{14} \; \text{g cm}^{-3}$.}
\label{snProfilesFig}
\end{figure*}

To ensure our bounds are robust
with respect to SN modeling, we use four representative models to derive the constraints. The models are SFHo-$18.6$, SFHo-$18.8$, SFHo-$20.0$, and LS220-$20.0$, where SFHo \cite{Steiner:2012rk} and LS220 \cite{Lattimer:1991nc} are the equations of states used at supernuclear densities, and $18.6$ \cite{Woosley:2002zz}, $18.8$ \cite{Sukhbold:2017cnt}, and $20.0$ \cite{Woosley:2007as} are the progenitor masses. More information of these models can be found in \cite{Bollig:2020xdr}. The temperature profiles and normalized number densities of particles in the PNS at $1\, \text{s}$ after core collapse are shown in Fig.~\ref{snProfilesFig}.%
\footnote{The full profile data can be found at the Garching Core Collapse Supernova Archive \cite{Garching}.}
For all $\vp$, we find that the SFHo-$18.8$ model leads to the most conservative bounds, and so we adopt this model when presenting our results.

In medium, the SM photon acquires nonzero effective plasma mass \cite{Braaten:1993jw, An:2013yfc} which is characterized by the real part of the photon polarization tensor
\begin{align}
\label{realPolarizationTensor}
\reP_\snL &=\frac{3 \omega_p^2}{v^2}(1-v^2) \left[\frac{1}{2v}\,\text{ln} \left(\frac{1+v}{1-v}\right)-1\right] \\
\reP_\snT &=\frac{2 \omega_p^2}{2v^2} \left[1-\frac{1-v^2}{2v}\,\text{ln} \left(\frac{1+v}{1-v}\right)\right].
\end{align}
Here $\omega_p^2 = 4\pi \alpha_{\text{EM}} n_e/\sqrt{m_{e}^2+(3\pi^2 n_e)^{2/3}}$ is the plasma frequency, $v$ is the photon velocity, and $\snT, \,\snL$ are the transverse and longitudinal polarizations of the photon each with a different in-medium dispersion relation. The polarization tensor also dynamically changes the coupling of $X$ with the SM particles charged under $U(1)_{\text{EM}}$. Explicitly,
\begin{align} 
\label{CoupInMediumSN}
g_X & \longrightarrow  g_{X,m} \;=\; \frac{m_X^2}{m_X^2-\Pi} g_X \; \equiv \; \xi_m g_X  &&(e,q,p)\\ 
g_X & \longrightarrow  g_X 
&&(\nu,n),
\end{align}
where
\begin{equation}
\label{mediumEffectOnCoup}
\xi^2_m \; = \; \frac{1}{\left(1- \reP/m_X^2 \right)^2+\left(\imP/m_X^2\right)^2} \, .
\end{equation}
The imaginary part of the polarization tensor is $\imP_{\snLT}=-\omega (\Gamma_{\text{abs} \mid \snLT }-\Gamma_{\text{prod} \mid \snLT })$, where $\Gamma_{\text{abs}}$ and $\Gamma_{\text{prod}}$ are the absorption width and production rate of SM photon at energy $\omega$ \cite{Weldon:1983jn}. In local thermal equilibrium at temperature $T$, these rates are related through detailed balance $\Gamma_{\text{prod}} =e^{-\omega/T}  \Gamma_{\text{abs}}$. 
The absorption rates of interest in a proto-neutron star are inverse bremsstrahlung, semi-Compton scattering, and $X$ decays into $e^{+}e^{-}$ and $\nu\bar{\nu}$. 

The bremsstrahlung rate is composed of $pn$, $pp$, and $nn$ collisions and is derived in App.~\ref{BremAppendix} using the soft radiation approximation,
\begin{widetext}
\begin{equation}
\Gamma^{ij}_{\text{ibr}|\snLT}=\frac{8 n_i n_j }{15 \pi^2  \omega^3 S_\snLT}\left(\frac{\pi T}{ m_N}\right)^{3/2}
\begin{cases}
\left[10q_{-}^2 \langle\sigma_{ij}^{(2)}\rangle +9q_{+}^2 \frac{Tv^2 }{m_N} \langle\sigma_{ij}^{(4)}\rangle
\right] &(\snT)\\
\left[
5q_{-}^2  \langle\sigma_{ij}^{(2)}\rangle  +6q_{+}^2 \frac{Tv^2 }{m_N}\langle\sigma_{ij}^{(4)}\rangle
\right](1-v^2) &(\snL) \, .
\end{cases}
\label{GammaBremIbr}
\end{equation}
\end{widetext}
Here $q_\pm= q_1\pm q_2$ with $q_1,q_2$ the charges of the two nucleons, $\langle\sigma_{ij}^{(2)}(T)\rangle=\frac{1}{2}\int^\infty_0 dx e^{-x}x^2\sigma_{ij}^{(2)}(xT)$ and
$\quad \langle\sigma_{ij}^{(4)}(T)\rangle=\frac{1}{6}\int^\infty_0 dx e^{-x}x^3\sigma_{ij}^{(4)}(xT)$ are the energy and angle-averaged nucleon-nucleon scattering cross section found in \cite{Rrapaj:2015wgs}. The production rate is identical with the restriction that the energy of $X$ must not exceed the energy of the collision. This takes $\int_0^\infty \rightarrow  \int_{\omega/T}^\infty$; for example $\langle\sigma_{ij}^{(2)}(T)\rangle \rightarrow \langle\sigma_{ij}^{(2)}(T,\omega)\rangle =\frac{1}{2}\int^\infty_{\omega/T} dx e^{-x}x^2\sigma_{ij}^{(2)}(xT)$.

When calculating the imaginary part of the polarization tensor of the SM photon $\imP_{\snLT}$, the charges in question are $q_p=e$ for the proton and $q_n=0$ for the neutron. When calculating the absorption rate for $X$, the proton charge is $q_{p, \rm med}=\xi_m q_p=\xi_m g_X (\cos\varphi+\sin\varphi)$ and the neutron charge $q_{n, \rm med}= q_n=g_X\sin\varphi$. 
In a general model, the $pn,\, pp$ and $nn$ charges in Eq.~(\ref{GammaBremIbr}) become
\begin{eqnarray}
(q_p\pm q_n)^2_{\rm med}&=& q_n^2 + \xi_m^2 \left[ (q_p^2 \pm 2q_p q_n) \mp 2 \frac{\reP}{m_X^2}q_p q_n\right]\nonumber\\
\label{bremInMediumCoup}
(q_p+ q_p)^2_{\rm med}&=& 4 \xi_m^2 q_p^2\\
(q_n+ q_n)^2_{\rm med}&=& 4 q_n^2 \, .\nonumber
\end{eqnarray}
When the couplings of the proton and neutron to $X$ are the same [$q_{-}^2=0$ in Eq.~(\ref{GammaBremIbr})], the bremsstrahlung rate arises only at the quadrupole order and is suppressed. When the couplings are different, the $pn$ bremsstrahlung rate dominates. In the dark photon case, there is no coupling to neutrons and we are left with two diagrams for $pn$ and four for $pp$. For $pp$ bremsstrahlung rate, $q_{-}=q_p-q_p=0$ and first contribution is suppressed by $T/m_N$ as compared to the $pn$ bremsstrahlung rate. Hence only the $pn$ bremsstrahlung rate is important in the dark photon limit. In the $B-L$ case with ${\rm Re} \, \Pi \ll m_X^2$, the couplings of $X$ to the proton and neutron are the same, and so the leading contribution to bremsstrahlung is of order $\mathcal{O}(T/m_N)$ requiring all of $pn,\, pp$, and $nn$ bremsstrahlung rates to be included [the $q_{+}^2\langle\sigma_{ij}^{(4)}\rangle$ terms]. However, when ${\rm Re} \, \Pi \gg m_X^2$, only the $pn$ bremsstrahlung rate (off of neutrons) is important since the in-medium proton charge is suppressed and is thus non-degenerate with the neutron charge. 
Similarly, in the chargephobic limit, only the $pn$ bremsstrahlung rate is important, though in this case, the bremsstrahlung is $X$ off neutrons (and not protons). 
Hence, the $pp$ and $nn$ bremsstrahlung are only important in the $B-L$ case when $m_X \gtrsim 10 \,\mev$.

\begin{figure*}
\centering
\includegraphics[width=0.325\textwidth]{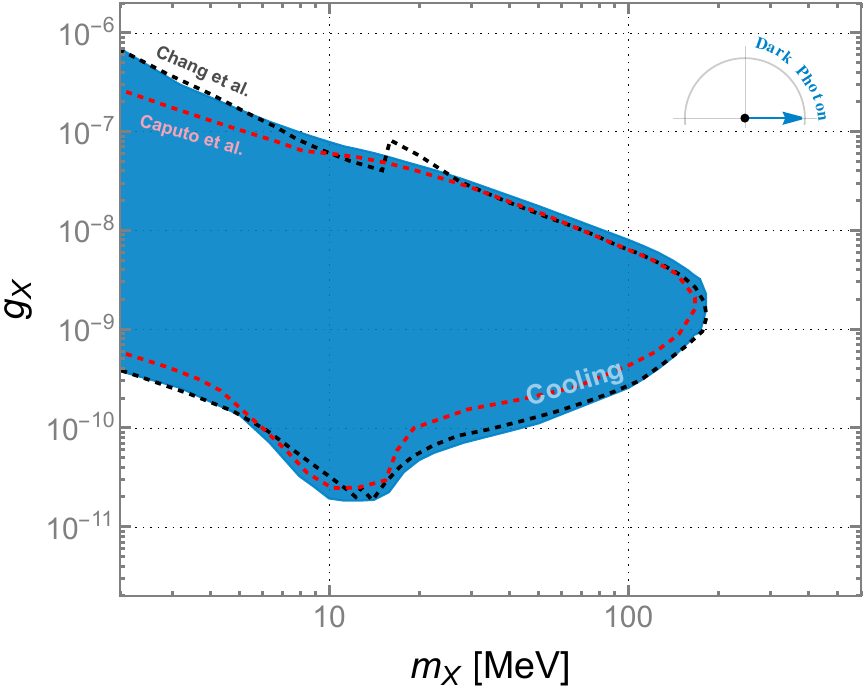}
\hfill
\includegraphics[width=0.325\textwidth]{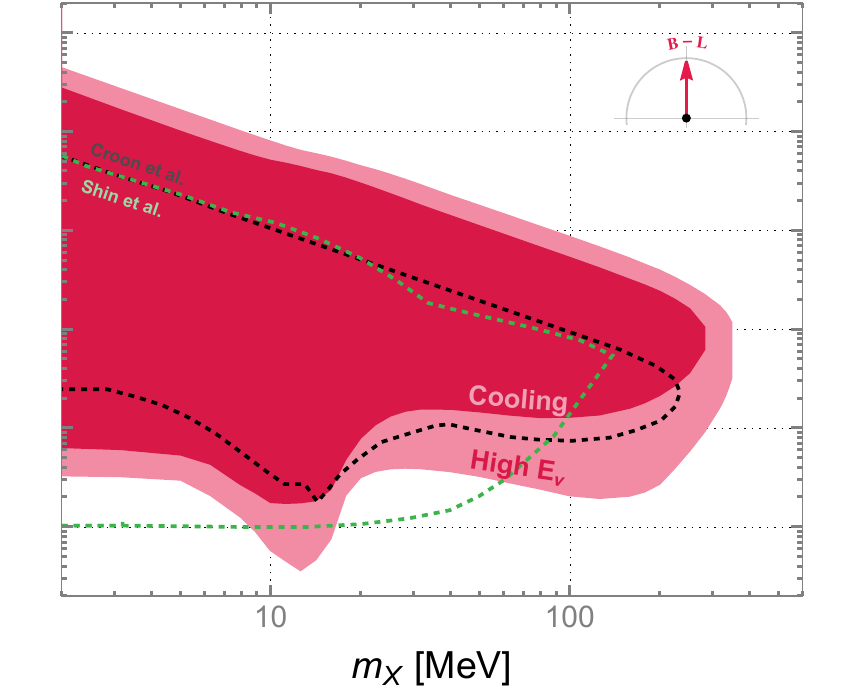}
\hfill
\includegraphics[width=0.325\textwidth]{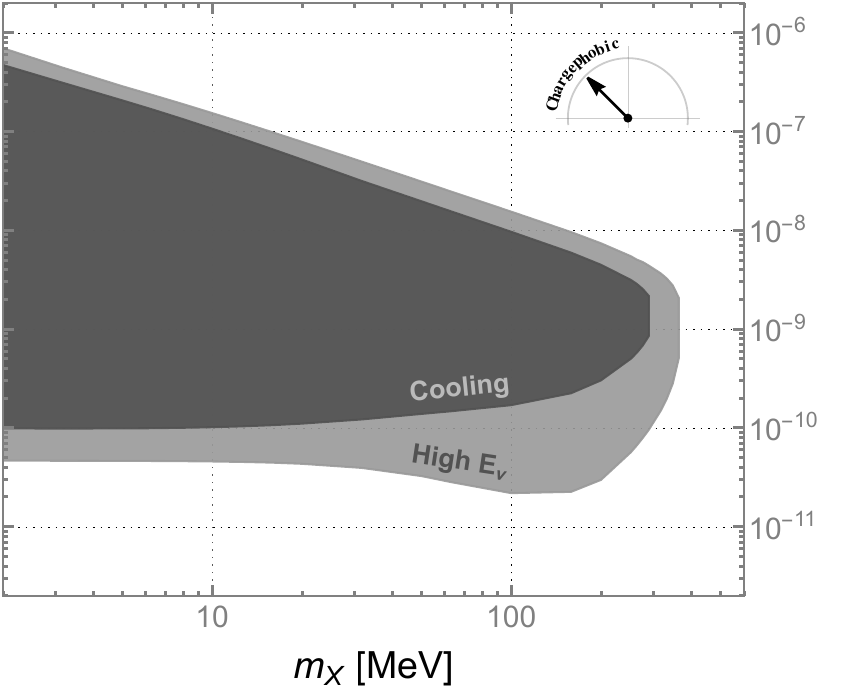}
 \caption{
 SN1987a constraints on the dark photon case $\vp=0$ (left, blue), the $B-L$ case $\vp=\pi/2$ (middle, red), and the chargephobic case $\vp=3\pi/4$ (right, gray). The colors are chosen to match the compass inset, with the darker and lighter shades corresponding to PNS cooling and non-observation of high energy neutrinos bounds respectively. In the dark photon case, the black and red dashed lines are cooling bounds taken from Refs. \cite{Chang:2016ntp} and \cite{Caputo:2025avc} with $R_{\rm max}= R_\nu$. In the $B-L$ case, the black and green dashed lines are the cooling bounds taken from Ref. \cite{Croon:2020lrf} ($R_{\rm max} = 100 \,\km$, and using a similar analysis to Ref. \cite{Chang:2016ntp}) and Ref. \cite{Shin:2021bvz} ($R_{\rm max} = 1000 \,\km$); our cooling constraint resolves the tension between these two previous bounds as discussed in the text.
 } 
 \label{snDPandBmLCnstrnts}			
\end{figure*}

The decay rates $X \rightarrow e^{+}e^{-}$ and $X \rightarrow \nu_i\bar{\nu}_i$ provide an important contribution to the optical depth. The widths are 
\begin{widetext}
\begin{equation}
\label{snDecaysEqn}
\Gamma_{f\bar{f}|  \snLT} \; = \; \mathcal{C}_f\frac{1}{4\pi}\frac{(q_f)_m^2 m_X^2}{\sqrt{\omega^2-m_X^2}}\int_{x_{-}}^{x_{+}} \frac{\textbf{m}_f(\omega x)^2_\snLT \, dx }{
\left[\exp\left(\frac{\mu_f-\omega x}{T}\right)+1\right]
\left[\exp\left(-\frac{\mu_f+\omega(1-x)}{T}\right)+1\right]},
\end{equation}
\end{widetext}
where $\mu_f$ is the chemical potential, and $x_\pm= \frac{1}{2}\left[ 1\pm \sqrt{(1-4(m_{f}^{\rm eff})^2/m_X^2)(1-m_X^2/\omega^2)}  \right]$ are the limits of integration, and $\mathcal{C}_f= 1 \;(1/2)$ for $\ee\; (\nu_i\bar{\nu}_i)$. We also have the dimensionless squared matrix elements $\textbf{m}_f(E_1)^2_\snT=(m_{f}^{\rm eff})^2/m_X^2+z(E_1)$ and $\textbf{m}(E_1)^2_  \snL=1-2z(E_1)$ where
\begin{equation}
\label{decayeezE1}
z(E_1)=\frac{E_1E_2}{m_X^2}-\frac{(E_1\omega-m_X^2/2)(E_2\omega-m_X^2/2)}{\omega^2-m_X^2},
\end{equation}
and $E_2=\omega-E_1$ \cite{Chang:2016ntp}. The two factors in the denominator of Eq.~(\ref{snDecaysEqn}) are due to Pauli blocking of $f$ and $\bar{f}$ respectively. In particular, $\mu_e/T\gg 1$ and so the second term reduces to unity, since the positron abundance is negligible in the PNS\@.  On the other hand, $\mu_{\nu_\tau}=0$, and so both terms contribute to blocking. In medium, the electrons also acquire an effective mass \cite{Braaten:1991hg} 
\begin{equation}
\label{inMediumElectronMassSN}
m_{e}^{\rm eff}=\frac{m_{e,0}}{2} +\sqrt{\frac{m_{e,0}^2}{4}+\frac{\alpha}{2 \pi}\left(\pi^2 T^2 +\mu_e^2 \right)}.
\end{equation}
This electron effective mass ranges between $0.6-11.3\; \mev$ within $R_\nu$ and the electron-positron decay channel is accessible only when $m_X\geq 2 m_{e}^{\rm eff}$. As a consequence, this channel can never be resonant since $m_X<2 m_{e}^{\rm eff}$ is always true when $m_X^2 = \reP_{\snLT}$. It is unclear if previous results in the literature have considered this effect. Furthermore, electron-positron coalescence can be safely ignored everywhere since positrons are so rare in the PNS\@. On the other hand, neutrino-antineutrino coalescence is only important in the high mass region of the $B-L$ case where the bremsstrahlung rate is suppressed. The PNS cooling constraints on the dark photon, $B-L$, and chargephobic cases are shown in Fig.~\ref{snDPandBmLCnstrnts}. In the $B-L$ and chargephobic case, Fig.~\ref{snDPandBmLCnstrnts} also shows constraints from the non-observation of high energy neutrinos which will be discussed in Sec. \ref{highEnuSec}.

In the dark photon case, we find excellent agreement with previous cooling bounds in Refs.~\cite{Chang:2016ntp, Caputo:2025avc}. The small discrepancy between the three bounds can be attributed to differences in the adopted CCSN model, and a small variation in the treatment of bremsstrahlung production rate.

In the $B-L$ case, there are several qualitative differences between our bound and the two previous bounds in Refs.~\cite{Croon:2020lrf, Shin:2021bvz} (where these two prior results also disagreed with each other in parts of the parameter space).  First, the upper bound is set by the choice of $R_{\rm max}$, the maximum radius where $X$ can be reabsorbed or decay without contributing to the cooling luminosity. Our choice of $R_{\rm max}=R_\nu \sim 25\,\km$, as advocated for in Ref. \cite{Lai:2024mse}, predictably results in a stronger upper bound when compared to the choice of  $R_{\rm max}\sim 1000 \, \km$ in Ref. \cite{Shin:2021bvz} due to the reduced trapping. We have explicitly checked that we reproduce the upper bound in Refs. \cite{Croon:2020lrf, Shin:2021bvz} for $R_{\rm max}\sim 1000 \, \km$.

Next we explain the features observed for the lower bounds of $B-L$ case.  For $m_X \gtrsim 20 \;\mev$, in-medium effects are negligible and the proton/neutron charges are degenerate resulting in quadrupole suppressed bremsstrahlung production and absorption rates. In this regime, neutrino coalescence is an important production mode. For $m_X \lesssim 10 \;\mev$, finite temperature effects alter the proton effective charge but not for the neutron. This alleviates the quadrupole suppression in $pn$ bremsstrahlung and the resulting production is two fold: $pn$ bremsstrahlung off of neutrons which is $m_X$ independent, and resonant production off of protons when $m_X^2 = \reP_{\snLT}$. The resonance around $m_X \sim 10 \;\mev$ in the lower bound of the $B-L$ case is the same as that in the dark photon case; this is to be expected since the resonant production can be understood as a thermal bath of photons slowly transitioning to $X$ independent of the details of the production process \cite{An:2013yfc}. 

The lower bound in Ref.~\cite{Shin:2021bvz} (dashed green) is an order of magnitude stronger than ours. This discrepancy can be attributed to the way the bremsstrahlung rates are implemented. Ref.~\cite{Shin:2021bvz} used one-pion exchange, which is known to overestimate this rate as compared to nucleon-nucleon bremsstrahlung data. This overestimate was first shown in \cite{Rrapaj:2015wgs} and more recently in \cite{Hardy:2024gwy}. Following the argument in Appendix D of \cite{Hardy:2024gwy}, we have used soft radiation approximation in deriving our bremsstrahlung rates Eq.~(\ref{GammaBremIbr}). The lower bound in Ref.~\cite{Croon:2020lrf} (dashed black) for $m_X\lesssim 10 \;\mev$ was obtained by recasting the constraint on a vector boson coupling to just baryon number. However, in this process they did not take into account finite temperature effects lifting the bremsstrahlung quadrupole suppression, and hence arrived at a bound that is weaker than ours.

In the chargephobic case, the dominant production mode is $pn$ bremsstrahlung off of neutrons which is both $m_X$ independent and oblivious to finite temperature corrections. This rate sets the lower bound. The upper bound is set by inverse bremsstrahlung and decays into neutrinos inside the PNS\@. Furthermore, with the absence of the $X \rightarrow \ee$ decay channel, there are no efficient processes to facilitate the absorption of $X$ outside the neutrinosphere. Consequently,  the constraint is insensitive $\rfar$. In terms of complexity, the chargephobic SN cooling constraint is the simplest to derive among the three cases that we considered.  
Nevertheless, this limit demonstrates the resilience of the SN cooling bound even when $X$ has no couplings to electrons and protons.


\subsubsection{High Energy Neutrinos}
\label{highEnuSec}

\phantomsection
\label{sec:cnst:highEnuSec1}

The observed neutrino signal from SN1987A shows no evidence of neutrinos with $E_\nu \gtrsim 75 \;\mev$ \cite{Kamiokande-II:1987idp, Bionta:1987qt}; again, in accordance with CCSN numerical simulations  \cite{Burrows:1986me, Burrows:1987zz,Fiorillo:2023frv}. However, the luminosity of $X$ contributing to cooling in Sec.~\ref{pnsCoolingSec}, can also contribute an additional flux of high energy neutrinos that could have been observed in the same terrestrial detectors. This can arise through the direct decay $X\rightarrow \nu_i\bar{\nu}_i$ or through $X\rightarrow \mu^{+}\mu^{-}\; (\pi^{+}\pi^{-})$ and subsequent neutrinos from muon/pion decay. The non-observation of high-energy neutrinos has been used to set constraints on various models \cite{Fiorillo:2022cdq, Akita:2022etk, Syvolap:2023trc, Akita:2023iwq, Syvolap:2024hdh, Telalovic:2024cot, Blinov:2025aha}. In this work, we adapt the analysis employed to Ref. \cite{Blinov:2025aha} to our model as follows. The differential $\nu_e$ flux at the detector is
\begin{widetext}
\begin{equation}
\label{snHighEdetFluxEq}
\frac{d\Phi_{\nu_e}}{dE_\nu}=\frac{1}{3} \frac{\text{BR}(\nu)}{4\pi R^2_{1987A}} \int_0^\infty d\omega  \frac{1}{\omega} \frac{\Theta(E_\nu -E_\nu^{\rm min}) \times \Theta(E_\nu^{\rm max}-E_\nu )}{E_\nu^{\rm max}-E_\nu^{\rm min}} \frac{dL_X}{d\omega},
\end{equation}
\end{widetext}
where $\text{BR}(\nu)$ is the branching fraction of $X$ to neutrinos, $R_{1987A}= 51.4 \, \text{kpc}$ is the distance between Earth and SN1987A, $dL_X/d\omega$ is the differential luminosity in Eq.~ (\ref{Luminosity}), and the Heaviside theta term is the probability a neutrino
with energy $E_\nu$ is obtained from $X$ with energy $\omega$. 
At the detector level, two detection channels are at play: inverse beta decay $\bar{\nu}_e+p \rightarrow n +e^{+}$, and scattering off oxygen $\bar{\nu}_e/\nu_e+{}^{16} \text{O} \rightarrow e^{\pm} +Y$; these cross sections can be found in Refs.~\cite{Marteau:1999zp, Kolbe:2002gk, Strumia:2003zx, Formaggio:2012cpf, Fiorillo:2022cdq}. For the IMB detector \cite{IMB:1988suc}, the expected number of high energy events is
\begin{equation}
\label{snHighEnumEventsEq}
N_{\rm ev}=\epsilon_{\rm det} \sum_i \int_0^{10\,{\rm s}} dt \int_{90 \,\mev}^{215 \,\mev} dE_\nu \frac{d\Phi_{\nu_e} }{dE_\nu} \sigma^i_{\rm det}(E_\nu) N_i,
\end{equation}
with $\epsilon_{\rm det}$ the energy-independent detection efficiency. Here, we consider an exposure time of $10\,\text{s}$ and a fiducial mass of $6.8\; \text{kton}$ corresponding to $N_{{}^{16}\text{O}}= 2.3 \times 10^{32}$ and $N_{{}^1\text{H}}= 4.5 \times 10^{32}$. The time integral is performed by evaluating the differential luminosity/flux at the set of time snapshots provided by the Garching Archive \cite{Garching}. Finally, assuming Poissonian statistics, we set $95 \, \%$ confidence level bounds on parameter space that yields $N_{\rm ev}\geq 3$. These constraints are shown in Fig.~\ref{snDPandBmLCnstrnts} for the $B-L$ and chargephobic limits.


\subsubsection{Electromagnetic-Based Supernova Signals}
\label{EMSSsec}

\phantomsection
\label{sec:cnst:snEM}

When $X \rightarrow \ee$ is accessible, such as in the dark photon case, core collapse supernovae are capable of constraining $X$ emission not only through cooling but also through several electromagnetic signatures labeled in Fig.~\ref{DPandBmLCurrentConstraintPlots} as LESNe, SN1987A $\gamma$, $511\, \kev$ line, and the fireball bounds PVO/GW170817 that were taken directly from Ref.~\cite{Caputo:2025avc}.

The LESNe constraint arises by requiring that the energy deposited in the mantle, due to dark photon decays into $\ee$, does not over-energize a sample of very low-energy Type II supernovae whose light curves imply explosion energies $\lesssim 0.1 \;\text{Bethe}$ \cite{Pastorello:2003tc, Spiro_2014, Sung:2019xie,  Caputo:2022mah, Fiorillo:2025yzf}. Instead of decaying in the mantle, if the dark photons escape the progenitor and subsequently decays into $\ee$ pairs and photons outside the star, this would yield a detectable prompt $\mev$ $\gamma$-ray burst signal; the absence of such signal in the Gamma-Ray Spectrometer (GRS) in SN1987A \cite{Oberauer:1993yr} sets the constraint labeled SN1987A $\gamma$. A similar constraint can be placed from stacking Type Ic SNe events as was done for ALPs \cite{Candon:2025ypl}; translating this constraint to vector mediators is left for future studies. When the decays of escaped $X$ are frequent enough that the produced $\ee$ and photons become optically thick and thermalize, they form a fireball \cite{Piran:1999kx}. The Pioneer Venus Orbiter, which was sensitive to photons in the $0.2-2\;\mev$ range consistent with the expected thermalized fireball spectrum, saw no significant $\gamma$-ray excess from SN1987A \cite{Diamond:2023scc, DeRocco:2019njg}, thereby setting the constraint fireball (PVO). Similarly, for the compact binary merger GW170817 \cite{LIGOScientific:2017vwq, Murguia-Berthier:2020tfs}, X-ray observations from CALET/CGBM, Konus-Wind, and Insight-HXMT/HE place an upper limit on any GRB-like fireball emission \cite{Diamond:2023cto}, setting the constraint fireball (GW170817).
The injection of positrons into the Milky Way through these decays can substantially enhance the diffuse $\gamma$-ray emission from the Galactic Center, which is tightly constrained by the INTEGRAL/SPI measurement of the $511\, \kev$ line \cite{Prantzos:2010wi, DeRocco:2019njg}.

We emphasize that we did not derive or recast these constraints, but rather simply adopted them directly from \cite{Caputo:2025avc} for the dark photon case in Fig.~\ref{DPandBmLCurrentConstraintPlots}. In the $B-L$ case, we expect these constraints to remain but with some slight changes compared to the dark photon case. The difference can be characterized as follows: the $X\rightarrow \ee$ branching fraction in $B-L$ is $40\,\%$ smaller than in the DP case which reduces the signal from these constraints. Furthermore, important differences arise for the dominant production modes for $X$ inside the PNS between the dark photon and $B-L$ cases as discussed in Sec.~\ref{pnsCoolingSec}. For example, an estimate of the lower bound of the LESNe constraint can be obtained by simply rescaling the lower bound of the cooling bound \cite{Caputo:2025avc}. Hence, we expect the qualitative shape of the LESNe bound in the $B-L$ to mimic that of the $B-L$ cooling bound in Fig.~\ref{snDPandBmLCnstrnts}. Recasting these constraints to $B-L$, and more general angles $\vp$, requires a dedicated analysis that is beyond the scope of our work. 
We have taken the liberty to overlay these dark photon bounds
from \cite{Caputo:2025avc}  as unlabeled light-shaded regions with dashed-lines in the $B-L$ case shown in  Fig.~\ref{DPandBmLCurrentConstraintPlots}.  
This is merely illustrative of where 
one might expect these constraints to appear.

In the chargephobic vector boson case, 
virtually all of aforementioned constraints vanish due to the absence of $X$ decays to charged leptons.  The one potential exception to this is the LESNe constraint, since this extends up to a dark photon mass of about $500$~MeV\@.
For chargephobic masses $m_X \gtrsim 200$~MeV, a very small branching fraction to charged leptons
($\sim 10^{-6}$)
develops from the renormalization-group evolution of the couplings.  However, we suspect this is much too small to lead to a constraint on its own.
Instead, for $m_X > 2 m_\pi$, 
a chargephobic vector boson has an increasing
branching fraction to $\pi^+\pi^-$ (few to $20\%$),
that may or may not affect the low energy supernova dynamics.  While we have not shown a constraint in Fig.~\ref{CPConstraintPlots}, we caution the reader that a small region of LESNe parameter space may be constrained, if the convolution of $X$ decay and $\pi$ decay leads to enough energy deposition.  This is an interesting question, but beyond the scope of our work.

 \subsection{Neutrino Experiments}
 \label{NeutrinoSec}

Neutrino experiments can be classified into four categories: neutrino-electron scattering (NES) experiments, coherent elastic neutrino-nucleus scattering ($\cevns$) experiments, experiments sensitive to collider neutrinos, and searches for invisible final states. The last of these relies on the assumption that the invisible final states can be treated as $X$ decaying to neutrinos. NES and $\cevns$ rates are mediated by the neutral and charged electroweak current interactions in the SM and obtain a new contribution through $X$ exchange. The resulting enhancement in these rates can be detected in experiments sensitive to the NES and $\cevns$ process, or conversely, the lack of signal can be interpreted as constraints on the $\{m_X, g_X\}$ parameter space. In this section, we describe how $X$ modifies each of these four classes of neutrino observables and outline the procedure used to translate them into constraints on the parameter space of a given $\vp$. 

In the SM, the electron-neutrino scattering amplitude consists of contributions from the charged and neutral currents. The SM differential cross section is
\begin{widetext}
 \begin{equation}
\label{neutrinoElectronSigmaSM}
\left[\frac{d\sigma}{dE_r}(\nu e^{-}\rightarrow \nu e^{-})\right]_{\text{SM}}=\;\frac{2G_F^2 m_e}{\pi E_\nu^2}
\left( a_1^2 E_\nu^2 +a_2^2 (E_\nu-E_r)^2 -a_1a_2 m_e E_r \right),
\end{equation}
\end{widetext}
where $G_F$ is Fermi coupling constant, $E_r$ is the recoil energy of the electron, $E_\nu$ is the energy of incoming neutrino, and $\{a_1,a_2\}=\text{sin}^2\theta_W\{1, 1\}+\frac{1}{2}A$ with $A=\{1,0\}$ for $\nu_e e$, $\{0,1\}$ for $\bar{\nu}_e e$, $\{-1,0\}$ for $\nu_\alpha e$, and $\{0,-1\}$ for $\bar{\nu}_\alpha e$. The incoming neutrino must have a minimum energy of $E_\nu^\text{min}=\frac{1}{2}\left(E_r+\sqrt{E_r^2+2 E_r m_e}\right)$ for a given recoil energy. The $X$ contributions to NES and the interference between $X$ exchange and the SM $Z$ exchange are given by
\begin{widetext}
 \begin{eqnarray}
\label{neutrinoElectronSigmaDP}
\left[\frac{d\sigma}{dE_r}\right]_{\text{X}}&=&\frac{g_X^4 m_e\sin^2{\varphi}\,(\cos{\varphi}+\sin{\varphi})^2}{4 \pi  E_\nu^2(m_X^2+2m_e E_r)^2}
\left(2 E_\nu^2 +E_r^2-2E_\nu E_r-m_e E_r \right),\\
\left[\frac{d\sigma}{dE_r}\right]_{\text{Int}}&=&\frac{g_X^2G_F m_e\sin{\varphi}(\cos{\varphi}+\sin{\varphi})}{2\sqrt{2} \pi  E_\nu^2(m_X^2+2m_e E_r)^2} 
\left(b_1(2 E_\nu^2-m_e E_r )+b_2(2 E_r^2-4E_\nu E_r) +\beta \right),
\label{neutrinoElectronSigmaInterference}
\end{eqnarray}
\end{widetext}
where $\beta=\sin^2{\theta_W}(8E_\nu^2-8E_\nu E_r-4m_e E_r+4E_r^2)$, and $\{b_1,b_2\}$ are $\{1,0\}$ for $\nu_e e$, $\{1,1\}$ for $\bar{\nu}_e e$, $\{-1,0\}$ for $\nu_\alpha e$, and $\{-1,-1\}$ for $\bar{\nu}_\alpha e$ \cite{Bilmis:2015lja}.

$\cevns$ occurs when the momentum transfer between the neutrino and the nucleus is comparable to inverse of the nuclear radius.  The differential cross section is 
\begin{widetext}
 \begin{equation}
\label{coherentCrossSectionGeneral}
\left[\frac{d\sigma}{dE_r}\right]_{\text{CE}\nu\text{NS}}=\;\frac{G_F^2 Q^2 }{2\pi } M F^2(2 M E_r)\left( 2- M\frac{E_r}{E_\nu^2}-2\frac{E_r}{E_\nu}+\frac{E_r^2}{E_\nu^2}
\right)
\end{equation}
\end{widetext}
where $M$ is the mass of the target nucleus and $F(Q^2)$ is the nuclear form factor. In the SM, the effective charge is $Q_\text{SM}=Z(\frac{1}{2}-2 \sin^2{\theta_W})- \frac{1}{2}N$ for a nucleus with $Z$ protons and $N$ neutrons. This charge is approximately $Q_\text{SM}\sim - \frac{1}{2}N$ since the proton weak charge is quite small. When we include the contribution from $X$, the effective charge \cite{Liao:2017} becomes
\begin{widetext}
 \begin{equation}
\label{coherentNSIcharge}
Q^2_{\text{SM}+X}\;=\; \left(Q_\text{SM} -\left[Z(\cos{\varphi}+\sin{\varphi}) +N \sin{\varphi}\right]\sin{\varphi}
\frac{g_X^2 }{2\sqrt{2}G_F(2 ME_r+m_X^2)}\right)^2.
\end{equation} 
\end{widetext}
The $\cevns$ related constraints are of particular interest in this paper as they are some of the few terrestrial experiment constraints that do not vanish in the chargephobic case.


\subsubsection{Coherent elastic neutrino-nucleus scattering Experiments}

\phantomsection
\label{sec:cnst:CEnuNS}

The first observation of $\cevns$ in 2017 by the COHERENT collaboration \cite{COHERENT:2017ipa} spurred an extensive experimental program aimed at precision measurements of this process \cite{Abdullah:2022zue}. The $\cevns$ rate is sensitive to modifications by BSM mediators coupled to neutrinos and nucleons and can be leveraged to set constraints on light vector mediators \cite{Dent:2016wcr, Shoemaker:2017lzs,  Papoulias:2017qdn, Liao:2017, Abdullah:2018ykz, Denton:2018xmq, CONNIE:2019xid, Cadeddu:2020nbr, Miranda:2020tif, CONUS:2021dwh,  delaVega:2021wpx, Coloma:2022avw, AtzoriCorona:2022moj, DeRomeri:2025csu, Chattaraj:2025fvx,TEXONO:2025sub}.  In our mass range of interest, the current leading $\cevns$ constraints on light vector mediators are placed by COHERENT \cite{COHERENT:2017ipa, COHERENT:2021xmm,COHERENT:2020iec} and CONUS+ \cite{Ackermann:2025obx}. These two constraints will be the focus of this section and will be treated as representative of the current $\cevns$ constraint landscape for vector mediators of masses $m_X>1\;\mev$. 

The COHERENT collaboration has measured the $\cevns$ process using two detectors, in a sodium doped CsI scintillator \cite{COHERENT:2017ipa, COHERENT:2021xmm} and in a liquid argon detector \cite{COHERENT:2020iec} with future plans expanding to more detector materials \cite{Konovalov:2024uhc}. These detectors are exposed to the neutrino emissions from the Spallation Neutron Source (SNS) at Oak Ridge National Laboratory. In the CsI detector \cite{COHERENT:2021xmm}, the SM best fit prediction models the observed data with a $\chi^2/\text{dof}=82.6/98$ and the observed data rejects the no-$\cevns$ hypothesis at $11.6 \;\sigma$. 
The $X$ exchange contribution to the 
$\cevns$ rate, Eq.~(\ref{coherentCrossSectionGeneral}), can spoil this agreement and lead to a constraint on the couplings of $X$. In this work, we focus on constraints obtained from the CsI data release only; the inclusion of argon data provides only slight improvements. We set constraints following the analysis in \cite{DeRomeri:2022twg} 
using a Poissonian least-squares function over event counts in $9$ energy and $11$ timing bins.

Three neutrino flavors are produced in the SNS; prompt mono-energetic $30\;\text{MeV}$ $\nu_\mu$ produced via the decay of stopped $\pi^{+}$,  and delayed $\bar{\nu}_\mu$ and $\nu_{e}$ produced from the decay of $\mu^{+}$ which was in turn produced from pion decay. Neutrino fluxes $d\Phi_\nu/dE_\nu$ and timing spectra $f_T^\nu(t)$ can be found in \cite{Picciau:2022xzi}. For a given nucleus $\mathcal{N}$, the expected number of events per energy $i$ and time $j$ bin is 
\begin{widetext}
\begin{align}
\label{coherentEnergyTimeBinExpectedNumber}
N_{i,j}^{\cevns , \;\mathcal{N}}
&= \sum_{\nu=\nu_e, \nu_\mu, \bar{\nu}_\mu}N_{\rm target} 
\int_{t^j}^{t^{j+1}} dt f^\nu_T(t, \alpha_6) \epsilon_T(t) \int_{E^j_{\rm nr}}^{E^{j+1}_{\rm nr}}d E_{\rm nr}\epsilon_E(E_{\rm nr},\alpha_7) \times \nonumber\\
&\int_0^{E_{\rm nr}^{'\rm max}}d E'_{\rm nr}P(E_{\rm nr},E'_{\rm nr}) \int_{E_\nu^{\rm min}(E'_{\rm nr})}^{E_{\nu}^{\rm max}} d E_\nu \frac{d\Phi_\nu}{d E_\nu}(E_\nu) \frac{d\sigma_{\nu \mathcal{N}}}{dE_r}(E_\nu,E'_{\rm nr},\alpha_4),
\end{align}
\end{widetext}
where $N_{\rm target}$ is the number of target atoms in the detector, $\epsilon_T (\epsilon_E)$ are the time (energy)-dependent detector efficiencies, and $E_{\rm nr}$ and $E'_{\rm nr}$ are the reconstructed and true nuclear recoil energy respectively related via the energy resolution function $P(E_{\rm nr},E'_{\rm nr})$. The first two integrals are bounded by bin edges, the third integral by $E_{\rm nr}^{'\rm max}= 2(E_{\nu}^{\rm max})^2/m_N$, and the fourth by $E_\nu^{\rm min}\simeq \sqrt{m_N E'_{\rm nr}/2}$ and $E_\nu^{\rm max}=m_\mu/2$. The total binned number of events is  given  by $N_{i,j}^{\cevns}=N_{i,j}^{\cevns , \;\text{Cs}}+N_{i,j}^{\cevns , \;\text{I}}$. 

The analysis is then performed using a Poissonian
least-squares function
\begin{widetext}
\begin{equation}
    \label{CsIChiSquaredEq}
    \chi^2_{\rm CsI} \; = \; 2 \sum_{i=1}^{9}\sum_{j=1}^{11}
    \left[
    N_{\rm th}^{\rm CsI}-N_{ij}^{\rm exp}+N_{ij}^{\rm exp}\ln\left(\frac{N_{ij}^{\rm exp}}{N_{\rm th}^{\rm CsI}}\right)
    \right]
    +\sum_{k=0}^{4}\left(\frac{\alpha_k}{\sigma_k}\right)^2,
\end{equation}
\begin{equation}
\label{CsITotalTheoryPredictionEqn}
N_{\rm th}^{\text{CsI} ,\cevns}=(1+\alpha_0)N_{i,j}^{\cevns}(\alpha_4,\alpha_6,\alpha_7) +(1+\alpha_1)N_{i,j}^{\rm BRN}(\alpha_6)+(1+\alpha_2)N_{i,j}^{\rm NIN}(\alpha_6)+(1+\alpha_3)N_{i,j}^{\rm SSB}. 
\end{equation}
\end{widetext}
We use the CsI unbinned data events provided by the COHERENT collaboration  \cite{COHERENT:2021xmm} to extract the measured binned event number $N_{ij}^{\rm exp}$ and to estimate the number of events $N_{i,j}^{\rm SSB}$ from the beam-uncorrelated steady-state background (SSB). We also include the prompt beam-related neutron (BRN) and neutrino-induced neutron (NIN) events. The nuisance parameters are as follows: $\alpha_0$ encodes the efficiency and flux uncertainties, $\alpha_1\; (\alpha_1)  \;[\alpha_3]$ correspond to the BRN (NIN) [SSB] uncertainties, $\alpha_4$ modifies the nuclear form factor $R_{A}=1.23 A^{1/3}(1+\alpha_4)$, $\alpha_6$ accommodates the uncertainty in beam timing $(\pm 250 \, \text{ns})$, and $\alpha_7$ allows for deviations in the energy-dependent detector efficiency ($\epsilon_E[x +\alpha_7]$ with $\alpha_7$ allowed to vary in the range $[-1,1]\times \,\text{PE}$)
\footnote{The nuisance parameter $\alpha_5$ is associated with the quenching factor. This parameter can be neglected in the region of the parameter space $(m_X >1\,\mev)$ where electron scattering does not play an important role. We drop this parameter but keep the numbering convention to match \cite{COHERENT:2021xmm}.}.
Finally, the uncertainties $\sigma_i$ associated with nuisance parameter $\alpha_i$ are given by $\{\alpha_0,\,\dots,\,\alpha_4\}= \{11,\; 25,\; 35,\; 2.1,\; 5,\; 3.8 \} \,\%$.

The CONUS+ collaboration recently achieved the first clear observation of reactor neutrino $\cevns$ \cite{Ackermann:2025obx}, with earlier evidence from Dresden-II \cite{Colaresi:2022obx} and prior CONUS runs \cite{CONUS:2021dwh}. Furthermore, many reactor-based experiments aim to lower detection thresholds and in doing so enhance sensitivity to light mediators; such as TEXONO \cite{TEXONO:2024vfk}, NUCLEUS \cite{NUCLEUS:2022zti}, RICOCHET \cite{Ricochet:2022pzj}, and CONNIE \cite{CONNIE:2024pwt}. Reactor-based $\cevns$ experiments probe neutrino energies of a few MeV, providing leading $\cevns$ sensitivity to light vector mediators below roughly $m_X \lesssim 10 \; \mev$. This mass range is dominated by $\dneff$ constraints discussed in Section \ref{neffSection}, however, the CONUS+ constraint still rules out a novel region of parameter space. Similar to the COHERENT analysis, we follow the procedure laid out in \cite{DeRomeri:2025csu, Chattaraj:2025fvx} to place constraints from the CONUS+ data release \cite{Ackermann:2025obx}. Note, the combined analysis of CONUS+, TEXONO, and COHERENT data \cite{AtzoriCorona:2025ygn} result in a slightly stronger bound; this is beyond the scope of our analysis and we do not attempt to replicate these results.

\begin{figure*}
\includegraphics[width=0.49\textwidth]{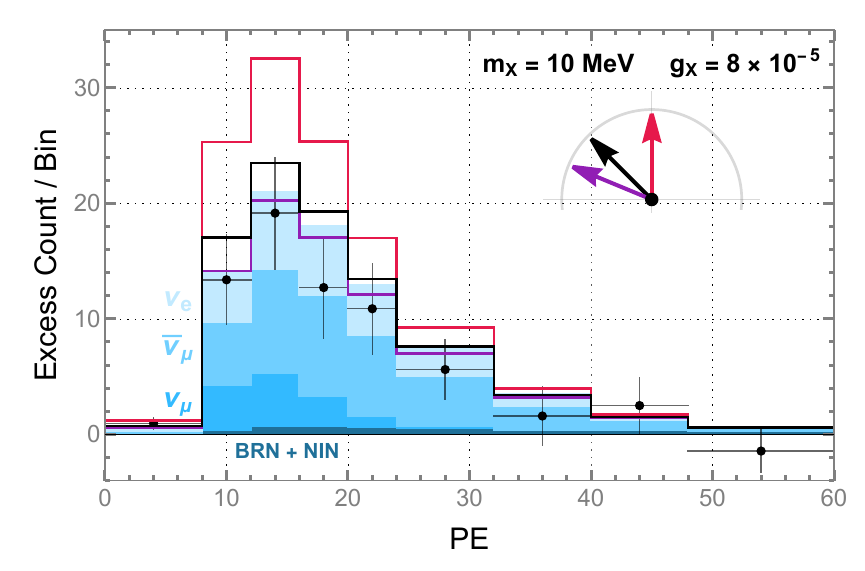}
\label{CoherentExcessPlot}
\hfill
\includegraphics[width=0.49\textwidth]{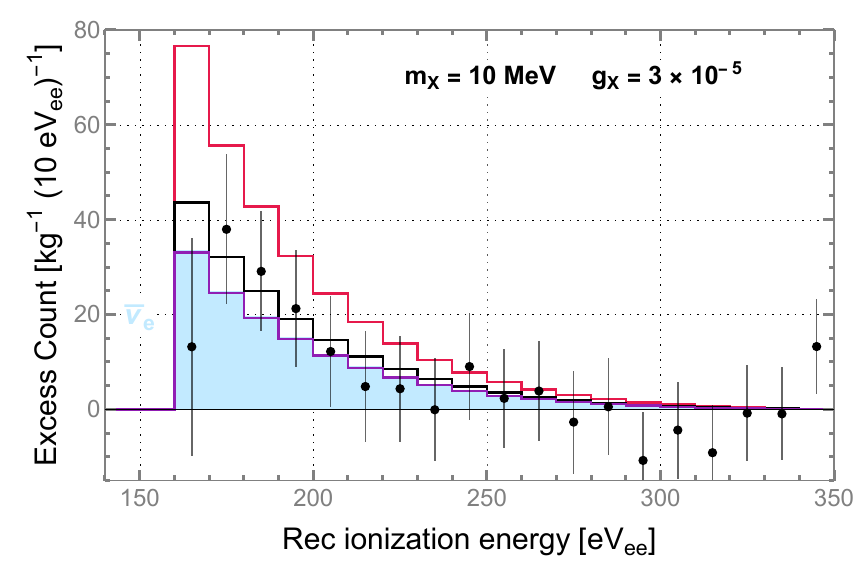}
\label{conusExcessPlot}
\caption{
(Left) COHERENT CsI detector data residual over SSB background \cite{COHERENT:2021xmm}. (Right) CONUS+ excess counts from three HPGe detectors (C2, C3, C5) \cite{Ackermann:2025obx}. The axes are chosen to match those in the original plots in the data release. The shaded steps represent the $\nu_i$ SM $\cevns$ prediction and the BRN and NIN backgrounds. The lines with no filling correspond to total excess count (SM + BSM) when $X$ with charge assignment of $B-L$, $\vp = 3\pi/4$, and $\vp = \pi-\text{tan}^{-1}(55/133)$ are considered. In all BSM cases, we have $m_X = 10 \, \mev$ with (right) $g_X = 8\times 10^{-5}$ and (left) $g_X = 3\times 10^{-5}$.} 
 \label{CEnuNSExcessPlots}	
\end{figure*}

\begin{figure*}
\centering
\includegraphics[width=0.325\textwidth]{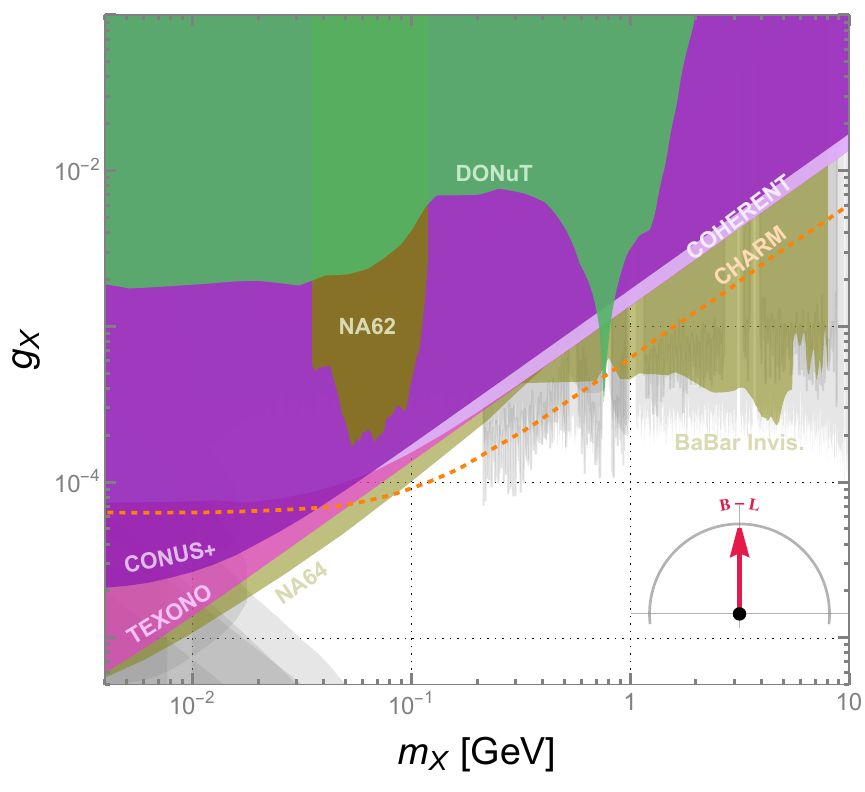}
\hfill
\includegraphics[width=0.325\textwidth]{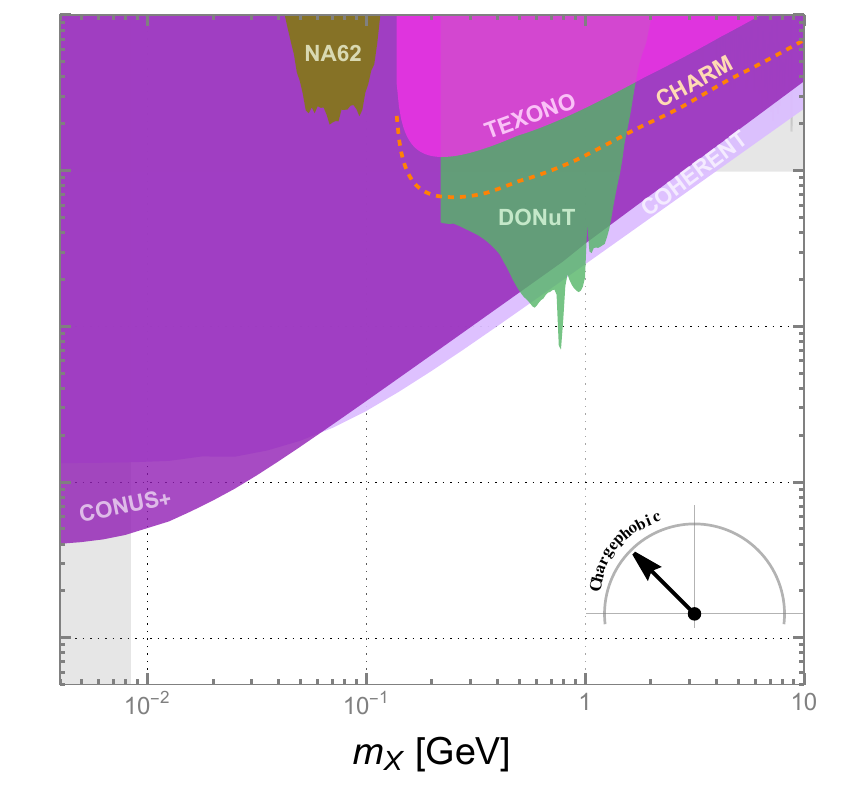}
\hfill
\includegraphics[width=0.325\textwidth]{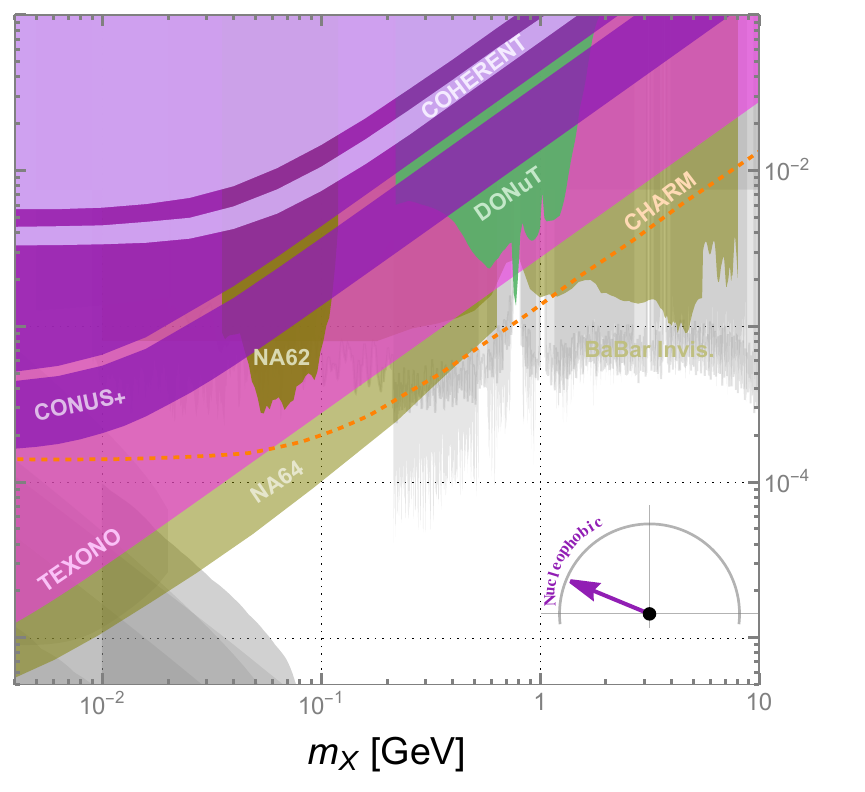}
 \caption{Constraints on (left) $B-L$ ($\vp=\pi/2$), (middle) chargephobic ($\vp=3\pi/4$), and (right) nucleophobic [$\vp=\pi-\text{tan}^{-1}(55/133)$] limits with a focus on the neutrino constraints discussed in Sec.~\ref{NeutrinoSec}. Other constraints in Fig.~\ref{DPandBmLCurrentConstraintPlots} are displayed in gray.
 } 
 \label{neutrinoCnstrntPlot}				
\end{figure*}

An example of the excess count calculation for $B-L$, chargephobic, and nucleophobic cases in COHERENT and CONUS+ detectors is shown in Fig. \ref{CEnuNSExcessPlots}. The nucleophobic case is chosen such that the effective charge of a cesium nucleus vanishes. This specific choice is arbitrary, but captures the typical extent to which $\cevns$ experiment sensitivity to $X$ is suppressed. The nucleus charge scales as $A$ in $B-L$, $\sim N$ in the SM and the chargephobic case, and $(N-(N_{\rm Cs}/Z_{\rm Cs})Z)$ in the nucleophobic case. This explains the extent to which each model contributes to the excess count as is evident in the BSM excess (BSM $-$ SM) in both experiments. The excess count does not vanish in the nucleophobic case since 
there are couplings to Ge and I nuclei.
These couplings are, however, suppressed: $\{x_{\rm Cs},\;x_{\rm I},\; x_{\rm Ge}\}_{\cancel{Cs}}= \{0,\; -0.4,\; -2\}$ when compared to $\{x_{\rm Cs},\;x_{\rm I},\; x_{\rm Ge},\; x_{\rm Ar}\}_{B-L}= \{133,\; 127,\; 72,\; 40\}$ for $B-L$. Furthermore, the negative nucleus charge in the nucleophobic case interferes destructively with the SM nucleus charge in Eq.~(\ref{coherentNSIcharge}). This can lead to excess counts lower than those in the SM as shown in Fig.~\ref{CEnuNSExcessPlots}. Finally, the COHERENT and CONUS+ constraints are shown in Fig.~\ref{neutrinoCnstrntPlot} where we emphasize constraints from the neutrino experiments discussed in Sec.~\ref{NeutrinoSec}. The aforementioned interference effect can also be seen through the unconstrained band in the COHERENT and CONUS+ nucleophobic bounds.

\phantomsection
\label{sec:cnst:COH}

Finally, to illustrate the potential of next-generation $\cevns$ experiments, we show the recasted projected COH-CryoCsI II sensitivity \cite{AtzoriCorona:2025ibl} in Fig. \ref{CPConstraintPlots} for the chargephobic limit. Here, the combination of a larger cryogenic CsI target, 
a lower nuclear recoil threshold, and an increased SNS beam power allows this setup to substantially improve upon current $\cevns$ bounds.


\subsubsection{Neutrino-electron Scattering Experiments}

\phantomsection
\label{sec:cnst:NES}

Experiments in this category are sensitive to deviations from the SM NES rate via the $X$ mediated NES channel. In our parameter space, the leading NES experimental constraints are set by Borexino Phase-II \cite{Bellini:2011rx, Borexino:2013zhu, BOREXINO:2014pcl, Borexino:2017rsf, BOREXINO:2018ohr, BOREXINO:2020aww}, TEXONO \cite{TEXONO:2009knm}, and CHARM-II \cite{VILAIN1994246,CHARM-II:1993phx}. 
TEXONO is a CsI(Tl) scintillating crystal array measuring $\bar{\nu}_ee^{-}$ elastic scattering with the $\bar{\nu}_e$ produced at the Kuo-Sheng Nuclear power station. The detector is exposed to an average $\bar{\nu}_e$ flux of $6.4\times 10^{12}\;\text{cm}^{-2}\,\text{s}^{-1}$ with standard reactor $\bar{\nu}_e$ spectrum \cite{Vogel:1989iv}. We derive the constraint from TEXONO following the analysis in \cite{Bilmis:2015lja, Lindner:2018kjo}. Borexino presents a comprehensive measurement of the solar neutrino spectrum from the pp chain in the energy range of $(0.19 - 16) \,\mev$. This data set in conjunction with standard solar model were used in \cite{Coloma:2022umy,Bilmis:2015lja,Harnik:2012ni} to set a constraint on the $B-L$ case. We adopt this constraint, without reproducing it, for the $B-L$ case and recast it using the appropriate charge scaling. For $m_X\gtrsim 3\; \mev$, TEXONO is the leading NES constraint whereas Borexino leads in the region below where the $\dneff$ constraint dominates.  Hence, we focus on TEXONO constraints when presenting our analysis in Fig.~\ref{neutrinoCnstrntPlot}. As for CHARM-II, it was pointed out in \cite{Bauer:2018onh} that the exact neutrino fluxes seem to be unknown in the experimental publication. This casts some doubt on the validity of neutrino rate calculation in \cite{Bilmis:2015lja, Lindner:2018kjo}. Following \cite{Bauer:2018onh}, we show the CHARM bound as a dashed line to remind the reader of this subtlety and we assume that the correct neutrino flux was used for the bound calculation.


\subsubsection{Invisible Searches}\label{invisSection}

\phantomsection
\label{sec:cnst:invis}

Invisible searches assume that the dark photon decays predominantly into a dark sector state $\chi$ via $X \to \chi \bar{\chi}$, which occurs when $m_X > 2m_\chi$ and the coupling $g_X$ is sufficiently small that visible decays are subdominant \cite{Fox:2011fx, NA64:2017vtt, BaBar:2017tiz,Belle-II:2019qfb, NA62:2019meo, NA64:2021xzo, NA64:2022yly, NA64:2023wbi}. The experimental signature is a visible SM particle accompanied by large missing energy and momentum. The same signature arises if $X$ instead decays mainly to neutrinos, $X \to \nu \bar{\nu}$. As a result, existing invisible searches can be recast 
using the neutrino branching fraction for a given $\vp$.%
\footnote{At the $J/\psi,\psi(2S), \, \psi(3770),$ and $\Upsilon(1S)-\Upsilon(4S)$ resonances, the $X \rightarrow \nu\bar{\nu}$ branching ratio can be non-trivially altered from the one reported in Fig.~\ref{BranchingFractionFig}. In these narrow regions, the invisible search constraints can lose sensitivity or even vanish. We have not included this effect and report the invisible search constraints as they are traditionally displayed in the literature \cite{Bauer:2018onh, Ilten:2018crw}.
}

In our analysis, we consider three invisible searches to provide constraints on $X$.  The BaBar invisible search uses $53 \;\text{fb}^{-1}$ of  $e^{+}e^{-}$  collision data to search for a single high-energy photon and a large missing momentum and energy via the process $e^{+}e^{-} \longrightarrow \gamma X$ followed by $X$ invisible decay \cite{BaBar:2017tiz}. The invisible search in $\text{NA}64$ uses an electron beam dump setup to look for single electron events with large missing energy from $e Z \rightarrow e Z X$ production followed by invisible $X$ decay \cite{NA64:2023wbi}. The $\text{NA}62$ invisible search uses the process $K^{+}\longrightarrow \pi^{+}\pi^{0}, \; \pi^{0}\longrightarrow \gamma X$ followed by an invisible decay of $X$ \cite{NA62:2019meo}. BaBar and NA64 set leading constraints on $B-L$ and nucleophobic cases as shown in Fig.~\ref{neutrinoCnstrntPlot}. NA62 is one of the few constraints that remain in the chargephobic case as it does not rely on a charged lepton coupling for either production or detection. The constraint is nonetheless significantly weakened due to the $\pi^{0}\longrightarrow \gamma X$ suppression in the chargephobic case as seen in Fig.~\ref{mesonDecayWidthPlot}. 
 

\subsubsection{Collider Neutrinos}\label{colliderNeutrinosSec}

\phantomsection
\label{sec:cnst:colliderNu}

Collider neutrino experiments \cite{DONuT:2007bsg, SHiP:2015vad, FASER:2019dxq, SHiP:2020sos} are sensitive to new physics that produces an additional forward flux of tau neutrinos on top of the SM $\nu_\tau$ component from $D_s$ and other heavy meson decays. In our model,  $X$ is produced via bremsstrahlung or meson decays and can subsequently decay as $X \rightarrow \nu_\tau \bar{\nu}_\tau$, thereby enhancing the flux incident on these detectors. To illustrate this, we use two representative experiments: the fixed target DONuT experiment at Fermilab as a current constraint and the forward LHC detector FASER$\nu$ for future projected sensitivity. In the $B-L$ case, we simply adopt these constraint from \cite{Kling:2020iar}. We  recast this constraint using Eq.~(\ref{recastingEquation}) and the LHCb production fractions in Figure 11 from \cite{Ilten:2018crw} for masses $m_X> 0.2\; \gev$. We do not recast or show the constraints for  $m_X< 0.2\; \gev$ in the chargephobic and nucleophobic cases in Fig.~\ref{neutrinoCnstrntPlot}; this region of parameter space is already well covered by NES and $\cevns$ experiments. 

There are several qualitative differences in the shape of the DONuT constraint between the $B-L$, chargephobic, and nucleophobic cases. For $B-L$, the resonance near $m_X \sim 0.8 \; \gev$ is driven by $\omega$–$X$ mixing, while the $\rho$–$X$ production mode is absent. In the chargephobic case, the $\rho$–$X$ channel is active, leading to a broader resonance, while the $\omega$–$X$ mixing is reduced by a factor of 8, as shown in Fig.~\ref{sigmaProdRatioRep}.  In the nucleophobic case, $\omega-X$ mixing is highly suppressed and so the resonance reach is reduced. The branching fraction to tau neutrinos is also affected by the width of the hadronic resonance; see the $B-L$ and chargephobic cases in Fig.~\ref{BranchingFractionFig}. The combination of these effects accounts for the shape of the DONuT constraint near the resonance.
 In our analysis, we used 
the LHCb production fractions \cite{Ilten:2018crw} to determine
the relative rates, however, they are not provided near the $\omega$ resonance.  We have used interpolations 
in the mass ranges where no results were shown in \cite{Ilten:2018crw}, and so our recasted DONuT resonant behavior in this mass range should be taken as approximate. A robust bound near the $\omega$ resonance requires a dedicated reanalysis of Ref.~\cite{Kling:2020iar} taking into account the production modes in Ref.~\cite{Kling:2025udr}, but this is beyond the scope of this work.

\subsection{Beam Dump \& Fixed Target Experiments}

\phantomsection
\label{sec:cnst:beamDump}

Electron and proton beam dump (BD) experiments direct high intensity beams onto thick targets, which can produce $X$ via bremsstrahlung (in $e$BD \& $pBD$) or meson decay/mixing (in $pBD$). A long-lived $X$ vector boson that is produced in this collision and survives the shielding can decay in a downstream decay volume, with its decay products registered in a detector located anywhere from meters to hundreds of meters away.  Decay modes of $X$ that lead to visible signals are typically $X \rightarrow \ell^{+}\ell^{-}$ and in particular $X \rightarrow \ee$, which appears as an electromagnetic shower. Accordingly, constraints set by these beam dump experiments vanish in the chargephobic case. A notable exception is the projected sensitivity of SHiP \cite{Alekhin:2015byh, SHiP:2020vbd, Zhou:2024aeu}, whose ability to fully reconstruct displaced $X$ decay vertices extends the signal reach to the broad range of hadronic decays in Fig.~\ref{VMDBranchingFractionFig}. This makes SHiP uniquely positioned to probe the chargephobic parameter space in the future.

In electron beam dumps, the production cross section of $X$ can be approximated using the Weizs\"{a}cker Williams approximation where the target nuclei $\mathcal{N}$ are replaced by an effective flux of photons \cite{Bjorken:1988as}. The bremsstrahlung process is simplified to semi-Compton scattering, $e\,\gamma \longrightarrow  e\, X$, and is given to a good approximation by 
\begin{equation}
\label{eBeamDumpCrossSection}
\frac{d\sigma_X}{dr_e}\simeq \frac{\alpha^2}{\pi}g_X^2(1+\sin{2 \varphi}) \,\xi \sqrt{1-\frac{m_X^2}{E_e^2}}\frac{1-r_e+\frac{r_e^2}{3}}{m_X^2\frac{1-r_e}{xre}+m_X^2 r_e}, 
\end{equation} 
where $r_e\equiv E_X/E_e$ is the fraction of electron beam's energy carried by $X$ and $\xi(E_e, m_X, Z, A)$ is the effective photon flux \cite{Bjorken:2009mm}.  The electron beam can interact with the shielding material which diminishes its energy from initial beam energy $E_0$ to $E_e$. The energy distribution of the electron beam as function of penetration depth $t$ \cite{Tsai:1986tx} is
\begin{equation}
\label{beamEnergyPenetrationDepth}
I_e(E_0,E_e,t)= \frac{1}{E_0}\frac{
\ln{\left(\frac{E_0}{E_e}\right)}^{\frac{4}{3}t-1}
}{
\Gamma(\frac{4}{3}t)
}.
\end{equation} 
The expected number of $e^{+}e^{-}$ events in the detector for an electron beam of $N_e$ electrons and energy $E_0$ is 
\begin{widetext}
\begin{align} 
\label{numElectronBeamDump}
N\simeq N_e \frac{N_0 X_0}{A}&\int^{E_0-m_e}_{m_X}dE_X\int^{E_0}_{E_X+m_e}dE_e\int^{T_\text{sh}}_{0}dt_{\text{sh}}\nonumber\\
 & \times\left[ I_e(E_0,E_e,t_{\text{sh}}) \frac{1}{E_e}\frac{d\sigma_X}{dr_e} 
  e^{-L_{\text{sh}}/l_X}
 \left(
1- e^{-L_{\text{dec}}/l_X}
 \right)  \right] 
 \,\text{Br}_{e^{+}e^{-}}.
\end{align}
\end{widetext}
Here $N_0$ is Avogadro’s number, $X_0$ and $\rho_{\text{sh}}$ are the unit radiation length and density of target material, $A$ is mass number of target, $T_\text{sh}=L_{\text{sh}} \rho_{\text{sh}} /X_0$   , and $l_X=\beta \gamma \tau$ is the $X$ decay length in the lab frame. The exponential terms ensure that $X$ must decay after the shielding  but before or within the detector where $L_{\text{dec}}$ is the length of the decay region. This term is obtained by integrating the differential decay probability $dP(l)/dl= e^{-l/l_X}/l_X$ from $L_{\text{sh}}$ and $L_{\text{tot}}=L_{\text{sh}}+L_{\text{dec}}$. A detailed explanation of computing this integral is provided in the appendix of \cite{Andreas:2013xxa}. Constraints are placed by excluding regions of the $\{m_X, g_X \}$ parameter space that result in excess events over the $95\%$ C.L. upper limit of the number of observed events in each experiment. Using this framework, we derive constraints from Orsay \cite{Davier:1989wz}, KEK \cite{Konaka:1986cb}, NA64 \cite{NA64:2018lsq, NA64:2019auh}, $\text{E}137$ \cite{Bjorken:1988as}, $\text{E}141$ \cite{Riordan:1987aw}, and $\text{E}774$ \cite{Bross:1989mp}.

In proton beam dumps, $X$ can be produced either via bremsstrahlung or through the decays of mesons ($\pi^{0},\,\eta,\,\eta'$) produced in the target. The bremsstrahlung process is analogous to the electron beam dump case, with the effective cross section for proton collision determined experimentally \cite{Cudell:2001pn}. Constraints on a dark photon have been derived in \cite{Blumlein:2011mv, Gninenko:2011uv, Gninenko:2012eq} on U70 \cite{Blumlein:2013cua}, NOMAD \cite{NOMAD:2001eyx},  PS191 \cite{Bernardi:1985ny}, CHARM \cite{Tsai:2019buq}, LSND \cite{Bauer:2018onh}, $\nu\text{Cal}$ \cite{Tsai:2019buq}, and FASER \cite{FASER:2023tle, FASER:2018eoc}. In our analysis, we recast these dark photon constraints to our model using 
Eq.~(\ref{recastingEquation}) and the methodology developed in \cite{Ilten:2018crw}. The same procedure is applied to fixed target experiments: APEX \cite{APEX:2011dww}, A1/MAMI \cite{A1:2011yso, Merkel:2014avp}, HPS \cite{Battaglieri:2014hga, Adrian:2022nkt}, and NA48/2 \cite{NA482:2015wmo}.  These experiments probe $m_X < 1\;\gev$ and rely on $X \to e^{+}e^{-}$ decay as their signal channel. In this mass range, the RGE induced electron charge in the chargephobic limit is at most $x_e \sim\mathcal{O}(10^{-3})$, so both the bremsstrahlung production channel and $\ee$ decay are suppressed by $x_e^{2}$. As a result, the expected event yields in the chargephobic case are negligible and so the constraints vanish.

\phantomsection
\label{sec:cnst:SHiP}

For the projected SHiP sensitivity, we recast the dark photon reach from Ref.~\cite{SHiP:2020vbd} using the LHCb production fractions \cite{Ilten:2018crw}. In the chargephobic case, the charged lepton decays are negligible and the SHiP sensitivity is effectively determined by the hadronic modes. Therefore, we find that SHiP gains sensitivity starting at $m_X \simeq 2m_\pi$, where the $X \to \pi^{+}\pi^{-}$ channel becomes kinematically accessible. For larger masses, multi-pion
and kaon final states also contribute to detection signal, 
see Fig.~\ref{VMDBranchingFractionFig}.
Above the pion threshold, a dominant fraction of the production and detection channels remains unsuppressed, so SHiP retains substantial reach in the chargephobic limit. Thus, SHiP is a uniquely powerful probe of the chargephobic case, complementing the $\cevns$ constraints and providing outstanding opportunities for future discovery.

As a final remark, recent studies of neutron initiated production at FASER/FASER2 show that a ``neutron beam dump'' can enhance
sensitivity to the protophobic model \cite{Dev:2023zts, Kling:2025udr}. These results still rely on $X \to \ell^{+}\ell^{-}$ decays for detection and as such do not set constraints on the chargephobic model.

\subsection{Collider Experiments}

In this section, we summarize the constraints obtained from collider searches at LHCb \cite{LHCb:2017trq}, CMS \cite{CMS:2019buh}, KLOE \cite{KLOE-2:2011hhj, KLOE-2:2012lii, Anastasi:2015qla, KLOE-2:2018kqf}, BESIII \cite{BESIII:2017fwv},  and BaBar \cite{BaBar:2014zli}. We recast these bounds to our general $\vp$ model using Eq.~(\ref{recastingEquation}), following the production, decay, and experimental efficiency rescaling procedures outlined in \cite{Ilten:2018crw}.
Note that there are projections Refs. \cite{Graham:2021ggy, Craik:2022riw}
(and other possible constraints \cite{Seto:2025mte})
for the improvements in the sensitivity to the dark photon by the LHCb experiment. However, we do not anticipate the large improvements in the dark photon parameter space $m_{A'} \lesssim 500$~MeV to extend to the chargephobic vector boson case given the very small branching fraction to charged leptons.

\phantomsection
\label{sec:cnst:LHC}

The LHCb dark photon search \cite{LHCb:2017trq} considers both prompt and displaced $X\longrightarrow \mumu$ decays in $pp$ collisions at $13$~ TeV using $1.6\;\text{fb}^{-1}$ of data. Because the analysis relies on the dimuon final state, the constraint weakens considerably on the chargephobic case. The constraint does not vanish, however, due to the loop induced charged lepton coupling $x_e \sim\mathcal{O}(10^{-2})$ at larger masses $m_X \gtrsim 1$~GeV that is generated by the RGE running as described in App.~\ref{RGEAppendix}. The relevant production modes include meson decays ($\eta \rightarrow X \gamma$,  $\omega \rightarrow X \pi^{0}$), vector meson mixing ($\omega \rightarrow X$,  $\rho \rightarrow X $, $\phi \rightarrow X$), and Drell–Yan ($q\bar{q}\rightarrow X$). The dominant production modes vary across the $m_X$ mass range. Consequently, each $\vp$ value can suppress/remove some but not all production mechanisms. For $m_X\lesssim 0.4\; \gev$ and $m_X\gtrsim 1\; \gev$, the leading channels are $\eta \rightarrow X \gamma$ ($\sim 65 \%$) and $u \bar{u} \rightarrow X$ ($\sim 70 \%$) which are suppressed/vanish at $\vp = \pi - \tanInv 2$. In the intermediate region $0.4-0.9\; \gev$, the leading channels are $\omega \rightarrow X\pi^0$ ($\sim 45 \%$) and $\rho \rightarrow X$ mixing ($\sim 65 \%$) both of which vanish for the $B-L$ case. A similar analysis is applied to the CMS constraint \cite{CMS:2019buh}. 


\phantomsection
\label{sec:cnst:KLOE}

The KLOE experiment conducted several searches for dark photon using data taken at $\text{DA}\Phi \text{NE}$ with $\sqrt{s}\simeq 1020 \,\text{MeV}$ \cite{KLOE-2:2011hhj, KLOE-2:2012lii, Anastasi:2015qla, KLOE-2:2018kqf}. These analyses considered the process $\ee \longrightarrow \gamma X$ with $X\longrightarrow \ee$,  $X\longrightarrow \mumu$, and $X\longrightarrow \pi^{+}\pi^{-}$, as well as the decay chain $\phi \longrightarrow X \eta$ with $X\longrightarrow \ee$. In both cases, 
constraints on the chargephobic 
parameter space are highly suppressed 
due to the $x_e^2$ suppression on the production and/or the decay.

%

\phantomsection
\label{sec:cnst:BaBar}

Finally, the BaBar dark photon search \cite{BaBar:2014zli} considered the process $\ee \longrightarrow \gamma X$ with $X\longrightarrow \ee$ and $X\longrightarrow  \mumu$ primarily at the $\Upsilon(4S)$ but also at $\Upsilon(3S)\; \&\; \Upsilon(2S)$ resonances. 
The reported constraint spans $0.02\;\text{GeV}<m_X<10.2\;\text{GeV}$ and has a sensitivity scaling as $x_e^4$ since both the production and decay proceed through electron coupling. Consequently, the bound vanishes for the chargephobic case even after including small RGE induced couplings. A similar suppression applies to BESIII \cite{BESIII:2017fwv}.

\subsection{Precision Constraints}

Precision measurements of SM observables provide powerful probes of
new light vectors, since even weak couplings can shift well measured
quantities away from their SM predictions. In this section, we discuss the constraints set from the electron/muon anomalous magnetic moments and electroweak precision tests. We also comment on the sensitivity of the proposed REDTOP experiment \cite{REDTOP:2022slw}. 

\phantomsection
\label{sec:cnst:gm2}

We first consider the one-loop contribution of $X$ to the anomalous magnetic moments $(g-2)_{\ell}$ of the electron and muon, $a_\ell^X$, computed following Ref. \cite{Pospelov:2008zw}. The $(g-2)_{e}$ constraint is set such that the effective shift of the fine structure constant, $\Delta \alpha= 2\pi a^{X}_e$, does not exceed $1\,\text{ppb}$ in line with the current $\mathcal{O}( 1 \,\text{ppb})$ spread among the most precise determinations of $\alpha$ \cite{Parker:2018vye, Morel:2020dww,  Fan:2022eto}; previous constraints were set at $15\,\text{ppb}$ \cite{Pospelov:2008zw} based on older $\alpha$ determinations including Ref. \cite{Clade:2006zz}. For the $(g-2)_{\mu}$, we use the updated SM prediction and final world average measurement of $a_\mu$ \cite{Aliberti:2025beg} 
and require  $a^X_\mu < (a_{\mu}^{\rm exp}-a_{\mu}^{\rm SM}+2\sigma)$ where $\sigma$ is the quoted uncertainty on this difference. These constraints are relevant in the dark photon case where they cover the region of parameter space where collider constraints are not sensitive due to masking out meson resonances. For other $\vp$, these constraints become subdominant to NES and $\cevns$ constraints. In particular, in chargephobic case, the absence of couplings to charged leptons  
when $m_X \lesssim 200$~MeV combined with the very small renormalization group induced couplings above this scale imply that both $(g-2)_{e,\mu}$ are negligible throughout the parameter space.  


\phantomsection
\label{sec:cnst:EWP}

At low energies, $X$ couples to the electromagnetic current, while at
electroweak scales, this becomes a coupling to the hypercharge current.
This causes a 
mass mixing between $X$ and the $Z$ boson and also a suppressed coupling to the $Z$ current, see App. \ref{generatingDarkMixingAppendix}. This mixing shifts the physical $Z$ mass and its couplings to fermions, leading to constraints from
electroweak precision observables. For a dark photon, constraints were derived in Refs.~\cite{Hook:2010tw, Curtin:2014cca, Loizos:2023xbj} by performing global fits to LEP, SLC, Tevatron, and LHC electroweak precision data. In our analysis, we adopt the $95\%\,\text{C.L.}$ exclusion 
curve based on the PDG world average value of $m_W$ (solid blue curve in Fig.~1 of Ref.~\cite{Loizos:2023xbj}) in the dark photon limit and recast this constraint for arbitrary $\vp$ using the relation $\epsilon e = g_X \cos\varphi$. We show this constraint, labeled EWP, in Figs.~\ref{DPandBmLCurrentConstraintPlots}-\ref{CPConstraintPlots}. 


\phantomsection
\label{sec:cnst:REDTOP}

Finally, REDTOP is a proposed high-intensity $\eta/\eta'$ factory that aims to collect a data sample of order $10^{14}$ $\eta$ and $10^{12}$ $\eta'$ mesons to study very rare decays and to search for BSM physics in the MeV-GeV range \cite{REDTOP:2022slw, Zielinski:2025wfa}. In our case, $X$ can be produced at REDTOP via $\eta,\eta'\to\gamma X$ and the detector is sensitive to the following subsequent decays $X \to e^{+}e^{-}$, $X \to \mu^{+}\mu^{-}$, and $X \to \pi^{+}\pi^{-}$. In particular, the hadronic channel $X \to \pi^{+}\pi^{-}$ remains open in the chargephobic limit. In Fig.~\ref{CPConstraintPlots}, we show the recasted projected REDTOP sensitivity with $10^{18}$ protons on target (solid black in Fig.~39 of Ref.~\cite{Gan:2020aco}).


\section{Discussion}

There is a huge effort exploring the space of possibilities for light vector boson mediators.  In this paper, we have demonstrated that when a vector boson $X^\mu$ couples to a specific ``chargephobic'' linear combination of the electromagnetic and $B-L$ currents, there are dramatic changes to the constraints within the $\{m_X, g_X\}$ parameter space.  The changes arise because a chargephobic vector boson does not have (tree-level) couplings to electrons and protons.  
The resulting sensitivity of most beam dump and collider based experiments is  negligible or highly suppressed because they rely on production from proton or electron beams and/or they rely on the decays to leptons that produce an easily visible signal.
Two classes of constraints on $X^\mu$ remain largely unaffected when the couplings to charged leptons and singly-charged baryons are absent:  supernova cooling constraints and the constraints on the contributions to $\Delta N_{\rm eff}$ in the early universe, due to the nonzero couplings to neutrinos and neutrons.

It is interesting to compare the constraints on the chargephobic case to those on other combinations of currents that are also more weakly constrained.  Two specific examples are:  $B - 3 L_\tau$ and baryon number by itself.
A vector boson coupled to
$B - 3 L_\tau$ also has highly suppressed constraints 
(e.g. \cite{Heeck:2018nzc,Bauer:2020itv})
from beam dump experiments and collider experiments due to the absence of interactions with the first two generations of charged leptons.
Moreover, because $B - 3 L_\tau$ 
does not couple to electron or
muon neutrinos, it is also 
unconstrained by 
COHERENT experiment results on 
neutrino-nucleus scattering.
However, as emphasized in
Refs.~\cite{Heeck:2018nzc,Coloma:2020gfv}
$B - 3 L_\tau$ is strongly constrained by the neutrino flavor-dependent non-standard interactions (NSI) 
that affect neutrino oscillations.
These turn out to be stronger than the constraints from COHERENT neutrino-nucleus scattering, and so for much of the parameter space we find that the chargephobic limit is \emph{more weakly constrained} than $B - 3 L_\tau$.

A vector boson coupled to just baryon number is problematic, because baryon number is a non-conserved (anomalous) current in the SM\@.  Refs.~\cite{Dror:2017ehi,Dror:2017nsg} examined this case in detail, 
finding that rare meson decay
mediated by the anomalous interaction
set several constraints on the parameter space that lead to constraints that are more restrictive than those on the chargephobic case for $m_X \lesssim 200$~MeV and $600 \; {\rm MeV}  \lesssim m_X \lesssim 4 \; {\rm GeV}$\@.  In other ranges the two cases are similar, and for larger $m_X$, chargephobic is more constrained due to LHCb searches that are (re-)activated due to the small renormalization group-generated couplings to the charged leptons (see App.~\ref{RGEAppendix}) as well as from electroweak precision constraints.

Several open questions remain:  one is the prospects for gaining \emph{improved} sensitivity to the chargephobic case.  In Fig.~\ref{CPConstraintPlots} we present a rough guide to the increased sensitivities for a handful of experiments, but there are no doubt more possibilities that are worth exploring.  One of our primary goals in the paper is to point out that relying on a charged lepton decay of a light gauge boson can be disastrous for an experiment's sensitivity.  SHiP \cite{SHiP:2015vad,SHiP:2020sos,SHiP:2020vbd} will be able to probe new regions of the chargephobic parameter space due to the significantly improved detector capabilities
that enable it to be sensitive to the hadronic decays of $X$ when $m_X \gtrsim 300$~MeV\@.  
Also, the proposed REDTOP experiment 
\cite{REDTOP:2022slw}
can gain 
sensitivity to chargephobic couplings due to rare decays of $\eta,\eta'$ mesons to vector bosons. 

There are also detailed questions one can ask regarding whether there are additional production or decay modes that experiments could be sensitive to.  The production of vector bosons in beam dump experiments occurs through bremsstrahlung ``coherently'' off the proton beam.  However, one might expect there is also a rate for incoherent contribution (emission from the quarks)\footnote{We thank Ryan Plestid for pointing this out to us.}, that is either form-factor suppressed or suppressed by the degree of inelastic scattering.
We also have seen that there are improved constraints on both the chargephobic case and $B-L$ by considering the absence of high energy neutrino events from SN1987A,
as was found for the case of $L_\mu - L_\tau$ \cite{Blinov:2025aha}.

It is also worth commenting on the very interesting ensemble of bounds on a dark photon arising from its production in supernovae and decay into $e^+e^-$, leading to significant observable signals \cite{Caputo:2025avc}.  
Most of these constraints do not apply to the chargephobic vector boson because of the dominance of the decay $X \rightarrow \nu\bar{\nu}$ throughout the mass range that these constraints apply. 
One needs to be a bit more careful with the low energy supernova constraint (LESNe), since for the dark photon case this extends to $m_{A'} \sim \; 500$~MeV\@.  
A chargephobic vector boson with mass $m_X > 2 m_\pi$
has an increasing
branching fraction to $\pi^+\pi^-$ (few to $20\%$),
that may or may not affect the low energy supernova dynamics.  It would be interesting to determine if a small region of chargephobic parameter space could be constrained, though it requires the convolution of $X$ decay and $\pi$ decay to lead to enough energy deposition to disrupt the supernova explosions.  This is an interesting question, but beyond the scope of our work.

 Finally, while we have focused much of the paper on the specific case of chargephobic, $\vp = 3\pi/4$,
our analysis pipeline allows us to consider any angle.  As we have pointed out in the paper, some angles have suppressed rates for specific processes where mixing with a particular QCD vector resonance does not occur.  We have also pointed out the ``nucleophobic'' 
(a related idea to ``xenophobic'' in Ref.~\cite{Feng:2013vod})
case where 
$\vp =\pi-\text{tan}^{-1}(Z_{\mathcal{N}}/A_{\mathcal{N}})$,
and the coherent interaction of the vector boson with the nucleus vanishes for a specific isotope.  Our interest in this particular case was to explore how much weaker the constraints from COHERENT and other neutrino-nucleus scattering experiments can be.  Of course when we consider mixing angle away from the chargephobic limit, constraints from neutrino-electron scattering reappear, as well the ensemble of beam dump and collider experiments, as we show in Fig.~\ref{neutrinoCnstrntPlot} (right).
Nevertheless, there could be a host of interesting phenomenology when a  nucleophobic gauge boson is fused with a model of dark matter -- there could be an interesting parameter space where dark matter - electron scattering experiments could have sensitivity in regions of parameter space what would otherwise have been considered ruled out by nuclear recoil (or Migdal) scattering.

\section*{Acknowledgments}

We thank Hans-Thomas Janka and Daniel Kresse for providing access to the Garching supernova model data used in this analysis. 
We thank 
Andrea Caputo,
Ryan Plestid, 
and Yotam Soreq 
for helpful discussions
and are grateful to 
Andrea Caputo,
Pierre Fayet,
Damiano Fiorillo
and Edoardo Vitagliano
for insightful comments on an earlier version of the paper. 
This work is supported in part by the U.S. Department of Energy under grant number DE-SC0011640.

\appendix

\section{Generating the Dark Mixing Angle}\label{generatingDarkMixingAppendix}

Throughout the paper, we utilized the low energy effective theory of $X^\mu$ coupled to a linear combination of 
the electromagnetic current and $B-L$ current, Eq.~\ref{interactionLagrangian}, that can be written as
$g_X \left[
  (\cos \vp) j^{\text{EM}}_\mu 
  + (\sin \vp) j^{B-L}_{\mu} 
\right] X^\mu$. 
We will not address the strength of the coupling, $g_X$.  In the case of dark photon, generating a small kinetic mixing has been explored \cite{Gherghetta:2019coi}; for $B-L$, a small coupling could arise from a gauged (and then spontaneously broken) $U(1)_{B-L}$ group that simply has a small gauge coupling. The question we wish to address is how a dark mixing angle $\vp$ could generated by UV physics.

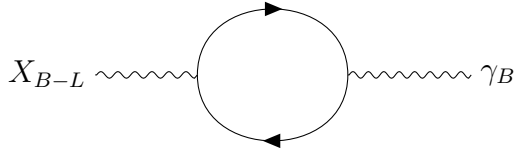
\begin{figure}
\begin{center}
\begin{tikzpicture}
  \begin{feynman}
    \vertex (XL) at (-3,0) {\large $X_{B-L}$};
    \vertex (G)  at ( 3,0) {\large $\gamma_B$};

    \vertex (L)  at (-1,0);
    \vertex (R)  at ( 1,0);

    \diagram*{
      (XL) -- [photon] (L),
      (R)  -- [photon] (G),

      (L) -- [fermion, half left] (R),

      (R) -- [fermion, half left] (L),
    };
  \end{feynman}
\end{tikzpicture}
\end{center}
    \caption{Contributions to the mixing of $B-L$ with hypercharge.}
    \label{fig:selfenergymixing}
\end{figure}

Here we consider extending the SM to include a gauged $U(1)_{B-L}$ at a high scale that is, at lower energies, spontaneously broken to generate a mass for the vector boson that couples to $B-L$. It is well known (e.g., \cite{Salvioni:2009jp,
Heeck:2014zfa}) that one-loop corrections arising from matter that transforms under both $U(1)_{B-L}$ and $U(1)_Y$ will lead to these gauge bosons mixing with each other, see Fig.~\ref{fig:selfenergymixing}.
In the case that is of most interest to us, $g_{B-L} \ll g_Y$, the hypercharge gauge boson remains virtually unchanged due to this mixing.  Instead, the gauge boson that (initially) couples to $B-L$, however, will have a sizable one-loop induced coupling to the hypercharge current. The contributions to the loop corrections include ``threshold'' or ``matching'' finite corrections from heavy fields that transform under both $U(1)_{B-L}$ and $U(1)_Y$, as well as renormalization group generated mixing from light fields (including SM fields).

For example, assuming that $X^\mu$ couples to just $B-L$ at a high scale $\mu_{\rm high}$, the leading renormalization group induced contribution to hypercharge (and ultimately electric charge) is
\begin{eqnarray}
    \epsilon e &=& 
    \epsilon g_Y \cos \theta_W \; \simeq \; - \frac{b_{\rm SM} \cos\theta_W}{16\pi^2} g_Y^2 g_{B-L} \ln \frac{\mu_{\rm high}}{\mu_{\rm low}} \, .
\label{eq:RGhighlow}
\end{eqnarray}
The coefficient $b_{\rm SM} = 32/3$ $(26/3)$ includes (does not include) three right-handed neutrinos. A sizable coupling of $X^\mu$ to the electromagnetic current is possible even with just the SM field content and a large amount of renormalization group evolution.  Moreover, it is interesting that the SM field content leads to $\vp$ evolving \emph{towards} the chargephobic limit ($\vp > \pi/2$).  We caution, however, that Eq.~(\ref{eq:RGhighlow}) is just the leading result; once $\epsilon g_Y$ is comparable to $g_{B-L}$ there are significant corrections from additional contributions in the RGE\@. To achieve an arbitrary angle $0 < \varphi < \pi$, additional matter beyond the SM is needed either as threshold corrections or as light(er) fields contributing to the renormalization group evolution.

The mixing with hypercharge generates the following terms in the lagrangian,
\begin{equation}
\mathcal{L}\supset \left(
\ve e j^\text{EM}_\mu
+g_{B-L}  j^{B-L}_\mu 
-\frac{\ve e}{\cos^2 \theta_w}\frac{\delta}{1-\delta} j^Z_\mu
\right)X^\mu,
\label{hyperchargeMixLagrangian}
\end{equation}
where $\delta= m_X^2/m_Z^2$; this expression follows from Appendix A of Ref.~\cite{Bauer:2018onh}, retaining terms up to order $\ve \delta$. The first two terms, the electromagnetic and $B-L$ currents, are the couplings central to this work, as discussed in Sec. \ref{sec:vectorbosonsEMBminusL}. The third term couples $X$ to the SM $Z$ current. For a particle whose electromagnetic or $B-L$ charge is of order one, the $Z$ current contribution is an $\mathcal{O}(\ve\delta)$ correction and is negligible throughout our parameter space, where $\delta< 1$. 
In the chargephobic case, the coupling to charged leptons vanishes at tree level; renormalization group evolution generates a small residual coupling (Appendix \ref{RGEAppendix}), which is overtaken by the $Z$ current contribution when $m_X$ approaches $m_Z$, as shown in Fig. \ref{chargeSumFigures}.

\section{Renormalization Group Equations}\label{RGEAppendix}

In the chargephobic limit, $X^\mu$ couples the linear combination $- j^{\rm EM}_\mu + j^{B-L}_\mu$, corresponding to the mixing angle $\varphi = 3\pi/4$, or equivalently $\epsilon e = - g_{B-L}$.
The mixing angle is not stable under renormalization group evolution. We can determine the running of $\varphi$ by first determining the RGEs of the individual couplings. RGEs have been considered before in similar contexts \cite{delAguila:1988jz, delAguila:1995rb,Babu:1996vt, Dienes:1996zr}. In our case, the RGEs are given by 
\begin{widetext}
\begin{eqnarray} 
\label{coupRGE1}
\mu\frac{de}{d\mu} &=& \frac{1}{12\pi^2} e^{3} \sum_i N_c^iQ_i^2\\ 
\label{coupRGE2}
\mu\frac{d \gbl}{d\mu} &=& \frac{1}{12\pi^2} \gbl \sum_i N_c^i\left[\gbl^{2}q_i^2 + (\ve e)^2  Q_i^2 + 2 \gbl (\ve e)Q_iq_i \right]\\ 
\label{coupRGE3}
\mu\frac{d\ve e}{d\mu} &=& \frac{1}{12\pi^2} \sum_i\left[
\gbl^{2} (\ve e) q_i^2 + (\ve e)^3 Q_i^2 + 2 e^2  (\ve e) Q_i^2 + 2 e^2 \gbl Q_iq_i + 2 \gbl (\ve e)^2 Q_iq_i
\right],
\end{eqnarray}
\end{widetext}
where $\mu$ is the renormalization scale and the charges of the matter fermions contributing to the RGEs
are given by $Q_i$ (EM) and $q_i$ ($B-L$) in Sec.~\ref{sec:vectorbosonsEMBminusL}.
Below $\sqrt{s} \sim 2\, \gev$, QCD is non-perturbative and it is mesons rather than quarks that determine thresholds in RGEs. For a detailed treatment see for example, \cite{Eidelman:1995ny, Jegerlehner:2006ju}. The precision obtained from this full treatment is not necessary to derive our constraints. Instead, we employ the following simplification: In the non-perturbative regime, $\sqrt{s} < 2\, \gev$, quarks are included perturbatively into the RGE at a scale corresponding to lightest meson that ``contains'' the quarks.
Namely, the up and down quark charges are added at the pion mass, the strange quark at the mass of the lightest Kaon, and the charm at the mass of the lightest D meson. 

\begin{figure*}
\centering
\begin{subfigure}[c]{0.72\textwidth}
  \centering
  \includegraphics[width=\linewidth]{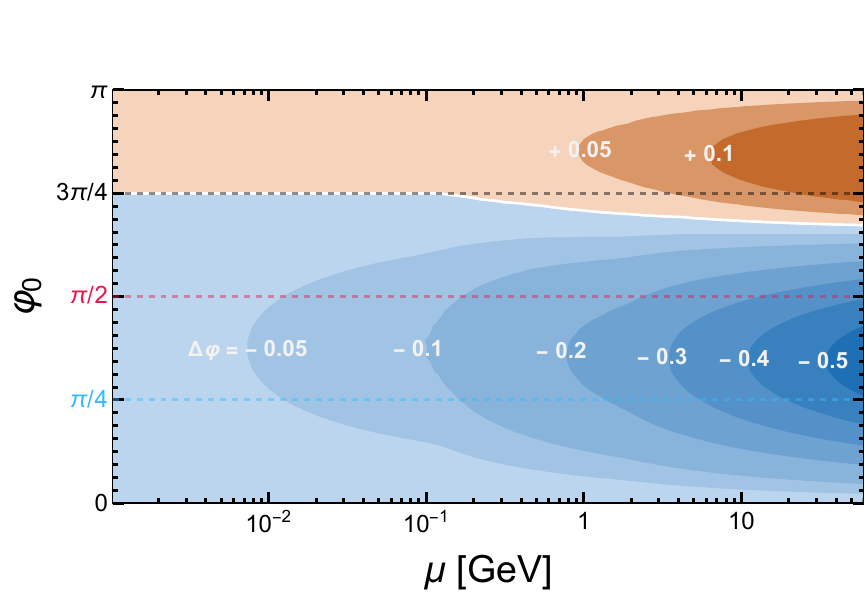}
\end{subfigure}
\hfill
\begin{subfigure}[c]{0.26\textwidth}
  \centering
  \includegraphics[height=\linewidth,angle=-90,origin=c]{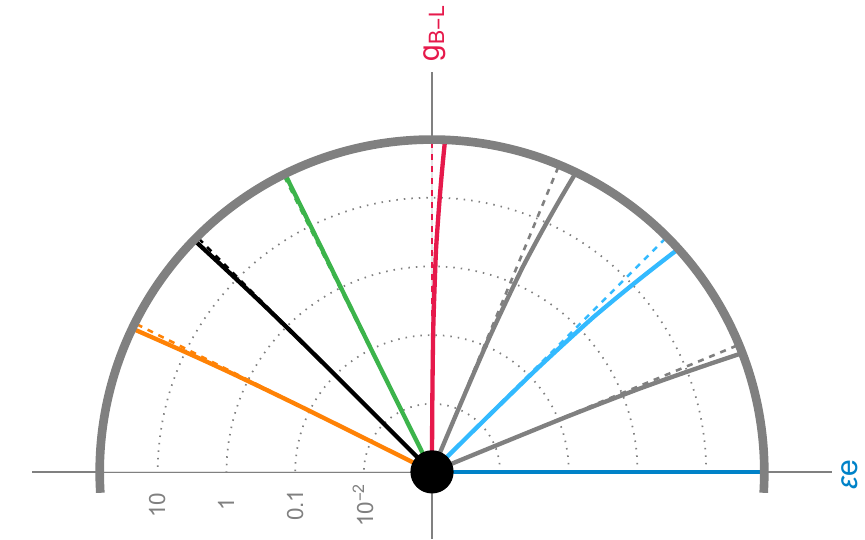}
\end{subfigure}
\caption{The running of the dark mixing angle with initial value $\vp_0$ at $\mu= m_e$. The plot on the left contains blue (orange) contours to indicate the negative (positive) deviation from initial dark mixing angle, $\Delta \vp = \vp(\mu) - \vp_0$. The white line represents $\Delta \vp = 0$ and we note that the chargephobic limit coincides with this line until the pion mass threshold. The plot on the right illustrates the shift using a compass plot as in Fig. \ref{angleParamSpaceFig} for selected dark mixing angles. Here, the radial direction also indicates the renormalization scale $\mu$. Explicitly, the center of the compass corresponds to $\mu= m_e$, the outer thick gray arc to $\mu= 60\,\gev$, and the dotted arcs
to intermediate values of $\mu$.}
\label{fig:angleevolution}
\end{figure*}

We can solve the RGEs, Eqs.~(\ref{coupRGE1})-(\ref{coupRGE3}), to obtain the running of the dark mixing angle 
as a function of the renormalization scale.  This is illustrated in Fig.~\ref{fig:angleevolution}, where we show the extent of $\varphi(\mu)$ evolution following any given initial value $\varphi_0$($\mu = m_e$) up to $\varphi(\mu = 60 \; {\rm GeV})$.  Overall, the shifts are relatively modest, which is evident in how the needles on the compass plot are slightly shifted shown in Fig.~\ref{fig:angleevolution}.  Notice that the angle $\varphi(\mu)$ can shift towards positive or negative values depending on the initial value.

\begin{figure*}
\centering
\includegraphics[width=0.9\textwidth]{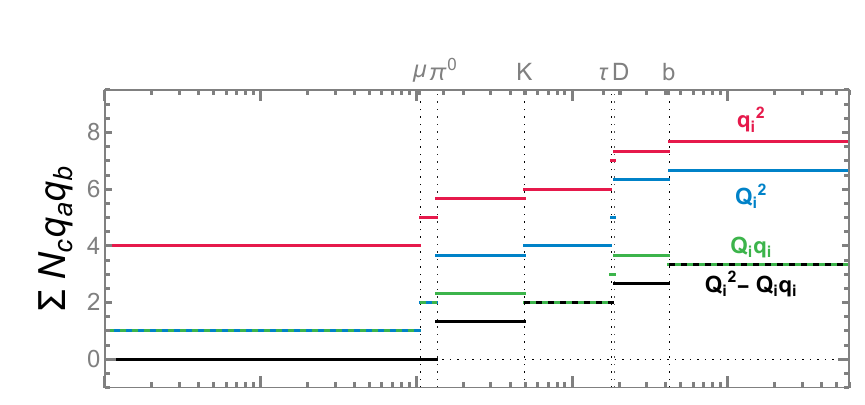}
\vfill
\includegraphics[width=0.9\textwidth]{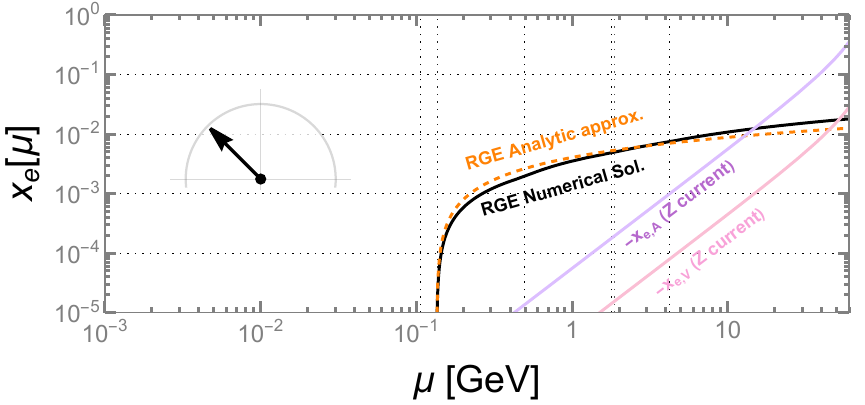}
 \caption{(Top) the squared charge sum of particles contributing to the RGE at threshold and (bottom)  $x_e$ running using the full numerical solution of Eq. \ref{coupRGE1}-\ref{coupRGE3} shown in black for $g_X = 10^{-2}$ or using the simple analytic approximation $(g_X \ll e)$ as discussed in the text (orange dashed). We also show the vector (pink) and axial (purple) contributions to the electron coupling from the $Z$ current induced by hypercharge mixing, see 
 Eq.~(\ref{hyperchargeMixLagrangian}).} 
 \label{chargeSumFigures}		
\end{figure*}

In the chargephobic model, we
numerically solve Eqs.~(\ref{coupRGE1})--(\ref{coupRGE3}) to obtain the running of $x_e$ as shown in Fig.~\ref{chargeSumFigures},
which also shows the 
sum of charges at each mass threshold for various configurations of charge assignments.
However, one can understand the results more easily by working in the limit 
$g_X = \sqrt{(\ve e)^2+g_{B-L}^2} \ll e$, 
and then the RGEs simplify to
\begin{eqnarray}
\mu\frac{de}{d\mu} &=& \frac{1}{12\pi^2} e^{3} \sum_i N_c^iQ_i^2\qquad(g_X \ll e \; \text{limit})\nonumber\\ 
\mu\frac{d\ve e}{d\mu} &=&\frac{1}{6\pi^2} 
 e^2\sum_i N_c^i\left[ (\ve e) Q_i^2+  \gbl Q_iq_i\right].
 \label{simplifiedRGEeqn}
\end{eqnarray}
Here, we have taken $\gbl$ to be constant since the leading contribution to its running is $\mathcal{O}(g_X^3)$ as compared with $\ve e$ that receives its leading contribution at order $\mathcal{O}(g_X e^2)$. Interestingly, leptons do not contribute to the running of $\ve e$.  This is clear because the lepton  charges satisfy $Q_l^2 = Q_l q_l$ (and neutrinos do not contribute $q_\nu=0$), and so at
the chargephobic mixing angle, $\ve e = -\gbl$, the contribution of leptons to the term in square  brackets in Eq.~(\ref{simplifiedRGEeqn}) vanishes.

The main driver of the $\ve e$ running is the $e$ running itself. A simple but useful analytic approximation for $\ve e [\mu]$ can be obtained by running $e$ starting at the pion threshold then suppressing the solution by $\ve$. The result is
\begin{align} 
\label{xeRGEsimpleSolElectro}
x_e [\mu] &=-\frac{1}{\sqrt{2}} \left(1-\frac{1}{\sqrt{1-\frac{e^2}{6\pi^2} \left(\frac{11}{3}\right) \log \left( \frac{\mu}{m_\pi}\right)}}\right)\nonumber\\
&\approx \frac{\sqrt{2}}{24\pi^2}\frac{11}{3} e^2 \log \left( \frac{\mu}{m_\pi}\right)
\quad\quad\quad
(\text{analytic approx.})
\end{align}
where the factor of $11/3$ is the squared charge sum of the electron, muon, up and down quarks all of which contribute to the running of $e$ above $\mu = m_\pi$. The running of $\ve e$ can be reformulated into the running of $\vp$. The deviation from $\vp_0$ is  
\begin{equation} 
\label{RGEreformulation}
\delta \vp \approx \frac{e^2}{24\pi^2} \left(\frac{11}{3}\right) \log \left( \frac{\mu}{m_\pi}\right).
\end{equation}
Notice that mixing angle receives a positive contribution at larger renormalization scales when starting from the chargephobic limit.  This means $X^\mu$ slightly shifts from equal couplings $g_X = |\ve e|/\sqrt{2} = g_{B-L}/\sqrt{2}$, to one where $X^\mu$ couples \emph{slightly} more to the to the electromagnetic current relative to $B-L$ ($\varphi$ evolves towards values slightly larger than  $3\pi/4$.)

Now let's briefly describe the effect of the renormalization group evolution of $\vp$ on various observables. 
For the purposes of this discussion, we take $\varphi = 3\pi/4$ at low energies, and then consider the effects of the shifts in the couplings of $X^\mu$ for observables that are sensitive to energies $\sqrt{s} \gtrsim m_\pi$ where deviations from chargephobic begin to occur, see Fig.~\ref{chargeSumFigures}. The dominant effect is the appearance of a small coupling of $X^\mu$ to charged leptons, since this is a critically important mode for $X^\mu$ to be observed.  The tail end of current beam dump experiment sensitivity extends into the region $m_\pi < m_X \lesssim 800$~MeV, however we find that the branching fraction $Br(X \rightarrow \ell^+\ell^-)$ where $x_e[m_X]$ remains small enough that these
experiments remain insensitive in this region.  Most of the supernova constraint region is also below the pion mass threshold, and so is not expected to change.  The one region that requires a more careful analysis is the low energy supernova constraints (the region labeled LESNe in Fig.~\ref{DPandBmLCurrentConstraintPlots} \cite{Caputo:2025avc}). However, it is likely that the hadronic decay of $X \rightarrow \pi^+\pi^-$ is the most important mode at larger chargephobic masses $m_X \gtrsim 300$~MeV, and this will require a detailed analysis to determine the sensitivity to a chargephobic vector boson.

The most prominent re-appearance of  experiments are LHCb constraints (shown in green in Fig.~\ref{CPConstraintPlots}) and the neutrino-electron scattering experiments TEXONO and CHARM (shown in pink and orange in Fig.~\ref{CPConstraintPlots}). However, these constraints are subdominant to $\cevns$ and EWP constraints respectively.

\section{Bremsstrahlung Interaction in Proto-Neutron Star}\label{BremAppendix}

The bremsstrahlung channels that contribute to absorption in a general $\varphi$ model are
\begin{widetext}
\begin{equation}
\label{bremDiagrams}
\scalebox{0.8}{
\begin{tikzpicture}[baseline=-5]
\begin{feynman}
\vertex (a) at (-1.5,0.4) {\(P_1\)};;
\vertex (b) at (1.5,0.4) {\(P_3\)};
\vertex (c) at (-1.5,-0.4)  {\(P_2\)};
\vertex (d) at (1.5,-0.4)  {\(P_4\)};
\vertex (x) at (0,0);
\vertex (y) at (-0.5,0.4);
\vertex (z) at (-1,1){\(K\)};
\diagram {
(a) -- (b),
(c) --  (d),
[blob] (x) , 
(y) --[boson] (z),
};
\end{feynman}
\end{tikzpicture}
}
\quad
\scalebox{0.8}{
\begin{tikzpicture}[baseline=-5]
\begin{feynman}
\vertex (a) at (-1.5,0.4) {\(P_1\)};;
\vertex (b) at (1.5,0.4) {\(P_3\)};
\vertex (c) at (-1.5,-0.4)  {\(P_2\)};
\vertex (d) at (1.5,-0.4)  {\(P_4\)};
\vertex (x) at (0,0);
\vertex (y) at (0.8,0.4);
\vertex (z) at (0.4,1){\(K\)};
\diagram {
(a) -- (b),
(c) --  (d),
[blob] (x) , 
(y) --[boson] (z),
};
\end{feynman}
\end{tikzpicture}
}
\quad
\scalebox{0.8}{
\begin{tikzpicture}[baseline=-5]
\begin{feynman}
\vertex (a) at (-1.5,0.4) {\(P_1\)};;
\vertex (b) at (1.5,0.4) {\(P_3\)};
\vertex (c) at (-1.5,-0.4)  {\(P_2\)};
\vertex (d) at (1.5,-0.4)  {\(P_4\)};
\vertex (x) at (0,0);
\vertex (y) at (-0.5,-0.4);
\vertex (z) at (-1,-1){\(K\)};
\diagram {
(a) -- (b),
(c) --  (d),
[blob] (x) , 
(y) --[boson] (z),
};
\end{feynman}
\end{tikzpicture}
} 
\quad
\scalebox{0.8}{
\begin{tikzpicture}[baseline=-5]
\begin{feynman}
\vertex (a) at (-1.5,0.4) {\(P_1\)};;
\vertex (b) at (1.5,0.4) {\(P_3\)};
\vertex (c) at (-1.5,-0.4)  {\(P_2\)};
\vertex (d) at (1.5,-0.4)  {\(P_4\)};
\vertex (x) at (0,0);
\vertex (y) at (0.8,-0.4);
\vertex (z) at (0.2,-1){\(K\)};
\diagram {
(a) -- (b),
(c) --  (d),
[blob] (x) , 
(y) --[boson] (z),
};
\end{feynman}
\end{tikzpicture}
},
\end{equation}
where $P_i = \{E_i, \vec{\beta}_i\}, \;\text{and} \;K=\{\omega, \vec{v}\}$ are the momenta of nucleons and $X$ and the shaded circle represents the nucleon-nucleon interaction which takes into account long-distance component arising from pion-exchanges and short-distance components that contribute to nucleon-nucleon scattering \cite{Rrapaj:2015wgs}. We will designate the charge $q_1,q_2$ to the upper and lower legs of the diagrams respectively. Following the derivation in \cite{Chang:2016ntp} with the necessary adjustments, the absorption width 
\begin{equation}
\label{GammaSRA1}
\Gamma_{\text{SRA}}\simeq \frac{n_i n_j}{256 \pi^5 S_\snLT \omega T^3 m_N^{7/2}} \int d \cos{\theta_{\text{CM}}} d^3\vec{p}_{+}d^3\vec{p}_{-}    \overline{  (\tilde{\varepsilon} \cdot J ) }^2 e^{-\frac{\vert\vec{p}_{+} \vert^2+\vert\vec{p}_{-} \vert^2}{4 m_N T} }\sqrt{T_{\text{CM}}} \frac{d \sigma_{ij}}{d \cos{\theta_{\text{CM}}}}, 
\end{equation}
\end{widetext}
where $S_\snL(S_\snT)=1(2)$ is the spin degeneracy of the polarization states, $n_i$ is the number density of nucleons, $m_N$ is the nucleon mass, $T_{\text{CM}}=(\vec{p}_1-\vec{p}_2)^2/4m_N$ is the center of mass kinetic energy, $\vec{p}_\pm=\vec{p}_1 \pm \vec{p}_2$, $\theta_{\text{CM}}$ is the scattering angle. Furthermore, $\tilde{\varepsilon}_\mu$ is the $X$ polarization, 
\begin{equation}
\label{quadrupoleCurrent}
J^\mu= \left[q_1 \left(\frac{P_1}{P_1 \cdot K}-\frac{P_3}{P_3 \cdot K} \right) +q_2 \left(\frac{P_2}{P_2 \cdot K}-\frac{P_4}{P_4 \cdot K} \right)     \right]^\mu,
\end{equation}
is the quadrupole current, and  $\overline{  (\tilde{\varepsilon} \cdot J ) }^2 \equiv \int d\Omega' \, \Sigma_{\text{pols.}}(\tilde{\varepsilon}_\mu J^\mu)^2$ integrated over the $X$ scattering angle. To leading order, this integral projected onto the transverse and longitudinal polarizations is 
\begin{widetext}
 \begin{equation}
\label{polarizationSumSolution}
\overline{  (\tilde{\varepsilon} \cdot J ) }^2_\snLT=\frac{8\pi}{15}\frac{T_\text{CM}}{m_N \omega^2}
\begin{cases}
\left[10q_{-}^2(1-\cos\theta)+3v^2\frac{T_\text{CM}}{m_N }q_{+}^2(1-\cos^2\theta)
\right] &(\snT)\\
\left[
5q_{-}^2(1-\cos\theta)+2v^2\frac{T_\tCM}{m_N }q_{+}^2(1-\cos^2\theta)
\right](1-v^2) &(\snL)
\end{cases}
\end{equation}
where $q_\pm= q_1\pm q_2$. The width in Eq. \ref{GammaSRA1} for the transverse mode becomes
\begin{equation}
\label{GammaSRA2}
\Gamma_{\text{SRA}|\snT}=\frac{ n_i n_j }{30 \pi^{1/2}\omega^3  m_N^3 T^{3/2} } \int \vert\vec{p}_{-}\vert^2 d\vert\vec{p}_{-}\vert  e^{-\vert\vec{p}_{-}\vert^2/4m_N T} T_\tCM^{3/2}
\left[
10q_{-}^2\sigma_{ij}^{(2)} +3v^2 \frac{T_\tCM}{m_N}q_{+}^2\sigma_{ij}^{(4)}
\right],
\end{equation}
\end{widetext}
where the $p_{+}$ integral was evaluated to $\int \vert\vec{p}_{+}\vert^2   e^{-\vert\vec{p}_{+}\vert^2/4m_N T} d\vert\vec{p}_{+}\vert=2\sqrt{\pi (m_N T)^3}$ and we defined 
\begin{eqnarray}
\sigma_{ij}^{(2)}(T_\tCM) &\equiv& \int d\cos{\theta_\tCM}(1-\cos{\theta_\tCM}) \frac{d\sigma_{ij}}{d\cos{\theta_\tCM}} \\ \sigma_{ij}^{(4)}(T_\tCM) &\equiv& \int d\cos{\theta_\tCM}(1-\cos^2{\theta_\tCM}) \frac{d\sigma_{ij}}{d\cos{\theta_\tCM}}. 
\end{eqnarray}
Finally, we change to a dimensionless variable $x=T_\tCM/T$ and do the $\vert p_{-}\vert$ integral and go through a similar process for the longitudinal mode to get inverse bremsstrahlung rate
\begin{widetext}
 \begin{equation}
\label{GammaBremIbrAppendix}
\Gamma_{\text{ibr}|\snLT}=\frac{8 n_i n_j }{15 \pi^2 \omega^3 S_\snLT}\left(\frac{\pi T}{ m_N}\right)^{3/2}
\begin{cases}
\left[10q_{-}^2 \langle\sigma_{ij}^{(2)}\rangle +9q_{+}^2 \frac{Tv^2 }{m_N} \langle\sigma_{ij}^{(4)}\rangle
\right] &(\snT)\\
\left[
5q_{-}^2  \langle\sigma_{ij}^{(2)}\rangle  +6q_{+}^2 \frac{Tv^2 }{m_N}\langle\sigma_{ij}^{(4)}\rangle
\right](1-v^2) &(\snL)
\end{cases}
\end{equation}
\end{widetext}
where 
\begin{eqnarray}
\langle\sigma_{ij}^{(2)}(T)\rangle &=& \frac{1}{2}\int^\infty_0 dx e^{-x}x^2\sigma_{ij}^{(2)}(xT) \\ 
\quad \langle\sigma_{ij}^{(4)}(T)\rangle &=& \frac{1}{6}\int^\infty_0 dx e^{-x}x^3\sigma_{ij}^{(4)}(xT) 
\end{eqnarray}
are the energy and angle-averaged nucleon nucleon scattering cross section found in \cite{Rrapaj:2015wgs}.  The calculation for production rate is identical with the restriction that the $X^\mu$ energy must not exceed the energy of the collision. This takes $\int_0^\infty \rightarrow  \int_{\omega/T}^\infty$; for example $\langle\sigma_{ij}^{(2)}(T)\rangle \rightarrow \langle\sigma_{ij}^{(2)}(T,\omega)\rangle =\frac{1}{2}\int^\infty_{\omega/T} dx e^{-x}x^2\sigma_{ij}^{(2)}(xT)$.

\bibliography{chargephobic}

\begin{thebibliography}{300}%
\makeatletter
\providecommand \@ifxundefined [1]{%
 \@ifx{#1\undefined}
}%
\providecommand \@ifnum [1]{%
 \ifnum #1\expandafter \@firstoftwo
 \else \expandafter \@secondoftwo
 \fi
}%
\providecommand \@ifx [1]{%
 \ifx #1\expandafter \@firstoftwo
 \else \expandafter \@secondoftwo
 \fi
}%
\providecommand \natexlab [1]{#1}%
\providecommand \enquote  [1]{``#1''}%
\providecommand \bibnamefont  [1]{#1}%
\providecommand \bibfnamefont [1]{#1}%
\providecommand \citenamefont [1]{#1}%
\providecommand \href@noop [0]{\@secondoftwo}%
\providecommand \href [0]{\begingroup \@sanitize@url \@href}%
\providecommand \@href[1]{\@@startlink{#1}\@@href}%
\providecommand \@@href[1]{\endgroup#1\@@endlink}%
\providecommand \@sanitize@url [0]{\catcode `\\12\catcode `\$12\catcode
  `\&12\catcode `\#12\catcode `\^12\catcode `\_12\catcode `\%12\relax}%
\providecommand \@@startlink[1]{}%
\providecommand \@@endlink[0]{}%
\providecommand \url  [0]{\begingroup\@sanitize@url \@url }%
\providecommand \@url [1]{\endgroup\@href {#1}{\urlprefix }}%
\providecommand \urlprefix  [0]{URL }%
\providecommand \Eprint [0]{\href }%
\providecommand \doibase [0]{https://doi.org/}%
\providecommand \selectlanguage [0]{\@gobble}%
\providecommand \bibinfo  [0]{\@secondoftwo}%
\providecommand \bibfield  [0]{\@secondoftwo}%
\providecommand \translation [1]{[#1]}%
\providecommand \BibitemOpen [0]{}%
\providecommand \bibitemStop [0]{}%
\providecommand \bibitemNoStop [0]{.\EOS\space}%
\providecommand \EOS [0]{\spacefactor3000\relax}%
\providecommand \BibitemShut  [1]{\csname bibitem#1\endcsname}%
\let\auto@bib@innerbib\@empty
\bibitem [{\citenamefont {Ilten}\ \emph {et~al.}(2018)\citenamefont {Ilten},
  \citenamefont {Soreq}, \citenamefont {Williams},\ and\ \citenamefont
  {Xue}}]{Ilten:2018crw}%
  \BibitemOpen
  \bibfield  {author} {\bibinfo {author} {\bibfnamefont {P.}~\bibnamefont
  {Ilten}}, \bibinfo {author} {\bibfnamefont {Y.}~\bibnamefont {Soreq}},
  \bibinfo {author} {\bibfnamefont {M.}~\bibnamefont {Williams}},\ and\
  \bibinfo {author} {\bibfnamefont {W.}~\bibnamefont {Xue}},\ }\bibfield
  {title} {\bibinfo {title} {{Serendipity in dark photon searches}},\ }\href
  {https://doi.org/10.1007/JHEP06(2018)004} {\bibfield  {journal} {\bibinfo
  {journal} {JHEP}\ }\textbf {\bibinfo {volume} {06}},\ \bibinfo {pages}
  {004}},\ \Eprint {https://arxiv.org/abs/1801.04847} {arXiv:1801.04847
  [hep-ph]} \BibitemShut {NoStop}%
\bibitem [{\citenamefont {Bauer}\ \emph
  {et~al.}(2018{\natexlab{a}})\citenamefont {Bauer}, \citenamefont
  {Foldenauer},\ and\ \citenamefont {Jaeckel}}]{Bauer:2018onh}%
  \BibitemOpen
  \bibfield  {author} {\bibinfo {author} {\bibfnamefont {M.}~\bibnamefont
  {Bauer}}, \bibinfo {author} {\bibfnamefont {P.}~\bibnamefont {Foldenauer}},\
  and\ \bibinfo {author} {\bibfnamefont {J.}~\bibnamefont {Jaeckel}},\
  }\bibfield  {title} {\bibinfo {title} {{Hunting All the Hidden Photons}},\
  }\href {https://doi.org/10.1007/JHEP07(2018)094} {\bibfield  {journal}
  {\bibinfo  {journal} {JHEP}\ }\textbf {\bibinfo {volume} {07}},\ \bibinfo
  {pages} {094}},\ \Eprint {https://arxiv.org/abs/1803.05466} {arXiv:1803.05466
  [hep-ph]} \BibitemShut {NoStop}%
\bibitem [{\citenamefont {Fabbrichesi}\ \emph {et~al.}(2020)\citenamefont
  {Fabbrichesi}, \citenamefont {Gabrielli},\ and\ \citenamefont
  {Lanfranchi}}]{Fabbrichesi:2020wbt}%
  \BibitemOpen
  \bibfield  {author} {\bibinfo {author} {\bibfnamefont {M.}~\bibnamefont
  {Fabbrichesi}}, \bibinfo {author} {\bibfnamefont {E.}~\bibnamefont
  {Gabrielli}},\ and\ \bibinfo {author} {\bibfnamefont {G.}~\bibnamefont
  {Lanfranchi}},\ }\bibfield  {title} {\bibinfo {title} {{The Dark Photon}}\
  }\href {https://doi.org/10.1007/978-3-030-62519-1}
  {10.1007/978-3-030-62519-1} (\bibinfo {year} {2020}),\ \Eprint
  {https://arxiv.org/abs/2005.01515} {arXiv:2005.01515 [hep-ph]} \BibitemShut
  {NoStop}%
\bibitem [{\citenamefont {Caputo}\ \emph {et~al.}(2021)\citenamefont {Caputo},
  \citenamefont {Millar}, \citenamefont {O'Hare},\ and\ \citenamefont
  {Vitagliano}}]{Caputo:2021eaa}%
  \BibitemOpen
  \bibfield  {author} {\bibinfo {author} {\bibfnamefont {A.}~\bibnamefont
  {Caputo}}, \bibinfo {author} {\bibfnamefont {A.~J.}\ \bibnamefont {Millar}},
  \bibinfo {author} {\bibfnamefont {C.~A.~J.}\ \bibnamefont {O'Hare}},\ and\
  \bibinfo {author} {\bibfnamefont {E.}~\bibnamefont {Vitagliano}},\ }\bibfield
   {title} {\bibinfo {title} {{Dark photon limits: A handbook}},\ }\href
  {https://doi.org/10.1103/PhysRevD.104.095029} {\bibfield  {journal} {\bibinfo
   {journal} {Phys. Rev. D}\ }\textbf {\bibinfo {volume} {104}},\ \bibinfo
  {pages} {095029} (\bibinfo {year} {2021})},\ \Eprint
  {https://arxiv.org/abs/2105.04565} {arXiv:2105.04565 [hep-ph]} \BibitemShut
  {NoStop}%
\bibitem [{\citenamefont {Caputo}\ \emph {et~al.}(2025)\citenamefont {Caputo},
  \citenamefont {Park},\ and\ \citenamefont {Yun}}]{Caputo:2025avc}%
  \BibitemOpen
  \bibfield  {author} {\bibinfo {author} {\bibfnamefont {A.}~\bibnamefont
  {Caputo}}, \bibinfo {author} {\bibfnamefont {J.}~\bibnamefont {Park}},\ and\
  \bibinfo {author} {\bibfnamefont {S.}~\bibnamefont {Yun}},\ }\bibfield
  {title} {\bibinfo {title} {{The Heavy Dark Photon Handbook: Cosmological and
  Astrophysical Bounds}},\ }\href@noop {} {\  (\bibinfo {year} {2025})},\
  \Eprint {https://arxiv.org/abs/2511.15785} {arXiv:2511.15785 [hep-ph]}
  \BibitemShut {NoStop}%
\bibitem [{\citenamefont {Davidson}(1979)}]{Davidson:1978pm}%
  \BibitemOpen
  \bibfield  {author} {\bibinfo {author} {\bibfnamefont {A.}~\bibnamefont
  {Davidson}},\ }\bibfield  {title} {\bibinfo {title} {{$B-L$ as the fourth
  color within an $\mathrm{SU}(2)_L \times \mathrm{U}(1)_R \times
  \mathrm{U}(1)$ model}},\ }\href {https://doi.org/10.1103/PhysRevD.20.776}
  {\bibfield  {journal} {\bibinfo  {journal} {Phys. Rev. D}\ }\textbf {\bibinfo
  {volume} {20}},\ \bibinfo {pages} {776} (\bibinfo {year} {1979})}\BibitemShut
  {NoStop}%
\bibitem [{\citenamefont {Marshak}\ and\ \citenamefont
  {Mohapatra}(1980)}]{Marshak:1979fm}%
  \BibitemOpen
  \bibfield  {author} {\bibinfo {author} {\bibfnamefont {R.~E.}\ \bibnamefont
  {Marshak}}\ and\ \bibinfo {author} {\bibfnamefont {R.~N.}\ \bibnamefont
  {Mohapatra}},\ }\bibfield  {title} {\bibinfo {title} {{Quark - Lepton
  Symmetry and B-L as the U(1) Generator of the Electroweak Symmetry Group}},\
  }\href {https://doi.org/10.1016/0370-2693(80)90436-0} {\bibfield  {journal}
  {\bibinfo  {journal} {Phys. Lett. B}\ }\textbf {\bibinfo {volume} {91}},\
  \bibinfo {pages} {222} (\bibinfo {year} {1980})}\BibitemShut {NoStop}%
\bibitem [{\citenamefont {Mohapatra}\ and\ \citenamefont
  {Marshak}(1980)}]{Mohapatra:1980qe}%
  \BibitemOpen
  \bibfield  {author} {\bibinfo {author} {\bibfnamefont {R.~N.}\ \bibnamefont
  {Mohapatra}}\ and\ \bibinfo {author} {\bibfnamefont {R.~E.}\ \bibnamefont
  {Marshak}},\ }\bibfield  {title} {\bibinfo {title} {{Local B-L Symmetry of
  Electroweak Interactions, Majorana Neutrinos and Neutron Oscillations}},\
  }\href {https://doi.org/10.1103/PhysRevLett.44.1316} {\bibfield  {journal}
  {\bibinfo  {journal} {Phys. Rev. Lett.}\ }\textbf {\bibinfo {volume} {44}},\
  \bibinfo {pages} {1316} (\bibinfo {year} {1980})},\ \bibinfo {note}
  {[Erratum: Phys.Rev.Lett. 44, 1643 (1980)]}\BibitemShut {NoStop}%
\bibitem [{\citenamefont {Fayet}(1980)}]{Fayet:1980ad}%
  \BibitemOpen
  \bibfield  {author} {\bibinfo {author} {\bibfnamefont {P.}~\bibnamefont
  {Fayet}},\ }\bibfield  {title} {\bibinfo {title} {{Effects of the Spin 1
  Partner of the Goldstino (Gravitino) on Neutral Current Phenomenology}},\
  }\href {https://doi.org/10.1016/0370-2693(80)90488-8} {\bibfield  {journal}
  {\bibinfo  {journal} {Phys. Lett. B}\ }\textbf {\bibinfo {volume} {95}},\
  \bibinfo {pages} {285} (\bibinfo {year} {1980})}\BibitemShut {NoStop}%
\bibitem [{\citenamefont {Fayet}(1981)}]{Fayet:1980rr}%
  \BibitemOpen
  \bibfield  {author} {\bibinfo {author} {\bibfnamefont {P.}~\bibnamefont
  {Fayet}},\ }\bibfield  {title} {\bibinfo {title} {{On the Search for a New
  Spin 1 Boson}},\ }\href {https://doi.org/10.1016/0550-3213(81)90122-X}
  {\bibfield  {journal} {\bibinfo  {journal} {Nucl. Phys. B}\ }\textbf
  {\bibinfo {volume} {187}},\ \bibinfo {pages} {184} (\bibinfo {year}
  {1981})}\BibitemShut {NoStop}%
\bibitem [{\citenamefont {Wetterich}(1981)}]{Wetterich:1981bx}%
  \BibitemOpen
  \bibfield  {author} {\bibinfo {author} {\bibfnamefont {C.}~\bibnamefont
  {Wetterich}},\ }\bibfield  {title} {\bibinfo {title} {{Neutrino Masses and
  the Scale of B-L Violation}},\ }\href
  {https://doi.org/10.1016/0550-3213(81)90279-0} {\bibfield  {journal}
  {\bibinfo  {journal} {Nucl. Phys. B}\ }\textbf {\bibinfo {volume} {187}},\
  \bibinfo {pages} {343} (\bibinfo {year} {1981})}\BibitemShut {NoStop}%
\bibitem [{\citenamefont {Fayet}(1986)}]{Fayet:1986rh}%
  \BibitemOpen
  \bibfield  {author} {\bibinfo {author} {\bibfnamefont {P.}~\bibnamefont
  {Fayet}},\ }\bibfield  {title} {\bibinfo {title} {{The Fifth Interaction in
  Grand Unified Theories: A New Force Acting Mostly on Neutrons and Particle
  Spins}},\ }\href {https://doi.org/10.1016/0370-2693(86)90271-6} {\bibfield
  {journal} {\bibinfo  {journal} {Phys. Lett. B}\ }\textbf {\bibinfo {volume}
  {172}},\ \bibinfo {pages} {363} (\bibinfo {year} {1986})}\BibitemShut
  {NoStop}%
\bibitem [{\citenamefont {Fayet}(1989)}]{Fayet:1989mq}%
  \BibitemOpen
  \bibfield  {author} {\bibinfo {author} {\bibfnamefont {P.}~\bibnamefont
  {Fayet}},\ }\bibfield  {title} {\bibinfo {title} {{The Fifth Force Charge as
  a Linear Combination of Baryonic, Leptonic (Or $B^-$l) and Electric
  Charges}},\ }\href {https://doi.org/10.1016/0370-2693(89)91294-X} {\bibfield
  {journal} {\bibinfo  {journal} {Phys. Lett. B}\ }\textbf {\bibinfo {volume}
  {227}},\ \bibinfo {pages} {127} (\bibinfo {year} {1989})}\BibitemShut
  {NoStop}%
\bibitem [{\citenamefont {Fayet}(1990)}]{Fayet:1990wx}%
  \BibitemOpen
  \bibfield  {author} {\bibinfo {author} {\bibfnamefont {P.}~\bibnamefont
  {Fayet}},\ }\bibfield  {title} {\bibinfo {title} {{Extra U(1)'s and New
  Forces}},\ }\href {https://doi.org/10.1016/0550-3213(90)90381-M} {\bibfield
  {journal} {\bibinfo  {journal} {Nucl. Phys. B}\ }\textbf {\bibinfo {volume}
  {347}},\ \bibinfo {pages} {743} (\bibinfo {year} {1990})}\BibitemShut
  {NoStop}%
\bibitem [{\citenamefont {Langacker}(2009)}]{Langacker:2008yv}%
  \BibitemOpen
  \bibfield  {author} {\bibinfo {author} {\bibfnamefont {P.}~\bibnamefont
  {Langacker}},\ }\bibfield  {title} {\bibinfo {title} {{The Physics of Heavy
  $Z^\prime$ Gauge Bosons}},\ }\href
  {https://doi.org/10.1103/RevModPhys.81.1199} {\bibfield  {journal} {\bibinfo
  {journal} {Rev. Mod. Phys.}\ }\textbf {\bibinfo {volume} {81}},\ \bibinfo
  {pages} {1199} (\bibinfo {year} {2009})},\ \Eprint
  {https://arxiv.org/abs/0801.1345} {arXiv:0801.1345 [hep-ph]} \BibitemShut
  {NoStop}%
\bibitem [{\citenamefont {Harnik}\ \emph {et~al.}(2012)\citenamefont {Harnik},
  \citenamefont {Kopp},\ and\ \citenamefont {Machado}}]{Harnik:2012ni}%
  \BibitemOpen
  \bibfield  {author} {\bibinfo {author} {\bibfnamefont {R.}~\bibnamefont
  {Harnik}}, \bibinfo {author} {\bibfnamefont {J.}~\bibnamefont {Kopp}},\ and\
  \bibinfo {author} {\bibfnamefont {P.~A.~N.}\ \bibnamefont {Machado}},\
  }\bibfield  {title} {\bibinfo {title} {{Exploring $\nu$ Signals in Dark
  Matter Detectors}},\ }\href {https://doi.org/10.1088/1475-7516/2012/07/026}
  {\bibfield  {journal} {\bibinfo  {journal} {JCAP}\ }\textbf {\bibinfo
  {volume} {1207}},\ \bibinfo {pages} {026}},\ \Eprint
  {https://arxiv.org/abs/1202.6073} {arXiv:1202.6073 [hep-ph]} \BibitemShut
  {NoStop}%
\bibitem [{\citenamefont {Heeck}(2014)}]{Heeck:2014zfa}%
  \BibitemOpen
  \bibfield  {author} {\bibinfo {author} {\bibfnamefont {J.}~\bibnamefont
  {Heeck}},\ }\bibfield  {title} {\bibinfo {title} {{Unbroken B \textendash{} L
  symmetry}},\ }\href {https://doi.org/10.1016/j.physletb.2014.10.067}
  {\bibfield  {journal} {\bibinfo  {journal} {Phys. Lett. B}\ }\textbf
  {\bibinfo {volume} {739}},\ \bibinfo {pages} {256} (\bibinfo {year}
  {2014})},\ \Eprint {https://arxiv.org/abs/1408.6845} {arXiv:1408.6845
  [hep-ph]} \BibitemShut {NoStop}%
\bibitem [{\citenamefont {Foot}(1991)}]{Foot:1990mn}%
  \BibitemOpen
  \bibfield  {author} {\bibinfo {author} {\bibfnamefont {R.}~\bibnamefont
  {Foot}},\ }\bibfield  {title} {\bibinfo {title} {{New Physics From Electric
  Charge Quantization?}},\ }\href {https://doi.org/10.1142/S0217732391000543}
  {\bibfield  {journal} {\bibinfo  {journal} {Mod. Phys. Lett. A}\ }\textbf
  {\bibinfo {volume} {6}},\ \bibinfo {pages} {527} (\bibinfo {year}
  {1991})}\BibitemShut {NoStop}%
\bibitem [{\citenamefont {He}\ \emph {et~al.}(1991)\citenamefont {He},
  \citenamefont {Joshi}, \citenamefont {Lew},\ and\ \citenamefont
  {Volkas}}]{He:1991qd}%
  \BibitemOpen
  \bibfield  {author} {\bibinfo {author} {\bibfnamefont {X.-G.}\ \bibnamefont
  {He}}, \bibinfo {author} {\bibfnamefont {G.~C.}\ \bibnamefont {Joshi}},
  \bibinfo {author} {\bibfnamefont {H.}~\bibnamefont {Lew}},\ and\ \bibinfo
  {author} {\bibfnamefont {R.~R.}\ \bibnamefont {Volkas}},\ }\bibfield  {title}
  {\bibinfo {title} {{Simplest Z-prime model}},\ }\href
  {https://doi.org/10.1103/PhysRevD.44.2118} {\bibfield  {journal} {\bibinfo
  {journal} {Phys. Rev. D}\ }\textbf {\bibinfo {volume} {44}},\ \bibinfo
  {pages} {2118} (\bibinfo {year} {1991})}\BibitemShut {NoStop}%
\bibitem [{\citenamefont {Foot}\ \emph {et~al.}(1994)\citenamefont {Foot},
  \citenamefont {He}, \citenamefont {Lew},\ and\ \citenamefont
  {Volkas}}]{Foot:1994vd}%
  \BibitemOpen
  \bibfield  {author} {\bibinfo {author} {\bibfnamefont {R.}~\bibnamefont
  {Foot}}, \bibinfo {author} {\bibfnamefont {X.~G.}\ \bibnamefont {He}},
  \bibinfo {author} {\bibfnamefont {H.}~\bibnamefont {Lew}},\ and\ \bibinfo
  {author} {\bibfnamefont {R.~R.}\ \bibnamefont {Volkas}},\ }\bibfield  {title}
  {\bibinfo {title} {{Model for a light Z-prime boson}},\ }\href
  {https://doi.org/10.1103/PhysRevD.50.4571} {\bibfield  {journal} {\bibinfo
  {journal} {Phys. Rev. D}\ }\textbf {\bibinfo {volume} {50}},\ \bibinfo
  {pages} {4571} (\bibinfo {year} {1994})},\ \Eprint
  {https://arxiv.org/abs/hep-ph/9401250} {arXiv:hep-ph/9401250} \BibitemShut
  {NoStop}%
\bibitem [{\citenamefont {Ma}(1998)}]{Ma:1997nq}%
  \BibitemOpen
  \bibfield  {author} {\bibinfo {author} {\bibfnamefont {E.}~\bibnamefont
  {Ma}},\ }\bibfield  {title} {\bibinfo {title} {{Gauged B - 3L(tau) and
  radiative neutrino masses}},\ }\href
  {https://doi.org/10.1016/S0370-2693(98)00599-1} {\bibfield  {journal}
  {\bibinfo  {journal} {Phys. Lett. B}\ }\textbf {\bibinfo {volume} {433}},\
  \bibinfo {pages} {74} (\bibinfo {year} {1998})},\ \Eprint
  {https://arxiv.org/abs/hep-ph/9709474} {arXiv:hep-ph/9709474} \BibitemShut
  {NoStop}%
\bibitem [{\citenamefont {Ma}\ and\ \citenamefont {Roy}(1998)}]{Ma:1998dp}%
  \BibitemOpen
  \bibfield  {author} {\bibinfo {author} {\bibfnamefont {E.}~\bibnamefont
  {Ma}}\ and\ \bibinfo {author} {\bibfnamefont {D.~P.}\ \bibnamefont {Roy}},\
  }\bibfield  {title} {\bibinfo {title} {{Phenomenology of the $B$ - 3L($\tau$)
  gauge boson}},\ }\href {https://doi.org/10.1103/PhysRevD.58.095005}
  {\bibfield  {journal} {\bibinfo  {journal} {Phys. Rev. D}\ }\textbf {\bibinfo
  {volume} {58}},\ \bibinfo {pages} {095005} (\bibinfo {year} {1998})},\
  \Eprint {https://arxiv.org/abs/hep-ph/9806210} {arXiv:hep-ph/9806210}
  \BibitemShut {NoStop}%
\bibitem [{\citenamefont {Appelquist}\ \emph {et~al.}(2003)\citenamefont
  {Appelquist}, \citenamefont {Dobrescu},\ and\ \citenamefont
  {Hopper}}]{Appelquist:2002mw}%
  \BibitemOpen
  \bibfield  {author} {\bibinfo {author} {\bibfnamefont {T.}~\bibnamefont
  {Appelquist}}, \bibinfo {author} {\bibfnamefont {B.~A.}\ \bibnamefont
  {Dobrescu}},\ and\ \bibinfo {author} {\bibfnamefont {A.~R.}\ \bibnamefont
  {Hopper}},\ }\bibfield  {title} {\bibinfo {title} {{Nonexotic Neutral Gauge
  Bosons}},\ }\href {https://doi.org/10.1103/PhysRevD.68.035012} {\bibfield
  {journal} {\bibinfo  {journal} {Phys. Rev. D}\ }\textbf {\bibinfo {volume}
  {68}},\ \bibinfo {pages} {035012} (\bibinfo {year} {2003})},\ \Eprint
  {https://arxiv.org/abs/hep-ph/0212073} {arXiv:hep-ph/0212073} \BibitemShut
  {NoStop}%
\bibitem [{\citenamefont {Boehm}\ and\ \citenamefont
  {Fayet}(2004)}]{Boehm:2003hm}%
  \BibitemOpen
  \bibfield  {author} {\bibinfo {author} {\bibfnamefont {C.}~\bibnamefont
  {Boehm}}\ and\ \bibinfo {author} {\bibfnamefont {P.}~\bibnamefont {Fayet}},\
  }\bibfield  {title} {\bibinfo {title} {{Scalar dark matter candidates}},\
  }\href {https://doi.org/10.1016/j.nuclphysb.2004.01.015} {\bibfield
  {journal} {\bibinfo  {journal} {Nucl. Phys. B}\ }\textbf {\bibinfo {volume}
  {683}},\ \bibinfo {pages} {219} (\bibinfo {year} {2004})},\ \Eprint
  {https://arxiv.org/abs/hep-ph/0305261} {arXiv:hep-ph/0305261} \BibitemShut
  {NoStop}%
\bibitem [{\citenamefont {Fayet}(2004)}]{Fayet:2004bw}%
  \BibitemOpen
  \bibfield  {author} {\bibinfo {author} {\bibfnamefont {P.}~\bibnamefont
  {Fayet}},\ }\bibfield  {title} {\bibinfo {title} {{Light spin 1/2 or spin 0
  dark matter particles}},\ }\href {https://doi.org/10.1103/PhysRevD.70.023514}
  {\bibfield  {journal} {\bibinfo  {journal} {Phys. Rev. D}\ }\textbf {\bibinfo
  {volume} {70}},\ \bibinfo {pages} {023514} (\bibinfo {year} {2004})},\
  \Eprint {https://arxiv.org/abs/hep-ph/0403226} {arXiv:hep-ph/0403226}
  \BibitemShut {NoStop}%
\bibitem [{\citenamefont {Fayet}(2006)}]{Fayet:2006sp}%
  \BibitemOpen
  \bibfield  {author} {\bibinfo {author} {\bibfnamefont {P.}~\bibnamefont
  {Fayet}},\ }\bibfield  {title} {\bibinfo {title} {{Constraints on Light Dark
  Matter and U bosons, from psi, Upsilon, K+, pi0, eta and eta-prime decays}},\
  }\href {https://doi.org/10.1103/PhysRevD.74.054034} {\bibfield  {journal}
  {\bibinfo  {journal} {Phys. Rev. D}\ }\textbf {\bibinfo {volume} {74}},\
  \bibinfo {pages} {054034} (\bibinfo {year} {2006})},\ \Eprint
  {https://arxiv.org/abs/hep-ph/0607318} {arXiv:hep-ph/0607318} \BibitemShut
  {NoStop}%
\bibitem [{\citenamefont {Fayet}(2007)}]{Fayet:2007ua}%
  \BibitemOpen
  \bibfield  {author} {\bibinfo {author} {\bibfnamefont {P.}~\bibnamefont
  {Fayet}},\ }\bibfield  {title} {\bibinfo {title} {{U-boson production in e+
  e- annihilations, psi and Upsilon decays, and Light Dark Matter}},\ }\href
  {https://doi.org/10.1103/PhysRevD.75.115017} {\bibfield  {journal} {\bibinfo
  {journal} {Phys. Rev. D}\ }\textbf {\bibinfo {volume} {75}},\ \bibinfo
  {pages} {115017} (\bibinfo {year} {2007})},\ \Eprint
  {https://arxiv.org/abs/hep-ph/0702176} {arXiv:hep-ph/0702176} \BibitemShut
  {NoStop}%
\bibitem [{\citenamefont {Salvioni}\ \emph {et~al.}(2010)\citenamefont
  {Salvioni}, \citenamefont {Strumia}, \citenamefont {Villadoro},\ and\
  \citenamefont {Zwirner}}]{Salvioni:2009jp}%
  \BibitemOpen
  \bibfield  {author} {\bibinfo {author} {\bibfnamefont {E.}~\bibnamefont
  {Salvioni}}, \bibinfo {author} {\bibfnamefont {A.}~\bibnamefont {Strumia}},
  \bibinfo {author} {\bibfnamefont {G.}~\bibnamefont {Villadoro}},\ and\
  \bibinfo {author} {\bibfnamefont {F.}~\bibnamefont {Zwirner}},\ }\bibfield
  {title} {\bibinfo {title} {{Non-universal minimal Z' models: present bounds
  and early LHC reach}},\ }\href {https://doi.org/10.1007/JHEP03(2010)010}
  {\bibfield  {journal} {\bibinfo  {journal} {JHEP}\ }\textbf {\bibinfo
  {volume} {03}},\ \bibinfo {pages} {010}},\ \Eprint
  {https://arxiv.org/abs/0911.1450} {arXiv:0911.1450 [hep-ph]} \BibitemShut
  {NoStop}%
\bibitem [{\citenamefont {Fayet}(2017)}]{Fayet:2016nyc}%
  \BibitemOpen
  \bibfield  {author} {\bibinfo {author} {\bibfnamefont {P.}~\bibnamefont
  {Fayet}},\ }\bibfield  {title} {\bibinfo {title} {{The light $U$ boson as the
  mediator of a new force, coupled to a combination of $Q,B,L$ and dark
  matter}},\ }\href {https://doi.org/10.1140/epjc/s10052-016-4568-9} {\bibfield
   {journal} {\bibinfo  {journal} {Eur. Phys. J. C}\ }\textbf {\bibinfo
  {volume} {77}},\ \bibinfo {pages} {53} (\bibinfo {year} {2017})},\ \Eprint
  {https://arxiv.org/abs/1611.05357} {arXiv:1611.05357 [hep-ph]} \BibitemShut
  {NoStop}%
\bibitem [{\citenamefont {Fayet}(2021)}]{Fayet:2020bmb}%
  \BibitemOpen
  \bibfield  {author} {\bibinfo {author} {\bibfnamefont {P.}~\bibnamefont
  {Fayet}},\ }\bibfield  {title} {\bibinfo {title} {{$U$ boson interpolating
  between a generalized dark photon or dark $Z$ , an axial boson, and an
  axionlike particle}},\ }\href {https://doi.org/10.1103/PhysRevD.103.035034}
  {\bibfield  {journal} {\bibinfo  {journal} {Phys. Rev. D}\ }\textbf {\bibinfo
  {volume} {103}},\ \bibinfo {pages} {035034} (\bibinfo {year} {2021})},\
  \Eprint {https://arxiv.org/abs/2010.04673} {arXiv:2010.04673 [hep-ph]}
  \BibitemShut {NoStop}%
\bibitem [{\citenamefont {Celis}\ \emph {et~al.}(2015)\citenamefont {Celis},
  \citenamefont {Fuentes-Martin}, \citenamefont {Jung},\ and\ \citenamefont
  {Serodio}}]{Celis:2015ara}%
  \BibitemOpen
  \bibfield  {author} {\bibinfo {author} {\bibfnamefont {A.}~\bibnamefont
  {Celis}}, \bibinfo {author} {\bibfnamefont {J.}~\bibnamefont
  {Fuentes-Martin}}, \bibinfo {author} {\bibfnamefont {M.}~\bibnamefont
  {Jung}},\ and\ \bibinfo {author} {\bibfnamefont {H.}~\bibnamefont
  {Serodio}},\ }\bibfield  {title} {\bibinfo {title} {{Family nonuniversal Z'
  models with protected flavor-changing interactions}},\ }\href
  {https://doi.org/10.1103/PhysRevD.92.015007} {\bibfield  {journal} {\bibinfo
  {journal} {Phys. Rev. D}\ }\textbf {\bibinfo {volume} {92}},\ \bibinfo
  {pages} {015007} (\bibinfo {year} {2015})},\ \Eprint
  {https://arxiv.org/abs/1505.03079} {arXiv:1505.03079 [hep-ph]} \BibitemShut
  {NoStop}%
\bibitem [{\citenamefont {Bauer}\ \emph
  {et~al.}(2018{\natexlab{b}})\citenamefont {Bauer}, \citenamefont
  {Diefenbacher}, \citenamefont {Plehn}, \citenamefont {Russell},\ and\
  \citenamefont {Camargo}}]{Bauer:2018egk}%
  \BibitemOpen
  \bibfield  {author} {\bibinfo {author} {\bibfnamefont {M.}~\bibnamefont
  {Bauer}}, \bibinfo {author} {\bibfnamefont {S.}~\bibnamefont {Diefenbacher}},
  \bibinfo {author} {\bibfnamefont {T.}~\bibnamefont {Plehn}}, \bibinfo
  {author} {\bibfnamefont {M.}~\bibnamefont {Russell}},\ and\ \bibinfo {author}
  {\bibfnamefont {D.~A.}\ \bibnamefont {Camargo}},\ }\bibfield  {title}
  {\bibinfo {title} {{Dark Matter in Anomaly-Free Gauge Extensions}},\ }\href
  {https://doi.org/10.21468/SciPostPhys.5.4.036} {\bibfield  {journal}
  {\bibinfo  {journal} {SciPost Phys.}\ }\textbf {\bibinfo {volume} {5}},\
  \bibinfo {pages} {036} (\bibinfo {year} {2018}{\natexlab{b}})},\ \Eprint
  {https://arxiv.org/abs/1805.01904} {arXiv:1805.01904 [hep-ph]} \BibitemShut
  {NoStop}%
\bibitem [{\citenamefont {Altmannshofer}\ \emph {et~al.}(2019)\citenamefont
  {Altmannshofer}, \citenamefont {Tammaro},\ and\ \citenamefont
  {Zupan}}]{Altmannshofer:2018xyo}%
  \BibitemOpen
  \bibfield  {author} {\bibinfo {author} {\bibfnamefont {W.}~\bibnamefont
  {Altmannshofer}}, \bibinfo {author} {\bibfnamefont {M.}~\bibnamefont
  {Tammaro}},\ and\ \bibinfo {author} {\bibfnamefont {J.}~\bibnamefont
  {Zupan}},\ }\bibfield  {title} {\bibinfo {title} {{Non-standard neutrino
  interactions and low energy experiments}},\ }\href
  {https://doi.org/10.1007/JHEP11(2021)113} {\bibfield  {journal} {\bibinfo
  {journal} {JHEP}\ }\textbf {\bibinfo {volume} {09}},\ \bibinfo {pages}
  {083}},\ \bibinfo {note} {[Erratum: JHEP 11, 113 (2021)]},\ \Eprint
  {https://arxiv.org/abs/1812.02778} {arXiv:1812.02778 [hep-ph]} \BibitemShut
  {NoStop}%
\bibitem [{\citenamefont {Heeck}\ \emph {et~al.}(2019)\citenamefont {Heeck},
  \citenamefont {Lindner}, \citenamefont {Rodejohann},\ and\ \citenamefont
  {Vogl}}]{Heeck:2018nzc}%
  \BibitemOpen
  \bibfield  {author} {\bibinfo {author} {\bibfnamefont {J.}~\bibnamefont
  {Heeck}}, \bibinfo {author} {\bibfnamefont {M.}~\bibnamefont {Lindner}},
  \bibinfo {author} {\bibfnamefont {W.}~\bibnamefont {Rodejohann}},\ and\
  \bibinfo {author} {\bibfnamefont {S.}~\bibnamefont {Vogl}},\ }\bibfield
  {title} {\bibinfo {title} {{Non-Standard Neutrino Interactions and Neutral
  Gauge Bosons}},\ }\href {https://doi.org/10.21468/SciPostPhys.6.3.038}
  {\bibfield  {journal} {\bibinfo  {journal} {SciPost Phys.}\ }\textbf
  {\bibinfo {volume} {6}},\ \bibinfo {pages} {038} (\bibinfo {year} {2019})},\
  \Eprint {https://arxiv.org/abs/1812.04067} {arXiv:1812.04067 [hep-ph]}
  \BibitemShut {NoStop}%
\bibitem [{\citenamefont {Escudero}\ \emph {et~al.}(2019)\citenamefont
  {Escudero}, \citenamefont {Hooper}, \citenamefont {Krnjaic},\ and\
  \citenamefont {Pierre}}]{Escudero:2019gzq}%
  \BibitemOpen
  \bibfield  {author} {\bibinfo {author} {\bibfnamefont {M.}~\bibnamefont
  {Escudero}}, \bibinfo {author} {\bibfnamefont {D.}~\bibnamefont {Hooper}},
  \bibinfo {author} {\bibfnamefont {G.}~\bibnamefont {Krnjaic}},\ and\ \bibinfo
  {author} {\bibfnamefont {M.}~\bibnamefont {Pierre}},\ }\bibfield  {title}
  {\bibinfo {title} {{Cosmology with A Very Light L$_{\mu}$ \ensuremath{-}
  L$_{\tau}$ Gauge Boson}},\ }\href {https://doi.org/10.1007/JHEP03(2019)071}
  {\bibfield  {journal} {\bibinfo  {journal} {JHEP}\ }\textbf {\bibinfo
  {volume} {03}},\ \bibinfo {pages} {071}},\ \Eprint
  {https://arxiv.org/abs/1901.02010} {arXiv:1901.02010 [hep-ph]} \BibitemShut
  {NoStop}%
\bibitem [{\citenamefont {Han}\ \emph {et~al.}(2019)\citenamefont {Han},
  \citenamefont {Liao}, \citenamefont {Liu},\ and\ \citenamefont
  {Marfatia}}]{Han:2019zkz}%
  \BibitemOpen
  \bibfield  {author} {\bibinfo {author} {\bibfnamefont {T.}~\bibnamefont
  {Han}}, \bibinfo {author} {\bibfnamefont {J.}~\bibnamefont {Liao}}, \bibinfo
  {author} {\bibfnamefont {H.}~\bibnamefont {Liu}},\ and\ \bibinfo {author}
  {\bibfnamefont {D.}~\bibnamefont {Marfatia}},\ }\bibfield  {title} {\bibinfo
  {title} {{Nonstandard neutrino interactions at COHERENT, DUNE, T2HK and
  LHC}},\ }\href {https://doi.org/10.1007/JHEP11(2019)028} {\bibfield
  {journal} {\bibinfo  {journal} {JHEP}\ }\textbf {\bibinfo {volume} {11}},\
  \bibinfo {pages} {028}},\ \Eprint {https://arxiv.org/abs/1910.03272}
  {arXiv:1910.03272 [hep-ph]} \BibitemShut {NoStop}%
\bibitem [{\citenamefont {Coloma}\ \emph {et~al.}(2021)\citenamefont {Coloma},
  \citenamefont {Gonzalez-Garcia},\ and\ \citenamefont
  {Maltoni}}]{Coloma:2020gfv}%
  \BibitemOpen
  \bibfield  {author} {\bibinfo {author} {\bibfnamefont {P.}~\bibnamefont
  {Coloma}}, \bibinfo {author} {\bibfnamefont {M.~C.}\ \bibnamefont
  {Gonzalez-Garcia}},\ and\ \bibinfo {author} {\bibfnamefont {M.}~\bibnamefont
  {Maltoni}},\ }\bibfield  {title} {\bibinfo {title} {{Neutrino oscillation
  constraints on U(1)' models: from non-standard interactions to long-range
  forces}},\ }\href {https://doi.org/10.1007/JHEP01(2021)114} {\bibfield
  {journal} {\bibinfo  {journal} {JHEP}\ }\textbf {\bibinfo {volume} {01}},\
  \bibinfo {pages} {114}},\ \bibinfo {note} {[Erratum: JHEP 11, 115 (2022)]},\
  \Eprint {https://arxiv.org/abs/2009.14220} {arXiv:2009.14220 [hep-ph]}
  \BibitemShut {NoStop}%
\bibitem [{\citenamefont {Bauer}\ \emph {et~al.}(2021)\citenamefont {Bauer},
  \citenamefont {Foldenauer},\ and\ \citenamefont {Mosny}}]{Bauer:2020itv}%
  \BibitemOpen
  \bibfield  {author} {\bibinfo {author} {\bibfnamefont {M.}~\bibnamefont
  {Bauer}}, \bibinfo {author} {\bibfnamefont {P.}~\bibnamefont {Foldenauer}},\
  and\ \bibinfo {author} {\bibfnamefont {M.}~\bibnamefont {Mosny}},\ }\bibfield
   {title} {\bibinfo {title} {{Flavor structure of anomaly-free hidden photon
  models}},\ }\href {https://doi.org/10.1103/PhysRevD.103.075024} {\bibfield
  {journal} {\bibinfo  {journal} {Phys. Rev. D}\ }\textbf {\bibinfo {volume}
  {103}},\ \bibinfo {pages} {075024} (\bibinfo {year} {2021})},\ \Eprint
  {https://arxiv.org/abs/2011.12973} {arXiv:2011.12973 [hep-ph]} \BibitemShut
  {NoStop}%
\bibitem [{\citenamefont {Ghosh}\ \emph {et~al.}(2024)\citenamefont {Ghosh},
  \citenamefont {Ghosh}, \citenamefont {Jeesun},\ and\ \citenamefont
  {Srivastava}}]{Ghosh:2024cxi}%
  \BibitemOpen
  \bibfield  {author} {\bibinfo {author} {\bibfnamefont {D.~K.}\ \bibnamefont
  {Ghosh}}, \bibinfo {author} {\bibfnamefont {P.}~\bibnamefont {Ghosh}},
  \bibinfo {author} {\bibfnamefont {S.}~\bibnamefont {Jeesun}},\ and\ \bibinfo
  {author} {\bibfnamefont {R.}~\bibnamefont {Srivastava}},\ }\bibfield  {title}
  {\bibinfo {title} {{Neff at CMB challenges U(1)X light gauge boson
  scenarios}},\ }\href {https://doi.org/10.1103/PhysRevD.110.075032} {\bibfield
   {journal} {\bibinfo  {journal} {Phys. Rev. D}\ }\textbf {\bibinfo {volume}
  {110}},\ \bibinfo {pages} {075032} (\bibinfo {year} {2024})},\ \Eprint
  {https://arxiv.org/abs/2404.10077} {arXiv:2404.10077 [hep-ph]} \BibitemShut
  {NoStop}%
\bibitem [{\citenamefont {Salvioni}\ \emph {et~al.}(2009)\citenamefont
  {Salvioni}, \citenamefont {Villadoro},\ and\ \citenamefont
  {Zwirner}}]{Salvioni:2009mt}%
  \BibitemOpen
  \bibfield  {author} {\bibinfo {author} {\bibfnamefont {E.}~\bibnamefont
  {Salvioni}}, \bibinfo {author} {\bibfnamefont {G.}~\bibnamefont
  {Villadoro}},\ and\ \bibinfo {author} {\bibfnamefont {F.}~\bibnamefont
  {Zwirner}},\ }\bibfield  {title} {\bibinfo {title} {{Minimal Z-prime models:
  Present bounds and early LHC reach}},\ }\href
  {https://doi.org/10.1088/1126-6708/2009/11/068} {\bibfield  {journal}
  {\bibinfo  {journal} {JHEP}\ }\textbf {\bibinfo {volume} {11}},\ \bibinfo
  {pages} {068}},\ \Eprint {https://arxiv.org/abs/0909.1320} {arXiv:0909.1320
  [hep-ph]} \BibitemShut {NoStop}%
\bibitem [{\citenamefont {Kribs}\ \emph {et~al.}(2022)\citenamefont {Kribs},
  \citenamefont {Lee},\ and\ \citenamefont {Martin}}]{Kribs:2022gri}%
  \BibitemOpen
  \bibfield  {author} {\bibinfo {author} {\bibfnamefont {G.~D.}\ \bibnamefont
  {Kribs}}, \bibinfo {author} {\bibfnamefont {G.}~\bibnamefont {Lee}},\ and\
  \bibinfo {author} {\bibfnamefont {A.}~\bibnamefont {Martin}},\ }\bibfield
  {title} {\bibinfo {title} {{Effective field theory of St\"uckelberg vector
  bosons}},\ }\href {https://doi.org/10.1103/PhysRevD.106.055020} {\bibfield
  {journal} {\bibinfo  {journal} {Phys. Rev. D}\ }\textbf {\bibinfo {volume}
  {106}},\ \bibinfo {pages} {055020} (\bibinfo {year} {2022})},\ \Eprint
  {https://arxiv.org/abs/2204.01755} {arXiv:2204.01755 [hep-ph]} \BibitemShut
  {NoStop}%
\bibitem [{\citenamefont {Dror}\ \emph
  {et~al.}(2017{\natexlab{a}})\citenamefont {Dror}, \citenamefont {Lasenby},\
  and\ \citenamefont {Pospelov}}]{Dror:2017ehi}%
  \BibitemOpen
  \bibfield  {author} {\bibinfo {author} {\bibfnamefont {J.~A.}\ \bibnamefont
  {Dror}}, \bibinfo {author} {\bibfnamefont {R.}~\bibnamefont {Lasenby}},\ and\
  \bibinfo {author} {\bibfnamefont {M.}~\bibnamefont {Pospelov}},\ }\bibfield
  {title} {\bibinfo {title} {{New constraints on light vectors coupled to
  anomalous currents}},\ }\href
  {https://doi.org/10.1103/PhysRevLett.119.141803} {\bibfield  {journal}
  {\bibinfo  {journal} {Phys. Rev. Lett.}\ }\textbf {\bibinfo {volume} {119}},\
  \bibinfo {pages} {141803} (\bibinfo {year} {2017}{\natexlab{a}})},\ \Eprint
  {https://arxiv.org/abs/1705.06726} {arXiv:1705.06726 [hep-ph]} \BibitemShut
  {NoStop}%
\bibitem [{\citenamefont {Dror}\ \emph
  {et~al.}(2017{\natexlab{b}})\citenamefont {Dror}, \citenamefont {Lasenby},\
  and\ \citenamefont {Pospelov}}]{Dror:2017nsg}%
  \BibitemOpen
  \bibfield  {author} {\bibinfo {author} {\bibfnamefont {J.~A.}\ \bibnamefont
  {Dror}}, \bibinfo {author} {\bibfnamefont {R.}~\bibnamefont {Lasenby}},\ and\
  \bibinfo {author} {\bibfnamefont {M.}~\bibnamefont {Pospelov}},\ }\bibfield
  {title} {\bibinfo {title} {{Dark forces coupled to nonconserved currents}},\
  }\href {https://doi.org/10.1103/PhysRevD.96.075036} {\bibfield  {journal}
  {\bibinfo  {journal} {Phys. Rev. D}\ }\textbf {\bibinfo {volume} {96}},\
  \bibinfo {pages} {075036} (\bibinfo {year} {2017}{\natexlab{b}})},\ \Eprint
  {https://arxiv.org/abs/1707.01503} {arXiv:1707.01503 [hep-ph]} \BibitemShut
  {NoStop}%
\bibitem [{Gar()}]{Garching}%
  \BibitemOpen
  \href {https://wwwmpa.mpa-garching.mpg.de/ccsnarchive/archive.html} {\bibinfo
  {title} {{Garching Core Collapse Supernova Archive}}}\BibitemShut {NoStop}%
\bibitem [{\citenamefont {Dror}\ \emph {et~al.}(2019)\citenamefont {Dror},
  \citenamefont {Lasenby},\ and\ \citenamefont {Pospelov}}]{Dror:2018wfl}%
  \BibitemOpen
  \bibfield  {author} {\bibinfo {author} {\bibfnamefont {J.~A.}\ \bibnamefont
  {Dror}}, \bibinfo {author} {\bibfnamefont {R.}~\bibnamefont {Lasenby}},\ and\
  \bibinfo {author} {\bibfnamefont {M.}~\bibnamefont {Pospelov}},\ }\bibfield
  {title} {\bibinfo {title} {{Light vectors coupled to bosonic currents}},\
  }\href {https://doi.org/10.1103/PhysRevD.99.055016} {\bibfield  {journal}
  {\bibinfo  {journal} {Phys. Rev. D}\ }\textbf {\bibinfo {volume} {99}},\
  \bibinfo {pages} {055016} (\bibinfo {year} {2019})},\ \Eprint
  {https://arxiv.org/abs/1811.00595} {arXiv:1811.00595 [hep-ph]} \BibitemShut
  {NoStop}%
\bibitem [{\citenamefont {Dror}(2020)}]{Dror:2020fbh}%
  \BibitemOpen
  \bibfield  {author} {\bibinfo {author} {\bibfnamefont {J.~A.}\ \bibnamefont
  {Dror}},\ }\bibfield  {title} {\bibinfo {title} {{Discovering leptonic forces
  using nonconserved currents}},\ }\href
  {https://doi.org/10.1103/PhysRevD.101.095013} {\bibfield  {journal} {\bibinfo
   {journal} {Phys. Rev. D}\ }\textbf {\bibinfo {volume} {101}},\ \bibinfo
  {pages} {095013} (\bibinfo {year} {2020})},\ \Eprint
  {https://arxiv.org/abs/2004.04750} {arXiv:2004.04750 [hep-ph]} \BibitemShut
  {NoStop}%
\bibitem [{\citenamefont {Michaels}\ and\ \citenamefont
  {Yu}(2021)}]{Michaels:2020fzj}%
  \BibitemOpen
  \bibfield  {author} {\bibinfo {author} {\bibfnamefont {L.}~\bibnamefont
  {Michaels}}\ and\ \bibinfo {author} {\bibfnamefont {F.}~\bibnamefont {Yu}},\
  }\bibfield  {title} {\bibinfo {title} {{Probing new $U(1)$ gauge symmetries
  via exotic $Z \to Z' \gamma$ decays}},\ }\href
  {https://doi.org/10.1007/JHEP03(2021)120} {\bibfield  {journal} {\bibinfo
  {journal} {JHEP}\ }\textbf {\bibinfo {volume} {03}},\ \bibinfo {pages}
  {120}},\ \Eprint {https://arxiv.org/abs/2010.00021} {arXiv:2010.00021
  [hep-ph]} \BibitemShut {NoStop}%
\bibitem [{\citenamefont {Esseili}\ and\ \citenamefont
  {Kribs}(2024)}]{Esseili:2023ldf}%
  \BibitemOpen
  \bibfield  {author} {\bibinfo {author} {\bibfnamefont {H.}~\bibnamefont
  {Esseili}}\ and\ \bibinfo {author} {\bibfnamefont {G.~D.}\ \bibnamefont
  {Kribs}},\ }\bibfield  {title} {\bibinfo {title} {{Cosmological implications
  of gauged U(1)$_{B-L}$ on \ensuremath{\Delta}N $_{eff}$ in the CMB and
  BBN}},\ }\href {https://doi.org/10.1088/1475-7516/2024/05/110} {\bibfield
  {journal} {\bibinfo  {journal} {JCAP}\ }\textbf {\bibinfo {volume} {05}},\
  \bibinfo {pages} {110}},\ \Eprint {https://arxiv.org/abs/2308.07955}
  {arXiv:2308.07955 [hep-ph]} \BibitemShut {NoStop}%
\bibitem [{\citenamefont {Feng}\ \emph {et~al.}(2013)\citenamefont {Feng},
  \citenamefont {Kumar},\ and\ \citenamefont {Sanford}}]{Feng:2013vod}%
  \BibitemOpen
  \bibfield  {author} {\bibinfo {author} {\bibfnamefont {J.~L.}\ \bibnamefont
  {Feng}}, \bibinfo {author} {\bibfnamefont {J.}~\bibnamefont {Kumar}},\ and\
  \bibinfo {author} {\bibfnamefont {D.}~\bibnamefont {Sanford}},\ }\bibfield
  {title} {\bibinfo {title} {{Xenophobic Dark Matter}},\ }\href
  {https://doi.org/10.1103/PhysRevD.88.015021} {\bibfield  {journal} {\bibinfo
  {journal} {Phys. Rev. D}\ }\textbf {\bibinfo {volume} {88}},\ \bibinfo
  {pages} {015021} (\bibinfo {year} {2013})},\ \Eprint
  {https://arxiv.org/abs/1306.2315} {arXiv:1306.2315 [hep-ph]} \BibitemShut
  {NoStop}%
\bibitem [{\citenamefont {Abreu}\ \emph {et~al.}(2024)\citenamefont {Abreu}
  \emph {et~al.}}]{FASER:2023tle}%
  \BibitemOpen
  \bibfield  {author} {\bibinfo {author} {\bibfnamefont {H.}~\bibnamefont
  {Abreu}} \emph {et~al.} (\bibinfo {collaboration} {FASER}),\ }\bibfield
  {title} {\bibinfo {title} {{Search for dark photons with the FASER detector
  at the LHC}},\ }\href {https://doi.org/10.1016/j.physletb.2023.138378}
  {\bibfield  {journal} {\bibinfo  {journal} {Phys. Lett. B}\ }\textbf
  {\bibinfo {volume} {848}},\ \bibinfo {pages} {138378} (\bibinfo {year}
  {2024})},\ \Eprint {https://arxiv.org/abs/2308.05587} {arXiv:2308.05587
  [hep-ex]} \BibitemShut {NoStop}%
\bibitem [{\citenamefont {Ariga}\ \emph {et~al.}(2019)\citenamefont {Ariga}
  \emph {et~al.}}]{FASER:2018eoc}%
  \BibitemOpen
  \bibfield  {author} {\bibinfo {author} {\bibfnamefont {A.}~\bibnamefont
  {Ariga}} \emph {et~al.} (\bibinfo {collaboration} {FASER}),\ }\bibfield
  {title} {\bibinfo {title} {{FASER\textquoteright{}s physics reach for
  long-lived particles}},\ }\href {https://doi.org/10.1103/PhysRevD.99.095011}
  {\bibfield  {journal} {\bibinfo  {journal} {Phys. Rev. D}\ }\textbf {\bibinfo
  {volume} {99}},\ \bibinfo {pages} {095011} (\bibinfo {year} {2019})},\
  \Eprint {https://arxiv.org/abs/1811.12522} {arXiv:1811.12522 [hep-ph]}
  \BibitemShut {NoStop}%
\bibitem [{\citenamefont {Archilli}\ \emph {et~al.}(2012)\citenamefont
  {Archilli} \emph {et~al.}}]{KLOE-2:2011hhj}%
  \BibitemOpen
  \bibfield  {author} {\bibinfo {author} {\bibfnamefont {F.}~\bibnamefont
  {Archilli}} \emph {et~al.} (\bibinfo {collaboration} {KLOE-2}),\ }\bibfield
  {title} {\bibinfo {title} {{Search for a vector gauge boson in $\phi$ meson
  decays with the KLOE detector}},\ }\href
  {https://doi.org/10.1016/j.physletb.2011.11.033} {\bibfield  {journal}
  {\bibinfo  {journal} {Phys. Lett. B}\ }\textbf {\bibinfo {volume} {706}},\
  \bibinfo {pages} {251} (\bibinfo {year} {2012})},\ \Eprint
  {https://arxiv.org/abs/1110.0411} {arXiv:1110.0411 [hep-ex]} \BibitemShut
  {NoStop}%
\bibitem [{\citenamefont {Fujiwara}\ \emph {et~al.}(1985)\citenamefont
  {Fujiwara}, \citenamefont {Kugo}, \citenamefont {Terao}, \citenamefont
  {Uehara},\ and\ \citenamefont {Yamawaki}}]{Fujiwara:1984mp}%
  \BibitemOpen
  \bibfield  {author} {\bibinfo {author} {\bibfnamefont {T.}~\bibnamefont
  {Fujiwara}}, \bibinfo {author} {\bibfnamefont {T.}~\bibnamefont {Kugo}},
  \bibinfo {author} {\bibfnamefont {H.}~\bibnamefont {Terao}}, \bibinfo
  {author} {\bibfnamefont {S.}~\bibnamefont {Uehara}},\ and\ \bibinfo {author}
  {\bibfnamefont {K.}~\bibnamefont {Yamawaki}},\ }\bibfield  {title} {\bibinfo
  {title} {{Nonabelian Anomaly and Vector Mesons as Dynamical Gauge Bosons of
  Hidden Local Symmetries}},\ }\href {https://doi.org/10.1143/PTP.73.926}
  {\bibfield  {journal} {\bibinfo  {journal} {Prog. Theor. Phys.}\ }\textbf
  {\bibinfo {volume} {73}},\ \bibinfo {pages} {926} (\bibinfo {year}
  {1985})}\BibitemShut {NoStop}%
\bibitem [{\citenamefont {Tulin}(2014)}]{Tulin:2014tya}%
  \BibitemOpen
  \bibfield  {author} {\bibinfo {author} {\bibfnamefont {S.}~\bibnamefont
  {Tulin}},\ }\bibfield  {title} {\bibinfo {title} {{New weakly-coupled forces
  hidden in low-energy QCD}},\ }\href
  {https://doi.org/10.1103/PhysRevD.89.114008} {\bibfield  {journal} {\bibinfo
  {journal} {Phys. Rev. D}\ }\textbf {\bibinfo {volume} {89}},\ \bibinfo
  {pages} {114008} (\bibinfo {year} {2014})},\ \Eprint
  {https://arxiv.org/abs/1404.4370} {arXiv:1404.4370 [hep-ph]} \BibitemShut
  {NoStop}%
\bibitem [{\citenamefont {Bilmis}\ \emph {et~al.}(2015)\citenamefont {Bilmis},
  \citenamefont {Turan}, \citenamefont {Aliev}, \citenamefont {Deniz},
  \citenamefont {Singh},\ and\ \citenamefont {Wong}}]{Bilmis:2015lja}%
  \BibitemOpen
  \bibfield  {author} {\bibinfo {author} {\bibfnamefont {S.}~\bibnamefont
  {Bilmis}}, \bibinfo {author} {\bibfnamefont {I.}~\bibnamefont {Turan}},
  \bibinfo {author} {\bibfnamefont {T.~M.}\ \bibnamefont {Aliev}}, \bibinfo
  {author} {\bibfnamefont {M.}~\bibnamefont {Deniz}}, \bibinfo {author}
  {\bibfnamefont {L.}~\bibnamefont {Singh}},\ and\ \bibinfo {author}
  {\bibfnamefont {H.~T.}\ \bibnamefont {Wong}},\ }\bibfield  {title} {\bibinfo
  {title} {{Constraints on Dark Photon from Neutrino-Electron Scattering
  Experiments}},\ }\href {https://doi.org/10.1103/PhysRevD.92.033009}
  {\bibfield  {journal} {\bibinfo  {journal} {Phys. Rev. D}\ }\textbf {\bibinfo
  {volume} {92}},\ \bibinfo {pages} {033009} (\bibinfo {year} {2015})},\
  \Eprint {https://arxiv.org/abs/1502.07763} {arXiv:1502.07763 [hep-ph]}
  \BibitemShut {NoStop}%
\bibitem [{\citenamefont {Lindner}\ \emph {et~al.}(2018)\citenamefont
  {Lindner}, \citenamefont {Queiroz}, \citenamefont {Rodejohann},\ and\
  \citenamefont {Xu}}]{Lindner:2018kjo}%
  \BibitemOpen
  \bibfield  {author} {\bibinfo {author} {\bibfnamefont {M.}~\bibnamefont
  {Lindner}}, \bibinfo {author} {\bibfnamefont {F.~S.}\ \bibnamefont
  {Queiroz}}, \bibinfo {author} {\bibfnamefont {W.}~\bibnamefont
  {Rodejohann}},\ and\ \bibinfo {author} {\bibfnamefont {X.-J.}\ \bibnamefont
  {Xu}},\ }\bibfield  {title} {\bibinfo {title} {{Neutrino-electron scattering:
  general constraints on Z' and dark photon models}},\ }\href
  {https://doi.org/10.1007/JHEP05(2018)098} {\bibfield  {journal} {\bibinfo
  {journal} {JHEP}\ }\textbf {\bibinfo {volume} {05}},\ \bibinfo {pages}
  {098}},\ \Eprint {https://arxiv.org/abs/1803.00060} {arXiv:1803.00060
  [hep-ph]} \BibitemShut {NoStop}%
\bibitem [{\citenamefont {De~Romeri}\ \emph {et~al.}(2023)\citenamefont
  {De~Romeri}, \citenamefont {Miranda}, \citenamefont {Papoulias},
  \citenamefont {Sanchez~Garcia}, \citenamefont {T{\'o}rtola},\ and\
  \citenamefont {Valle}}]{DeRomeri:2022twg}%
  \BibitemOpen
  \bibfield  {author} {\bibinfo {author} {\bibfnamefont {V.}~\bibnamefont
  {De~Romeri}}, \bibinfo {author} {\bibfnamefont {O.~G.}\ \bibnamefont
  {Miranda}}, \bibinfo {author} {\bibfnamefont {D.~K.}\ \bibnamefont
  {Papoulias}}, \bibinfo {author} {\bibfnamefont {G.}~\bibnamefont
  {Sanchez~Garcia}}, \bibinfo {author} {\bibfnamefont {M.}~\bibnamefont
  {T{\'o}rtola}},\ and\ \bibinfo {author} {\bibfnamefont {J.~W.~F.}\
  \bibnamefont {Valle}},\ }\bibfield  {title} {\bibinfo {title} {{Physics
  implications of a combined analysis of COHERENT CsI and LAr data}},\ }\href
  {https://doi.org/10.1007/JHEP04(2023)035} {\bibfield  {journal} {\bibinfo
  {journal} {JHEP}\ }\textbf {\bibinfo {volume} {04}},\ \bibinfo {pages}
  {035}},\ \Eprint {https://arxiv.org/abs/2211.11905} {arXiv:2211.11905
  [hep-ph]} \BibitemShut {NoStop}%
\bibitem [{\citenamefont {et~al. (Particle
  Data~Group)}(2020)}]{10.1093/ptep/ptaa104}%
  \BibitemOpen
  \bibfield  {author} {\bibinfo {author} {\bibfnamefont {P.~Z.}\ \bibnamefont
  {et~al. (Particle Data~Group)}},\ }\bibfield  {title} {\bibinfo {title}
  {{Review of Particle Physics}},\ }\bibfield  {journal} {\bibinfo  {journal}
  {Progress of Theoretical and Experimental Physics}\ }\textbf {\bibinfo
  {volume} {2020}},\ \href {https://doi.org/10.1093/ptep/ptaa104}
  {10.1093/ptep/ptaa104} (\bibinfo {year} {2020}),\ \bibinfo {note} {083C01},\
  \Eprint
  {https://arxiv.org/abs/https://academic.oup.com/ptep/article-pdf/2020/8/083C01/34673722/ptaa104.pdf}
  {https://academic.oup.com/ptep/article-pdf/2020/8/083C01/34673722/ptaa104.pdf}
  \BibitemShut {NoStop}%
\bibitem [{\citenamefont {Ezhela}\ \emph {et~al.}(2003)\citenamefont {Ezhela},
  \citenamefont {Lugovsky},\ and\ \citenamefont {Zenin}}]{Ezhela:2003pp}%
  \BibitemOpen
  \bibfield  {author} {\bibinfo {author} {\bibfnamefont {V.~V.}\ \bibnamefont
  {Ezhela}}, \bibinfo {author} {\bibfnamefont {S.~B.}\ \bibnamefont
  {Lugovsky}},\ and\ \bibinfo {author} {\bibfnamefont {O.~V.}\ \bibnamefont
  {Zenin}},\ }\bibfield  {title} {\bibinfo {title} {{Hadronic part of the muon
  g-2 estimated on the $\sigma$**2003(tot)($e^+ e^- \rightarrow$ hadrons)
  evaluated data compilation}},\ }\href@noop {} {\  (\bibinfo {year} {2003})},\
  \Eprint {https://arxiv.org/abs/hep-ph/0312114} {arXiv:hep-ph/0312114}
  \BibitemShut {NoStop}%
\bibitem [{\citenamefont {Lees}\ \emph
  {et~al.}(2012{\natexlab{a}})\citenamefont {Lees} \emph
  {et~al.}}]{BaBar:2012bdw}%
  \BibitemOpen
  \bibfield  {author} {\bibinfo {author} {\bibfnamefont {J.~P.}\ \bibnamefont
  {Lees}} \emph {et~al.} (\bibinfo {collaboration} {BaBar}),\ }\bibfield
  {title} {\bibinfo {title} {{Precise Measurement of the $e^+ e^- \to
  \pi^+\pi^- (\gamma)$ Cross Section with the Initial-State Radiation Method at
  BABAR}},\ }\href {https://doi.org/10.1103/PhysRevD.86.032013} {\bibfield
  {journal} {\bibinfo  {journal} {Phys. Rev. D}\ }\textbf {\bibinfo {volume}
  {86}},\ \bibinfo {pages} {032013} (\bibinfo {year} {2012}{\natexlab{a}})},\
  \Eprint {https://arxiv.org/abs/1205.2228} {arXiv:1205.2228 [hep-ex]}
  \BibitemShut {NoStop}%
\bibitem [{\citenamefont {Aubert}\ \emph {et~al.}(2004)\citenamefont {Aubert}
  \emph {et~al.}}]{BaBar:2004ytv}%
  \BibitemOpen
  \bibfield  {author} {\bibinfo {author} {\bibfnamefont {B.}~\bibnamefont
  {Aubert}} \emph {et~al.} (\bibinfo {collaboration} {BaBar}),\ }\bibfield
  {title} {\bibinfo {title} {{Study of $e^+e^- \to \pi^+ \pi^- \pi^0$ process
  using initial state radiation with BaBar}},\ }\href
  {https://doi.org/10.1103/PhysRevD.70.072004} {\bibfield  {journal} {\bibinfo
  {journal} {Phys. Rev. D}\ }\textbf {\bibinfo {volume} {70}},\ \bibinfo
  {pages} {072004} (\bibinfo {year} {2004})},\ \Eprint
  {https://arxiv.org/abs/hep-ex/0408078} {arXiv:hep-ex/0408078} \BibitemShut
  {NoStop}%
\bibitem [{\citenamefont {Lees}\ \emph {et~al.}(2013)\citenamefont {Lees} \emph
  {et~al.}}]{BaBar:2013jqz}%
  \BibitemOpen
  \bibfield  {author} {\bibinfo {author} {\bibfnamefont {J.~P.}\ \bibnamefont
  {Lees}} \emph {et~al.} (\bibinfo {collaboration} {BaBar}),\ }\bibfield
  {title} {\bibinfo {title} {{Precision measurement of the $e^+e^- →
  K^+K^-(\gamma)$ cross section with the initial-state radiation method at
  BABAR}},\ }\href {https://doi.org/10.1103/PhysRevD.88.032013} {\bibfield
  {journal} {\bibinfo  {journal} {Phys. Rev. D}\ }\textbf {\bibinfo {volume}
  {88}},\ \bibinfo {pages} {032013} (\bibinfo {year} {2013})},\ \Eprint
  {https://arxiv.org/abs/1306.3600} {arXiv:1306.3600 [hep-ex]} \BibitemShut
  {NoStop}%
\bibitem [{\citenamefont {Aubert}\ \emph {et~al.}(2008)\citenamefont {Aubert}
  \emph {et~al.}}]{BaBar:2007ceh}%
  \BibitemOpen
  \bibfield  {author} {\bibinfo {author} {\bibfnamefont {B.}~\bibnamefont
  {Aubert}} \emph {et~al.} (\bibinfo {collaboration} {BaBar}),\ }\bibfield
  {title} {\bibinfo {title} {{Measurements of $e^{+} e^{-} \to K^{+} K^{-}
  \eta$, $K^{+} K^{-} \pi^0$ and $K^0_{s} K^\pm \pi^\mp$ cross- sections using
  initial state radiation events}},\ }\href
  {https://doi.org/10.1103/PhysRevD.77.092002} {\bibfield  {journal} {\bibinfo
  {journal} {Phys. Rev. D}\ }\textbf {\bibinfo {volume} {77}},\ \bibinfo
  {pages} {092002} (\bibinfo {year} {2008})},\ \Eprint
  {https://arxiv.org/abs/0710.4451} {arXiv:0710.4451 [hep-ex]} \BibitemShut
  {NoStop}%
\bibitem [{\citenamefont {Lees}\ \emph
  {et~al.}(2012{\natexlab{b}})\citenamefont {Lees} \emph
  {et~al.}}]{BaBar:2012sxt}%
  \BibitemOpen
  \bibfield  {author} {\bibinfo {author} {\bibfnamefont {J.~P.}\ \bibnamefont
  {Lees}} \emph {et~al.} (\bibinfo {collaboration} {BaBar}),\ }\bibfield
  {title} {\bibinfo {title} {{Initial-State Radiation Measurement of the
  $e^+e^- -> \pi^+\pi^-\pi^+\pi^-$ Cross Section}},\ }\href
  {https://doi.org/10.1103/PhysRevD.85.112009} {\bibfield  {journal} {\bibinfo
  {journal} {Phys. Rev. D}\ }\textbf {\bibinfo {volume} {85}},\ \bibinfo
  {pages} {112009} (\bibinfo {year} {2012}{\natexlab{b}})},\ \Eprint
  {https://arxiv.org/abs/1201.5677} {arXiv:1201.5677 [hep-ex]} \BibitemShut
  {NoStop}%
\bibitem [{\citenamefont {Lees}\ \emph
  {et~al.}(2017{\natexlab{a}})\citenamefont {Lees} \emph
  {et~al.}}]{BaBar:2017zmc}%
  \BibitemOpen
  \bibfield  {author} {\bibinfo {author} {\bibfnamefont {J.~P.}\ \bibnamefont
  {Lees}} \emph {et~al.} (\bibinfo {collaboration} {BaBar}),\ }\bibfield
  {title} {\bibinfo {title} {{Measurement of the
  ${e}^{+}{e}^{{-}}{\rightarrow}{{\pi}}^{+}{{\pi}}^{{-}}{{\pi}}^{0}{{\pi}}^{0}$
  cross section using initial-state radiation at BABAR}},\ }\href
  {https://doi.org/10.1103/PhysRevD.96.092009} {\bibfield  {journal} {\bibinfo
  {journal} {Phys. Rev. D}\ }\textbf {\bibinfo {volume} {96}},\ \bibinfo
  {pages} {092009} (\bibinfo {year} {2017}{\natexlab{a}})},\ \Eprint
  {https://arxiv.org/abs/1709.01171} {arXiv:1709.01171 [hep-ex]} \BibitemShut
  {NoStop}%
\bibitem [{\citenamefont {Achasov}\ \emph {et~al.}(2003)\citenamefont {Achasov}
  \emph {et~al.}}]{Achasov:2003ir}%
  \BibitemOpen
  \bibfield  {author} {\bibinfo {author} {\bibfnamefont {M.~N.}\ \bibnamefont
  {Achasov}} \emph {et~al.},\ }\bibfield  {title} {\bibinfo {title} {{Study of
  the process e+ e- ---{\ensuremath{>}} pi+ pi- pi0 in the energy region
  s**(1/2) below 0.98-GeV}},\ }\href
  {https://doi.org/10.1103/PhysRevD.68.052006} {\bibfield  {journal} {\bibinfo
  {journal} {Phys. Rev. D}\ }\textbf {\bibinfo {volume} {68}},\ \bibinfo
  {pages} {052006} (\bibinfo {year} {2003})},\ \Eprint
  {https://arxiv.org/abs/hep-ex/0305049} {arXiv:hep-ex/0305049} \BibitemShut
  {NoStop}%
\bibitem [{\citenamefont {Achasov}\ \emph {et~al.}(2002)\citenamefont {Achasov}
  \emph {et~al.}}]{Achasov:2002ud}%
  \BibitemOpen
  \bibfield  {author} {\bibinfo {author} {\bibfnamefont {M.~N.}\ \bibnamefont
  {Achasov}} \emph {et~al.},\ }\bibfield  {title} {\bibinfo {title} {{Study of
  the process e+ e- ---{\ensuremath{>}} pi+ pi- pi0 in the energy region
  s**(1/2) from 0.98-GeV to 1.38-GeV}},\ }\href
  {https://doi.org/10.1103/PhysRevD.66.032001} {\bibfield  {journal} {\bibinfo
  {journal} {Phys. Rev. D}\ }\textbf {\bibinfo {volume} {66}},\ \bibinfo
  {pages} {032001} (\bibinfo {year} {2002})},\ \Eprint
  {https://arxiv.org/abs/hep-ex/0201040} {arXiv:hep-ex/0201040} \BibitemShut
  {NoStop}%
\bibitem [{\citenamefont {Sirunyan}\ \emph {et~al.}(2020)\citenamefont
  {Sirunyan} \emph {et~al.}}]{CMS:2019buh}%
  \BibitemOpen
  \bibfield  {author} {\bibinfo {author} {\bibfnamefont {A.~M.}\ \bibnamefont
  {Sirunyan}} \emph {et~al.} (\bibinfo {collaboration} {CMS}),\ }\bibfield
  {title} {\bibinfo {title} {{Search for a Narrow Resonance Lighter than 200
  GeV Decaying to a Pair of Muons in Proton-Proton Collisions at $\sqrt{s} =13$
  TeV}},\ }\href {https://doi.org/10.1103/PhysRevLett.124.131802} {\bibfield
  {journal} {\bibinfo  {journal} {Phys. Rev. Lett.}\ }\textbf {\bibinfo
  {volume} {124}},\ \bibinfo {pages} {131802} (\bibinfo {year} {2020})},\
  \Eprint {https://arxiv.org/abs/1912.04776} {arXiv:1912.04776 [hep-ex]}
  \BibitemShut {NoStop}%
\bibitem [{\citenamefont {Aghanim}\ \emph
  {et~al.}(2020{\natexlab{a}})\citenamefont {Aghanim} \emph
  {et~al.}}]{Planck:2018nkj}%
  \BibitemOpen
  \bibfield  {author} {\bibinfo {author} {\bibfnamefont {N.}~\bibnamefont
  {Aghanim}} \emph {et~al.} (\bibinfo {collaboration} {Planck}),\ }\bibfield
  {title} {\bibinfo {title} {{Planck 2018 results. I. Overview and the
  cosmological legacy of Planck}},\ }\href
  {https://doi.org/10.1051/0004-6361/201833880} {\bibfield  {journal} {\bibinfo
   {journal} {Astron. Astrophys.}\ }\textbf {\bibinfo {volume} {641}},\
  \bibinfo {pages} {A1} (\bibinfo {year} {2020}{\natexlab{a}})},\ \Eprint
  {https://arxiv.org/abs/1807.06205} {arXiv:1807.06205 [astro-ph.CO]}
  \BibitemShut {NoStop}%
\bibitem [{\citenamefont {Aghanim}\ \emph
  {et~al.}(2020{\natexlab{b}})\citenamefont {Aghanim} \emph
  {et~al.}}]{Planck:2018vyg}%
  \BibitemOpen
  \bibfield  {author} {\bibinfo {author} {\bibfnamefont {N.}~\bibnamefont
  {Aghanim}} \emph {et~al.} (\bibinfo {collaboration} {Planck}),\ }\bibfield
  {title} {\bibinfo {title} {{Planck 2018 results. VI. Cosmological
  parameters}},\ }\href {https://doi.org/10.1051/0004-6361/201833910}
  {\bibfield  {journal} {\bibinfo  {journal} {Astron. Astrophys.}\ }\textbf
  {\bibinfo {volume} {641}},\ \bibinfo {pages} {A6} (\bibinfo {year}
  {2020}{\natexlab{b}})},\ \bibinfo {note} {[Erratum: Astron.Astrophys. 652, C4
  (2021)]},\ \Eprint {https://arxiv.org/abs/1807.06209} {arXiv:1807.06209
  [astro-ph.CO]} \BibitemShut {NoStop}%
\bibitem [{\citenamefont {Calabrese}\ \emph {et~al.}(2025)\citenamefont
  {Calabrese} \emph {et~al.}}]{AtacamaCosmologyTelescope:2025nti}%
  \BibitemOpen
  \bibfield  {author} {\bibinfo {author} {\bibfnamefont {E.}~\bibnamefont
  {Calabrese}} \emph {et~al.} (\bibinfo {collaboration} {Atacama Cosmology
  Telescope}),\ }\bibfield  {title} {\bibinfo {title} {{The Atacama Cosmology
  Telescope: DR6 constraints on extended cosmological models}},\ }\href
  {https://doi.org/10.1088/1475-7516/2025/11/063} {\bibfield  {journal}
  {\bibinfo  {journal} {JCAP}\ }\textbf {\bibinfo {volume} {11}},\ \bibinfo
  {pages} {063}},\ \Eprint {https://arxiv.org/abs/2503.14454} {arXiv:2503.14454
  [astro-ph.CO]} \BibitemShut {NoStop}%
\bibitem [{\citenamefont {Cyburt}\ \emph {et~al.}(2016)\citenamefont {Cyburt},
  \citenamefont {Fields}, \citenamefont {Olive},\ and\ \citenamefont
  {Yeh}}]{Cyburt:2015mya}%
  \BibitemOpen
  \bibfield  {author} {\bibinfo {author} {\bibfnamefont {R.~H.}\ \bibnamefont
  {Cyburt}}, \bibinfo {author} {\bibfnamefont {B.~D.}\ \bibnamefont {Fields}},
  \bibinfo {author} {\bibfnamefont {K.~A.}\ \bibnamefont {Olive}},\ and\
  \bibinfo {author} {\bibfnamefont {T.-H.}\ \bibnamefont {Yeh}},\ }\bibfield
  {title} {\bibinfo {title} {{Big Bang Nucleosynthesis: 2015}},\ }\href
  {https://doi.org/10.1103/RevModPhys.88.015004} {\bibfield  {journal}
  {\bibinfo  {journal} {Rev. Mod. Phys.}\ }\textbf {\bibinfo {volume} {88}},\
  \bibinfo {pages} {015004} (\bibinfo {year} {2016})},\ \Eprint
  {https://arxiv.org/abs/1505.01076} {arXiv:1505.01076 [astro-ph.CO]}
  \BibitemShut {NoStop}%
\bibitem [{\citenamefont {Yeh}\ \emph {et~al.}(2021)\citenamefont {Yeh},
  \citenamefont {Olive},\ and\ \citenamefont {Fields}}]{Yeh:2020mgl}%
  \BibitemOpen
  \bibfield  {author} {\bibinfo {author} {\bibfnamefont {T.-H.}\ \bibnamefont
  {Yeh}}, \bibinfo {author} {\bibfnamefont {K.~A.}\ \bibnamefont {Olive}},\
  and\ \bibinfo {author} {\bibfnamefont {B.~D.}\ \bibnamefont {Fields}},\
  }\bibfield  {title} {\bibinfo {title} {{The impact of new $d(p,\gamma)$3
  rates on Big Bang Nucleosynthesis}},\ }\href
  {https://doi.org/10.1088/1475-7516/2021/03/046} {\bibfield  {journal}
  {\bibinfo  {journal} {JCAP}\ }\textbf {\bibinfo {volume} {03}},\ \bibinfo
  {pages} {046}},\ \Eprint {https://arxiv.org/abs/2011.13874} {arXiv:2011.13874
  [astro-ph.CO]} \BibitemShut {NoStop}%
\bibitem [{\citenamefont {Yeh}\ \emph {et~al.}(2022)\citenamefont {Yeh},
  \citenamefont {Shelton}, \citenamefont {Olive},\ and\ \citenamefont
  {Fields}}]{Yeh:2022heq}%
  \BibitemOpen
  \bibfield  {author} {\bibinfo {author} {\bibfnamefont {T.-H.}\ \bibnamefont
  {Yeh}}, \bibinfo {author} {\bibfnamefont {J.}~\bibnamefont {Shelton}},
  \bibinfo {author} {\bibfnamefont {K.~A.}\ \bibnamefont {Olive}},\ and\
  \bibinfo {author} {\bibfnamefont {B.~D.}\ \bibnamefont {Fields}},\ }\bibfield
   {title} {\bibinfo {title} {{Probing physics beyond the standard model:
  limits from BBN and the CMB independently and combined}},\ }\href
  {https://doi.org/10.1088/1475-7516/2022/10/046} {\bibfield  {journal}
  {\bibinfo  {journal} {JCAP}\ }\textbf {\bibinfo {volume} {10}},\ \bibinfo
  {pages} {046}},\ \Eprint {https://arxiv.org/abs/2207.13133} {arXiv:2207.13133
  [astro-ph.CO]} \BibitemShut {NoStop}%
\bibitem [{\citenamefont {Workman}\ \emph {et~al.}(2022)\citenamefont {Workman}
  \emph {et~al.}}]{Workman:2022ynf}%
  \BibitemOpen
  \bibfield  {author} {\bibinfo {author} {\bibfnamefont {R.~L.}\ \bibnamefont
  {Workman}} \emph {et~al.} (\bibinfo {collaboration} {Particle Data Group}),\
  }\bibfield  {title} {\bibinfo {title} {{Review of Particle Physics}},\ }\href
  {https://doi.org/10.1093/ptep/ptac097} {\bibfield  {journal} {\bibinfo
  {journal} {PTEP}\ }\textbf {\bibinfo {volume} {2022}},\ \bibinfo {pages}
  {083C01} (\bibinfo {year} {2022})}\BibitemShut {NoStop}%
\bibitem [{\citenamefont {de~Salas}\ and\ \citenamefont
  {Pastor}(2016)}]{deSalas:2016ztq}%
  \BibitemOpen
  \bibfield  {author} {\bibinfo {author} {\bibfnamefont {P.~F.}\ \bibnamefont
  {de~Salas}}\ and\ \bibinfo {author} {\bibfnamefont {S.}~\bibnamefont
  {Pastor}},\ }\bibfield  {title} {\bibinfo {title} {{Relic neutrino decoupling
  with flavour oscillations revisited}},\ }\href
  {https://doi.org/10.1088/1475-7516/2016/07/051} {\bibfield  {journal}
  {\bibinfo  {journal} {JCAP}\ }\textbf {\bibinfo {volume} {07}},\ \bibinfo
  {pages} {051}},\ \Eprint {https://arxiv.org/abs/1606.06986} {arXiv:1606.06986
  [hep-ph]} \BibitemShut {NoStop}%
\bibitem [{\citenamefont {Mangano}\ \emph {et~al.}(2005)\citenamefont
  {Mangano}, \citenamefont {Miele}, \citenamefont {Pastor}, \citenamefont
  {Pinto}, \citenamefont {Pisanti},\ and\ \citenamefont
  {Serpico}}]{Mangano:2005cc}%
  \BibitemOpen
  \bibfield  {author} {\bibinfo {author} {\bibfnamefont {G.}~\bibnamefont
  {Mangano}}, \bibinfo {author} {\bibfnamefont {G.}~\bibnamefont {Miele}},
  \bibinfo {author} {\bibfnamefont {S.}~\bibnamefont {Pastor}}, \bibinfo
  {author} {\bibfnamefont {T.}~\bibnamefont {Pinto}}, \bibinfo {author}
  {\bibfnamefont {O.}~\bibnamefont {Pisanti}},\ and\ \bibinfo {author}
  {\bibfnamefont {P.~D.}\ \bibnamefont {Serpico}},\ }\bibfield  {title}
  {\bibinfo {title} {{Relic neutrino decoupling including flavor
  oscillations}},\ }\href {https://doi.org/10.1016/j.nuclphysb.2005.09.041}
  {\bibfield  {journal} {\bibinfo  {journal} {Nucl. Phys. B}\ }\textbf
  {\bibinfo {volume} {729}},\ \bibinfo {pages} {221} (\bibinfo {year}
  {2005})},\ \Eprint {https://arxiv.org/abs/hep-ph/0506164}
  {arXiv:hep-ph/0506164} \BibitemShut {NoStop}%
\bibitem [{\citenamefont {Dolgov}(2002)}]{Dolgov:2002wy}%
  \BibitemOpen
  \bibfield  {author} {\bibinfo {author} {\bibfnamefont {A.~D.}\ \bibnamefont
  {Dolgov}},\ }\bibfield  {title} {\bibinfo {title} {{Neutrinos in
  cosmology}},\ }\href {https://doi.org/10.1016/S0370-1573(02)00139-4}
  {\bibfield  {journal} {\bibinfo  {journal} {Phys. Rept.}\ }\textbf {\bibinfo
  {volume} {370}},\ \bibinfo {pages} {333} (\bibinfo {year} {2002})},\ \Eprint
  {https://arxiv.org/abs/hep-ph/0202122} {arXiv:hep-ph/0202122} \BibitemShut
  {NoStop}%
\bibitem [{\citenamefont {Dicus}\ \emph {et~al.}(1982)\citenamefont {Dicus},
  \citenamefont {Kolb}, \citenamefont {Gleeson}, \citenamefont {Sudarshan},
  \citenamefont {Teplitz},\ and\ \citenamefont {Turner}}]{Dicus:1982bz}%
  \BibitemOpen
  \bibfield  {author} {\bibinfo {author} {\bibfnamefont {D.~A.}\ \bibnamefont
  {Dicus}}, \bibinfo {author} {\bibfnamefont {E.~W.}\ \bibnamefont {Kolb}},
  \bibinfo {author} {\bibfnamefont {A.~M.}\ \bibnamefont {Gleeson}}, \bibinfo
  {author} {\bibfnamefont {E.~C.~G.}\ \bibnamefont {Sudarshan}}, \bibinfo
  {author} {\bibfnamefont {V.~L.}\ \bibnamefont {Teplitz}},\ and\ \bibinfo
  {author} {\bibfnamefont {M.~S.}\ \bibnamefont {Turner}},\ }\bibfield  {title}
  {\bibinfo {title} {{Primordial Nucleosynthesis Including Radiative, Coulomb,
  and Finite Temperature Corrections to Weak Rates}},\ }\href
  {https://doi.org/10.1103/PhysRevD.26.2694} {\bibfield  {journal} {\bibinfo
  {journal} {Phys. Rev. D}\ }\textbf {\bibinfo {volume} {26}},\ \bibinfo
  {pages} {2694} (\bibinfo {year} {1982})}\BibitemShut {NoStop}%
\bibitem [{\citenamefont {Hannestad}\ and\ \citenamefont
  {Madsen}(1995)}]{Hannestad:1995rs}%
  \BibitemOpen
  \bibfield  {author} {\bibinfo {author} {\bibfnamefont {S.}~\bibnamefont
  {Hannestad}}\ and\ \bibinfo {author} {\bibfnamefont {J.}~\bibnamefont
  {Madsen}},\ }\bibfield  {title} {\bibinfo {title} {{Neutrino decoupling in
  the early universe}},\ }\href {https://doi.org/10.1103/PhysRevD.52.1764}
  {\bibfield  {journal} {\bibinfo  {journal} {Phys. Rev. D}\ }\textbf {\bibinfo
  {volume} {52}},\ \bibinfo {pages} {1764} (\bibinfo {year} {1995})},\ \Eprint
  {https://arxiv.org/abs/astro-ph/9506015} {arXiv:astro-ph/9506015}
  \BibitemShut {NoStop}%
\bibitem [{\citenamefont {Dodelson}\ and\ \citenamefont
  {Turner}(1992)}]{Dodelson:1992km}%
  \BibitemOpen
  \bibfield  {author} {\bibinfo {author} {\bibfnamefont {S.}~\bibnamefont
  {Dodelson}}\ and\ \bibinfo {author} {\bibfnamefont {M.~S.}\ \bibnamefont
  {Turner}},\ }\bibfield  {title} {\bibinfo {title} {{Nonequilibrium neutrino
  statistical mechanics in the expanding universe}},\ }\href
  {https://doi.org/10.1103/PhysRevD.46.3372} {\bibfield  {journal} {\bibinfo
  {journal} {Phys. Rev. D}\ }\textbf {\bibinfo {volume} {46}},\ \bibinfo
  {pages} {3372} (\bibinfo {year} {1992})}\BibitemShut {NoStop}%
\bibitem [{\citenamefont {Dolgov}\ \emph {et~al.}(1997)\citenamefont {Dolgov},
  \citenamefont {Hansen},\ and\ \citenamefont {Semikoz}}]{Dolgov:1997mb}%
  \BibitemOpen
  \bibfield  {author} {\bibinfo {author} {\bibfnamefont {A.~D.}\ \bibnamefont
  {Dolgov}}, \bibinfo {author} {\bibfnamefont {S.~H.}\ \bibnamefont {Hansen}},\
  and\ \bibinfo {author} {\bibfnamefont {D.~V.}\ \bibnamefont {Semikoz}},\
  }\bibfield  {title} {\bibinfo {title} {{Nonequilibrium corrections to the
  spectra of massless neutrinos in the early universe}},\ }\href
  {https://doi.org/10.1016/S0550-3213(97)00479-3} {\bibfield  {journal}
  {\bibinfo  {journal} {Nucl. Phys. B}\ }\textbf {\bibinfo {volume} {503}},\
  \bibinfo {pages} {426} (\bibinfo {year} {1997})},\ \Eprint
  {https://arxiv.org/abs/hep-ph/9703315} {arXiv:hep-ph/9703315} \BibitemShut
  {NoStop}%
\bibitem [{\citenamefont {Esposito}\ \emph {et~al.}(2000)\citenamefont
  {Esposito}, \citenamefont {Miele}, \citenamefont {Pastor}, \citenamefont
  {Peloso},\ and\ \citenamefont {Pisanti}}]{Esposito:2000hi}%
  \BibitemOpen
  \bibfield  {author} {\bibinfo {author} {\bibfnamefont {S.}~\bibnamefont
  {Esposito}}, \bibinfo {author} {\bibfnamefont {G.}~\bibnamefont {Miele}},
  \bibinfo {author} {\bibfnamefont {S.}~\bibnamefont {Pastor}}, \bibinfo
  {author} {\bibfnamefont {M.}~\bibnamefont {Peloso}},\ and\ \bibinfo {author}
  {\bibfnamefont {O.}~\bibnamefont {Pisanti}},\ }\bibfield  {title} {\bibinfo
  {title} {{Nonequilibrium spectra of degenerate relic neutrinos}},\ }\href
  {https://doi.org/10.1016/S0550-3213(00)00554-X} {\bibfield  {journal}
  {\bibinfo  {journal} {Nucl. Phys. B}\ }\textbf {\bibinfo {volume} {590}},\
  \bibinfo {pages} {539} (\bibinfo {year} {2000})},\ \Eprint
  {https://arxiv.org/abs/astro-ph/0005573} {arXiv:astro-ph/0005573}
  \BibitemShut {NoStop}%
\bibitem [{\citenamefont {Mangano}\ \emph {et~al.}(2002)\citenamefont
  {Mangano}, \citenamefont {Miele}, \citenamefont {Pastor},\ and\ \citenamefont
  {Peloso}}]{Mangano:2001iu}%
  \BibitemOpen
  \bibfield  {author} {\bibinfo {author} {\bibfnamefont {G.}~\bibnamefont
  {Mangano}}, \bibinfo {author} {\bibfnamefont {G.}~\bibnamefont {Miele}},
  \bibinfo {author} {\bibfnamefont {S.}~\bibnamefont {Pastor}},\ and\ \bibinfo
  {author} {\bibfnamefont {M.}~\bibnamefont {Peloso}},\ }\bibfield  {title}
  {\bibinfo {title} {{A Precision calculation of the effective number of
  cosmological neutrinos}},\ }\href
  {https://doi.org/10.1016/S0370-2693(02)01622-2} {\bibfield  {journal}
  {\bibinfo  {journal} {Phys. Lett. B}\ }\textbf {\bibinfo {volume} {534}},\
  \bibinfo {pages} {8} (\bibinfo {year} {2002})},\ \Eprint
  {https://arxiv.org/abs/astro-ph/0111408} {arXiv:astro-ph/0111408}
  \BibitemShut {NoStop}%
\bibitem [{\citenamefont {Birrell}\ \emph {et~al.}(2014)\citenamefont
  {Birrell}, \citenamefont {Yang},\ and\ \citenamefont
  {Rafelski}}]{Birrell:2014uka}%
  \BibitemOpen
  \bibfield  {author} {\bibinfo {author} {\bibfnamefont {J.}~\bibnamefont
  {Birrell}}, \bibinfo {author} {\bibfnamefont {C.-T.}\ \bibnamefont {Yang}},\
  and\ \bibinfo {author} {\bibfnamefont {J.}~\bibnamefont {Rafelski}},\
  }\bibfield  {title} {\bibinfo {title} {{Relic Neutrino Freeze-out: Dependence
  on Natural Constants}},\ }\href
  {https://doi.org/10.1016/j.nuclphysb.2014.11.020} {\bibfield  {journal}
  {\bibinfo  {journal} {Nucl. Phys. B}\ }\textbf {\bibinfo {volume} {890}},\
  \bibinfo {pages} {481} (\bibinfo {year} {2014})},\ \Eprint
  {https://arxiv.org/abs/1406.1759} {arXiv:1406.1759 [nucl-th]} \BibitemShut
  {NoStop}%
\bibitem [{\citenamefont {Grohs}\ \emph {et~al.}(2016)\citenamefont {Grohs},
  \citenamefont {Fuller}, \citenamefont {Kishimoto}, \citenamefont {Paris},\
  and\ \citenamefont {Vlasenko}}]{Grohs:2015tfy}%
  \BibitemOpen
  \bibfield  {author} {\bibinfo {author} {\bibfnamefont {E.}~\bibnamefont
  {Grohs}}, \bibinfo {author} {\bibfnamefont {G.~M.}\ \bibnamefont {Fuller}},
  \bibinfo {author} {\bibfnamefont {C.~T.}\ \bibnamefont {Kishimoto}}, \bibinfo
  {author} {\bibfnamefont {M.~W.}\ \bibnamefont {Paris}},\ and\ \bibinfo
  {author} {\bibfnamefont {A.}~\bibnamefont {Vlasenko}},\ }\bibfield  {title}
  {\bibinfo {title} {{Neutrino energy transport in weak decoupling and big bang
  nucleosynthesis}},\ }\href {https://doi.org/10.1103/PhysRevD.93.083522}
  {\bibfield  {journal} {\bibinfo  {journal} {Phys. Rev. D}\ }\textbf {\bibinfo
  {volume} {93}},\ \bibinfo {pages} {083522} (\bibinfo {year} {2016})},\
  \Eprint {https://arxiv.org/abs/1512.02205} {arXiv:1512.02205 [astro-ph.CO]}
  \BibitemShut {NoStop}%
\bibitem [{\citenamefont {Cielo}\ \emph {et~al.}(2023)\citenamefont {Cielo},
  \citenamefont {Escudero}, \citenamefont {Mangano},\ and\ \citenamefont
  {Pisanti}}]{Cielo:2023bqp}%
  \BibitemOpen
  \bibfield  {author} {\bibinfo {author} {\bibfnamefont {M.}~\bibnamefont
  {Cielo}}, \bibinfo {author} {\bibfnamefont {M.}~\bibnamefont {Escudero}},
  \bibinfo {author} {\bibfnamefont {G.}~\bibnamefont {Mangano}},\ and\ \bibinfo
  {author} {\bibfnamefont {O.}~\bibnamefont {Pisanti}},\ }\bibfield  {title}
  {\bibinfo {title} {{Neff in the Standard Model at NLO is 3.043}},\
  }\href@noop {} {\  (\bibinfo {year} {2023})},\ \Eprint
  {https://arxiv.org/abs/2306.05460} {arXiv:2306.05460 [hep-ph]} \BibitemShut
  {NoStop}%
\bibitem [{\citenamefont {Akita}\ and\ \citenamefont
  {Yamaguchi}(2020)}]{Akita:2020szl}%
  \BibitemOpen
  \bibfield  {author} {\bibinfo {author} {\bibfnamefont {K.}~\bibnamefont
  {Akita}}\ and\ \bibinfo {author} {\bibfnamefont {M.}~\bibnamefont
  {Yamaguchi}},\ }\bibfield  {title} {\bibinfo {title} {{A precision
  calculation of relic neutrino decoupling}},\ }\href
  {https://doi.org/10.1088/1475-7516/2020/08/012} {\bibfield  {journal}
  {\bibinfo  {journal} {JCAP}\ }\textbf {\bibinfo {volume} {08}},\ \bibinfo
  {pages} {012}},\ \Eprint {https://arxiv.org/abs/2005.07047} {arXiv:2005.07047
  [hep-ph]} \BibitemShut {NoStop}%
\bibitem [{\citenamefont {Froustey}\ \emph {et~al.}(2020)\citenamefont
  {Froustey}, \citenamefont {Pitrou},\ and\ \citenamefont
  {Volpe}}]{Froustey:2020mcq}%
  \BibitemOpen
  \bibfield  {author} {\bibinfo {author} {\bibfnamefont {J.}~\bibnamefont
  {Froustey}}, \bibinfo {author} {\bibfnamefont {C.}~\bibnamefont {Pitrou}},\
  and\ \bibinfo {author} {\bibfnamefont {M.~C.}\ \bibnamefont {Volpe}},\
  }\bibfield  {title} {\bibinfo {title} {{Neutrino decoupling including flavour
  oscillations and primordial nucleosynthesis}},\ }\href
  {https://doi.org/10.1088/1475-7516/2020/12/015} {\bibfield  {journal}
  {\bibinfo  {journal} {JCAP}\ }\textbf {\bibinfo {volume} {12}},\ \bibinfo
  {pages} {015}},\ \Eprint {https://arxiv.org/abs/2008.01074} {arXiv:2008.01074
  [hep-ph]} \BibitemShut {NoStop}%
\bibitem [{\citenamefont {Bennett}\ \emph {et~al.}(2021)\citenamefont
  {Bennett}, \citenamefont {Buldgen}, \citenamefont {De~Salas}, \citenamefont
  {Drewes}, \citenamefont {Gariazzo}, \citenamefont {Pastor},\ and\
  \citenamefont {Wong}}]{Bennett:2020zkv}%
  \BibitemOpen
  \bibfield  {author} {\bibinfo {author} {\bibfnamefont {J.~J.}\ \bibnamefont
  {Bennett}}, \bibinfo {author} {\bibfnamefont {G.}~\bibnamefont {Buldgen}},
  \bibinfo {author} {\bibfnamefont {P.~F.}\ \bibnamefont {De~Salas}}, \bibinfo
  {author} {\bibfnamefont {M.}~\bibnamefont {Drewes}}, \bibinfo {author}
  {\bibfnamefont {S.}~\bibnamefont {Gariazzo}}, \bibinfo {author}
  {\bibfnamefont {S.}~\bibnamefont {Pastor}},\ and\ \bibinfo {author}
  {\bibfnamefont {Y.~Y.~Y.}\ \bibnamefont {Wong}},\ }\bibfield  {title}
  {\bibinfo {title} {{Towards a precision calculation of $N_{\rm eff}$ in the
  Standard Model II: Neutrino decoupling in the presence of flavour
  oscillations and finite-temperature QED}},\ }\href
  {https://doi.org/10.1088/1475-7516/2021/04/073} {\bibfield  {journal}
  {\bibinfo  {journal} {JCAP}\ }\textbf {\bibinfo {volume} {04}},\ \bibinfo
  {pages} {073}},\ \Eprint {https://arxiv.org/abs/2012.02726} {arXiv:2012.02726
  [hep-ph]} \BibitemShut {NoStop}%
\bibitem [{\citenamefont {Sarkar}(1996)}]{Sarkar:1995dd}%
  \BibitemOpen
  \bibfield  {author} {\bibinfo {author} {\bibfnamefont {S.}~\bibnamefont
  {Sarkar}},\ }\bibfield  {title} {\bibinfo {title} {{Big bang nucleosynthesis
  and physics beyond the standard model}},\ }\href
  {https://doi.org/10.1088/0034-4885/59/12/001} {\bibfield  {journal} {\bibinfo
   {journal} {Rept. Prog. Phys.}\ }\textbf {\bibinfo {volume} {59}},\ \bibinfo
  {pages} {1493} (\bibinfo {year} {1996})},\ \Eprint
  {https://arxiv.org/abs/hep-ph/9602260} {arXiv:hep-ph/9602260} \BibitemShut
  {NoStop}%
\bibitem [{\citenamefont {Iocco}\ \emph {et~al.}(2009)\citenamefont {Iocco},
  \citenamefont {Mangano}, \citenamefont {Miele}, \citenamefont {Pisanti},\
  and\ \citenamefont {Serpico}}]{Iocco:2008va}%
  \BibitemOpen
  \bibfield  {author} {\bibinfo {author} {\bibfnamefont {F.}~\bibnamefont
  {Iocco}}, \bibinfo {author} {\bibfnamefont {G.}~\bibnamefont {Mangano}},
  \bibinfo {author} {\bibfnamefont {G.}~\bibnamefont {Miele}}, \bibinfo
  {author} {\bibfnamefont {O.}~\bibnamefont {Pisanti}},\ and\ \bibinfo {author}
  {\bibfnamefont {P.~D.}\ \bibnamefont {Serpico}},\ }\bibfield  {title}
  {\bibinfo {title} {{Primordial Nucleosynthesis: from precision cosmology to
  fundamental physics}},\ }\href
  {https://doi.org/10.1016/j.physrep.2009.02.002} {\bibfield  {journal}
  {\bibinfo  {journal} {Phys. Rept.}\ }\textbf {\bibinfo {volume} {472}},\
  \bibinfo {pages} {1} (\bibinfo {year} {2009})},\ \Eprint
  {https://arxiv.org/abs/0809.0631} {arXiv:0809.0631 [astro-ph]} \BibitemShut
  {NoStop}%
\bibitem [{\citenamefont {Pospelov}\ and\ \citenamefont
  {Pradler}(2010)}]{Pospelov:2010hj}%
  \BibitemOpen
  \bibfield  {author} {\bibinfo {author} {\bibfnamefont {M.}~\bibnamefont
  {Pospelov}}\ and\ \bibinfo {author} {\bibfnamefont {J.}~\bibnamefont
  {Pradler}},\ }\bibfield  {title} {\bibinfo {title} {{Big Bang Nucleosynthesis
  as a Probe of New Physics}},\ }\href
  {https://doi.org/10.1146/annurev.nucl.012809.104521} {\bibfield  {journal}
  {\bibinfo  {journal} {Ann. Rev. Nucl. Part. Sci.}\ }\textbf {\bibinfo
  {volume} {60}},\ \bibinfo {pages} {539} (\bibinfo {year} {2010})},\ \Eprint
  {https://arxiv.org/abs/1011.1054} {arXiv:1011.1054 [hep-ph]} \BibitemShut
  {NoStop}%
\bibitem [{\citenamefont {Blennow}\ \emph {et~al.}(2012)\citenamefont
  {Blennow}, \citenamefont {Fernandez-Martinez}, \citenamefont {Mena},
  \citenamefont {Redondo},\ and\ \citenamefont {Serra}}]{Blennow:2012de}%
  \BibitemOpen
  \bibfield  {author} {\bibinfo {author} {\bibfnamefont {M.}~\bibnamefont
  {Blennow}}, \bibinfo {author} {\bibfnamefont {E.}~\bibnamefont
  {Fernandez-Martinez}}, \bibinfo {author} {\bibfnamefont {O.}~\bibnamefont
  {Mena}}, \bibinfo {author} {\bibfnamefont {J.}~\bibnamefont {Redondo}},\ and\
  \bibinfo {author} {\bibfnamefont {P.}~\bibnamefont {Serra}},\ }\bibfield
  {title} {\bibinfo {title} {{Asymmetric Dark Matter and Dark Radiation}},\
  }\href {https://doi.org/10.1088/1475-7516/2012/07/022} {\bibfield  {journal}
  {\bibinfo  {journal} {JCAP}\ }\textbf {\bibinfo {volume} {07}},\ \bibinfo
  {pages} {022}},\ \Eprint {https://arxiv.org/abs/1203.5803} {arXiv:1203.5803
  [hep-ph]} \BibitemShut {NoStop}%
\bibitem [{\citenamefont {Boehm}\ \emph {et~al.}(2012)\citenamefont {Boehm},
  \citenamefont {Dolan},\ and\ \citenamefont {McCabe}}]{Boehm:2012gr}%
  \BibitemOpen
  \bibfield  {author} {\bibinfo {author} {\bibfnamefont {C.}~\bibnamefont
  {Boehm}}, \bibinfo {author} {\bibfnamefont {M.~J.}\ \bibnamefont {Dolan}},\
  and\ \bibinfo {author} {\bibfnamefont {C.}~\bibnamefont {McCabe}},\
  }\bibfield  {title} {\bibinfo {title} {{Increasing Neff with particles in
  thermal equilibrium with neutrinos}},\ }\href
  {https://doi.org/10.1088/1475-7516/2012/12/027} {\bibfield  {journal}
  {\bibinfo  {journal} {JCAP}\ }\textbf {\bibinfo {volume} {12}},\ \bibinfo
  {pages} {027}},\ \Eprint {https://arxiv.org/abs/1207.0497} {arXiv:1207.0497
  [astro-ph.CO]} \BibitemShut {NoStop}%
\bibitem [{\citenamefont {Boehm}\ \emph {et~al.}(2013)\citenamefont {Boehm},
  \citenamefont {Dolan},\ and\ \citenamefont {McCabe}}]{Boehm:2013jpa}%
  \BibitemOpen
  \bibfield  {author} {\bibinfo {author} {\bibfnamefont {C.}~\bibnamefont
  {Boehm}}, \bibinfo {author} {\bibfnamefont {M.~J.}\ \bibnamefont {Dolan}},\
  and\ \bibinfo {author} {\bibfnamefont {C.}~\bibnamefont {McCabe}},\
  }\bibfield  {title} {\bibinfo {title} {{A Lower Bound on the Mass of Cold
  Thermal Dark Matter from Planck}},\ }\href
  {https://doi.org/10.1088/1475-7516/2013/08/041} {\bibfield  {journal}
  {\bibinfo  {journal} {JCAP}\ }\textbf {\bibinfo {volume} {08}},\ \bibinfo
  {pages} {041}},\ \Eprint {https://arxiv.org/abs/1303.6270} {arXiv:1303.6270
  [hep-ph]} \BibitemShut {NoStop}%
\bibitem [{\citenamefont {Brust}\ \emph {et~al.}(2013)\citenamefont {Brust},
  \citenamefont {Kaplan},\ and\ \citenamefont {Walters}}]{Brust:2013ova}%
  \BibitemOpen
  \bibfield  {author} {\bibinfo {author} {\bibfnamefont {C.}~\bibnamefont
  {Brust}}, \bibinfo {author} {\bibfnamefont {D.~E.}\ \bibnamefont {Kaplan}},\
  and\ \bibinfo {author} {\bibfnamefont {M.~T.}\ \bibnamefont {Walters}},\
  }\bibfield  {title} {\bibinfo {title} {{New Light Species and the CMB}},\
  }\href {https://doi.org/10.1007/JHEP12(2013)058} {\bibfield  {journal}
  {\bibinfo  {journal} {JHEP}\ }\textbf {\bibinfo {volume} {12}},\ \bibinfo
  {pages} {058}},\ \Eprint {https://arxiv.org/abs/1303.5379} {arXiv:1303.5379
  [hep-ph]} \BibitemShut {NoStop}%
\bibitem [{\citenamefont {Vogel}\ and\ \citenamefont
  {Redondo}(2014)}]{Vogel:2013raa}%
  \BibitemOpen
  \bibfield  {author} {\bibinfo {author} {\bibfnamefont {H.}~\bibnamefont
  {Vogel}}\ and\ \bibinfo {author} {\bibfnamefont {J.}~\bibnamefont
  {Redondo}},\ }\bibfield  {title} {\bibinfo {title} {{Dark Radiation
  constraints on minicharged particles in models with a hidden photon}},\
  }\href {https://doi.org/10.1088/1475-7516/2014/02/029} {\bibfield  {journal}
  {\bibinfo  {journal} {JCAP}\ }\textbf {\bibinfo {volume} {02}},\ \bibinfo
  {pages} {029}},\ \Eprint {https://arxiv.org/abs/1311.2600} {arXiv:1311.2600
  [hep-ph]} \BibitemShut {NoStop}%
\bibitem [{\citenamefont {Fradette}\ \emph {et~al.}(2014)\citenamefont
  {Fradette}, \citenamefont {Pospelov}, \citenamefont {Pradler},\ and\
  \citenamefont {Ritz}}]{Fradette:2014sza}%
  \BibitemOpen
  \bibfield  {author} {\bibinfo {author} {\bibfnamefont {A.}~\bibnamefont
  {Fradette}}, \bibinfo {author} {\bibfnamefont {M.}~\bibnamefont {Pospelov}},
  \bibinfo {author} {\bibfnamefont {J.}~\bibnamefont {Pradler}},\ and\ \bibinfo
  {author} {\bibfnamefont {A.}~\bibnamefont {Ritz}},\ }\bibfield  {title}
  {\bibinfo {title} {{Cosmological Constraints on Very Dark Photons}},\ }\href
  {https://doi.org/10.1103/PhysRevD.90.035022} {\bibfield  {journal} {\bibinfo
  {journal} {Phys. Rev. D}\ }\textbf {\bibinfo {volume} {90}},\ \bibinfo
  {pages} {035022} (\bibinfo {year} {2014})},\ \Eprint
  {https://arxiv.org/abs/1407.0993} {arXiv:1407.0993 [hep-ph]} \BibitemShut
  {NoStop}%
\bibitem [{\citenamefont {Nollett}\ and\ \citenamefont
  {Steigman}(2015)}]{Nollett:2014lwa}%
  \BibitemOpen
  \bibfield  {author} {\bibinfo {author} {\bibfnamefont {K.~M.}\ \bibnamefont
  {Nollett}}\ and\ \bibinfo {author} {\bibfnamefont {G.}~\bibnamefont
  {Steigman}},\ }\bibfield  {title} {\bibinfo {title} {{BBN And The CMB
  Constrain Neutrino Coupled Light WIMPs}},\ }\href
  {https://doi.org/10.1103/PhysRevD.91.083505} {\bibfield  {journal} {\bibinfo
  {journal} {Phys. Rev. D}\ }\textbf {\bibinfo {volume} {91}},\ \bibinfo
  {pages} {083505} (\bibinfo {year} {2015})},\ \Eprint
  {https://arxiv.org/abs/1411.6005} {arXiv:1411.6005 [astro-ph.CO]}
  \BibitemShut {NoStop}%
\bibitem [{\citenamefont {Buen-Abad}\ \emph {et~al.}(2015)\citenamefont
  {Buen-Abad}, \citenamefont {Marques-Tavares},\ and\ \citenamefont
  {Schmaltz}}]{Buen-Abad:2015ova}%
  \BibitemOpen
  \bibfield  {author} {\bibinfo {author} {\bibfnamefont {M.~A.}\ \bibnamefont
  {Buen-Abad}}, \bibinfo {author} {\bibfnamefont {G.}~\bibnamefont
  {Marques-Tavares}},\ and\ \bibinfo {author} {\bibfnamefont {M.}~\bibnamefont
  {Schmaltz}},\ }\bibfield  {title} {\bibinfo {title} {{Non-Abelian dark matter
  and dark radiation}},\ }\href {https://doi.org/10.1103/PhysRevD.92.023531}
  {\bibfield  {journal} {\bibinfo  {journal} {Phys. Rev. D}\ }\textbf {\bibinfo
  {volume} {92}},\ \bibinfo {pages} {023531} (\bibinfo {year} {2015})},\
  \Eprint {https://arxiv.org/abs/1505.03542} {arXiv:1505.03542 [hep-ph]}
  \BibitemShut {NoStop}%
\bibitem [{\citenamefont {Chacko}\ \emph {et~al.}(2015)\citenamefont {Chacko},
  \citenamefont {Cui}, \citenamefont {Hong},\ and\ \citenamefont
  {Okui}}]{Chacko:2015noa}%
  \BibitemOpen
  \bibfield  {author} {\bibinfo {author} {\bibfnamefont {Z.}~\bibnamefont
  {Chacko}}, \bibinfo {author} {\bibfnamefont {Y.}~\bibnamefont {Cui}},
  \bibinfo {author} {\bibfnamefont {S.}~\bibnamefont {Hong}},\ and\ \bibinfo
  {author} {\bibfnamefont {T.}~\bibnamefont {Okui}},\ }\bibfield  {title}
  {\bibinfo {title} {{Hidden dark matter sector, dark radiation, and the
  CMB}},\ }\href {https://doi.org/10.1103/PhysRevD.92.055033} {\bibfield
  {journal} {\bibinfo  {journal} {Phys. Rev. D}\ }\textbf {\bibinfo {volume}
  {92}},\ \bibinfo {pages} {055033} (\bibinfo {year} {2015})},\ \Eprint
  {https://arxiv.org/abs/1505.04192} {arXiv:1505.04192 [hep-ph]} \BibitemShut
  {NoStop}%
\bibitem [{\citenamefont {Wilkinson}\ \emph {et~al.}(2016)\citenamefont
  {Wilkinson}, \citenamefont {Vincent}, \citenamefont {B\oe{}hm},\ and\
  \citenamefont {McCabe}}]{Wilkinson:2016gsy}%
  \BibitemOpen
  \bibfield  {author} {\bibinfo {author} {\bibfnamefont {R.~J.}\ \bibnamefont
  {Wilkinson}}, \bibinfo {author} {\bibfnamefont {A.~C.}\ \bibnamefont
  {Vincent}}, \bibinfo {author} {\bibfnamefont {C.}~\bibnamefont {B\oe{}hm}},\
  and\ \bibinfo {author} {\bibfnamefont {C.}~\bibnamefont {McCabe}},\
  }\bibfield  {title} {\bibinfo {title} {{Ruling out the light weakly
  interacting massive particle explanation of the Galactic 511 keV line}},\
  }\href {https://doi.org/10.1103/PhysRevD.94.103525} {\bibfield  {journal}
  {\bibinfo  {journal} {Phys. Rev. D}\ }\textbf {\bibinfo {volume} {94}},\
  \bibinfo {pages} {103525} (\bibinfo {year} {2016})},\ \Eprint
  {https://arxiv.org/abs/1602.01114} {arXiv:1602.01114 [astro-ph.CO]}
  \BibitemShut {NoStop}%
\bibitem [{\citenamefont {Huang}\ \emph {et~al.}(2018)\citenamefont {Huang},
  \citenamefont {Ohlsson},\ and\ \citenamefont {Zhou}}]{Huang:2017egl}%
  \BibitemOpen
  \bibfield  {author} {\bibinfo {author} {\bibfnamefont {G.-y.}\ \bibnamefont
  {Huang}}, \bibinfo {author} {\bibfnamefont {T.}~\bibnamefont {Ohlsson}},\
  and\ \bibinfo {author} {\bibfnamefont {S.}~\bibnamefont {Zhou}},\ }\bibfield
  {title} {\bibinfo {title} {{Observational Constraints on Secret Neutrino
  Interactions from Big Bang Nucleosynthesis}},\ }\href
  {https://doi.org/10.1103/PhysRevD.97.075009} {\bibfield  {journal} {\bibinfo
  {journal} {Phys. Rev. D}\ }\textbf {\bibinfo {volume} {97}},\ \bibinfo
  {pages} {075009} (\bibinfo {year} {2018})},\ \Eprint
  {https://arxiv.org/abs/1712.04792} {arXiv:1712.04792 [hep-ph]} \BibitemShut
  {NoStop}%
\bibitem [{\citenamefont {Escudero}(2019)}]{Escudero:2018mvt}%
  \BibitemOpen
  \bibfield  {author} {\bibinfo {author} {\bibfnamefont {M.}~\bibnamefont
  {Escudero}},\ }\bibfield  {title} {\bibinfo {title} {{Neutrino decoupling
  beyond the Standard Model: CMB constraints on the Dark Matter mass with a
  fast and precise $N_{\rm eff}$ evaluation}},\ }\href
  {https://doi.org/10.1088/1475-7516/2019/02/007} {\bibfield  {journal}
  {\bibinfo  {journal} {JCAP}\ }\textbf {\bibinfo {volume} {02}},\ \bibinfo
  {pages} {007}},\ \Eprint {https://arxiv.org/abs/1812.05605} {arXiv:1812.05605
  [hep-ph]} \BibitemShut {NoStop}%
\bibitem [{\citenamefont {Abazajian}\ and\ \citenamefont
  {Heeck}(2019)}]{Abazajian:2019oqj}%
  \BibitemOpen
  \bibfield  {author} {\bibinfo {author} {\bibfnamefont {K.~N.}\ \bibnamefont
  {Abazajian}}\ and\ \bibinfo {author} {\bibfnamefont {J.}~\bibnamefont
  {Heeck}},\ }\bibfield  {title} {\bibinfo {title} {{Observing Dirac neutrinos
  in the cosmic microwave background}},\ }\href
  {https://doi.org/10.1103/PhysRevD.100.075027} {\bibfield  {journal} {\bibinfo
   {journal} {Phys. Rev. D}\ }\textbf {\bibinfo {volume} {100}},\ \bibinfo
  {pages} {075027} (\bibinfo {year} {2019})},\ \Eprint
  {https://arxiv.org/abs/1908.03286} {arXiv:1908.03286 [hep-ph]} \BibitemShut
  {NoStop}%
\bibitem [{\citenamefont {Ibe}\ \emph {et~al.}(2020)\citenamefont {Ibe},
  \citenamefont {Kobayashi}, \citenamefont {Nakayama},\ and\ \citenamefont
  {Shirai}}]{Ibe:2019gpv}%
  \BibitemOpen
  \bibfield  {author} {\bibinfo {author} {\bibfnamefont {M.}~\bibnamefont
  {Ibe}}, \bibinfo {author} {\bibfnamefont {S.}~\bibnamefont {Kobayashi}},
  \bibinfo {author} {\bibfnamefont {Y.}~\bibnamefont {Nakayama}},\ and\
  \bibinfo {author} {\bibfnamefont {S.}~\bibnamefont {Shirai}},\ }\bibfield
  {title} {\bibinfo {title} {{Cosmological constraint on dark photon from
  N$_{eff}$}},\ }\href {https://doi.org/10.1007/JHEP04(2020)009} {\bibfield
  {journal} {\bibinfo  {journal} {JHEP}\ }\textbf {\bibinfo {volume} {04}},\
  \bibinfo {pages} {009}},\ \Eprint {https://arxiv.org/abs/1912.12152}
  {arXiv:1912.12152 [hep-ph]} \BibitemShut {NoStop}%
\bibitem [{\citenamefont {Escudero~Abenza}(2020)}]{EscuderoAbenza:2020cmq}%
  \BibitemOpen
  \bibfield  {author} {\bibinfo {author} {\bibfnamefont {M.}~\bibnamefont
  {Escudero~Abenza}},\ }\bibfield  {title} {\bibinfo {title} {{ Precision Early
  Universe Thermodynamics Made Simple: $N_{\rm eff}$ and Neutrino Decoupling in
  the Standard Model and Beyond}},\ }\href
  {https://doi.org/10.1088/1475-7516/2020/05/048} {\bibfield  {journal}
  {\bibinfo  {journal} {JCAP}\ }\textbf {\bibinfo {volume} {05}},\ \bibinfo
  {pages} {048}},\ \Eprint {https://arxiv.org/abs/2001.04466} {arXiv:2001.04466
  [hep-ph]} \BibitemShut {NoStop}%
\bibitem [{\citenamefont {Coffey}\ \emph {et~al.}(2020)\citenamefont {Coffey},
  \citenamefont {Forestell}, \citenamefont {Morrissey},\ and\ \citenamefont
  {White}}]{Coffey:2020oir}%
  \BibitemOpen
  \bibfield  {author} {\bibinfo {author} {\bibfnamefont {J.}~\bibnamefont
  {Coffey}}, \bibinfo {author} {\bibfnamefont {L.}~\bibnamefont {Forestell}},
  \bibinfo {author} {\bibfnamefont {D.~E.}\ \bibnamefont {Morrissey}},\ and\
  \bibinfo {author} {\bibfnamefont {G.}~\bibnamefont {White}},\ }\bibfield
  {title} {\bibinfo {title} {{Cosmological Bounds on sub-GeV Dark Vector Bosons
  from Electromagnetic Energy Injection}},\ }\href
  {https://doi.org/10.1007/JHEP07(2020)179} {\bibfield  {journal} {\bibinfo
  {journal} {JHEP}\ }\textbf {\bibinfo {volume} {07}},\ \bibinfo {pages}
  {179}},\ \Eprint {https://arxiv.org/abs/2003.02273} {arXiv:2003.02273
  [hep-ph]} \BibitemShut {NoStop}%
\bibitem [{\citenamefont {Luo}\ \emph {et~al.}(2020)\citenamefont {Luo},
  \citenamefont {Rodejohann},\ and\ \citenamefont {Xu}}]{Luo:2020sho}%
  \BibitemOpen
  \bibfield  {author} {\bibinfo {author} {\bibfnamefont {X.}~\bibnamefont
  {Luo}}, \bibinfo {author} {\bibfnamefont {W.}~\bibnamefont {Rodejohann}},\
  and\ \bibinfo {author} {\bibfnamefont {X.-J.}\ \bibnamefont {Xu}},\
  }\bibfield  {title} {\bibinfo {title} {{Dirac neutrinos and $N_{{\rm
  eff}}$}},\ }\href {https://doi.org/10.1088/1475-7516/2020/06/058} {\bibfield
  {journal} {\bibinfo  {journal} {JCAP}\ }\textbf {\bibinfo {volume} {06}},\
  \bibinfo {pages} {058}},\ \Eprint {https://arxiv.org/abs/2005.01629}
  {arXiv:2005.01629 [hep-ph]} \BibitemShut {NoStop}%
\bibitem [{\citenamefont {Luo}\ \emph {et~al.}(2021)\citenamefont {Luo},
  \citenamefont {Rodejohann},\ and\ \citenamefont {Xu}}]{Luo:2020fdt}%
  \BibitemOpen
  \bibfield  {author} {\bibinfo {author} {\bibfnamefont {X.}~\bibnamefont
  {Luo}}, \bibinfo {author} {\bibfnamefont {W.}~\bibnamefont {Rodejohann}},\
  and\ \bibinfo {author} {\bibfnamefont {X.-J.}\ \bibnamefont {Xu}},\
  }\bibfield  {title} {\bibinfo {title} {{Dirac neutrinos and N$_{eff}$. Part
  II. The freeze-in case}},\ }\href
  {https://doi.org/10.1088/1475-7516/2021/03/082} {\bibfield  {journal}
  {\bibinfo  {journal} {JCAP}\ }\textbf {\bibinfo {volume} {03}},\ \bibinfo
  {pages} {082}},\ \Eprint {https://arxiv.org/abs/2011.13059} {arXiv:2011.13059
  [hep-ph]} \BibitemShut {NoStop}%
\bibitem [{\citenamefont {Adshead}\ \emph {et~al.}(2022)\citenamefont
  {Adshead}, \citenamefont {Ralegankar},\ and\ \citenamefont
  {Shelton}}]{Adshead:2022ovo}%
  \BibitemOpen
  \bibfield  {author} {\bibinfo {author} {\bibfnamefont {P.}~\bibnamefont
  {Adshead}}, \bibinfo {author} {\bibfnamefont {P.}~\bibnamefont
  {Ralegankar}},\ and\ \bibinfo {author} {\bibfnamefont {J.}~\bibnamefont
  {Shelton}},\ }\bibfield  {title} {\bibinfo {title} {{Dark radiation
  constraints on portal interactions with hidden sectors}},\ }\href
  {https://doi.org/10.1088/1475-7516/2022/09/056} {\bibfield  {journal}
  {\bibinfo  {journal} {JCAP}\ }\textbf {\bibinfo {volume} {09}},\ \bibinfo
  {pages} {056}},\ \Eprint {https://arxiv.org/abs/2206.13530} {arXiv:2206.13530
  [hep-ph]} \BibitemShut {NoStop}%
\bibitem [{\citenamefont {Eijima}\ \emph {et~al.}(2022)\citenamefont {Eijima},
  \citenamefont {Seto},\ and\ \citenamefont {Shimomura}}]{Eijima:2022dec}%
  \BibitemOpen
  \bibfield  {author} {\bibinfo {author} {\bibfnamefont {S.}~\bibnamefont
  {Eijima}}, \bibinfo {author} {\bibfnamefont {O.}~\bibnamefont {Seto}},\ and\
  \bibinfo {author} {\bibfnamefont {T.}~\bibnamefont {Shimomura}},\ }\bibfield
  {title} {\bibinfo {title} {{Revisiting sterile neutrino dark matter in gauged
  U(1)B-L model}},\ }\href {https://doi.org/10.1103/PhysRevD.106.103513}
  {\bibfield  {journal} {\bibinfo  {journal} {Phys. Rev. D}\ }\textbf {\bibinfo
  {volume} {106}},\ \bibinfo {pages} {103513} (\bibinfo {year} {2022})},\
  \Eprint {https://arxiv.org/abs/2207.01775} {arXiv:2207.01775 [hep-ph]}
  \BibitemShut {NoStop}%
\bibitem [{\citenamefont {Sandner}\ \emph {et~al.}(2023)\citenamefont
  {Sandner}, \citenamefont {Escudero},\ and\ \citenamefont
  {Witte}}]{Sandner:2023ptm}%
  \BibitemOpen
  \bibfield  {author} {\bibinfo {author} {\bibfnamefont {S.}~\bibnamefont
  {Sandner}}, \bibinfo {author} {\bibfnamefont {M.}~\bibnamefont {Escudero}},\
  and\ \bibinfo {author} {\bibfnamefont {S.~J.}\ \bibnamefont {Witte}},\
  }\bibfield  {title} {\bibinfo {title} {{Precision CMB constraints on eV-scale
  bosons coupled to neutrinos}},\ }\href@noop {} {\  (\bibinfo {year}
  {2023})},\ \Eprint {https://arxiv.org/abs/2305.01692} {arXiv:2305.01692
  [hep-ph]} \BibitemShut {NoStop}%
\bibitem [{\citenamefont {Hall}\ \emph {et~al.}(2010)\citenamefont {Hall},
  \citenamefont {Jedamzik}, \citenamefont {March-Russell},\ and\ \citenamefont
  {West}}]{Hall:2009bx}%
  \BibitemOpen
  \bibfield  {author} {\bibinfo {author} {\bibfnamefont {L.~J.}\ \bibnamefont
  {Hall}}, \bibinfo {author} {\bibfnamefont {K.}~\bibnamefont {Jedamzik}},
  \bibinfo {author} {\bibfnamefont {J.}~\bibnamefont {March-Russell}},\ and\
  \bibinfo {author} {\bibfnamefont {S.~M.}\ \bibnamefont {West}},\ }\bibfield
  {title} {\bibinfo {title} {{Freeze-In Production of FIMP Dark Matter}},\
  }\href {https://doi.org/10.1007/JHEP03(2010)080} {\bibfield  {journal}
  {\bibinfo  {journal} {JHEP}\ }\textbf {\bibinfo {volume} {03}},\ \bibinfo
  {pages} {080}},\ \Eprint {https://arxiv.org/abs/0911.1120} {arXiv:0911.1120
  [hep-ph]} \BibitemShut {NoStop}%
\bibitem [{\citenamefont {Li}\ and\ \citenamefont {Xu}(2023)}]{Li:2023puz}%
  \BibitemOpen
  \bibfield  {author} {\bibinfo {author} {\bibfnamefont {S.-P.}\ \bibnamefont
  {Li}}\ and\ \bibinfo {author} {\bibfnamefont {X.-J.}\ \bibnamefont {Xu}},\
  }\bibfield  {title} {\bibinfo {title} {{N$_{eff}$ constraints on light
  mediators coupled to neutrinos: the dilution-resistant effect}},\ }\href
  {https://doi.org/10.1007/JHEP10(2023)012} {\bibfield  {journal} {\bibinfo
  {journal} {JHEP}\ }\textbf {\bibinfo {volume} {10}},\ \bibinfo {pages}
  {012}},\ \Eprint {https://arxiv.org/abs/2307.13967} {arXiv:2307.13967
  [hep-ph]} \BibitemShut {NoStop}%
\bibitem [{\citenamefont {Aver}\ \emph {et~al.}(2015)\citenamefont {Aver},
  \citenamefont {Olive},\ and\ \citenamefont {Skillman}}]{Aver:2015iza}%
  \BibitemOpen
  \bibfield  {author} {\bibinfo {author} {\bibfnamefont {E.}~\bibnamefont
  {Aver}}, \bibinfo {author} {\bibfnamefont {K.~A.}\ \bibnamefont {Olive}},\
  and\ \bibinfo {author} {\bibfnamefont {E.~D.}\ \bibnamefont {Skillman}},\
  }\bibfield  {title} {\bibinfo {title} {{The effects of He I
  \ensuremath{\lambda}10830 on helium abundance determinations}},\ }\href
  {https://doi.org/10.1088/1475-7516/2015/07/011} {\bibfield  {journal}
  {\bibinfo  {journal} {JCAP}\ }\textbf {\bibinfo {volume} {07}},\ \bibinfo
  {pages} {011}},\ \Eprint {https://arxiv.org/abs/1503.08146} {arXiv:1503.08146
  [astro-ph.CO]} \BibitemShut {NoStop}%
\bibitem [{\citenamefont {Pettini}\ and\ \citenamefont
  {Cooke}(2012)}]{Pettini:2012ph}%
  \BibitemOpen
  \bibfield  {author} {\bibinfo {author} {\bibfnamefont {M.}~\bibnamefont
  {Pettini}}\ and\ \bibinfo {author} {\bibfnamefont {R.}~\bibnamefont
  {Cooke}},\ }\bibfield  {title} {\bibinfo {title} {{A new, precise measurement
  of the primordial abundance of Deuterium}},\ }\href
  {https://doi.org/10.1111/j.1365-2966.2012.21665.x} {\bibfield  {journal}
  {\bibinfo  {journal} {Mon. Not. Roy. Astron. Soc.}\ }\textbf {\bibinfo
  {volume} {425}},\ \bibinfo {pages} {2477} (\bibinfo {year} {2012})},\ \Eprint
  {https://arxiv.org/abs/1205.3785} {arXiv:1205.3785 [astro-ph.CO]}
  \BibitemShut {NoStop}%
\bibitem [{\citenamefont {Cooke}\ \emph {et~al.}(2014)\citenamefont {Cooke},
  \citenamefont {Pettini}, \citenamefont {Jorgenson}, \citenamefont {Murphy},\
  and\ \citenamefont {Steidel}}]{Cooke:2013cba}%
  \BibitemOpen
  \bibfield  {author} {\bibinfo {author} {\bibfnamefont {R.}~\bibnamefont
  {Cooke}}, \bibinfo {author} {\bibfnamefont {M.}~\bibnamefont {Pettini}},
  \bibinfo {author} {\bibfnamefont {R.~A.}\ \bibnamefont {Jorgenson}}, \bibinfo
  {author} {\bibfnamefont {M.~T.}\ \bibnamefont {Murphy}},\ and\ \bibinfo
  {author} {\bibfnamefont {C.~C.}\ \bibnamefont {Steidel}},\ }\bibfield
  {title} {\bibinfo {title} {{Precision measures of the primordial abundance of
  deuterium}},\ }\href {https://doi.org/10.1088/0004-637X/781/1/31} {\bibfield
  {journal} {\bibinfo  {journal} {Astrophys. J.}\ }\textbf {\bibinfo {volume}
  {781}},\ \bibinfo {pages} {31} (\bibinfo {year} {2014})},\ \Eprint
  {https://arxiv.org/abs/1308.3240} {arXiv:1308.3240 [astro-ph.CO]}
  \BibitemShut {NoStop}%
\bibitem [{\citenamefont {Riemer-S\o{}rensen}\ \emph
  {et~al.}(2015)\citenamefont {Riemer-S\o{}rensen}, \citenamefont {Webb},
  \citenamefont {Crighton}, \citenamefont {Dumont}, \citenamefont {Ali},
  \citenamefont {Kotu\v{s}}, \citenamefont {Bainbridge}, \citenamefont
  {Murphy},\ and\ \citenamefont {Carswell}}]{Riemer-Sorensen:2014aoa}%
  \BibitemOpen
  \bibfield  {author} {\bibinfo {author} {\bibfnamefont {S.}~\bibnamefont
  {Riemer-S\o{}rensen}}, \bibinfo {author} {\bibfnamefont {J.~K.}\ \bibnamefont
  {Webb}}, \bibinfo {author} {\bibfnamefont {N.}~\bibnamefont {Crighton}},
  \bibinfo {author} {\bibfnamefont {V.}~\bibnamefont {Dumont}}, \bibinfo
  {author} {\bibfnamefont {K.}~\bibnamefont {Ali}}, \bibinfo {author}
  {\bibfnamefont {S.}~\bibnamefont {Kotu\v{s}}}, \bibinfo {author}
  {\bibfnamefont {M.}~\bibnamefont {Bainbridge}}, \bibinfo {author}
  {\bibfnamefont {M.~T.}\ \bibnamefont {Murphy}},\ and\ \bibinfo {author}
  {\bibfnamefont {R.}~\bibnamefont {Carswell}},\ }\bibfield  {title} {\bibinfo
  {title} {{A robust deuterium abundance; Re-measurement of the z=3.256
  absorption system towards the quasar PKS1937-1009}},\ }\href
  {https://doi.org/10.1093/mnras/stu2599} {\bibfield  {journal} {\bibinfo
  {journal} {Mon. Not. Roy. Astron. Soc.}\ }\textbf {\bibinfo {volume} {447}},\
  \bibinfo {pages} {2925} (\bibinfo {year} {2015})},\ \Eprint
  {https://arxiv.org/abs/1412.4043} {arXiv:1412.4043 [astro-ph.CO]}
  \BibitemShut {NoStop}%
\bibitem [{\citenamefont {Cooke}\ \emph {et~al.}(2016)\citenamefont {Cooke},
  \citenamefont {Pettini}, \citenamefont {Nollett},\ and\ \citenamefont
  {Jorgenson}}]{Cooke:2016rky}%
  \BibitemOpen
  \bibfield  {author} {\bibinfo {author} {\bibfnamefont {R.~J.}\ \bibnamefont
  {Cooke}}, \bibinfo {author} {\bibfnamefont {M.}~\bibnamefont {Pettini}},
  \bibinfo {author} {\bibfnamefont {K.~M.}\ \bibnamefont {Nollett}},\ and\
  \bibinfo {author} {\bibfnamefont {R.}~\bibnamefont {Jorgenson}},\ }\bibfield
  {title} {\bibinfo {title} {{The primordial deuterium abundance of the most
  metal-poor damped Ly$\alpha$ system}},\ }\href
  {https://doi.org/10.3847/0004-637X/830/2/148} {\bibfield  {journal} {\bibinfo
   {journal} {Astrophys. J.}\ }\textbf {\bibinfo {volume} {830}},\ \bibinfo
  {pages} {148} (\bibinfo {year} {2016})},\ \Eprint
  {https://arxiv.org/abs/1607.03900} {arXiv:1607.03900 [astro-ph.CO]}
  \BibitemShut {NoStop}%
\bibitem [{\citenamefont {Balashev}\ \emph {et~al.}(2016)\citenamefont
  {Balashev}, \citenamefont {Zavarygin}, \citenamefont {Ivanchik},
  \citenamefont {Telikova},\ and\ \citenamefont
  {Varshalovich}}]{Balashev:2015hoe}%
  \BibitemOpen
  \bibfield  {author} {\bibinfo {author} {\bibfnamefont {S.~A.}\ \bibnamefont
  {Balashev}}, \bibinfo {author} {\bibfnamefont {E.~O.}\ \bibnamefont
  {Zavarygin}}, \bibinfo {author} {\bibfnamefont {A.~V.}\ \bibnamefont
  {Ivanchik}}, \bibinfo {author} {\bibfnamefont {K.~N.}\ \bibnamefont
  {Telikova}},\ and\ \bibinfo {author} {\bibfnamefont {D.~A.}\ \bibnamefont
  {Varshalovich}},\ }\bibfield  {title} {\bibinfo {title} {{The primordial
  deuterium abundance: subDLA system at $z_{\rm abs}=2.437$ towards the QSO J
  1444+2919}},\ }\href {https://doi.org/10.1093/mnras/stw356} {\bibfield
  {journal} {\bibinfo  {journal} {Mon. Not. Roy. Astron. Soc.}\ }\textbf
  {\bibinfo {volume} {458}},\ \bibinfo {pages} {2188} (\bibinfo {year}
  {2016})},\ \Eprint {https://arxiv.org/abs/1511.01797} {arXiv:1511.01797
  [astro-ph.GA]} \BibitemShut {NoStop}%
\bibitem [{\citenamefont {Riemer-S\o{}rensen}\ \emph
  {et~al.}(2017)\citenamefont {Riemer-S\o{}rensen}, \citenamefont {Kotu\v{s}},
  \citenamefont {Webb}, \citenamefont {Ali}, \citenamefont {Dumont},
  \citenamefont {Murphy},\ and\ \citenamefont
  {Carswell}}]{Riemer-Sorensen:2017pey}%
  \BibitemOpen
  \bibfield  {author} {\bibinfo {author} {\bibfnamefont {S.}~\bibnamefont
  {Riemer-S\o{}rensen}}, \bibinfo {author} {\bibfnamefont {S.}~\bibnamefont
  {Kotu\v{s}}}, \bibinfo {author} {\bibfnamefont {J.~K.}\ \bibnamefont {Webb}},
  \bibinfo {author} {\bibfnamefont {K.}~\bibnamefont {Ali}}, \bibinfo {author}
  {\bibfnamefont {V.}~\bibnamefont {Dumont}}, \bibinfo {author} {\bibfnamefont
  {M.~T.}\ \bibnamefont {Murphy}},\ and\ \bibinfo {author} {\bibfnamefont
  {R.~F.}\ \bibnamefont {Carswell}},\ }\bibfield  {title} {\bibinfo {title} {{A
  precise deuterium abundance: remeasurement of the z = 3.572 absorption system
  towards the quasar PKS1937\textendash{}101}},\ }\href
  {https://doi.org/10.1093/mnras/stx681} {\bibfield  {journal} {\bibinfo
  {journal} {Mon. Not. Roy. Astron. Soc.}\ }\textbf {\bibinfo {volume} {468}},\
  \bibinfo {pages} {3239} (\bibinfo {year} {2017})},\ \Eprint
  {https://arxiv.org/abs/1703.06656} {arXiv:1703.06656 [astro-ph.CO]}
  \BibitemShut {NoStop}%
\bibitem [{\citenamefont {Zavarygin}\ \emph {et~al.}(2018)\citenamefont
  {Zavarygin}, \citenamefont {Webb}, \citenamefont {Riemer-S\o{}rensen},\ and\
  \citenamefont {Dumont}}]{Zavarygin:2018ara}%
  \BibitemOpen
  \bibfield  {author} {\bibinfo {author} {\bibfnamefont {E.~O.}\ \bibnamefont
  {Zavarygin}}, \bibinfo {author} {\bibfnamefont {J.~K.}\ \bibnamefont {Webb}},
  \bibinfo {author} {\bibfnamefont {S.}~\bibnamefont {Riemer-S\o{}rensen}},\
  and\ \bibinfo {author} {\bibfnamefont {V.}~\bibnamefont {Dumont}},\
  }\bibfield  {title} {\bibinfo {title} {{Primordial deuterium abundance at
  $z_{abs}$ = 2:504 towards Q1009+2956}},\ }\href
  {https://doi.org/10.1088/1742-6596/1038/1/012012} {\bibfield  {journal}
  {\bibinfo  {journal} {J. Phys. Conf. Ser.}\ }\textbf {\bibinfo {volume}
  {1038}},\ \bibinfo {pages} {012012} (\bibinfo {year} {2018})},\ \Eprint
  {https://arxiv.org/abs/1801.04704} {arXiv:1801.04704 [astro-ph.CO]}
  \BibitemShut {NoStop}%
\bibitem [{\citenamefont {Cooke}\ \emph {et~al.}(2018)\citenamefont {Cooke},
  \citenamefont {Pettini},\ and\ \citenamefont {Steidel}}]{Cooke:2017cwo}%
  \BibitemOpen
  \bibfield  {author} {\bibinfo {author} {\bibfnamefont {R.~J.}\ \bibnamefont
  {Cooke}}, \bibinfo {author} {\bibfnamefont {M.}~\bibnamefont {Pettini}},\
  and\ \bibinfo {author} {\bibfnamefont {C.~C.}\ \bibnamefont {Steidel}},\
  }\bibfield  {title} {\bibinfo {title} {{One Percent Determination of the
  Primordial Deuterium Abundance}},\ }\href
  {https://doi.org/10.3847/1538-4357/aaab53} {\bibfield  {journal} {\bibinfo
  {journal} {Astrophys. J.}\ }\textbf {\bibinfo {volume} {855}},\ \bibinfo
  {pages} {102} (\bibinfo {year} {2018})},\ \Eprint
  {https://arxiv.org/abs/1710.11129} {arXiv:1710.11129 [astro-ph.CO]}
  \BibitemShut {NoStop}%
\bibitem [{\citenamefont {Aver}\ \emph {et~al.}(2021)\citenamefont {Aver},
  \citenamefont {Berg}, \citenamefont {Olive}, \citenamefont {Pogge},
  \citenamefont {Salzer},\ and\ \citenamefont {Skillman}}]{Aver:2020fon}%
  \BibitemOpen
  \bibfield  {author} {\bibinfo {author} {\bibfnamefont {E.}~\bibnamefont
  {Aver}}, \bibinfo {author} {\bibfnamefont {D.~A.}\ \bibnamefont {Berg}},
  \bibinfo {author} {\bibfnamefont {K.~A.}\ \bibnamefont {Olive}}, \bibinfo
  {author} {\bibfnamefont {R.~W.}\ \bibnamefont {Pogge}}, \bibinfo {author}
  {\bibfnamefont {J.~J.}\ \bibnamefont {Salzer}},\ and\ \bibinfo {author}
  {\bibfnamefont {E.~D.}\ \bibnamefont {Skillman}},\ }\bibfield  {title}
  {\bibinfo {title} {{Improving helium abundance determinations with Leo P as a
  case study}},\ }\href {https://doi.org/10.1088/1475-7516/2021/03/027}
  {\bibfield  {journal} {\bibinfo  {journal} {JCAP}\ }\textbf {\bibinfo
  {volume} {03}},\ \bibinfo {pages} {027}},\ \Eprint
  {https://arxiv.org/abs/2010.04180} {arXiv:2010.04180 [astro-ph.CO]}
  \BibitemShut {NoStop}%
\bibitem [{\citenamefont {Valerdi}\ \emph {et~al.}(2019)\citenamefont
  {Valerdi}, \citenamefont {Peimbert}, \citenamefont {Peimbert},\ and\
  \citenamefont {Sixtos}}]{Valerdi:2019beb}%
  \BibitemOpen
  \bibfield  {author} {\bibinfo {author} {\bibfnamefont {M.}~\bibnamefont
  {Valerdi}}, \bibinfo {author} {\bibfnamefont {A.}~\bibnamefont {Peimbert}},
  \bibinfo {author} {\bibfnamefont {M.}~\bibnamefont {Peimbert}},\ and\
  \bibinfo {author} {\bibfnamefont {A.}~\bibnamefont {Sixtos}},\ }\bibfield
  {title} {\bibinfo {title} {{Determination of the Primordial Helium Abundance
  Based on NGC 346, an H ii Region of the Small Magellanic Cloud}},\ }\href
  {https://doi.org/10.3847/1538-4357/ab14e4} {\bibfield  {journal} {\bibinfo
  {journal} {Astrophys. J.}\ }\textbf {\bibinfo {volume} {876}},\ \bibinfo
  {pages} {98} (\bibinfo {year} {2019})},\ \Eprint
  {https://arxiv.org/abs/1904.01594} {arXiv:1904.01594 [astro-ph.GA]}
  \BibitemShut {NoStop}%
\bibitem [{\citenamefont {Fern\'andez}\ \emph {et~al.}(2019)\citenamefont
  {Fern\'andez}, \citenamefont {Terlevich}, \citenamefont {D\'\i{}az},\ and\
  \citenamefont {Terlevich}}]{Fernandez:2019hds}%
  \BibitemOpen
  \bibfield  {author} {\bibinfo {author} {\bibfnamefont {V.}~\bibnamefont
  {Fern\'andez}}, \bibinfo {author} {\bibfnamefont {E.}~\bibnamefont
  {Terlevich}}, \bibinfo {author} {\bibfnamefont {A.~I.}\ \bibnamefont
  {D\'\i{}az}},\ and\ \bibinfo {author} {\bibfnamefont {R.}~\bibnamefont
  {Terlevich}},\ }\bibfield  {title} {\bibinfo {title} {{A Bayesian direct
  method implementation to fit emission line spectra: Application to the
  primordial He abundance determination}},\ }\href
  {https://doi.org/10.1093/mnras/stz1433} {\bibfield  {journal} {\bibinfo
  {journal} {Mon. Not. Roy. Astron. Soc.}\ }\textbf {\bibinfo {volume} {487}},\
  \bibinfo {pages} {3221} (\bibinfo {year} {2019})},\ \Eprint
  {https://arxiv.org/abs/1905.09215} {arXiv:1905.09215 [astro-ph.GA]}
  \BibitemShut {NoStop}%
\bibitem [{\citenamefont {Kurichin}\ \emph {et~al.}(2021)\citenamefont
  {Kurichin}, \citenamefont {Kislitsyn}, \citenamefont {Klimenko},
  \citenamefont {Balashev},\ and\ \citenamefont {Ivanchik}}]{Kurichin:2021ppm}%
  \BibitemOpen
  \bibfield  {author} {\bibinfo {author} {\bibfnamefont {O.~A.}\ \bibnamefont
  {Kurichin}}, \bibinfo {author} {\bibfnamefont {P.~A.}\ \bibnamefont
  {Kislitsyn}}, \bibinfo {author} {\bibfnamefont {V.~V.}\ \bibnamefont
  {Klimenko}}, \bibinfo {author} {\bibfnamefont {S.~A.}\ \bibnamefont
  {Balashev}},\ and\ \bibinfo {author} {\bibfnamefont {A.~V.}\ \bibnamefont
  {Ivanchik}},\ }\bibfield  {title} {\bibinfo {title} {{A new determination of
  the primordial helium abundance using the analyses of H II region spectra
  from SDSS}},\ }\href {https://doi.org/10.1093/mnras/stab215} {\bibfield
  {journal} {\bibinfo  {journal} {Mon. Not. Roy. Astron. Soc.}\ }\textbf
  {\bibinfo {volume} {502}},\ \bibinfo {pages} {3045} (\bibinfo {year}
  {2021})},\ \Eprint {https://arxiv.org/abs/2101.09127} {arXiv:2101.09127
  [astro-ph.CO]} \BibitemShut {NoStop}%
\bibitem [{\citenamefont {Hsyu}\ \emph {et~al.}(2020)\citenamefont {Hsyu},
  \citenamefont {Cooke}, \citenamefont {Prochaska},\ and\ \citenamefont
  {Bolte}}]{Hsyu:2020uqb}%
  \BibitemOpen
  \bibfield  {author} {\bibinfo {author} {\bibfnamefont {T.}~\bibnamefont
  {Hsyu}}, \bibinfo {author} {\bibfnamefont {R.~J.}\ \bibnamefont {Cooke}},
  \bibinfo {author} {\bibfnamefont {J.~X.}\ \bibnamefont {Prochaska}},\ and\
  \bibinfo {author} {\bibfnamefont {M.}~\bibnamefont {Bolte}},\ }\bibfield
  {title} {\bibinfo {title} {{The PHLEK Survey: A New Determination of the
  Primordial Helium Abundance}},\ }\href
  {https://doi.org/10.3847/1538-4357/ab91af} {\bibfield  {journal} {\bibinfo
  {journal} {Astrophys. J.}\ }\textbf {\bibinfo {volume} {896}},\ \bibinfo
  {pages} {77} (\bibinfo {year} {2020})},\ \Eprint
  {https://arxiv.org/abs/2005.12290} {arXiv:2005.12290 [astro-ph.GA]}
  \BibitemShut {NoStop}%
\bibitem [{\citenamefont {Hirata}\ \emph {et~al.}(1987)\citenamefont {Hirata}
  \emph {et~al.}}]{Kamiokande-II:1987idp}%
  \BibitemOpen
  \bibfield  {author} {\bibinfo {author} {\bibfnamefont {K.}~\bibnamefont
  {Hirata}} \emph {et~al.} (\bibinfo {collaboration} {Kamiokande-II}),\
  }\bibfield  {title} {\bibinfo {title} {{Observation of a Neutrino Burst from
  the Supernova SN 1987a}},\ }\href
  {https://doi.org/10.1103/PhysRevLett.58.1490} {\bibfield  {journal} {\bibinfo
   {journal} {Phys. Rev. Lett.}\ }\textbf {\bibinfo {volume} {58}},\ \bibinfo
  {pages} {1490} (\bibinfo {year} {1987})}\BibitemShut {NoStop}%
\bibitem [{\citenamefont {Bionta}\ \emph {et~al.}(1987)\citenamefont {Bionta}
  \emph {et~al.}}]{Bionta:1987qt}%
  \BibitemOpen
  \bibfield  {author} {\bibinfo {author} {\bibfnamefont {R.~M.}\ \bibnamefont
  {Bionta}} \emph {et~al.},\ }\bibfield  {title} {\bibinfo {title}
  {{Observation of a Neutrino Burst in Coincidence with Supernova SN 1987a in
  the Large Magellanic Cloud}},\ }\href
  {https://doi.org/10.1103/PhysRevLett.58.1494} {\bibfield  {journal} {\bibinfo
   {journal} {Phys. Rev. Lett.}\ }\textbf {\bibinfo {volume} {58}},\ \bibinfo
  {pages} {1494} (\bibinfo {year} {1987})}\BibitemShut {NoStop}%
\bibitem [{\citenamefont {Burrows}\ and\ \citenamefont
  {Lattimer}(1986)}]{Burrows:1986me}%
  \BibitemOpen
  \bibfield  {author} {\bibinfo {author} {\bibfnamefont {A.}~\bibnamefont
  {Burrows}}\ and\ \bibinfo {author} {\bibfnamefont {J.~M.}\ \bibnamefont
  {Lattimer}},\ }\bibfield  {title} {\bibinfo {title} {{The birth of neutron
  stars}},\ }\href {https://doi.org/10.1086/164405} {\bibfield  {journal}
  {\bibinfo  {journal} {Astrophys. J.}\ }\textbf {\bibinfo {volume} {307}},\
  \bibinfo {pages} {178} (\bibinfo {year} {1986})}\BibitemShut {NoStop}%
\bibitem [{\citenamefont {Burrows}\ and\ \citenamefont
  {Lattimer}(1987)}]{Burrows:1987zz}%
  \BibitemOpen
  \bibfield  {author} {\bibinfo {author} {\bibfnamefont {A.}~\bibnamefont
  {Burrows}}\ and\ \bibinfo {author} {\bibfnamefont {J.~M.}\ \bibnamefont
  {Lattimer}},\ }\bibfield  {title} {\bibinfo {title} {{Neutrinos from SN
  1987A}},\ }\href {https://doi.org/10.1086/184938} {\bibfield  {journal}
  {\bibinfo  {journal} {Astrophys. J. Lett.}\ }\textbf {\bibinfo {volume}
  {318}},\ \bibinfo {pages} {L63} (\bibinfo {year} {1987})}\BibitemShut
  {NoStop}%
\bibitem [{\citenamefont {Fiorillo}\ \emph
  {et~al.}(2023{\natexlab{a}})\citenamefont {Fiorillo}, \citenamefont
  {Heinlein}, \citenamefont {Janka}, \citenamefont {Raffelt}, \citenamefont
  {Vitagliano},\ and\ \citenamefont {Bollig}}]{Fiorillo:2023frv}%
  \BibitemOpen
  \bibfield  {author} {\bibinfo {author} {\bibfnamefont {D.~F.~G.}\
  \bibnamefont {Fiorillo}}, \bibinfo {author} {\bibfnamefont {M.}~\bibnamefont
  {Heinlein}}, \bibinfo {author} {\bibfnamefont {H.-T.}\ \bibnamefont {Janka}},
  \bibinfo {author} {\bibfnamefont {G.}~\bibnamefont {Raffelt}}, \bibinfo
  {author} {\bibfnamefont {E.}~\bibnamefont {Vitagliano}},\ and\ \bibinfo
  {author} {\bibfnamefont {R.}~\bibnamefont {Bollig}},\ }\bibfield  {title}
  {\bibinfo {title} {{Supernova simulations confront SN 1987A neutrinos}},\
  }\href {https://doi.org/10.1103/PhysRevD.108.083040} {\bibfield  {journal}
  {\bibinfo  {journal} {Phys. Rev. D}\ }\textbf {\bibinfo {volume} {108}},\
  \bibinfo {pages} {083040} (\bibinfo {year} {2023}{\natexlab{a}})},\ \Eprint
  {https://arxiv.org/abs/2308.01403} {arXiv:2308.01403 [astro-ph.HE]}
  \BibitemShut {NoStop}%
\bibitem [{\citenamefont {Raffelt}(1990)}]{Raffelt:1990yz}%
  \BibitemOpen
  \bibfield  {author} {\bibinfo {author} {\bibfnamefont {G.~G.}\ \bibnamefont
  {Raffelt}},\ }\bibfield  {title} {\bibinfo {title} {{Astrophysical methods to
  constrain axions and other novel particle phenomena}},\ }\href
  {https://doi.org/10.1016/0370-1573(90)90054-6} {\bibfield  {journal}
  {\bibinfo  {journal} {Phys. Rept.}\ }\textbf {\bibinfo {volume} {198}},\
  \bibinfo {pages} {1} (\bibinfo {year} {1990})}\BibitemShut {NoStop}%
\bibitem [{\citenamefont {Raffelt}(1996)}]{Raffelt:1996wa}%
  \BibitemOpen
  \bibfield  {author} {\bibinfo {author} {\bibfnamefont {G.~G.}\ \bibnamefont
  {Raffelt}},\ }\href@noop {} {\emph {\bibinfo {title} {{Stars as laboratories
  for fundamental physics}: {The astrophysics of neutrinos, axions, and other
  weakly interacting particles}}}}\ (\bibinfo {year} {1996})\BibitemShut
  {NoStop}%
\bibitem [{\citenamefont {Burrows}(2018)}]{Burrows:2018qjy}%
  \BibitemOpen
  \bibfield  {author} {\bibinfo {author} {\bibfnamefont {A.}~\bibnamefont
  {Burrows}},\ }\bibfield  {title} {\bibinfo {title} {{A Brief History of the
  Co-evolution of Supernova Theory with Neutrino Physics}}\ }(\bibinfo {year}
  {2018})\ \Eprint {https://arxiv.org/abs/1812.05612} {arXiv:1812.05612
  [astro-ph.SR]} \BibitemShut {NoStop}%
\bibitem [{\citenamefont {Janka}\ \emph {et~al.}(2007)\citenamefont {Janka},
  \citenamefont {Langanke}, \citenamefont {Marek}, \citenamefont
  {Martinez-Pinedo},\ and\ \citenamefont {Mueller}}]{Janka:2006fh}%
  \BibitemOpen
  \bibfield  {author} {\bibinfo {author} {\bibfnamefont {H.-T.}\ \bibnamefont
  {Janka}}, \bibinfo {author} {\bibfnamefont {K.}~\bibnamefont {Langanke}},
  \bibinfo {author} {\bibfnamefont {A.}~\bibnamefont {Marek}}, \bibinfo
  {author} {\bibfnamefont {G.}~\bibnamefont {Martinez-Pinedo}},\ and\ \bibinfo
  {author} {\bibfnamefont {B.}~\bibnamefont {Mueller}},\ }\bibfield  {title}
  {\bibinfo {title} {{Theory of Core-Collapse Supernovae}},\ }\href
  {https://doi.org/10.1016/j.physrep.2007.02.002} {\bibfield  {journal}
  {\bibinfo  {journal} {Phys. Rept.}\ }\textbf {\bibinfo {volume} {442}},\
  \bibinfo {pages} {38} (\bibinfo {year} {2007})},\ \Eprint
  {https://arxiv.org/abs/astro-ph/0612072} {arXiv:astro-ph/0612072}
  \BibitemShut {NoStop}%
\bibitem [{\citenamefont {Fiorillo}\ \emph
  {et~al.}(2023{\natexlab{b}})\citenamefont {Fiorillo}, \citenamefont
  {Raffelt},\ and\ \citenamefont {Vitagliano}}]{Fiorillo:2022cdq}%
  \BibitemOpen
  \bibfield  {author} {\bibinfo {author} {\bibfnamefont {D.~F.~G.}\
  \bibnamefont {Fiorillo}}, \bibinfo {author} {\bibfnamefont {G.~G.}\
  \bibnamefont {Raffelt}},\ and\ \bibinfo {author} {\bibfnamefont
  {E.}~\bibnamefont {Vitagliano}},\ }\bibfield  {title} {\bibinfo {title}
  {{Strong Supernova 1987A Constraints on Bosons Decaying to Neutrinos}},\
  }\href {https://doi.org/10.1103/PhysRevLett.131.021001} {\bibfield  {journal}
  {\bibinfo  {journal} {Phys. Rev. Lett.}\ }\textbf {\bibinfo {volume} {131}},\
  \bibinfo {pages} {021001} (\bibinfo {year} {2023}{\natexlab{b}})},\ \Eprint
  {https://arxiv.org/abs/2209.11773} {arXiv:2209.11773 [hep-ph]} \BibitemShut
  {NoStop}%
\bibitem [{\citenamefont {Akita}\ \emph {et~al.}(2022)\citenamefont {Akita},
  \citenamefont {Im},\ and\ \citenamefont {Masud}}]{Akita:2022etk}%
  \BibitemOpen
  \bibfield  {author} {\bibinfo {author} {\bibfnamefont {K.}~\bibnamefont
  {Akita}}, \bibinfo {author} {\bibfnamefont {S.~H.}\ \bibnamefont {Im}},\ and\
  \bibinfo {author} {\bibfnamefont {M.}~\bibnamefont {Masud}},\ }\bibfield
  {title} {\bibinfo {title} {{Probing non-standard neutrino interactions with a
  light boson from next galactic and diffuse supernova neutrinos}},\ }\href
  {https://doi.org/10.1007/JHEP12(2022)050} {\bibfield  {journal} {\bibinfo
  {journal} {JHEP}\ }\textbf {\bibinfo {volume} {12}},\ \bibinfo {pages}
  {050}},\ \Eprint {https://arxiv.org/abs/2206.06852} {arXiv:2206.06852
  [hep-ph]} \BibitemShut {NoStop}%
\bibitem [{\citenamefont {Syvolap}(2023)}]{Syvolap:2023trc}%
  \BibitemOpen
  \bibfield  {author} {\bibinfo {author} {\bibfnamefont {V.}~\bibnamefont
  {Syvolap}},\ }\bibfield  {title} {\bibinfo {title} {{Testing heavy neutral
  leptons produced in the supernovae explosions with future neutrino
  detectors}},\ }\href@noop {} {\  (\bibinfo {year} {2023})},\ \Eprint
  {https://arxiv.org/abs/2301.07052} {arXiv:2301.07052 [hep-ph]} \BibitemShut
  {NoStop}%
\bibitem [{\citenamefont {Akita}\ \emph {et~al.}(2024)\citenamefont {Akita},
  \citenamefont {Im}, \citenamefont {Masud},\ and\ \citenamefont
  {Yun}}]{Akita:2023iwq}%
  \BibitemOpen
  \bibfield  {author} {\bibinfo {author} {\bibfnamefont {K.}~\bibnamefont
  {Akita}}, \bibinfo {author} {\bibfnamefont {S.~H.}\ \bibnamefont {Im}},
  \bibinfo {author} {\bibfnamefont {M.}~\bibnamefont {Masud}},\ and\ \bibinfo
  {author} {\bibfnamefont {S.}~\bibnamefont {Yun}},\ }\bibfield  {title}
  {\bibinfo {title} {{Limits on heavy neutral leptons, Z' bosons and majorons
  from high-energy supernova neutrinos}},\ }\href
  {https://doi.org/10.1007/JHEP07(2024)057} {\bibfield  {journal} {\bibinfo
  {journal} {JHEP}\ }\textbf {\bibinfo {volume} {07}},\ \bibinfo {pages}
  {057}},\ \Eprint {https://arxiv.org/abs/2312.13627} {arXiv:2312.13627
  [hep-ph]} \BibitemShut {NoStop}%
\bibitem [{\citenamefont {Syvolap}\ and\ \citenamefont
  {Ruchayskiy}(2024)}]{Syvolap:2024hdh}%
  \BibitemOpen
  \bibfield  {author} {\bibinfo {author} {\bibfnamefont {V.}~\bibnamefont
  {Syvolap}}\ and\ \bibinfo {author} {\bibfnamefont {O.}~\bibnamefont
  {Ruchayskiy}},\ }\bibfield  {title} {\bibinfo {title} {{High-energy neutrino
  signals from supernova explosions: A new window into dark photon parameter
  space}},\ }\href {https://doi.org/10.1103/PhysRevD.110.115043} {\bibfield
  {journal} {\bibinfo  {journal} {Phys. Rev. D}\ }\textbf {\bibinfo {volume}
  {110}},\ \bibinfo {pages} {115043} (\bibinfo {year} {2024})},\ \Eprint
  {https://arxiv.org/abs/2404.19191} {arXiv:2404.19191 [hep-ph]} \BibitemShut
  {NoStop}%
\bibitem [{\citenamefont {Telalovic}\ \emph {et~al.}(2024)\citenamefont
  {Telalovic}, \citenamefont {Fiorillo}, \citenamefont
  {Mart{\'\i}nez-Mirav{\'e}}, \citenamefont {Vitagliano},\ and\ \citenamefont
  {Bustamante}}]{Telalovic:2024cot}%
  \BibitemOpen
  \bibfield  {author} {\bibinfo {author} {\bibfnamefont {B.}~\bibnamefont
  {Telalovic}}, \bibinfo {author} {\bibfnamefont {D.~F.~G.}\ \bibnamefont
  {Fiorillo}}, \bibinfo {author} {\bibfnamefont {P.}~\bibnamefont
  {Mart{\'\i}nez-Mirav{\'e}}}, \bibinfo {author} {\bibfnamefont
  {E.}~\bibnamefont {Vitagliano}},\ and\ \bibinfo {author} {\bibfnamefont
  {M.}~\bibnamefont {Bustamante}},\ }\bibfield  {title} {\bibinfo {title} {{The
  next galactic supernova can uncover mass and couplings of particles decaying
  to neutrinos}},\ }\href {https://doi.org/10.1088/1475-7516/2024/11/011}
  {\bibfield  {journal} {\bibinfo  {journal} {JCAP}\ }\textbf {\bibinfo
  {volume} {11}},\ \bibinfo {pages} {011}},\ \Eprint
  {https://arxiv.org/abs/2406.15506} {arXiv:2406.15506 [hep-ph]} \BibitemShut
  {NoStop}%
\bibitem [{\citenamefont {Blinov}\ \emph {et~al.}(2025)\citenamefont {Blinov},
  \citenamefont {Fox}, \citenamefont {Kelly}, \citenamefont {Plestid},\ and\
  \citenamefont {Zhou}}]{Blinov:2025aha}%
  \BibitemOpen
  \bibfield  {author} {\bibinfo {author} {\bibfnamefont {N.}~\bibnamefont
  {Blinov}}, \bibinfo {author} {\bibfnamefont {P.~J.}\ \bibnamefont {Fox}},
  \bibinfo {author} {\bibfnamefont {K.~J.}\ \bibnamefont {Kelly}}, \bibinfo
  {author} {\bibfnamefont {R.}~\bibnamefont {Plestid}},\ and\ \bibinfo {author}
  {\bibfnamefont {T.}~\bibnamefont {Zhou}},\ }\bibfield  {title} {\bibinfo
  {title} {{$L_\mu-L_\tau$ gauge bosons in beam dumps and supernovae}},\
  }\href@noop {} {\  (\bibinfo {year} {2025})},\ \Eprint
  {https://arxiv.org/abs/2511.09619} {arXiv:2511.09619 [hep-ph]} \BibitemShut
  {NoStop}%
\bibitem [{\citenamefont {Pastorello}\ \emph {et~al.}(2004)\citenamefont
  {Pastorello} \emph {et~al.}}]{Pastorello:2003tc}%
  \BibitemOpen
  \bibfield  {author} {\bibinfo {author} {\bibfnamefont {A.}~\bibnamefont
  {Pastorello}} \emph {et~al.},\ }\bibfield  {title} {\bibinfo {title} {{Low
  luminosity type II supernovae: spectroscopic and photometric evolution}},\
  }\href {https://doi.org/10.1111/j.1365-2966.2004.07173.x} {\bibfield
  {journal} {\bibinfo  {journal} {Mon. Not. Roy. Astron. Soc.}\ }\textbf
  {\bibinfo {volume} {347}},\ \bibinfo {pages} {74} (\bibinfo {year} {2004})},\
  \Eprint {https://arxiv.org/abs/astro-ph/0309264} {arXiv:astro-ph/0309264}
  \BibitemShut {NoStop}%
\bibitem [{\citenamefont {Prantzos}\ \emph {et~al.}(2011)\citenamefont
  {Prantzos} \emph {et~al.}}]{Prantzos:2010wi}%
  \BibitemOpen
  \bibfield  {author} {\bibinfo {author} {\bibfnamefont {N.}~\bibnamefont
  {Prantzos}} \emph {et~al.},\ }\bibfield  {title} {\bibinfo {title} {{The 511
  keV emission from positron annihilation in the Galaxy}},\ }\href
  {https://doi.org/10.1103/RevModPhys.83.1001} {\bibfield  {journal} {\bibinfo
  {journal} {Rev. Mod. Phys.}\ }\textbf {\bibinfo {volume} {83}},\ \bibinfo
  {pages} {1001} (\bibinfo {year} {2011})},\ \Eprint
  {https://arxiv.org/abs/1009.4620} {arXiv:1009.4620 [astro-ph.HE]}
  \BibitemShut {NoStop}%
\bibitem [{\citenamefont {Spiro}\ \emph {et~al.}(2014)\citenamefont {Spiro}
  \emph {et~al.}}]{Spiro_2014}%
  \BibitemOpen
  \bibfield  {author} {\bibinfo {author} {\bibfnamefont {S.}~\bibnamefont
  {Spiro}} \emph {et~al.},\ }\bibfield  {title} {\bibinfo {title} {Low
  luminosity type ii supernovae – ii. pointing towards moderate mass
  precursors},\ }\href {https://doi.org/10.1093/mnras/stu156} {\bibfield
  {journal} {\bibinfo  {journal} {Monthly Notices of the Royal Astronomical
  Society}\ }\textbf {\bibinfo {volume} {439}},\ \bibinfo {pages} {2873–2892}
  (\bibinfo {year} {2014})}\BibitemShut {NoStop}%
\bibitem [{\citenamefont {Sung}\ \emph {et~al.}(2019)\citenamefont {Sung},
  \citenamefont {Tu},\ and\ \citenamefont {Wu}}]{Sung:2019xie}%
  \BibitemOpen
  \bibfield  {author} {\bibinfo {author} {\bibfnamefont {A.}~\bibnamefont
  {Sung}}, \bibinfo {author} {\bibfnamefont {H.}~\bibnamefont {Tu}},\ and\
  \bibinfo {author} {\bibfnamefont {M.-R.}\ \bibnamefont {Wu}},\ }\bibfield
  {title} {\bibinfo {title} {{New constraint from supernova explosions on light
  particles beyond the Standard Model}},\ }\href
  {https://doi.org/10.1103/PhysRevD.99.121305} {\bibfield  {journal} {\bibinfo
  {journal} {Phys. Rev. D}\ }\textbf {\bibinfo {volume} {99}},\ \bibinfo
  {pages} {121305} (\bibinfo {year} {2019})},\ \Eprint
  {https://arxiv.org/abs/1903.07923} {arXiv:1903.07923 [hep-ph]} \BibitemShut
  {NoStop}%
\bibitem [{\citenamefont {Caputo}\ \emph
  {et~al.}(2022{\natexlab{a}})\citenamefont {Caputo}, \citenamefont {Janka},
  \citenamefont {Raffelt},\ and\ \citenamefont {Vitagliano}}]{Caputo:2022mah}%
  \BibitemOpen
  \bibfield  {author} {\bibinfo {author} {\bibfnamefont {A.}~\bibnamefont
  {Caputo}}, \bibinfo {author} {\bibfnamefont {H.-T.}\ \bibnamefont {Janka}},
  \bibinfo {author} {\bibfnamefont {G.}~\bibnamefont {Raffelt}},\ and\ \bibinfo
  {author} {\bibfnamefont {E.}~\bibnamefont {Vitagliano}},\ }\bibfield  {title}
  {\bibinfo {title} {{Low-Energy Supernovae Severely Constrain Radiative
  Particle Decays}},\ }\href {https://doi.org/10.1103/PhysRevLett.128.221103}
  {\bibfield  {journal} {\bibinfo  {journal} {Phys. Rev. Lett.}\ }\textbf
  {\bibinfo {volume} {128}},\ \bibinfo {pages} {221103} (\bibinfo {year}
  {2022}{\natexlab{a}})},\ \Eprint {https://arxiv.org/abs/2201.09890}
  {arXiv:2201.09890 [astro-ph.HE]} \BibitemShut {NoStop}%
\bibitem [{\citenamefont {Diamond}\ \emph {et~al.}(2024)\citenamefont
  {Diamond}, \citenamefont {Fiorillo}, \citenamefont {Marques-Tavares},
  \citenamefont {Tamborra},\ and\ \citenamefont
  {Vitagliano}}]{Diamond:2023cto}%
  \BibitemOpen
  \bibfield  {author} {\bibinfo {author} {\bibfnamefont {M.}~\bibnamefont
  {Diamond}}, \bibinfo {author} {\bibfnamefont {D.~F.~G.}\ \bibnamefont
  {Fiorillo}}, \bibinfo {author} {\bibfnamefont {G.}~\bibnamefont
  {Marques-Tavares}}, \bibinfo {author} {\bibfnamefont {I.}~\bibnamefont
  {Tamborra}},\ and\ \bibinfo {author} {\bibfnamefont {E.}~\bibnamefont
  {Vitagliano}},\ }\bibfield  {title} {\bibinfo {title} {{Multimessenger
  Constraints on Radiatively Decaying Axions from GW170817}},\ }\href
  {https://doi.org/10.1103/PhysRevLett.132.101004} {\bibfield  {journal}
  {\bibinfo  {journal} {Phys. Rev. Lett.}\ }\textbf {\bibinfo {volume} {132}},\
  \bibinfo {pages} {101004} (\bibinfo {year} {2024})},\ \Eprint
  {https://arxiv.org/abs/2305.10327} {arXiv:2305.10327 [hep-ph]} \BibitemShut
  {NoStop}%
\bibitem [{\citenamefont {Fiorillo}\ \emph {et~al.}(2025)\citenamefont
  {Fiorillo}, \citenamefont {Pitik},\ and\ \citenamefont
  {Vitagliano}}]{Fiorillo:2025yzf}%
  \BibitemOpen
  \bibfield  {author} {\bibinfo {author} {\bibfnamefont {D.~F.~G.}\
  \bibnamefont {Fiorillo}}, \bibinfo {author} {\bibfnamefont {T.}~\bibnamefont
  {Pitik}},\ and\ \bibinfo {author} {\bibfnamefont {E.}~\bibnamefont
  {Vitagliano}},\ }\bibfield  {title} {\bibinfo {title} {{Energy Transfer by
  Feebly Interacting Particles in Supernovae: The Trapping Regime}},\ }\href
  {https://doi.org/10.1103/cz94-dqxt} {\bibfield  {journal} {\bibinfo
  {journal} {Phys. Rev. Lett.}\ }\textbf {\bibinfo {volume} {135}},\ \bibinfo
  {pages} {071005} (\bibinfo {year} {2025})},\ \Eprint
  {https://arxiv.org/abs/2503.13653} {arXiv:2503.13653 [hep-ph]} \BibitemShut
  {NoStop}%
\bibitem [{\citenamefont {Cand{\'o}n}\ \emph {et~al.}(2025)\citenamefont
  {Cand{\'o}n}, \citenamefont {Fiorillo}, \citenamefont {Janka}, \citenamefont
  {van Baal},\ and\ \citenamefont {Vitagliano}}]{Candon:2025ypl}%
  \BibitemOpen
  \bibfield  {author} {\bibinfo {author} {\bibfnamefont {F.~R.}\ \bibnamefont
  {Cand{\'o}n}}, \bibinfo {author} {\bibfnamefont {D.~F.~G.}\ \bibnamefont
  {Fiorillo}}, \bibinfo {author} {\bibfnamefont {H.-T.}\ \bibnamefont {Janka}},
  \bibinfo {author} {\bibfnamefont {B.~F.~A.}\ \bibnamefont {van Baal}},\ and\
  \bibinfo {author} {\bibfnamefont {E.}~\bibnamefont {Vitagliano}},\ }\bibfield
   {title} {\bibinfo {title} {{Small Progenitors, Large Couplings: Type Ic
  Supernova Constraints on Radiatively Decaying Particles}},\ }\href@noop {} {\
   (\bibinfo {year} {2025})},\ \Eprint {https://arxiv.org/abs/2509.18253}
  {arXiv:2509.18253 [hep-ph]} \BibitemShut {NoStop}%
\bibitem [{\citenamefont {Chang}\ \emph {et~al.}(2017)\citenamefont {Chang},
  \citenamefont {Essig},\ and\ \citenamefont {McDermott}}]{Chang:2016ntp}%
  \BibitemOpen
  \bibfield  {author} {\bibinfo {author} {\bibfnamefont {J.~H.}\ \bibnamefont
  {Chang}}, \bibinfo {author} {\bibfnamefont {R.}~\bibnamefont {Essig}},\ and\
  \bibinfo {author} {\bibfnamefont {S.~D.}\ \bibnamefont {McDermott}},\
  }\bibfield  {title} {\bibinfo {title} {{Revisiting Supernova 1987A
  Constraints on Dark Photons}},\ }\href
  {https://doi.org/10.1007/JHEP01(2017)107} {\bibfield  {journal} {\bibinfo
  {journal} {JHEP}\ }\textbf {\bibinfo {volume} {01}},\ \bibinfo {pages}
  {107}},\ \Eprint {https://arxiv.org/abs/1611.03864} {arXiv:1611.03864
  [hep-ph]} \BibitemShut {NoStop}%
\bibitem [{\citenamefont {Hardy}\ and\ \citenamefont
  {Lasenby}(2017)}]{Hardy:2016kme}%
  \BibitemOpen
  \bibfield  {author} {\bibinfo {author} {\bibfnamefont {E.}~\bibnamefont
  {Hardy}}\ and\ \bibinfo {author} {\bibfnamefont {R.}~\bibnamefont
  {Lasenby}},\ }\bibfield  {title} {\bibinfo {title} {{Stellar cooling bounds
  on new light particles: plasma mixing effects}},\ }\href
  {https://doi.org/10.1007/JHEP02(2017)033} {\bibfield  {journal} {\bibinfo
  {journal} {JHEP}\ }\textbf {\bibinfo {volume} {02}},\ \bibinfo {pages}
  {033}},\ \Eprint {https://arxiv.org/abs/1611.05852} {arXiv:1611.05852
  [hep-ph]} \BibitemShut {NoStop}%
\bibitem [{\citenamefont {Mori}\ \emph {et~al.}(2026)\citenamefont {Mori},
  \citenamefont {Takiwaki},\ and\ \citenamefont {Kohri}}]{Mori:2025baz}%
  \BibitemOpen
  \bibfield  {author} {\bibinfo {author} {\bibfnamefont {K.}~\bibnamefont
  {Mori}}, \bibinfo {author} {\bibfnamefont {T.}~\bibnamefont {Takiwaki}},\
  and\ \bibinfo {author} {\bibfnamefont {K.}~\bibnamefont {Kohri}},\ }\bibfield
   {title} {\bibinfo {title} {{Verifying the failing supernova constraint on
  dark photons with two-dimensional hydrodynamic simulations}},\ }\href
  {https://doi.org/10.1103/d4zp-bxbn} {\bibfield  {journal} {\bibinfo
  {journal} {Phys. Rev. D}\ }\textbf {\bibinfo {volume} {113}},\ \bibinfo
  {pages} {L041303} (\bibinfo {year} {2026})},\ \Eprint
  {https://arxiv.org/abs/2511.18122} {arXiv:2511.18122 [astro-ph.HE]}
  \BibitemShut {NoStop}%
\bibitem [{\citenamefont {Bjorken}\ \emph {et~al.}(2009)\citenamefont
  {Bjorken}, \citenamefont {Essig}, \citenamefont {Schuster},\ and\
  \citenamefont {Toro}}]{Bjorken:2009mm}%
  \BibitemOpen
  \bibfield  {author} {\bibinfo {author} {\bibfnamefont {J.~D.}\ \bibnamefont
  {Bjorken}}, \bibinfo {author} {\bibfnamefont {R.}~\bibnamefont {Essig}},
  \bibinfo {author} {\bibfnamefont {P.}~\bibnamefont {Schuster}},\ and\
  \bibinfo {author} {\bibfnamefont {N.}~\bibnamefont {Toro}},\ }\bibfield
  {title} {\bibinfo {title} {{New Fixed-Target Experiments to Search for Dark
  Gauge Forces}},\ }\href {https://doi.org/10.1103/PhysRevD.80.075018}
  {\bibfield  {journal} {\bibinfo  {journal} {Phys. Rev. D}\ }\textbf {\bibinfo
  {volume} {80}},\ \bibinfo {pages} {075018} (\bibinfo {year} {2009})},\
  \Eprint {https://arxiv.org/abs/0906.0580} {arXiv:0906.0580 [hep-ph]}
  \BibitemShut {NoStop}%
\bibitem [{\citenamefont {Dent}\ \emph {et~al.}(2012)\citenamefont {Dent},
  \citenamefont {Ferrer},\ and\ \citenamefont {Krauss}}]{Dent:2012mx}%
  \BibitemOpen
  \bibfield  {author} {\bibinfo {author} {\bibfnamefont {J.~B.}\ \bibnamefont
  {Dent}}, \bibinfo {author} {\bibfnamefont {F.}~\bibnamefont {Ferrer}},\ and\
  \bibinfo {author} {\bibfnamefont {L.~M.}\ \bibnamefont {Krauss}},\ }\bibfield
   {title} {\bibinfo {title} {{Constraints on Light Hidden Sector Gauge Bosons
  from Supernova Cooling}},\ }\href@noop {} {\  (\bibinfo {year} {2012})},\
  \Eprint {https://arxiv.org/abs/1201.2683} {arXiv:1201.2683 [astro-ph.CO]}
  \BibitemShut {NoStop}%
\bibitem [{\citenamefont {Kazanas}\ \emph {et~al.}(2014)\citenamefont
  {Kazanas}, \citenamefont {Mohapatra}, \citenamefont {Nussinov}, \citenamefont
  {Teplitz},\ and\ \citenamefont {Zhang}}]{Kazanas:2014mca}%
  \BibitemOpen
  \bibfield  {author} {\bibinfo {author} {\bibfnamefont {D.}~\bibnamefont
  {Kazanas}}, \bibinfo {author} {\bibfnamefont {R.~N.}\ \bibnamefont
  {Mohapatra}}, \bibinfo {author} {\bibfnamefont {S.}~\bibnamefont {Nussinov}},
  \bibinfo {author} {\bibfnamefont {V.~L.}\ \bibnamefont {Teplitz}},\ and\
  \bibinfo {author} {\bibfnamefont {Y.}~\bibnamefont {Zhang}},\ }\bibfield
  {title} {\bibinfo {title} {{Supernova Bounds on the Dark Photon Using its
  Electromagnetic Decay}},\ }\href
  {https://doi.org/10.1016/j.nuclphysb.2014.11.009} {\bibfield  {journal}
  {\bibinfo  {journal} {Nucl. Phys. B}\ }\textbf {\bibinfo {volume} {890}},\
  \bibinfo {pages} {17} (\bibinfo {year} {2014})},\ \Eprint
  {https://arxiv.org/abs/1410.0221} {arXiv:1410.0221 [hep-ph]} \BibitemShut
  {NoStop}%
\bibitem [{\citenamefont {Rrapaj}\ and\ \citenamefont
  {Reddy}(2016)}]{Rrapaj:2015wgs}%
  \BibitemOpen
  \bibfield  {author} {\bibinfo {author} {\bibfnamefont {E.}~\bibnamefont
  {Rrapaj}}\ and\ \bibinfo {author} {\bibfnamefont {S.}~\bibnamefont {Reddy}},\
  }\bibfield  {title} {\bibinfo {title} {{Nucleon-nucleon bremsstrahlung of
  dark gauge bosons and revised supernova constraints}},\ }\href
  {https://doi.org/10.1103/PhysRevC.94.045805} {\bibfield  {journal} {\bibinfo
  {journal} {Phys. Rev. C}\ }\textbf {\bibinfo {volume} {94}},\ \bibinfo
  {pages} {045805} (\bibinfo {year} {2016})},\ \Eprint
  {https://arxiv.org/abs/1511.09136} {arXiv:1511.09136 [nucl-th]} \BibitemShut
  {NoStop}%
\bibitem [{\citenamefont {Croon}\ \emph {et~al.}(2021)\citenamefont {Croon},
  \citenamefont {Elor}, \citenamefont {Leane},\ and\ \citenamefont
  {McDermott}}]{Croon:2020lrf}%
  \BibitemOpen
  \bibfield  {author} {\bibinfo {author} {\bibfnamefont {D.}~\bibnamefont
  {Croon}}, \bibinfo {author} {\bibfnamefont {G.}~\bibnamefont {Elor}},
  \bibinfo {author} {\bibfnamefont {R.~K.}\ \bibnamefont {Leane}},\ and\
  \bibinfo {author} {\bibfnamefont {S.~D.}\ \bibnamefont {McDermott}},\
  }\bibfield  {title} {\bibinfo {title} {{Supernova Muons: New Constraints on
  $Z$' Bosons, Axions and ALPs}},\ }\href
  {https://doi.org/10.1007/JHEP01(2021)107} {\bibfield  {journal} {\bibinfo
  {journal} {JHEP}\ }\textbf {\bibinfo {volume} {01}},\ \bibinfo {pages}
  {107}},\ \Eprint {https://arxiv.org/abs/2006.13942} {arXiv:2006.13942
  [hep-ph]} \BibitemShut {NoStop}%
\bibitem [{\citenamefont {Shin}\ and\ \citenamefont
  {Yun}(2022)}]{Shin:2021bvz}%
  \BibitemOpen
  \bibfield  {author} {\bibinfo {author} {\bibfnamefont {C.~S.}\ \bibnamefont
  {Shin}}\ and\ \bibinfo {author} {\bibfnamefont {S.}~\bibnamefont {Yun}},\
  }\bibfield  {title} {\bibinfo {title} {{Dark gauge boson production from
  neutron stars via nucleon-nucleon bremsstrahlung}},\ }\href
  {https://doi.org/10.1007/JHEP02(2022)133} {\bibfield  {journal} {\bibinfo
  {journal} {JHEP}\ }\textbf {\bibinfo {volume} {02}},\ \bibinfo {pages}
  {133}},\ \Eprint {https://arxiv.org/abs/2110.03362} {arXiv:2110.03362
  [hep-ph]} \BibitemShut {NoStop}%
\bibitem [{\citenamefont {Caputo}\ \emph
  {et~al.}(2022{\natexlab{b}})\citenamefont {Caputo}, \citenamefont {Raffelt},\
  and\ \citenamefont {Vitagliano}}]{Caputo:2021rux}%
  \BibitemOpen
  \bibfield  {author} {\bibinfo {author} {\bibfnamefont {A.}~\bibnamefont
  {Caputo}}, \bibinfo {author} {\bibfnamefont {G.}~\bibnamefont {Raffelt}},\
  and\ \bibinfo {author} {\bibfnamefont {E.}~\bibnamefont {Vitagliano}},\
  }\bibfield  {title} {\bibinfo {title} {{Muonic boson limits: Supernova
  redux}},\ }\href {https://doi.org/10.1103/PhysRevD.105.035022} {\bibfield
  {journal} {\bibinfo  {journal} {Phys. Rev. D}\ }\textbf {\bibinfo {volume}
  {105}},\ \bibinfo {pages} {035022} (\bibinfo {year} {2022}{\natexlab{b}})},\
  \Eprint {https://arxiv.org/abs/2109.03244} {arXiv:2109.03244 [hep-ph]}
  \BibitemShut {NoStop}%
\bibitem [{\citenamefont {Caputo}\ \emph
  {et~al.}(2022{\natexlab{c}})\citenamefont {Caputo}, \citenamefont {Raffelt},\
  and\ \citenamefont {Vitagliano}}]{Caputo:2022rca}%
  \BibitemOpen
  \bibfield  {author} {\bibinfo {author} {\bibfnamefont {A.}~\bibnamefont
  {Caputo}}, \bibinfo {author} {\bibfnamefont {G.}~\bibnamefont {Raffelt}},\
  and\ \bibinfo {author} {\bibfnamefont {E.}~\bibnamefont {Vitagliano}},\
  }\bibfield  {title} {\bibinfo {title} {{Radiative transfer in stars by feebly
  interacting bosons}},\ }\href {https://doi.org/10.1088/1475-7516/2022/08/045}
  {\bibfield  {journal} {\bibinfo  {journal} {JCAP}\ }\textbf {\bibinfo
  {volume} {08}}\bibfield  {number} {\bibinfo  {number} { (08)},\ \bibinfo
  {pages} {045}},\ }\Eprint {https://arxiv.org/abs/2204.11862}
  {arXiv:2204.11862 [astro-ph.SR]} \BibitemShut {NoStop}%
\bibitem [{\citenamefont {Fiorillo}\ \emph
  {et~al.}(2024{\natexlab{a}})\citenamefont {Fiorillo}, \citenamefont
  {Raffelt},\ and\ \citenamefont {Vitagliano}}]{Fiorillo:2023cas}%
  \BibitemOpen
  \bibfield  {author} {\bibinfo {author} {\bibfnamefont {D.~F.~G.}\
  \bibnamefont {Fiorillo}}, \bibinfo {author} {\bibfnamefont {G.~G.}\
  \bibnamefont {Raffelt}},\ and\ \bibinfo {author} {\bibfnamefont
  {E.}~\bibnamefont {Vitagliano}},\ }\bibfield  {title} {\bibinfo {title}
  {{Supernova emission of secretly interacting neutrino fluid: Theoretical
  foundations}},\ }\href {https://doi.org/10.1103/PhysRevD.109.023017}
  {\bibfield  {journal} {\bibinfo  {journal} {Phys. Rev. D}\ }\textbf {\bibinfo
  {volume} {109}},\ \bibinfo {pages} {023017} (\bibinfo {year}
  {2024}{\natexlab{a}})},\ \Eprint {https://arxiv.org/abs/2307.15122}
  {arXiv:2307.15122 [hep-ph]} \BibitemShut {NoStop}%
\bibitem [{\citenamefont {Fiorillo}\ \emph
  {et~al.}(2024{\natexlab{b}})\citenamefont {Fiorillo}, \citenamefont
  {Raffelt},\ and\ \citenamefont {Vitagliano}}]{Fiorillo:2023ytr}%
  \BibitemOpen
  \bibfield  {author} {\bibinfo {author} {\bibfnamefont {D.~F.~G.}\
  \bibnamefont {Fiorillo}}, \bibinfo {author} {\bibfnamefont {G.~G.}\
  \bibnamefont {Raffelt}},\ and\ \bibinfo {author} {\bibfnamefont
  {E.}~\bibnamefont {Vitagliano}},\ }\bibfield  {title} {\bibinfo {title}
  {{Large Neutrino Secret Interactions Have a Small Impact on Supernovae}},\
  }\href {https://doi.org/10.1103/PhysRevLett.132.021002} {\bibfield  {journal}
  {\bibinfo  {journal} {Phys. Rev. Lett.}\ }\textbf {\bibinfo {volume} {132}},\
  \bibinfo {pages} {021002} (\bibinfo {year} {2024}{\natexlab{b}})},\ \Eprint
  {https://arxiv.org/abs/2307.15115} {arXiv:2307.15115 [hep-ph]} \BibitemShut
  {NoStop}%
\bibitem [{\citenamefont {Lai}\ \emph {et~al.}(2024)\citenamefont {Lai},
  \citenamefont {Leung},\ and\ \citenamefont {Lin}}]{Lai:2024mse}%
  \BibitemOpen
  \bibfield  {author} {\bibinfo {author} {\bibfnamefont {K.-C.}\ \bibnamefont
  {Lai}}, \bibinfo {author} {\bibfnamefont {C.~S.~J.}\ \bibnamefont {Leung}},\
  and\ \bibinfo {author} {\bibfnamefont {G.-L.}\ \bibnamefont {Lin}},\
  }\bibfield  {title} {\bibinfo {title} {{SN1987A constraints to BSM models
  with extra neutral bosons near the trapping regime:
  U(1)L{\ensuremath{\mu}}-L{\ensuremath{\tau}} model as an illustrative
  example}},\ }\href {https://doi.org/10.1103/PhysRevD.110.103023} {\bibfield
  {journal} {\bibinfo  {journal} {Phys. Rev. D}\ }\textbf {\bibinfo {volume}
  {110}},\ \bibinfo {pages} {103023} (\bibinfo {year} {2024})},\ \Eprint
  {https://arxiv.org/abs/2401.16023} {arXiv:2401.16023 [hep-ph]} \BibitemShut
  {NoStop}%
\bibitem [{\citenamefont {Steiner}\ \emph {et~al.}(2013)\citenamefont
  {Steiner}, \citenamefont {Hempel},\ and\ \citenamefont
  {Fischer}}]{Steiner:2012rk}%
  \BibitemOpen
  \bibfield  {author} {\bibinfo {author} {\bibfnamefont {A.~W.}\ \bibnamefont
  {Steiner}}, \bibinfo {author} {\bibfnamefont {M.}~\bibnamefont {Hempel}},\
  and\ \bibinfo {author} {\bibfnamefont {T.}~\bibnamefont {Fischer}},\
  }\bibfield  {title} {\bibinfo {title} {{Core-collapse supernova equations of
  state based on neutron star observations}},\ }\href
  {https://doi.org/10.1088/0004-637X/774/1/17} {\bibfield  {journal} {\bibinfo
  {journal} {Astrophys. J.}\ }\textbf {\bibinfo {volume} {774}},\ \bibinfo
  {pages} {17} (\bibinfo {year} {2013})},\ \Eprint
  {https://arxiv.org/abs/1207.2184} {arXiv:1207.2184 [astro-ph.SR]}
  \BibitemShut {NoStop}%
\bibitem [{\citenamefont {Lattimer}\ and\ \citenamefont
  {Swesty}(1991)}]{Lattimer:1991nc}%
  \BibitemOpen
  \bibfield  {author} {\bibinfo {author} {\bibfnamefont {J.~M.}\ \bibnamefont
  {Lattimer}}\ and\ \bibinfo {author} {\bibfnamefont {F.~D.}\ \bibnamefont
  {Swesty}},\ }\bibfield  {title} {\bibinfo {title} {{A Generalized equation of
  state for hot, dense matter}},\ }\href
  {https://doi.org/10.1016/0375-9474(91)90452-C} {\bibfield  {journal}
  {\bibinfo  {journal} {Nucl. Phys. A}\ }\textbf {\bibinfo {volume} {535}},\
  \bibinfo {pages} {331} (\bibinfo {year} {1991})}\BibitemShut {NoStop}%
\bibitem [{\citenamefont {Woosley}\ \emph {et~al.}(2002)\citenamefont
  {Woosley}, \citenamefont {Heger},\ and\ \citenamefont
  {Weaver}}]{Woosley:2002zz}%
  \BibitemOpen
  \bibfield  {author} {\bibinfo {author} {\bibfnamefont {S.~E.}\ \bibnamefont
  {Woosley}}, \bibinfo {author} {\bibfnamefont {A.}~\bibnamefont {Heger}},\
  and\ \bibinfo {author} {\bibfnamefont {T.~A.}\ \bibnamefont {Weaver}},\
  }\bibfield  {title} {\bibinfo {title} {{The evolution and explosion of
  massive stars}},\ }\href {https://doi.org/10.1103/RevModPhys.74.1015}
  {\bibfield  {journal} {\bibinfo  {journal} {Rev. Mod. Phys.}\ }\textbf
  {\bibinfo {volume} {74}},\ \bibinfo {pages} {1015} (\bibinfo {year}
  {2002})}\BibitemShut {NoStop}%
\bibitem [{\citenamefont {Sukhbold}\ \emph {et~al.}(2018)\citenamefont
  {Sukhbold}, \citenamefont {Woosley},\ and\ \citenamefont
  {Heger}}]{Sukhbold:2017cnt}%
  \BibitemOpen
  \bibfield  {author} {\bibinfo {author} {\bibfnamefont {T.}~\bibnamefont
  {Sukhbold}}, \bibinfo {author} {\bibfnamefont {S.}~\bibnamefont {Woosley}},\
  and\ \bibinfo {author} {\bibfnamefont {A.}~\bibnamefont {Heger}},\ }\bibfield
   {title} {\bibinfo {title} {{A High-resolution Study of Presupernova Core
  Structure}},\ }\href {https://doi.org/10.3847/1538-4357/aac2da} {\bibfield
  {journal} {\bibinfo  {journal} {Astrophys. J.}\ }\textbf {\bibinfo {volume}
  {860}},\ \bibinfo {pages} {93} (\bibinfo {year} {2018})},\ \Eprint
  {https://arxiv.org/abs/1710.03243} {arXiv:1710.03243 [astro-ph.HE]}
  \BibitemShut {NoStop}%
\bibitem [{\citenamefont {Woosley}\ and\ \citenamefont
  {Heger}(2007)}]{Woosley:2007as}%
  \BibitemOpen
  \bibfield  {author} {\bibinfo {author} {\bibfnamefont {S.~E.}\ \bibnamefont
  {Woosley}}\ and\ \bibinfo {author} {\bibfnamefont {A.}~\bibnamefont
  {Heger}},\ }\bibfield  {title} {\bibinfo {title} {{Nucleosynthesis and
  Remnants in Massive Stars of Solar Metallicity}},\ }\href
  {https://doi.org/10.1016/j.physrep.2007.02.009} {\bibfield  {journal}
  {\bibinfo  {journal} {Phys. Rept.}\ }\textbf {\bibinfo {volume} {442}},\
  \bibinfo {pages} {269} (\bibinfo {year} {2007})},\ \Eprint
  {https://arxiv.org/abs/astro-ph/0702176} {arXiv:astro-ph/0702176}
  \BibitemShut {NoStop}%
\bibitem [{\citenamefont {Bollig}\ \emph {et~al.}(2020)\citenamefont {Bollig},
  \citenamefont {DeRocco}, \citenamefont {Graham},\ and\ \citenamefont
  {Janka}}]{Bollig:2020xdr}%
  \BibitemOpen
  \bibfield  {author} {\bibinfo {author} {\bibfnamefont {R.}~\bibnamefont
  {Bollig}}, \bibinfo {author} {\bibfnamefont {W.}~\bibnamefont {DeRocco}},
  \bibinfo {author} {\bibfnamefont {P.~W.}\ \bibnamefont {Graham}},\ and\
  \bibinfo {author} {\bibfnamefont {H.-T.}\ \bibnamefont {Janka}},\ }\bibfield
  {title} {\bibinfo {title} {{Muons in Supernovae: Implications for the
  Axion-Muon Coupling}},\ }\href
  {https://doi.org/10.1103/PhysRevLett.125.051104} {\bibfield  {journal}
  {\bibinfo  {journal} {Phys. Rev. Lett.}\ }\textbf {\bibinfo {volume} {125}},\
  \bibinfo {pages} {051104} (\bibinfo {year} {2020})},\ \bibinfo {note}
  {[Erratum: Phys.Rev.Lett. 126, 189901 (2021)]},\ \Eprint
  {https://arxiv.org/abs/2005.07141} {arXiv:2005.07141 [hep-ph]} \BibitemShut
  {NoStop}%
\bibitem [{\citenamefont {Braaten}\ and\ \citenamefont
  {Segel}(1993)}]{Braaten:1993jw}%
  \BibitemOpen
  \bibfield  {author} {\bibinfo {author} {\bibfnamefont {E.}~\bibnamefont
  {Braaten}}\ and\ \bibinfo {author} {\bibfnamefont {D.}~\bibnamefont
  {Segel}},\ }\bibfield  {title} {\bibinfo {title} {{Neutrino energy loss from
  the plasma process at all temperatures and densities}},\ }\href
  {https://doi.org/10.1103/PhysRevD.48.1478} {\bibfield  {journal} {\bibinfo
  {journal} {Phys. Rev. D}\ }\textbf {\bibinfo {volume} {48}},\ \bibinfo
  {pages} {1478} (\bibinfo {year} {1993})},\ \Eprint
  {https://arxiv.org/abs/hep-ph/9302213} {arXiv:hep-ph/9302213} \BibitemShut
  {NoStop}%
\bibitem [{\citenamefont {An}\ \emph {et~al.}(2013)\citenamefont {An},
  \citenamefont {Pospelov},\ and\ \citenamefont {Pradler}}]{An:2013yfc}%
  \BibitemOpen
  \bibfield  {author} {\bibinfo {author} {\bibfnamefont {H.}~\bibnamefont
  {An}}, \bibinfo {author} {\bibfnamefont {M.}~\bibnamefont {Pospelov}},\ and\
  \bibinfo {author} {\bibfnamefont {J.}~\bibnamefont {Pradler}},\ }\bibfield
  {title} {\bibinfo {title} {{New stellar constraints on dark photons}},\
  }\href {https://doi.org/10.1016/j.physletb.2013.07.008} {\bibfield  {journal}
  {\bibinfo  {journal} {Phys. Lett. B}\ }\textbf {\bibinfo {volume} {725}},\
  \bibinfo {pages} {190} (\bibinfo {year} {2013})},\ \Eprint
  {https://arxiv.org/abs/1302.3884} {arXiv:1302.3884 [hep-ph]} \BibitemShut
  {NoStop}%
\bibitem [{\citenamefont {Weldon}(1983)}]{Weldon:1983jn}%
  \BibitemOpen
  \bibfield  {author} {\bibinfo {author} {\bibfnamefont {H.~A.}\ \bibnamefont
  {Weldon}},\ }\bibfield  {title} {\bibinfo {title} {{Simple Rules for
  Discontinuities in Finite Temperature Field Theory}},\ }\href
  {https://doi.org/10.1103/PhysRevD.28.2007} {\bibfield  {journal} {\bibinfo
  {journal} {Phys. Rev. D}\ }\textbf {\bibinfo {volume} {28}},\ \bibinfo
  {pages} {2007} (\bibinfo {year} {1983})}\BibitemShut {NoStop}%
\bibitem [{\citenamefont {Braaten}(1991)}]{Braaten:1991hg}%
  \BibitemOpen
  \bibfield  {author} {\bibinfo {author} {\bibfnamefont {E.}~\bibnamefont
  {Braaten}},\ }\bibfield  {title} {\bibinfo {title} {{Neutrino emissivity of
  an ultrarelativistic plasma from positron and plasmino annihilation}},\
  }\href@noop {} {\  (\bibinfo {year} {1991})}\BibitemShut {NoStop}%
\bibitem [{\citenamefont {Hardy}\ \emph {et~al.}(2024)\citenamefont {Hardy},
  \citenamefont {Sokolov},\ and\ \citenamefont {Stubbs}}]{Hardy:2024gwy}%
  \BibitemOpen
  \bibfield  {author} {\bibinfo {author} {\bibfnamefont {E.}~\bibnamefont
  {Hardy}}, \bibinfo {author} {\bibfnamefont {A.}~\bibnamefont {Sokolov}},\
  and\ \bibinfo {author} {\bibfnamefont {H.}~\bibnamefont {Stubbs}},\
  }\bibfield  {title} {\bibinfo {title} {{Supernova bounds on new scalars from
  resonant and soft emission}},\ }\href@noop {} {\  (\bibinfo {year} {2024})},\
  \Eprint {https://arxiv.org/abs/2410.17347} {arXiv:2410.17347 [hep-ph]}
  \BibitemShut {NoStop}%
\bibitem [{\citenamefont {Marteau}\ \emph {et~al.}(1999)\citenamefont
  {Marteau}, \citenamefont {Delorme},\ and\ \citenamefont
  {Ericson}}]{Marteau:1999zp}%
  \BibitemOpen
  \bibfield  {author} {\bibinfo {author} {\bibfnamefont {J.}~\bibnamefont
  {Marteau}}, \bibinfo {author} {\bibfnamefont {J.}~\bibnamefont {Delorme}},\
  and\ \bibinfo {author} {\bibfnamefont {M.}~\bibnamefont {Ericson}},\
  }\bibfield  {title} {\bibinfo {title} {{Neutrino oxygen interactions: Role of
  nuclear physics in the atmospheric neutrino anomaly}},\ }in\ \href@noop {}
  {\emph {\bibinfo {booktitle} {{34th Rencontres de Moriond: Electroweak
  Interactions and Unified Theories}}}}\ (\bibinfo {year} {1999})\ pp.\
  \bibinfo {pages} {121--126},\ \Eprint {https://arxiv.org/abs/hep-ph/9906449}
  {arXiv:hep-ph/9906449} \BibitemShut {NoStop}%
\bibitem [{\citenamefont {Kolbe}\ \emph {et~al.}(2002)\citenamefont {Kolbe},
  \citenamefont {Langanke},\ and\ \citenamefont {Vogel}}]{Kolbe:2002gk}%
  \BibitemOpen
  \bibfield  {author} {\bibinfo {author} {\bibfnamefont {E.}~\bibnamefont
  {Kolbe}}, \bibinfo {author} {\bibfnamefont {K.}~\bibnamefont {Langanke}},\
  and\ \bibinfo {author} {\bibfnamefont {P.}~\bibnamefont {Vogel}},\ }\bibfield
   {title} {\bibinfo {title} {{Estimates of weak and electromagnetic nuclear
  decay signatures for neutrino reactions in Super-Kamiokande}},\ }\href
  {https://doi.org/10.1103/PhysRevD.66.013007} {\bibfield  {journal} {\bibinfo
  {journal} {Phys. Rev. D}\ }\textbf {\bibinfo {volume} {66}},\ \bibinfo
  {pages} {013007} (\bibinfo {year} {2002})}\BibitemShut {NoStop}%
\bibitem [{\citenamefont {Strumia}\ and\ \citenamefont
  {Vissani}(2003)}]{Strumia:2003zx}%
  \BibitemOpen
  \bibfield  {author} {\bibinfo {author} {\bibfnamefont {A.}~\bibnamefont
  {Strumia}}\ and\ \bibinfo {author} {\bibfnamefont {F.}~\bibnamefont
  {Vissani}},\ }\bibfield  {title} {\bibinfo {title} {{Precise quasielastic
  neutrino/nucleon cross-section}},\ }\href
  {https://doi.org/10.1016/S0370-2693(03)00616-6} {\bibfield  {journal}
  {\bibinfo  {journal} {Phys. Lett. B}\ }\textbf {\bibinfo {volume} {564}},\
  \bibinfo {pages} {42} (\bibinfo {year} {2003})},\ \Eprint
  {https://arxiv.org/abs/astro-ph/0302055} {arXiv:astro-ph/0302055}
  \BibitemShut {NoStop}%
\bibitem [{\citenamefont {Formaggio}\ and\ \citenamefont
  {Zeller}(2012)}]{Formaggio:2012cpf}%
  \BibitemOpen
  \bibfield  {author} {\bibinfo {author} {\bibfnamefont {J.~A.}\ \bibnamefont
  {Formaggio}}\ and\ \bibinfo {author} {\bibfnamefont {G.~P.}\ \bibnamefont
  {Zeller}},\ }\bibfield  {title} {\bibinfo {title} {{From eV to EeV: Neutrino
  Cross Sections Across Energy Scales}},\ }\href
  {https://doi.org/10.1103/RevModPhys.84.1307} {\bibfield  {journal} {\bibinfo
  {journal} {Rev. Mod. Phys.}\ }\textbf {\bibinfo {volume} {84}},\ \bibinfo
  {pages} {1307} (\bibinfo {year} {2012})},\ \Eprint
  {https://arxiv.org/abs/1305.7513} {arXiv:1305.7513 [hep-ex]} \BibitemShut
  {NoStop}%
\bibitem [{\citenamefont {Bratton}\ \emph {et~al.}(1988)\citenamefont {Bratton}
  \emph {et~al.}}]{IMB:1988suc}%
  \BibitemOpen
  \bibfield  {author} {\bibinfo {author} {\bibfnamefont {C.~B.}\ \bibnamefont
  {Bratton}} \emph {et~al.} (\bibinfo {collaboration} {IMB}),\ }\bibfield
  {title} {\bibinfo {title} {{Angular Distribution of Events From Sn1987a}},\
  }\href {https://doi.org/10.1103/PhysRevD.37.3361} {\bibfield  {journal}
  {\bibinfo  {journal} {Phys. Rev. D}\ }\textbf {\bibinfo {volume} {37}},\
  \bibinfo {pages} {3361} (\bibinfo {year} {1988})}\BibitemShut {NoStop}%
\bibitem [{\citenamefont {Oberauer}\ \emph {et~al.}(1993)\citenamefont
  {Oberauer}, \citenamefont {Hagner}, \citenamefont {Raffelt},\ and\
  \citenamefont {Rieger}}]{Oberauer:1993yr}%
  \BibitemOpen
  \bibfield  {author} {\bibinfo {author} {\bibfnamefont {L.}~\bibnamefont
  {Oberauer}}, \bibinfo {author} {\bibfnamefont {C.}~\bibnamefont {Hagner}},
  \bibinfo {author} {\bibfnamefont {G.}~\bibnamefont {Raffelt}},\ and\ \bibinfo
  {author} {\bibfnamefont {E.}~\bibnamefont {Rieger}},\ }\bibfield  {title}
  {\bibinfo {title} {{Supernova bounds on neutrino radiative decays}},\ }\href
  {https://doi.org/10.1016/0927-6505(93)90004-W} {\bibfield  {journal}
  {\bibinfo  {journal} {Astropart. Phys.}\ }\textbf {\bibinfo {volume} {1}},\
  \bibinfo {pages} {377} (\bibinfo {year} {1993})}\BibitemShut {NoStop}%
\bibitem [{\citenamefont {Piran}(1999)}]{Piran:1999kx}%
  \BibitemOpen
  \bibfield  {author} {\bibinfo {author} {\bibfnamefont {T.}~\bibnamefont
  {Piran}},\ }\bibfield  {title} {\bibinfo {title} {{Gamma-ray bursts and the
  fireball model}},\ }\href {https://doi.org/10.1016/S0370-1573(98)00127-6}
  {\bibfield  {journal} {\bibinfo  {journal} {Phys. Rept.}\ }\textbf {\bibinfo
  {volume} {314}},\ \bibinfo {pages} {575} (\bibinfo {year} {1999})},\ \Eprint
  {https://arxiv.org/abs/astro-ph/9810256} {arXiv:astro-ph/9810256}
  \BibitemShut {NoStop}%
\bibitem [{\citenamefont {Diamond}\ \emph {et~al.}(2023)\citenamefont
  {Diamond}, \citenamefont {Fiorillo}, \citenamefont {Marques-Tavares},\ and\
  \citenamefont {Vitagliano}}]{Diamond:2023scc}%
  \BibitemOpen
  \bibfield  {author} {\bibinfo {author} {\bibfnamefont {M.}~\bibnamefont
  {Diamond}}, \bibinfo {author} {\bibfnamefont {D.~F.~G.}\ \bibnamefont
  {Fiorillo}}, \bibinfo {author} {\bibfnamefont {G.}~\bibnamefont
  {Marques-Tavares}},\ and\ \bibinfo {author} {\bibfnamefont {E.}~\bibnamefont
  {Vitagliano}},\ }\bibfield  {title} {\bibinfo {title} {{Axion-sourced
  fireballs from supernovae}},\ }\href
  {https://doi.org/10.1103/PhysRevD.107.103029} {\bibfield  {journal} {\bibinfo
   {journal} {Phys. Rev. D}\ }\textbf {\bibinfo {volume} {107}},\ \bibinfo
  {pages} {103029} (\bibinfo {year} {2023})},\ \bibinfo {note} {[Erratum:
  Phys.Rev.D 108, 049902 (2023)]},\ \Eprint {https://arxiv.org/abs/2303.11395}
  {arXiv:2303.11395 [hep-ph]} \BibitemShut {NoStop}%
\bibitem [{\citenamefont {DeRocco}\ \emph {et~al.}(2019)\citenamefont
  {DeRocco}, \citenamefont {Graham}, \citenamefont {Kasen}, \citenamefont
  {Marques-Tavares},\ and\ \citenamefont {Rajendran}}]{DeRocco:2019njg}%
  \BibitemOpen
  \bibfield  {author} {\bibinfo {author} {\bibfnamefont {W.}~\bibnamefont
  {DeRocco}}, \bibinfo {author} {\bibfnamefont {P.~W.}\ \bibnamefont {Graham}},
  \bibinfo {author} {\bibfnamefont {D.}~\bibnamefont {Kasen}}, \bibinfo
  {author} {\bibfnamefont {G.}~\bibnamefont {Marques-Tavares}},\ and\ \bibinfo
  {author} {\bibfnamefont {S.}~\bibnamefont {Rajendran}},\ }\bibfield  {title}
  {\bibinfo {title} {{Observable signatures of dark photons from supernovae}},\
  }\href {https://doi.org/10.1007/JHEP02(2019)171} {\bibfield  {journal}
  {\bibinfo  {journal} {JHEP}\ }\textbf {\bibinfo {volume} {02}},\ \bibinfo
  {pages} {171}},\ \Eprint {https://arxiv.org/abs/1901.08596} {arXiv:1901.08596
  [hep-ph]} \BibitemShut {NoStop}%
\bibitem [{\citenamefont {Abbott}\ \emph {et~al.}(2017)\citenamefont {Abbott}
  \emph {et~al.}}]{LIGOScientific:2017vwq}%
  \BibitemOpen
  \bibfield  {author} {\bibinfo {author} {\bibfnamefont {B.~P.}\ \bibnamefont
  {Abbott}} \emph {et~al.} (\bibinfo {collaboration} {LIGO Scientific,
  Virgo}),\ }\bibfield  {title} {\bibinfo {title} {{GW170817: Observation of
  Gravitational Waves from a Binary Neutron Star Inspiral}},\ }\href
  {https://doi.org/10.1103/PhysRevLett.119.161101} {\bibfield  {journal}
  {\bibinfo  {journal} {Phys. Rev. Lett.}\ }\textbf {\bibinfo {volume} {119}},\
  \bibinfo {pages} {161101} (\bibinfo {year} {2017})},\ \Eprint
  {https://arxiv.org/abs/1710.05832} {arXiv:1710.05832 [gr-qc]} \BibitemShut
  {NoStop}%
\bibitem [{\citenamefont {Murguia-Berthier}\ \emph {et~al.}(2021)\citenamefont
  {Murguia-Berthier}, \citenamefont {Ramirez-Ruiz}, \citenamefont {De~Colle},
  \citenamefont {Janiuk}, \citenamefont {Rosswog},\ and\ \citenamefont
  {Lee}}]{Murguia-Berthier:2020tfs}%
  \BibitemOpen
  \bibfield  {author} {\bibinfo {author} {\bibfnamefont {A.}~\bibnamefont
  {Murguia-Berthier}}, \bibinfo {author} {\bibfnamefont {E.}~\bibnamefont
  {Ramirez-Ruiz}}, \bibinfo {author} {\bibfnamefont {F.}~\bibnamefont
  {De~Colle}}, \bibinfo {author} {\bibfnamefont {A.}~\bibnamefont {Janiuk}},
  \bibinfo {author} {\bibfnamefont {S.}~\bibnamefont {Rosswog}},\ and\ \bibinfo
  {author} {\bibfnamefont {W.~H.}\ \bibnamefont {Lee}},\ }\bibfield  {title}
  {\bibinfo {title} {{The Fate of the Merger Remnant in GW170817 and its
  Imprint on the Jet Structure}},\ }\href
  {https://doi.org/10.3847/1538-4357/abd08e} {\bibfield  {journal} {\bibinfo
  {journal} {Astrophys. J.}\ }\textbf {\bibinfo {volume} {908}},\ \bibinfo
  {pages} {152} (\bibinfo {year} {2021})},\ \Eprint
  {https://arxiv.org/abs/2007.12245} {arXiv:2007.12245 [astro-ph.HE]}
  \BibitemShut {NoStop}%
\bibitem [{\citenamefont {Liao}\ and\ \citenamefont
  {Marfatia}(2017)}]{Liao:2017}%
  \BibitemOpen
  \bibfield  {author} {\bibinfo {author} {\bibfnamefont {J.}~\bibnamefont
  {Liao}}\ and\ \bibinfo {author} {\bibfnamefont {D.}~\bibnamefont
  {Marfatia}},\ }\bibfield  {title} {\bibinfo {title} {Coherent constraints on
  nonstandard neutrino interactions},\ }\href
  {https://doi.org/10.1016/j.physletb.2017.10.046} {\bibfield  {journal}
  {\bibinfo  {journal} {Physics Letters B}\ }\textbf {\bibinfo {volume}
  {775}},\ \bibinfo {pages} {54–57} (\bibinfo {year} {2017})}\BibitemShut
  {NoStop}%
\bibitem [{\citenamefont {Akimov}\ \emph {et~al.}(2017)\citenamefont {Akimov}
  \emph {et~al.}}]{COHERENT:2017ipa}%
  \BibitemOpen
  \bibfield  {author} {\bibinfo {author} {\bibfnamefont {D.}~\bibnamefont
  {Akimov}} \emph {et~al.} (\bibinfo {collaboration} {COHERENT}),\ }\bibfield
  {title} {\bibinfo {title} {{Observation of Coherent Elastic Neutrino-Nucleus
  Scattering}},\ }\href {https://doi.org/10.1126/science.aao0990} {\bibfield
  {journal} {\bibinfo  {journal} {Science}\ }\textbf {\bibinfo {volume}
  {357}},\ \bibinfo {pages} {1123} (\bibinfo {year} {2017})},\ \Eprint
  {https://arxiv.org/abs/1708.01294} {arXiv:1708.01294 [nucl-ex]} \BibitemShut
  {NoStop}%
\bibitem [{\citenamefont {Abdullah}\ \emph {et~al.}(2022)\citenamefont
  {Abdullah} \emph {et~al.}}]{Abdullah:2022zue}%
  \BibitemOpen
  \bibfield  {author} {\bibinfo {author} {\bibfnamefont {M.}~\bibnamefont
  {Abdullah}} \emph {et~al.},\ }\bibfield  {title} {\bibinfo {title} {{Coherent
  elastic neutrino-nucleus scattering: Terrestrial and astrophysical
  applications}},\ }\href@noop {} {\  (\bibinfo {year} {2022})},\ \Eprint
  {https://arxiv.org/abs/2203.07361} {arXiv:2203.07361 [hep-ph]} \BibitemShut
  {NoStop}%
\bibitem [{\citenamefont {Dent}\ \emph {et~al.}(2017)\citenamefont {Dent},
  \citenamefont {Dutta}, \citenamefont {Liao}, \citenamefont {Newstead},
  \citenamefont {Strigari},\ and\ \citenamefont {Walker}}]{Dent:2016wcr}%
  \BibitemOpen
  \bibfield  {author} {\bibinfo {author} {\bibfnamefont {J.~B.}\ \bibnamefont
  {Dent}}, \bibinfo {author} {\bibfnamefont {B.}~\bibnamefont {Dutta}},
  \bibinfo {author} {\bibfnamefont {S.}~\bibnamefont {Liao}}, \bibinfo {author}
  {\bibfnamefont {J.~L.}\ \bibnamefont {Newstead}}, \bibinfo {author}
  {\bibfnamefont {L.~E.}\ \bibnamefont {Strigari}},\ and\ \bibinfo {author}
  {\bibfnamefont {J.~W.}\ \bibnamefont {Walker}},\ }\bibfield  {title}
  {\bibinfo {title} {{Probing light mediators at ultralow threshold energies
  with coherent elastic neutrino-nucleus scattering}},\ }\href
  {https://doi.org/10.1103/PhysRevD.96.095007} {\bibfield  {journal} {\bibinfo
  {journal} {Phys. Rev. D}\ }\textbf {\bibinfo {volume} {96}},\ \bibinfo
  {pages} {095007} (\bibinfo {year} {2017})},\ \Eprint
  {https://arxiv.org/abs/1612.06350} {arXiv:1612.06350 [hep-ph]} \BibitemShut
  {NoStop}%
\bibitem [{\citenamefont {Shoemaker}(2017)}]{Shoemaker:2017lzs}%
  \BibitemOpen
  \bibfield  {author} {\bibinfo {author} {\bibfnamefont {I.~M.}\ \bibnamefont
  {Shoemaker}},\ }\bibfield  {title} {\bibinfo {title} {{COHERENT search
  strategy for beyond standard model neutrino interactions}},\ }\href
  {https://doi.org/10.1103/PhysRevD.95.115028} {\bibfield  {journal} {\bibinfo
  {journal} {Phys. Rev. D}\ }\textbf {\bibinfo {volume} {95}},\ \bibinfo
  {pages} {115028} (\bibinfo {year} {2017})},\ \Eprint
  {https://arxiv.org/abs/1703.05774} {arXiv:1703.05774 [hep-ph]} \BibitemShut
  {NoStop}%
\bibitem [{\citenamefont {Papoulias}\ and\ \citenamefont
  {Kosmas}(2018)}]{Papoulias:2017qdn}%
  \BibitemOpen
  \bibfield  {author} {\bibinfo {author} {\bibfnamefont {D.~K.}\ \bibnamefont
  {Papoulias}}\ and\ \bibinfo {author} {\bibfnamefont {T.~S.}\ \bibnamefont
  {Kosmas}},\ }\bibfield  {title} {\bibinfo {title} {{COHERENT constraints to
  conventional and exotic neutrino physics}},\ }\href
  {https://doi.org/10.1103/PhysRevD.97.033003} {\bibfield  {journal} {\bibinfo
  {journal} {Phys. Rev. D}\ }\textbf {\bibinfo {volume} {97}},\ \bibinfo
  {pages} {033003} (\bibinfo {year} {2018})},\ \Eprint
  {https://arxiv.org/abs/1711.09773} {arXiv:1711.09773 [hep-ph]} \BibitemShut
  {NoStop}%
\bibitem [{\citenamefont {Abdullah}\ \emph {et~al.}(2018)\citenamefont
  {Abdullah}, \citenamefont {Dent}, \citenamefont {Dutta}, \citenamefont
  {Kane}, \citenamefont {Liao},\ and\ \citenamefont
  {Strigari}}]{Abdullah:2018ykz}%
  \BibitemOpen
  \bibfield  {author} {\bibinfo {author} {\bibfnamefont {M.}~\bibnamefont
  {Abdullah}}, \bibinfo {author} {\bibfnamefont {J.~B.}\ \bibnamefont {Dent}},
  \bibinfo {author} {\bibfnamefont {B.}~\bibnamefont {Dutta}}, \bibinfo
  {author} {\bibfnamefont {G.~L.}\ \bibnamefont {Kane}}, \bibinfo {author}
  {\bibfnamefont {S.}~\bibnamefont {Liao}},\ and\ \bibinfo {author}
  {\bibfnamefont {L.~E.}\ \bibnamefont {Strigari}},\ }\bibfield  {title}
  {\bibinfo {title} {{Coherent elastic neutrino nucleus scattering as a probe
  of a Z' through kinetic and mass mixing effects}},\ }\href
  {https://doi.org/10.1103/PhysRevD.98.015005} {\bibfield  {journal} {\bibinfo
  {journal} {Phys. Rev. D}\ }\textbf {\bibinfo {volume} {98}},\ \bibinfo
  {pages} {015005} (\bibinfo {year} {2018})},\ \Eprint
  {https://arxiv.org/abs/1803.01224} {arXiv:1803.01224 [hep-ph]} \BibitemShut
  {NoStop}%
\bibitem [{\citenamefont {Denton}\ \emph {et~al.}(2018)\citenamefont {Denton},
  \citenamefont {Farzan},\ and\ \citenamefont {Shoemaker}}]{Denton:2018xmq}%
  \BibitemOpen
  \bibfield  {author} {\bibinfo {author} {\bibfnamefont {P.~B.}\ \bibnamefont
  {Denton}}, \bibinfo {author} {\bibfnamefont {Y.}~\bibnamefont {Farzan}},\
  and\ \bibinfo {author} {\bibfnamefont {I.~M.}\ \bibnamefont {Shoemaker}},\
  }\bibfield  {title} {\bibinfo {title} {{Testing large non-standard neutrino
  interactions with arbitrary mediator mass after COHERENT data}},\ }\href
  {https://doi.org/10.1007/JHEP07(2018)037} {\bibfield  {journal} {\bibinfo
  {journal} {JHEP}\ }\textbf {\bibinfo {volume} {07}},\ \bibinfo {pages}
  {037}},\ \Eprint {https://arxiv.org/abs/1804.03660} {arXiv:1804.03660
  [hep-ph]} \BibitemShut {NoStop}%
\bibitem [{\citenamefont {Aguilar-Arevalo}\ \emph {et~al.}(2020)\citenamefont
  {Aguilar-Arevalo} \emph {et~al.}}]{CONNIE:2019xid}%
  \BibitemOpen
  \bibfield  {author} {\bibinfo {author} {\bibfnamefont {A.}~\bibnamefont
  {Aguilar-Arevalo}} \emph {et~al.} (\bibinfo {collaboration} {CONNIE}),\
  }\bibfield  {title} {\bibinfo {title} {{Search for light mediators in the
  low-energy data of the CONNIE reactor neutrino experiment}},\ }\href
  {https://doi.org/10.1007/JHEP04(2020)054} {\bibfield  {journal} {\bibinfo
  {journal} {JHEP}\ }\textbf {\bibinfo {volume} {04}},\ \bibinfo {pages}
  {054}},\ \Eprint {https://arxiv.org/abs/1910.04951} {arXiv:1910.04951
  [hep-ex]} \BibitemShut {NoStop}%
\bibitem [{\citenamefont {Cadeddu}\ \emph {et~al.}(2021)\citenamefont
  {Cadeddu}, \citenamefont {Cargioli}, \citenamefont {Dordei}, \citenamefont
  {Giunti}, \citenamefont {Li}, \citenamefont {Picciau},\ and\ \citenamefont
  {Zhang}}]{Cadeddu:2020nbr}%
  \BibitemOpen
  \bibfield  {author} {\bibinfo {author} {\bibfnamefont {M.}~\bibnamefont
  {Cadeddu}}, \bibinfo {author} {\bibfnamefont {N.}~\bibnamefont {Cargioli}},
  \bibinfo {author} {\bibfnamefont {F.}~\bibnamefont {Dordei}}, \bibinfo
  {author} {\bibfnamefont {C.}~\bibnamefont {Giunti}}, \bibinfo {author}
  {\bibfnamefont {Y.~F.}\ \bibnamefont {Li}}, \bibinfo {author} {\bibfnamefont
  {E.}~\bibnamefont {Picciau}},\ and\ \bibinfo {author} {\bibfnamefont {Y.~Y.}\
  \bibnamefont {Zhang}},\ }\bibfield  {title} {\bibinfo {title} {{Constraints
  on light vector mediators through coherent elastic neutrino nucleus
  scattering data from COHERENT}},\ }\href
  {https://doi.org/10.1007/JHEP01(2021)116} {\bibfield  {journal} {\bibinfo
  {journal} {JHEP}\ }\textbf {\bibinfo {volume} {01}},\ \bibinfo {pages}
  {116}},\ \Eprint {https://arxiv.org/abs/2008.05022} {arXiv:2008.05022
  [hep-ph]} \BibitemShut {NoStop}%
\bibitem [{\citenamefont {Miranda}\ \emph {et~al.}(2020)\citenamefont
  {Miranda}, \citenamefont {Papoulias}, \citenamefont {Sanchez~Garcia},
  \citenamefont {Sanders}, \citenamefont {T{\'o}rtola},\ and\ \citenamefont
  {Valle}}]{Miranda:2020tif}%
  \BibitemOpen
  \bibfield  {author} {\bibinfo {author} {\bibfnamefont {O.~G.}\ \bibnamefont
  {Miranda}}, \bibinfo {author} {\bibfnamefont {D.~K.}\ \bibnamefont
  {Papoulias}}, \bibinfo {author} {\bibfnamefont {G.}~\bibnamefont
  {Sanchez~Garcia}}, \bibinfo {author} {\bibfnamefont {O.}~\bibnamefont
  {Sanders}}, \bibinfo {author} {\bibfnamefont {M.}~\bibnamefont
  {T{\'o}rtola}},\ and\ \bibinfo {author} {\bibfnamefont {J.~W.~F.}\
  \bibnamefont {Valle}},\ }\bibfield  {title} {\bibinfo {title} {{Implications
  of the first detection of coherent elastic neutrino-nucleus scattering
  (CEvNS) with Liquid Argon}},\ }\href
  {https://doi.org/10.1007/JHEP05(2020)130} {\bibfield  {journal} {\bibinfo
  {journal} {JHEP}\ }\textbf {\bibinfo {volume} {05}},\ \bibinfo {pages}
  {130}},\ \bibinfo {note} {[Erratum: JHEP 01, 067 (2021)]},\ \Eprint
  {https://arxiv.org/abs/2003.12050} {arXiv:2003.12050 [hep-ph]} \BibitemShut
  {NoStop}%
\bibitem [{\citenamefont {Bonet}\ \emph {et~al.}(2022)\citenamefont {Bonet}
  \emph {et~al.}}]{CONUS:2021dwh}%
  \BibitemOpen
  \bibfield  {author} {\bibinfo {author} {\bibfnamefont {H.}~\bibnamefont
  {Bonet}} \emph {et~al.} (\bibinfo {collaboration} {CONUS}),\ }\bibfield
  {title} {\bibinfo {title} {{Novel constraints on neutrino physics beyond the
  standard model from the CONUS experiment}},\ }\href
  {https://doi.org/10.1007/JHEP05(2022)085} {\bibfield  {journal} {\bibinfo
  {journal} {JHEP}\ }\textbf {\bibinfo {volume} {05}},\ \bibinfo {pages}
  {085}},\ \Eprint {https://arxiv.org/abs/2110.02174} {arXiv:2110.02174
  [hep-ph]} \BibitemShut {NoStop}%
\bibitem [{\citenamefont {de~la Vega}\ \emph {et~al.}(2021)\citenamefont {de~la
  Vega}, \citenamefont {Flores}, \citenamefont {Nath},\ and\ \citenamefont
  {Peinado}}]{delaVega:2021wpx}%
  \BibitemOpen
  \bibfield  {author} {\bibinfo {author} {\bibfnamefont {L.~M.~G.}\
  \bibnamefont {de~la Vega}}, \bibinfo {author} {\bibfnamefont {L.~J.}\
  \bibnamefont {Flores}}, \bibinfo {author} {\bibfnamefont {N.}~\bibnamefont
  {Nath}},\ and\ \bibinfo {author} {\bibfnamefont {E.}~\bibnamefont
  {Peinado}},\ }\bibfield  {title} {\bibinfo {title} {{Complementarity between
  dark matter direct searches and CE{\ensuremath{\nu}}NS experiments in U(1)'
  models}},\ }\href {https://doi.org/10.1007/JHEP09(2021)146} {\bibfield
  {journal} {\bibinfo  {journal} {JHEP}\ }\textbf {\bibinfo {volume} {09}},\
  \bibinfo {pages} {146}},\ \Eprint {https://arxiv.org/abs/2107.04037}
  {arXiv:2107.04037 [hep-ph]} \BibitemShut {NoStop}%
\bibitem [{\citenamefont {Coloma}\ \emph
  {et~al.}(2022{\natexlab{a}})\citenamefont {Coloma}, \citenamefont {Esteban},
  \citenamefont {Gonzalez-Garcia}, \citenamefont {Larizgoitia}, \citenamefont
  {Monrabal},\ and\ \citenamefont {Palomares-Ruiz}}]{Coloma:2022avw}%
  \BibitemOpen
  \bibfield  {author} {\bibinfo {author} {\bibfnamefont {P.}~\bibnamefont
  {Coloma}}, \bibinfo {author} {\bibfnamefont {I.}~\bibnamefont {Esteban}},
  \bibinfo {author} {\bibfnamefont {M.~C.}\ \bibnamefont {Gonzalez-Garcia}},
  \bibinfo {author} {\bibfnamefont {L.}~\bibnamefont {Larizgoitia}}, \bibinfo
  {author} {\bibfnamefont {F.}~\bibnamefont {Monrabal}},\ and\ \bibinfo
  {author} {\bibfnamefont {S.}~\bibnamefont {Palomares-Ruiz}},\ }\bibfield
  {title} {\bibinfo {title} {{Bounds on new physics with data of the Dresden-II
  reactor experiment and COHERENT}},\ }\href
  {https://doi.org/10.1007/JHEP05(2022)037} {\bibfield  {journal} {\bibinfo
  {journal} {JHEP}\ }\textbf {\bibinfo {volume} {05}},\ \bibinfo {pages}
  {037}},\ \Eprint {https://arxiv.org/abs/2202.10829} {arXiv:2202.10829
  [hep-ph]} \BibitemShut {NoStop}%
\bibitem [{\citenamefont {Atzori~Corona}\ \emph {et~al.}(2022)\citenamefont
  {Atzori~Corona}, \citenamefont {Cadeddu}, \citenamefont {Cargioli},
  \citenamefont {Dordei}, \citenamefont {Giunti}, \citenamefont {Li},
  \citenamefont {Picciau}, \citenamefont {Ternes},\ and\ \citenamefont
  {Zhang}}]{AtzoriCorona:2022moj}%
  \BibitemOpen
  \bibfield  {author} {\bibinfo {author} {\bibfnamefont {M.}~\bibnamefont
  {Atzori~Corona}}, \bibinfo {author} {\bibfnamefont {M.}~\bibnamefont
  {Cadeddu}}, \bibinfo {author} {\bibfnamefont {N.}~\bibnamefont {Cargioli}},
  \bibinfo {author} {\bibfnamefont {F.}~\bibnamefont {Dordei}}, \bibinfo
  {author} {\bibfnamefont {C.}~\bibnamefont {Giunti}}, \bibinfo {author}
  {\bibfnamefont {Y.~F.}\ \bibnamefont {Li}}, \bibinfo {author} {\bibfnamefont
  {E.}~\bibnamefont {Picciau}}, \bibinfo {author} {\bibfnamefont {C.~A.}\
  \bibnamefont {Ternes}},\ and\ \bibinfo {author} {\bibfnamefont {Y.~Y.}\
  \bibnamefont {Zhang}},\ }\bibfield  {title} {\bibinfo {title} {{Probing light
  mediators and (g {\ensuremath{-}} 2)$_{\mu}$ through detection of coherent
  elastic neutrino nucleus scattering at COHERENT}},\ }\href
  {https://doi.org/10.1007/JHEP05(2022)109} {\bibfield  {journal} {\bibinfo
  {journal} {JHEP}\ }\textbf {\bibinfo {volume} {05}},\ \bibinfo {pages}
  {109}},\ \Eprint {https://arxiv.org/abs/2202.11002} {arXiv:2202.11002
  [hep-ph]} \BibitemShut {NoStop}%
\bibitem [{\citenamefont {De~Romeri}\ \emph {et~al.}(2025)\citenamefont
  {De~Romeri}, \citenamefont {Papoulias},\ and\ \citenamefont
  {Sanchez~Garcia}}]{DeRomeri:2025csu}%
  \BibitemOpen
  \bibfield  {author} {\bibinfo {author} {\bibfnamefont {V.}~\bibnamefont
  {De~Romeri}}, \bibinfo {author} {\bibfnamefont {D.~K.}\ \bibnamefont
  {Papoulias}},\ and\ \bibinfo {author} {\bibfnamefont {G.}~\bibnamefont
  {Sanchez~Garcia}},\ }\bibfield  {title} {\bibinfo {title} {{Implications of
  the first CONUS+ measurement of coherent elastic neutrino-nucleus
  scattering}},\ }\href {https://doi.org/10.1103/PhysRevD.111.075025}
  {\bibfield  {journal} {\bibinfo  {journal} {Phys. Rev. D}\ }\textbf {\bibinfo
  {volume} {111}},\ \bibinfo {pages} {075025} (\bibinfo {year} {2025})},\
  \Eprint {https://arxiv.org/abs/2501.17843} {arXiv:2501.17843 [hep-ph]}
  \BibitemShut {NoStop}%
\bibitem [{\citenamefont {Chattaraj}\ \emph {et~al.}(2025)\citenamefont
  {Chattaraj}, \citenamefont {Majumdar},\ and\ \citenamefont
  {Srivastava}}]{Chattaraj:2025fvx}%
  \BibitemOpen
  \bibfield  {author} {\bibinfo {author} {\bibfnamefont {A.}~\bibnamefont
  {Chattaraj}}, \bibinfo {author} {\bibfnamefont {A.}~\bibnamefont
  {Majumdar}},\ and\ \bibinfo {author} {\bibfnamefont {R.}~\bibnamefont
  {Srivastava}},\ }\bibfield  {title} {\bibinfo {title} {{Probing standard
  model and beyond with reactor CE{\ensuremath{\nu}}NS data of CONUS+
  experiment}},\ }\href {https://doi.org/10.1016/j.physletb.2025.139438}
  {\bibfield  {journal} {\bibinfo  {journal} {Phys. Lett. B}\ }\textbf
  {\bibinfo {volume} {864}},\ \bibinfo {pages} {139438} (\bibinfo {year}
  {2025})},\ \Eprint {https://arxiv.org/abs/2501.12441} {arXiv:2501.12441
  [hep-ph]} \BibitemShut {NoStop}%
\bibitem [{\citenamefont {Karada{\u{g}}}\ \emph {et~al.}(2025)\citenamefont
  {Karada{\u{g}}} \emph {et~al.}}]{TEXONO:2025sub}%
  \BibitemOpen
  \bibfield  {author} {\bibinfo {author} {\bibfnamefont {S.}~\bibnamefont
  {Karada{\u{g}}}} \emph {et~al.} (\bibinfo {collaboration} {TEXONO}),\
  }\bibfield  {title} {\bibinfo {title} {{Constraints on new physics with light
  mediators and generalized neutrino interactions via coherent elastic neutrino
  nucleus scattering}},\ }\href {https://doi.org/10.1103/63xf-t6fz} {\bibfield
  {journal} {\bibinfo  {journal} {Phys. Rev. D}\ }\textbf {\bibinfo {volume}
  {112}},\ \bibinfo {pages} {035038} (\bibinfo {year} {2025})},\ \Eprint
  {https://arxiv.org/abs/2502.20007} {arXiv:2502.20007 [hep-ex]} \BibitemShut
  {NoStop}%
\bibitem [{\citenamefont {Akimov}\ \emph {et~al.}(2022)\citenamefont {Akimov}
  \emph {et~al.}}]{COHERENT:2021xmm}%
  \BibitemOpen
  \bibfield  {author} {\bibinfo {author} {\bibfnamefont {D.}~\bibnamefont
  {Akimov}} \emph {et~al.} (\bibinfo {collaboration} {COHERENT}),\ }\bibfield
  {title} {\bibinfo {title} {{Measurement of the Coherent Elastic
  Neutrino-Nucleus Scattering Cross Section on CsI by COHERENT}},\ }\href
  {https://doi.org/10.1103/PhysRevLett.129.081801} {\bibfield  {journal}
  {\bibinfo  {journal} {Phys. Rev. Lett.}\ }\textbf {\bibinfo {volume} {129}},\
  \bibinfo {pages} {081801} (\bibinfo {year} {2022})},\ \Eprint
  {https://arxiv.org/abs/2110.07730} {arXiv:2110.07730 [hep-ex]} \BibitemShut
  {NoStop}%
\bibitem [{\citenamefont {Akimov}\ \emph {et~al.}(2021)\citenamefont {Akimov}
  \emph {et~al.}}]{COHERENT:2020iec}%
  \BibitemOpen
  \bibfield  {author} {\bibinfo {author} {\bibfnamefont {D.}~\bibnamefont
  {Akimov}} \emph {et~al.} (\bibinfo {collaboration} {COHERENT}),\ }\bibfield
  {title} {\bibinfo {title} {{First Measurement of Coherent Elastic
  Neutrino-Nucleus Scattering on Argon}},\ }\href
  {https://doi.org/10.1103/PhysRevLett.126.012002} {\bibfield  {journal}
  {\bibinfo  {journal} {Phys. Rev. Lett.}\ }\textbf {\bibinfo {volume} {126}},\
  \bibinfo {pages} {012002} (\bibinfo {year} {2021})},\ \Eprint
  {https://arxiv.org/abs/2003.10630} {arXiv:2003.10630 [nucl-ex]} \BibitemShut
  {NoStop}%
\bibitem [{\citenamefont {Ackermann}\ \emph {et~al.}(2025)\citenamefont
  {Ackermann} \emph {et~al.}}]{Ackermann:2025obx}%
  \BibitemOpen
  \bibfield  {author} {\bibinfo {author} {\bibfnamefont {N.}~\bibnamefont
  {Ackermann}} \emph {et~al.},\ }\bibfield  {title} {\bibinfo {title} {{Direct
  observation of coherent elastic antineutrino{\textendash}nucleus
  scattering}},\ }\href {https://doi.org/10.1038/s41586-025-09322-2} {\bibfield
   {journal} {\bibinfo  {journal} {Nature}\ }\textbf {\bibinfo {volume}
  {643}},\ \bibinfo {pages} {1229} (\bibinfo {year} {2025})},\ \Eprint
  {https://arxiv.org/abs/2501.05206} {arXiv:2501.05206 [hep-ex]} \BibitemShut
  {NoStop}%
\bibitem [{\citenamefont {Konovalov}(2024)}]{Konovalov:2024uhc}%
  \BibitemOpen
  \bibfield  {author} {\bibinfo {author} {\bibfnamefont {A.~M.}\ \bibnamefont
  {Konovalov}},\ }\bibfield  {title} {\bibinfo {title} {{Status of the COHERENT
  Experiment and New Physics Opportunities at the SNS Facility}},\ }\href
  {https://doi.org/10.3103/S0027134924701686} {\bibfield  {journal} {\bibinfo
  {journal} {Moscow Univ. Phys. Bull.}\ }\textbf {\bibinfo {volume} {79}},\
  \bibinfo {pages} {220} (\bibinfo {year} {2024})}\BibitemShut {NoStop}%
\bibitem [{\citenamefont {Picciau}(2022)}]{Picciau:2022xzi}%
  \BibitemOpen
  \bibfield  {author} {\bibinfo {author} {\bibfnamefont {E.}~\bibnamefont
  {Picciau}},\ }\emph {\bibinfo {title} {{Low-energy signatures in
  DarkSide-50experiment and neutrino scattering processes}}},\ \href@noop {}
  {Ph.D. thesis},\ \bibinfo  {school} {Universit{\`a} degli Studi di Cagliari,
  Italy, Cagliari U.} (\bibinfo {year} {2022})\BibitemShut {NoStop}%
\bibitem [{\citenamefont {Colaresi}\ \emph {et~al.}(2022)\citenamefont
  {Colaresi}, \citenamefont {Collar}, \citenamefont {Hossbach}, \citenamefont
  {Lewis},\ and\ \citenamefont {Yocum}}]{Colaresi:2022obx}%
  \BibitemOpen
  \bibfield  {author} {\bibinfo {author} {\bibfnamefont {J.}~\bibnamefont
  {Colaresi}}, \bibinfo {author} {\bibfnamefont {J.~I.}\ \bibnamefont
  {Collar}}, \bibinfo {author} {\bibfnamefont {T.~W.}\ \bibnamefont
  {Hossbach}}, \bibinfo {author} {\bibfnamefont {C.~M.}\ \bibnamefont
  {Lewis}},\ and\ \bibinfo {author} {\bibfnamefont {K.~M.}\ \bibnamefont
  {Yocum}},\ }\bibfield  {title} {\bibinfo {title} {{Measurement of Coherent
  Elastic Neutrino-Nucleus Scattering from Reactor Antineutrinos}},\ }\href
  {https://doi.org/10.1103/PhysRevLett.129.211802} {\bibfield  {journal}
  {\bibinfo  {journal} {Phys. Rev. Lett.}\ }\textbf {\bibinfo {volume} {129}},\
  \bibinfo {pages} {211802} (\bibinfo {year} {2022})},\ \Eprint
  {https://arxiv.org/abs/2202.09672} {arXiv:2202.09672 [hep-ex]} \BibitemShut
  {NoStop}%
\bibitem [{\citenamefont {Karmakar}\ \emph {et~al.}(2025)\citenamefont
  {Karmakar} \emph {et~al.}}]{TEXONO:2024vfk}%
  \BibitemOpen
  \bibfield  {author} {\bibinfo {author} {\bibfnamefont {S.}~\bibnamefont
  {Karmakar}} \emph {et~al.} (\bibinfo {collaboration} {TEXONO}),\ }\bibfield
  {title} {\bibinfo {title} {{New Limits on the Coherent Neutrino-Nucleus
  Elastic Scattering Cross Section at the Kuo-Sheng Reactor-Neutrino
  Laboratory}},\ }\href {https://doi.org/10.1103/PhysRevLett.134.121802}
  {\bibfield  {journal} {\bibinfo  {journal} {Phys. Rev. Lett.}\ }\textbf
  {\bibinfo {volume} {134}},\ \bibinfo {pages} {121802} (\bibinfo {year}
  {2025})},\ \Eprint {https://arxiv.org/abs/2411.18812} {arXiv:2411.18812
  [nucl-ex]} \BibitemShut {NoStop}%
\bibitem [{\citenamefont {Goupy}\ \emph {et~al.}(2023)\citenamefont {Goupy}
  \emph {et~al.}}]{NUCLEUS:2022zti}%
  \BibitemOpen
  \bibfield  {author} {\bibinfo {author} {\bibfnamefont {C.}~\bibnamefont
  {Goupy}} \emph {et~al.} (\bibinfo {collaboration} {NUCLEUS}),\ }\bibfield
  {title} {\bibinfo {title} {{Exploring coherent elastic neutrino-nucleus
  scattering of reactor neutrinos with the NUCLEUS experiment}},\ }\href
  {https://doi.org/10.21468/SciPostPhysProc.12.053} {\bibfield  {journal}
  {\bibinfo  {journal} {SciPost Phys. Proc.}\ }\textbf {\bibinfo {volume}
  {12}},\ \bibinfo {pages} {053} (\bibinfo {year} {2023})},\ \Eprint
  {https://arxiv.org/abs/2211.04189} {arXiv:2211.04189 [physics.ins-det]}
  \BibitemShut {NoStop}%
\bibitem [{\citenamefont {Augier}\ \emph {et~al.}(2023)\citenamefont {Augier}
  \emph {et~al.}}]{Ricochet:2022pzj}%
  \BibitemOpen
  \bibfield  {author} {\bibinfo {author} {\bibfnamefont {C.}~\bibnamefont
  {Augier}} \emph {et~al.} (\bibinfo {collaboration} {Ricochet}),\ }\bibfield
  {title} {\bibinfo {title} {{Fast neutron background characterization of the
  future Ricochet experiment at the ILL research nuclear reactor}},\ }\href
  {https://doi.org/10.1140/epjc/s10052-022-11150-x} {\bibfield  {journal}
  {\bibinfo  {journal} {Eur. Phys. J. C}\ }\textbf {\bibinfo {volume} {83}},\
  \bibinfo {pages} {20} (\bibinfo {year} {2023})},\ \Eprint
  {https://arxiv.org/abs/2208.01760} {arXiv:2208.01760 [astro-ph.IM]}
  \BibitemShut {NoStop}%
\bibitem [{\citenamefont {Aguilar-Arevalo}\ \emph {et~al.}(2024)\citenamefont
  {Aguilar-Arevalo} \emph {et~al.}}]{CONNIE:2024pwt}%
  \BibitemOpen
  \bibfield  {author} {\bibinfo {author} {\bibfnamefont {A.~A.}\ \bibnamefont
  {Aguilar-Arevalo}} \emph {et~al.} (\bibinfo {collaboration} {CONNIE}),\
  }\bibfield  {title} {\bibinfo {title} {{Searches for CE{\ensuremath{\nu}}NS
  and Physics beyond the Standard Model using Skipper-CCDs at CONNIE}},\
  }\href@noop {} {\  (\bibinfo {year} {2024})},\ \Eprint
  {https://arxiv.org/abs/2403.15976} {arXiv:2403.15976 [hep-ex]} \BibitemShut
  {NoStop}%
\bibitem [{\citenamefont {Atzori~Corona}\ \emph
  {et~al.}(2025{\natexlab{a}})\citenamefont {Atzori~Corona}, \citenamefont
  {Cadeddu}, \citenamefont {Cargioli}, \citenamefont {Dordei},\ and\
  \citenamefont {Giunti}}]{AtzoriCorona:2025ygn}%
  \BibitemOpen
  \bibfield  {author} {\bibinfo {author} {\bibfnamefont {M.}~\bibnamefont
  {Atzori~Corona}}, \bibinfo {author} {\bibfnamefont {M.}~\bibnamefont
  {Cadeddu}}, \bibinfo {author} {\bibfnamefont {N.}~\bibnamefont {Cargioli}},
  \bibinfo {author} {\bibfnamefont {F.}~\bibnamefont {Dordei}},\ and\ \bibinfo
  {author} {\bibfnamefont {C.}~\bibnamefont {Giunti}},\ }\bibfield  {title}
  {\bibinfo {title} {{Reactor antineutrinos CE{\ensuremath{\nu}}NS on
  germanium: CONUS+ and TEXONO as a new gateway to SM and BSM physics}},\
  }\href {https://doi.org/10.1103/n563-8v8d} {\bibfield  {journal} {\bibinfo
  {journal} {Phys. Rev. D}\ }\textbf {\bibinfo {volume} {112}},\ \bibinfo
  {pages} {015007} (\bibinfo {year} {2025}{\natexlab{a}})},\ \Eprint
  {https://arxiv.org/abs/2501.18550} {arXiv:2501.18550 [hep-ph]} \BibitemShut
  {NoStop}%
\bibitem [{\citenamefont {Atzori~Corona}\ \emph
  {et~al.}(2025{\natexlab{b}})\citenamefont {Atzori~Corona}, \citenamefont
  {Cadeddu}, \citenamefont {Cargioli}, \citenamefont {Dordei}, \citenamefont
  {Giunti},\ and\ \citenamefont {Pavarani}}]{AtzoriCorona:2025ibl}%
  \BibitemOpen
  \bibfield  {author} {\bibinfo {author} {\bibfnamefont {M.}~\bibnamefont
  {Atzori~Corona}}, \bibinfo {author} {\bibfnamefont {M.}~\bibnamefont
  {Cadeddu}}, \bibinfo {author} {\bibfnamefont {N.}~\bibnamefont {Cargioli}},
  \bibinfo {author} {\bibfnamefont {F.}~\bibnamefont {Dordei}}, \bibinfo
  {author} {\bibfnamefont {C.}~\bibnamefont {Giunti}},\ and\ \bibinfo {author}
  {\bibfnamefont {R.}~\bibnamefont {Pavarani}},\ }\bibfield  {title} {\bibinfo
  {title} {{Toward precision physics tests with future COHERENT detectors}},\
  }\href@noop {} {\  (\bibinfo {year} {2025}{\natexlab{b}})},\ \Eprint
  {https://arxiv.org/abs/2509.04205} {arXiv:2509.04205 [physics.ins-det]}
  \BibitemShut {NoStop}%
\bibitem [{\citenamefont {Bellini}\ \emph {et~al.}(2011)\citenamefont {Bellini}
  \emph {et~al.}}]{Bellini:2011rx}%
  \BibitemOpen
  \bibfield  {author} {\bibinfo {author} {\bibfnamefont {G.}~\bibnamefont
  {Bellini}} \emph {et~al.},\ }\bibfield  {title} {\bibinfo {title} {{Precision
  measurement of the 7Be solar neutrino interaction rate in Borexino}},\ }\href
  {https://doi.org/10.1103/PhysRevLett.107.141302} {\bibfield  {journal}
  {\bibinfo  {journal} {Phys. Rev. Lett.}\ }\textbf {\bibinfo {volume} {107}},\
  \bibinfo {pages} {141302} (\bibinfo {year} {2011})},\ \Eprint
  {https://arxiv.org/abs/1104.1816} {arXiv:1104.1816 [hep-ex]} \BibitemShut
  {NoStop}%
\bibitem [{\citenamefont {Bellini}\ \emph
  {et~al.}(2014{\natexlab{a}})\citenamefont {Bellini} \emph
  {et~al.}}]{Borexino:2013zhu}%
  \BibitemOpen
  \bibfield  {author} {\bibinfo {author} {\bibfnamefont {G.}~\bibnamefont
  {Bellini}} \emph {et~al.} (\bibinfo {collaboration} {Borexino}),\ }\bibfield
  {title} {\bibinfo {title} {{Final results of Borexino Phase-I on low energy
  solar neutrino spectroscopy}},\ }\href
  {https://doi.org/10.1103/PhysRevD.89.112007} {\bibfield  {journal} {\bibinfo
  {journal} {Phys. Rev. D}\ }\textbf {\bibinfo {volume} {89}},\ \bibinfo
  {pages} {112007} (\bibinfo {year} {2014}{\natexlab{a}})},\ \Eprint
  {https://arxiv.org/abs/1308.0443} {arXiv:1308.0443 [hep-ex]} \BibitemShut
  {NoStop}%
\bibitem [{\citenamefont {Bellini}\ \emph
  {et~al.}(2014{\natexlab{b}})\citenamefont {Bellini} \emph
  {et~al.}}]{BOREXINO:2014pcl}%
  \BibitemOpen
  \bibfield  {author} {\bibinfo {author} {\bibfnamefont {G.}~\bibnamefont
  {Bellini}} \emph {et~al.} (\bibinfo {collaboration} {BOREXINO}),\ }\bibfield
  {title} {\bibinfo {title} {{Neutrinos from the primary
  proton{\textendash}proton fusion process in the Sun}},\ }\href
  {https://doi.org/10.1038/nature13702} {\bibfield  {journal} {\bibinfo
  {journal} {Nature}\ }\textbf {\bibinfo {volume} {512}},\ \bibinfo {pages}
  {383} (\bibinfo {year} {2014}{\natexlab{b}})}\BibitemShut {NoStop}%
\bibitem [{\citenamefont {Agostini}\ \emph {et~al.}(2019)\citenamefont
  {Agostini} \emph {et~al.}}]{Borexino:2017rsf}%
  \BibitemOpen
  \bibfield  {author} {\bibinfo {author} {\bibfnamefont {M.}~\bibnamefont
  {Agostini}} \emph {et~al.} (\bibinfo {collaboration} {Borexino}),\ }\bibfield
   {title} {\bibinfo {title} {{First Simultaneous Precision Spectroscopy of
  $pp$, $^7$Be, and $pep$ Solar Neutrinos with Borexino Phase-II}},\ }\href
  {https://doi.org/10.1103/PhysRevD.100.082004} {\bibfield  {journal} {\bibinfo
   {journal} {Phys. Rev. D}\ }\textbf {\bibinfo {volume} {100}},\ \bibinfo
  {pages} {082004} (\bibinfo {year} {2019})},\ \Eprint
  {https://arxiv.org/abs/1707.09279} {arXiv:1707.09279 [hep-ex]} \BibitemShut
  {NoStop}%
\bibitem [{\citenamefont {Agostini}\ \emph {et~al.}(2018)\citenamefont
  {Agostini} \emph {et~al.}}]{BOREXINO:2018ohr}%
  \BibitemOpen
  \bibfield  {author} {\bibinfo {author} {\bibfnamefont {M.}~\bibnamefont
  {Agostini}} \emph {et~al.} (\bibinfo {collaboration} {BOREXINO}),\ }\bibfield
   {title} {\bibinfo {title} {{Comprehensive measurement of $pp$-chain solar
  neutrinos}},\ }\href {https://doi.org/10.1038/s41586-018-0624-y} {\bibfield
  {journal} {\bibinfo  {journal} {Nature}\ }\textbf {\bibinfo {volume} {562}},\
  \bibinfo {pages} {505} (\bibinfo {year} {2018})}\BibitemShut {NoStop}%
\bibitem [{\citenamefont {Agostini}\ \emph {et~al.}(2020)\citenamefont
  {Agostini} \emph {et~al.}}]{BOREXINO:2020aww}%
  \BibitemOpen
  \bibfield  {author} {\bibinfo {author} {\bibfnamefont {M.}~\bibnamefont
  {Agostini}} \emph {et~al.} (\bibinfo {collaboration} {BOREXINO}),\ }\bibfield
   {title} {\bibinfo {title} {{Experimental evidence of neutrinos produced in
  the CNO fusion cycle in the Sun}},\ }\href
  {https://doi.org/10.1038/s41586-020-2934-0} {\bibfield  {journal} {\bibinfo
  {journal} {Nature}\ }\textbf {\bibinfo {volume} {587}},\ \bibinfo {pages}
  {577} (\bibinfo {year} {2020})},\ \Eprint {https://arxiv.org/abs/2006.15115}
  {arXiv:2006.15115 [hep-ex]} \BibitemShut {NoStop}%
\bibitem [{\citenamefont {Deniz}\ \emph {et~al.}(2010)\citenamefont {Deniz}
  \emph {et~al.}}]{TEXONO:2009knm}%
  \BibitemOpen
  \bibfield  {author} {\bibinfo {author} {\bibfnamefont {M.}~\bibnamefont
  {Deniz}} \emph {et~al.} (\bibinfo {collaboration} {TEXONO}),\ }\bibfield
  {title} {\bibinfo {title} {{Measurement of Nu(e)-bar -Electron Scattering
  Cross-Section with a CsI(Tl) Scintillating Crystal Array at the Kuo-Sheng
  Nuclear Power Reactor}},\ }\href {https://doi.org/10.1103/PhysRevD.81.072001}
  {\bibfield  {journal} {\bibinfo  {journal} {Phys. Rev. D}\ }\textbf {\bibinfo
  {volume} {81}},\ \bibinfo {pages} {072001} (\bibinfo {year} {2010})},\
  \Eprint {https://arxiv.org/abs/0911.1597} {arXiv:0911.1597 [hep-ex]}
  \BibitemShut {NoStop}%
\bibitem [{\citenamefont {et~al.}(1994)}]{VILAIN1994246}%
  \BibitemOpen
  \bibfield  {author} {\bibinfo {author} {\bibfnamefont {P.~V.}\ \bibnamefont
  {et~al.}},\ }\bibfield  {title} {\bibinfo {title} {Precision measurement of
  electroweak parameters from the scattering of muon-neutrinos on electrons},\
  }\href {https://doi.org/https://doi.org/10.1016/0370-2693(94)91421-4}
  {\bibfield  {journal} {\bibinfo  {journal} {Physics Letters B}\ }\textbf
  {\bibinfo {volume} {335}},\ \bibinfo {pages} {246} (\bibinfo {year}
  {1994})}\BibitemShut {NoStop}%
\bibitem [{\citenamefont {Vilain}\ \emph {et~al.}(1993)\citenamefont {Vilain}
  \emph {et~al.}}]{CHARM-II:1993phx}%
  \BibitemOpen
  \bibfield  {author} {\bibinfo {author} {\bibfnamefont {P.}~\bibnamefont
  {Vilain}} \emph {et~al.} (\bibinfo {collaboration} {CHARM-II}),\ }\bibfield
  {title} {\bibinfo {title} {{Measurement of differential cross-sections for
  muon-neutrino electron scattering}},\ }\href
  {https://doi.org/10.1016/0370-2693(93)90408-A} {\bibfield  {journal}
  {\bibinfo  {journal} {Phys. Lett. B}\ }\textbf {\bibinfo {volume} {302}},\
  \bibinfo {pages} {351} (\bibinfo {year} {1993})}\BibitemShut {NoStop}%
\bibitem [{\citenamefont {Vogel}\ and\ \citenamefont
  {Engel}(1989)}]{Vogel:1989iv}%
  \BibitemOpen
  \bibfield  {author} {\bibinfo {author} {\bibfnamefont {P.}~\bibnamefont
  {Vogel}}\ and\ \bibinfo {author} {\bibfnamefont {J.}~\bibnamefont {Engel}},\
  }\bibfield  {title} {\bibinfo {title} {{Neutrino Electromagnetic
  Form-Factors}},\ }\href {https://doi.org/10.1103/PhysRevD.39.3378} {\bibfield
   {journal} {\bibinfo  {journal} {Phys. Rev. D}\ }\textbf {\bibinfo {volume}
  {39}},\ \bibinfo {pages} {3378} (\bibinfo {year} {1989})}\BibitemShut
  {NoStop}%
\bibitem [{\citenamefont {Coloma}\ \emph
  {et~al.}(2022{\natexlab{b}})\citenamefont {Coloma}, \citenamefont
  {Gonzalez-Garcia}, \citenamefont {Maltoni}, \citenamefont {Pinheiro},\ and\
  \citenamefont {Urrea}}]{Coloma:2022umy}%
  \BibitemOpen
  \bibfield  {author} {\bibinfo {author} {\bibfnamefont {P.}~\bibnamefont
  {Coloma}}, \bibinfo {author} {\bibfnamefont {M.~C.}\ \bibnamefont
  {Gonzalez-Garcia}}, \bibinfo {author} {\bibfnamefont {M.}~\bibnamefont
  {Maltoni}}, \bibinfo {author} {\bibfnamefont {J.~a.~P.}\ \bibnamefont
  {Pinheiro}},\ and\ \bibinfo {author} {\bibfnamefont {S.}~\bibnamefont
  {Urrea}},\ }\bibfield  {title} {\bibinfo {title} {{Constraining new physics
  with Borexino Phase-II spectral data}},\ }\href
  {https://doi.org/10.1007/JHEP07(2022)138} {\bibfield  {journal} {\bibinfo
  {journal} {JHEP}\ }\textbf {\bibinfo {volume} {07}},\ \bibinfo {pages}
  {138}},\ \bibinfo {note} {[Erratum: JHEP 11, 138 (2022)]},\ \Eprint
  {https://arxiv.org/abs/2204.03011} {arXiv:2204.03011 [hep-ph]} \BibitemShut
  {NoStop}%
\bibitem [{\citenamefont {Fox}\ \emph {et~al.}(2011)\citenamefont {Fox},
  \citenamefont {Harnik}, \citenamefont {Kopp},\ and\ \citenamefont
  {Tsai}}]{Fox:2011fx}%
  \BibitemOpen
  \bibfield  {author} {\bibinfo {author} {\bibfnamefont {P.~J.}\ \bibnamefont
  {Fox}}, \bibinfo {author} {\bibfnamefont {R.}~\bibnamefont {Harnik}},
  \bibinfo {author} {\bibfnamefont {J.}~\bibnamefont {Kopp}},\ and\ \bibinfo
  {author} {\bibfnamefont {Y.}~\bibnamefont {Tsai}},\ }\bibfield  {title}
  {\bibinfo {title} {{LEP Shines Light on Dark Matter}},\ }\href
  {https://doi.org/10.1103/PhysRevD.84.014028} {\bibfield  {journal} {\bibinfo
  {journal} {Phys. Rev. D}\ }\textbf {\bibinfo {volume} {84}},\ \bibinfo
  {pages} {014028} (\bibinfo {year} {2011})},\ \Eprint
  {https://arxiv.org/abs/1103.0240} {arXiv:1103.0240 [hep-ph]} \BibitemShut
  {NoStop}%
\bibitem [{\citenamefont {Banerjee}\ \emph
  {et~al.}(2018{\natexlab{a}})\citenamefont {Banerjee} \emph
  {et~al.}}]{NA64:2017vtt}%
  \BibitemOpen
  \bibfield  {author} {\bibinfo {author} {\bibfnamefont {D.}~\bibnamefont
  {Banerjee}} \emph {et~al.} (\bibinfo {collaboration} {NA64}),\ }\bibfield
  {title} {\bibinfo {title} {{Search for vector mediator of Dark Matter
  production in invisible decay mode}},\ }\href
  {https://doi.org/10.1103/PhysRevD.97.072002} {\bibfield  {journal} {\bibinfo
  {journal} {Phys. Rev. D}\ }\textbf {\bibinfo {volume} {97}},\ \bibinfo
  {pages} {072002} (\bibinfo {year} {2018}{\natexlab{a}})},\ \Eprint
  {https://arxiv.org/abs/1710.00971} {arXiv:1710.00971 [hep-ex]} \BibitemShut
  {NoStop}%
\bibitem [{\citenamefont {Lees}\ \emph
  {et~al.}(2017{\natexlab{b}})\citenamefont {Lees} \emph
  {et~al.}}]{BaBar:2017tiz}%
  \BibitemOpen
  \bibfield  {author} {\bibinfo {author} {\bibfnamefont {J.~P.}\ \bibnamefont
  {Lees}} \emph {et~al.} (\bibinfo {collaboration} {BaBar}),\ }\bibfield
  {title} {\bibinfo {title} {{Search for Invisible Decays of a Dark Photon
  Produced in ${e}^{+}{e}^{-}$ Collisions at BaBar}},\ }\href
  {https://doi.org/10.1103/PhysRevLett.119.131804} {\bibfield  {journal}
  {\bibinfo  {journal} {Phys. Rev. Lett.}\ }\textbf {\bibinfo {volume} {119}},\
  \bibinfo {pages} {131804} (\bibinfo {year} {2017}{\natexlab{b}})},\ \Eprint
  {https://arxiv.org/abs/1702.03327} {arXiv:1702.03327 [hep-ex]} \BibitemShut
  {NoStop}%
\bibitem [{\citenamefont {Adachi}\ \emph {et~al.}(2020)\citenamefont {Adachi}
  \emph {et~al.}}]{Belle-II:2019qfb}%
  \BibitemOpen
  \bibfield  {author} {\bibinfo {author} {\bibfnamefont {I.}~\bibnamefont
  {Adachi}} \emph {et~al.} (\bibinfo {collaboration} {Belle-II}),\ }\bibfield
  {title} {\bibinfo {title} {{Search for an Invisibly Decaying $Z^{\prime}$
  Boson at Belle II in $e^+ e^- \to \mu^+ \mu^- (e^{\pm} \mu^{\mp})$ Plus
  Missing Energy Final States}},\ }\href
  {https://doi.org/10.1103/PhysRevLett.124.141801} {\bibfield  {journal}
  {\bibinfo  {journal} {Phys. Rev. Lett.}\ }\textbf {\bibinfo {volume} {124}},\
  \bibinfo {pages} {141801} (\bibinfo {year} {2020})},\ \Eprint
  {https://arxiv.org/abs/1912.11276} {arXiv:1912.11276 [hep-ex]} \BibitemShut
  {NoStop}%
\bibitem [{\citenamefont {Cortina~Gil}\ \emph {et~al.}(2019)\citenamefont
  {Cortina~Gil} \emph {et~al.}}]{NA62:2019meo}%
  \BibitemOpen
  \bibfield  {author} {\bibinfo {author} {\bibfnamefont {E.}~\bibnamefont
  {Cortina~Gil}} \emph {et~al.} (\bibinfo {collaboration} {NA62}),\ }\bibfield
  {title} {\bibinfo {title} {{Search for production of an invisible dark photon
  in $\pi^0$ decays}},\ }\href {https://doi.org/10.1007/JHEP05(2019)182}
  {\bibfield  {journal} {\bibinfo  {journal} {JHEP}\ }\textbf {\bibinfo
  {volume} {05}},\ \bibinfo {pages} {182}},\ \Eprint
  {https://arxiv.org/abs/1903.08767} {arXiv:1903.08767 [hep-ex]} \BibitemShut
  {NoStop}%
\bibitem [{\citenamefont {Andreev}\ \emph {et~al.}(2021)\citenamefont {Andreev}
  \emph {et~al.}}]{NA64:2021xzo}%
  \BibitemOpen
  \bibfield  {author} {\bibinfo {author} {\bibfnamefont {Y.~M.}\ \bibnamefont
  {Andreev}} \emph {et~al.} (\bibinfo {collaboration} {NA64}),\ }\bibfield
  {title} {\bibinfo {title} {{Constraints on New Physics in Electron $g-2$ from
  a Search for Invisible Decays of a Scalar, Pseudoscalar, Vector, and Axial
  Vector}},\ }\href {https://doi.org/10.1103/PhysRevLett.126.211802} {\bibfield
   {journal} {\bibinfo  {journal} {Phys. Rev. Lett.}\ }\textbf {\bibinfo
  {volume} {126}},\ \bibinfo {pages} {211802} (\bibinfo {year} {2021})},\
  \Eprint {https://arxiv.org/abs/2102.01885} {arXiv:2102.01885 [hep-ex]}
  \BibitemShut {NoStop}%
\bibitem [{\citenamefont {Andreev}\ \emph {et~al.}(2022)\citenamefont {Andreev}
  \emph {et~al.}}]{NA64:2022yly}%
  \BibitemOpen
  \bibfield  {author} {\bibinfo {author} {\bibfnamefont {Y.~M.}\ \bibnamefont
  {Andreev}} \emph {et~al.} (\bibinfo {collaboration} {NA64}),\ }\bibfield
  {title} {\bibinfo {title} {{Search for a New B-L Z' Gauge Boson with the NA64
  Experiment at CERN}},\ }\href
  {https://doi.org/10.1103/PhysRevLett.129.161801} {\bibfield  {journal}
  {\bibinfo  {journal} {Phys. Rev. Lett.}\ }\textbf {\bibinfo {volume} {129}},\
  \bibinfo {pages} {161801} (\bibinfo {year} {2022})},\ \Eprint
  {https://arxiv.org/abs/2207.09979} {arXiv:2207.09979 [hep-ex]} \BibitemShut
  {NoStop}%
\bibitem [{\citenamefont {Andreev}\ \emph {et~al.}(2023)\citenamefont {Andreev}
  \emph {et~al.}}]{NA64:2023wbi}%
  \BibitemOpen
  \bibfield  {author} {\bibinfo {author} {\bibfnamefont {Y.~M.}\ \bibnamefont
  {Andreev}} \emph {et~al.} (\bibinfo {collaboration} {NA64}),\ }\bibfield
  {title} {\bibinfo {title} {{Search for Light Dark Matter with NA64 at
  CERN}},\ }\href {https://doi.org/10.1103/PhysRevLett.131.161801} {\bibfield
  {journal} {\bibinfo  {journal} {Phys. Rev. Lett.}\ }\textbf {\bibinfo
  {volume} {131}},\ \bibinfo {pages} {161801} (\bibinfo {year} {2023})},\
  \Eprint {https://arxiv.org/abs/2307.02404} {arXiv:2307.02404 [hep-ex]}
  \BibitemShut {NoStop}%
\bibitem [{\citenamefont {Kodama}\ \emph {et~al.}(2008)\citenamefont {Kodama}
  \emph {et~al.}}]{DONuT:2007bsg}%
  \BibitemOpen
  \bibfield  {author} {\bibinfo {author} {\bibfnamefont {K.}~\bibnamefont
  {Kodama}} \emph {et~al.} (\bibinfo {collaboration} {DONuT}),\ }\bibfield
  {title} {\bibinfo {title} {{Final tau-neutrino results from the DONuT
  experiment}},\ }\href {https://doi.org/10.1103/PhysRevD.78.052002} {\bibfield
   {journal} {\bibinfo  {journal} {Phys. Rev. D}\ }\textbf {\bibinfo {volume}
  {78}},\ \bibinfo {pages} {052002} (\bibinfo {year} {2008})},\ \Eprint
  {https://arxiv.org/abs/0711.0728} {arXiv:0711.0728 [hep-ex]} \BibitemShut
  {NoStop}%
\bibitem [{\citenamefont {Anelli}\ \emph {et~al.}(2015)\citenamefont {Anelli}
  \emph {et~al.}}]{SHiP:2015vad}%
  \BibitemOpen
  \bibfield  {author} {\bibinfo {author} {\bibfnamefont {M.}~\bibnamefont
  {Anelli}} \emph {et~al.} (\bibinfo {collaboration} {SHiP}),\ }\bibfield
  {title} {\bibinfo {title} {{A facility to Search for Hidden Particles (SHiP)
  at the CERN SPS}},\ }\href@noop {} {\  (\bibinfo {year} {2015})},\ \Eprint
  {https://arxiv.org/abs/1504.04956} {arXiv:1504.04956 [physics.ins-det]}
  \BibitemShut {NoStop}%
\bibitem [{\citenamefont {Abreu}\ \emph {et~al.}(2020)\citenamefont {Abreu}
  \emph {et~al.}}]{FASER:2019dxq}%
  \BibitemOpen
  \bibfield  {author} {\bibinfo {author} {\bibfnamefont {H.}~\bibnamefont
  {Abreu}} \emph {et~al.} (\bibinfo {collaboration} {FASER}),\ }\bibfield
  {title} {\bibinfo {title} {{Detecting and Studying High-Energy Collider
  Neutrinos with FASER at the LHC}},\ }\href
  {https://doi.org/10.1140/epjc/s10052-020-7631-5} {\bibfield  {journal}
  {\bibinfo  {journal} {Eur. Phys. J. C}\ }\textbf {\bibinfo {volume} {80}},\
  \bibinfo {pages} {61} (\bibinfo {year} {2020})},\ \Eprint
  {https://arxiv.org/abs/1908.02310} {arXiv:1908.02310 [hep-ex]} \BibitemShut
  {NoStop}%
\bibitem [{\citenamefont {Ahdida}\ \emph {et~al.}(2020)\citenamefont {Ahdida}
  \emph {et~al.}}]{SHiP:2020sos}%
  \BibitemOpen
  \bibfield  {author} {\bibinfo {author} {\bibfnamefont {C.}~\bibnamefont
  {Ahdida}} \emph {et~al.} (\bibinfo {collaboration} {SHiP}),\ }\bibfield
  {title} {\bibinfo {title} {{SND@LHC}},\ }\href@noop {} {\  (\bibinfo {year}
  {2020})},\ \Eprint {https://arxiv.org/abs/2002.08722} {arXiv:2002.08722
  [physics.ins-det]} \BibitemShut {NoStop}%
\bibitem [{\citenamefont {Kling}(2020)}]{Kling:2020iar}%
  \BibitemOpen
  \bibfield  {author} {\bibinfo {author} {\bibfnamefont {F.}~\bibnamefont
  {Kling}},\ }\bibfield  {title} {\bibinfo {title} {{Probing light gauge bosons
  in tau neutrino experiments}},\ }\href
  {https://doi.org/10.1103/PhysRevD.102.015007} {\bibfield  {journal} {\bibinfo
   {journal} {Phys. Rev. D}\ }\textbf {\bibinfo {volume} {102}},\ \bibinfo
  {pages} {015007} (\bibinfo {year} {2020})},\ \Eprint
  {https://arxiv.org/abs/2005.03594} {arXiv:2005.03594 [hep-ph]} \BibitemShut
  {NoStop}%
\bibitem [{\citenamefont {Kling}\ \emph {et~al.}(2025)\citenamefont {Kling},
  \citenamefont {Reimitz},\ and\ \citenamefont {Ritz}}]{Kling:2025udr}%
  \BibitemOpen
  \bibfield  {author} {\bibinfo {author} {\bibfnamefont {F.}~\bibnamefont
  {Kling}}, \bibinfo {author} {\bibfnamefont {P.}~\bibnamefont {Reimitz}},\
  and\ \bibinfo {author} {\bibfnamefont {A.}~\bibnamefont {Ritz}},\ }\bibfield
  {title} {\bibinfo {title} {{Dark Vector Boson Bremsstrahlung: New Form
  Factors for a Broader Class of Models}},\ }\href@noop {} {\  (\bibinfo {year}
  {2025})},\ \Eprint {https://arxiv.org/abs/2509.09437} {arXiv:2509.09437
  [hep-ph]} \BibitemShut {NoStop}%
\bibitem [{\citenamefont {Alekhin}\ \emph {et~al.}(2016)\citenamefont {Alekhin}
  \emph {et~al.}}]{Alekhin:2015byh}%
  \BibitemOpen
  \bibfield  {author} {\bibinfo {author} {\bibfnamefont {S.}~\bibnamefont
  {Alekhin}} \emph {et~al.},\ }\bibfield  {title} {\bibinfo {title} {{A
  facility to Search for Hidden Particles at the CERN SPS: the SHiP physics
  case}},\ }\href {https://doi.org/10.1088/0034-4885/79/12/124201} {\bibfield
  {journal} {\bibinfo  {journal} {Rept. Prog. Phys.}\ }\textbf {\bibinfo
  {volume} {79}},\ \bibinfo {pages} {124201} (\bibinfo {year} {2016})},\
  \Eprint {https://arxiv.org/abs/1504.04855} {arXiv:1504.04855 [hep-ph]}
  \BibitemShut {NoStop}%
\bibitem [{\citenamefont {Ahdida}\ \emph {et~al.}(2021)\citenamefont {Ahdida}
  \emph {et~al.}}]{SHiP:2020vbd}%
  \BibitemOpen
  \bibfield  {author} {\bibinfo {author} {\bibfnamefont {C.}~\bibnamefont
  {Ahdida}} \emph {et~al.} (\bibinfo {collaboration} {SHiP}),\ }\bibfield
  {title} {\bibinfo {title} {{Sensitivity of the SHiP experiment to dark
  photons decaying to a pair of charged particles}},\ }\href
  {https://doi.org/10.1140/epjc/s10052-021-09224-3} {\bibfield  {journal}
  {\bibinfo  {journal} {Eur. Phys. J. C}\ }\textbf {\bibinfo {volume} {81}},\
  \bibinfo {pages} {451} (\bibinfo {year} {2021})},\ \Eprint
  {https://arxiv.org/abs/2011.05115} {arXiv:2011.05115 [hep-ex]} \BibitemShut
  {NoStop}%
\bibitem [{\citenamefont {Zhou}\ \emph {et~al.}(2025)\citenamefont {Zhou},
  \citenamefont {Plestid}, \citenamefont {Kelly}, \citenamefont {Blinov},\ and\
  \citenamefont {Fox}}]{Zhou:2024aeu}%
  \BibitemOpen
  \bibfield  {author} {\bibinfo {author} {\bibfnamefont {T.}~\bibnamefont
  {Zhou}}, \bibinfo {author} {\bibfnamefont {R.}~\bibnamefont {Plestid}},
  \bibinfo {author} {\bibfnamefont {K.~J.}\ \bibnamefont {Kelly}}, \bibinfo
  {author} {\bibfnamefont {N.}~\bibnamefont {Blinov}},\ and\ \bibinfo {author}
  {\bibfnamefont {P.~J.}\ \bibnamefont {Fox}},\ }\bibfield  {title} {\bibinfo
  {title} {{Long-lived vectors from electromagnetic cascades at SHiP}},\ }\href
  {https://doi.org/10.1007/JHEP02(2025)107} {\bibfield  {journal} {\bibinfo
  {journal} {JHEP}\ }\textbf {\bibinfo {volume} {02}},\ \bibinfo {pages}
  {107}},\ \Eprint {https://arxiv.org/abs/2412.01880} {arXiv:2412.01880
  [hep-ph]} \BibitemShut {NoStop}%
\bibitem [{\citenamefont {Bjorken}\ \emph {et~al.}(1988)\citenamefont
  {Bjorken}, \citenamefont {Ecklund}, \citenamefont {Nelson}, \citenamefont
  {Abashian}, \citenamefont {Church}, \citenamefont {Lu}, \citenamefont {Mo},
  \citenamefont {Nunamaker},\ and\ \citenamefont {Rassmann}}]{Bjorken:1988as}%
  \BibitemOpen
  \bibfield  {author} {\bibinfo {author} {\bibfnamefont {J.~D.}\ \bibnamefont
  {Bjorken}}, \bibinfo {author} {\bibfnamefont {S.}~\bibnamefont {Ecklund}},
  \bibinfo {author} {\bibfnamefont {W.~R.}\ \bibnamefont {Nelson}}, \bibinfo
  {author} {\bibfnamefont {A.}~\bibnamefont {Abashian}}, \bibinfo {author}
  {\bibfnamefont {C.}~\bibnamefont {Church}}, \bibinfo {author} {\bibfnamefont
  {B.}~\bibnamefont {Lu}}, \bibinfo {author} {\bibfnamefont {L.~W.}\
  \bibnamefont {Mo}}, \bibinfo {author} {\bibfnamefont {T.~A.}\ \bibnamefont
  {Nunamaker}},\ and\ \bibinfo {author} {\bibfnamefont {P.}~\bibnamefont
  {Rassmann}},\ }\bibfield  {title} {\bibinfo {title} {{Search for Neutral
  Metastable Penetrating Particles Produced in the SLAC Beam Dump}},\ }\href
  {https://doi.org/10.1103/PhysRevD.38.3375} {\bibfield  {journal} {\bibinfo
  {journal} {Phys. Rev. D}\ }\textbf {\bibinfo {volume} {38}},\ \bibinfo
  {pages} {3375} (\bibinfo {year} {1988})}\BibitemShut {NoStop}%
\bibitem [{\citenamefont {Tsai}(1986)}]{Tsai:1986tx}%
  \BibitemOpen
  \bibfield  {author} {\bibinfo {author} {\bibfnamefont {Y.-S.}\ \bibnamefont
  {Tsai}},\ }\bibfield  {title} {\bibinfo {title} {{AXION BREMSSTRAHLUNG BY AN
  ELECTRON BEAM}},\ }\href {https://doi.org/10.1103/PhysRevD.34.1326}
  {\bibfield  {journal} {\bibinfo  {journal} {Phys. Rev. D}\ }\textbf {\bibinfo
  {volume} {34}},\ \bibinfo {pages} {1326} (\bibinfo {year}
  {1986})}\BibitemShut {NoStop}%
\bibitem [{\citenamefont {Andreas}(2013)}]{Andreas:2013xxa}%
  \BibitemOpen
  \bibfield  {author} {\bibinfo {author} {\bibfnamefont {S.}~\bibnamefont
  {Andreas}},\ }\emph {\bibinfo {title} {{Light Weakly Interacting Particles:
  Constraints and Connection to Dark Matter}}},\ \href
  {https://doi.org/10.3204/DESY-THESIS-2013-024} {Ph.D. thesis},\ \bibinfo
  {school} {Hamburg U.} (\bibinfo {year} {2013})\BibitemShut {NoStop}%
\bibitem [{\citenamefont {Davier}\ and\ \citenamefont
  {Nguyen~Ngoc}(1989)}]{Davier:1989wz}%
  \BibitemOpen
  \bibfield  {author} {\bibinfo {author} {\bibfnamefont {M.}~\bibnamefont
  {Davier}}\ and\ \bibinfo {author} {\bibfnamefont {H.}~\bibnamefont
  {Nguyen~Ngoc}},\ }\bibfield  {title} {\bibinfo {title} {{An Unambiguous
  Search for a Light Higgs Boson}},\ }\href
  {https://doi.org/10.1016/0370-2693(89)90174-3} {\bibfield  {journal}
  {\bibinfo  {journal} {Phys. Lett. B}\ }\textbf {\bibinfo {volume} {229}},\
  \bibinfo {pages} {150} (\bibinfo {year} {1989})}\BibitemShut {NoStop}%
\bibitem [{\citenamefont {Konaka}\ \emph {et~al.}(1986)\citenamefont {Konaka}
  \emph {et~al.}}]{Konaka:1986cb}%
  \BibitemOpen
  \bibfield  {author} {\bibinfo {author} {\bibfnamefont {A.}~\bibnamefont
  {Konaka}} \emph {et~al.},\ }\bibfield  {title} {\bibinfo {title} {{Search for
  Neutral Particles in Electron Beam Dump Experiment}},\ }\href
  {https://doi.org/10.1103/PhysRevLett.57.659} {\bibfield  {journal} {\bibinfo
  {journal} {Phys. Rev. Lett.}\ }\textbf {\bibinfo {volume} {57}},\ \bibinfo
  {pages} {659} (\bibinfo {year} {1986})}\BibitemShut {NoStop}%
\bibitem [{\citenamefont {Banerjee}\ \emph
  {et~al.}(2018{\natexlab{b}})\citenamefont {Banerjee} \emph
  {et~al.}}]{NA64:2018lsq}%
  \BibitemOpen
  \bibfield  {author} {\bibinfo {author} {\bibfnamefont {D.}~\bibnamefont
  {Banerjee}} \emph {et~al.} (\bibinfo {collaboration} {NA64}),\ }\bibfield
  {title} {\bibinfo {title} {{Search for a Hypothetical 16.7 MeV Gauge Boson
  and Dark Photons in the NA64 Experiment at CERN}},\ }\href
  {https://doi.org/10.1103/PhysRevLett.120.231802} {\bibfield  {journal}
  {\bibinfo  {journal} {Phys. Rev. Lett.}\ }\textbf {\bibinfo {volume} {120}},\
  \bibinfo {pages} {231802} (\bibinfo {year} {2018}{\natexlab{b}})},\ \Eprint
  {https://arxiv.org/abs/1803.07748} {arXiv:1803.07748 [hep-ex]} \BibitemShut
  {NoStop}%
\bibitem [{\citenamefont {Banerjee}\ \emph {et~al.}(2020)\citenamefont
  {Banerjee} \emph {et~al.}}]{NA64:2019auh}%
  \BibitemOpen
  \bibfield  {author} {\bibinfo {author} {\bibfnamefont {D.}~\bibnamefont
  {Banerjee}} \emph {et~al.} (\bibinfo {collaboration} {NA64}),\ }\bibfield
  {title} {\bibinfo {title} {{Improved limits on a hypothetical X(16.7) boson
  and a dark photon decaying into $e^+e^-$ pairs}},\ }\href
  {https://doi.org/10.1103/PhysRevD.101.071101} {\bibfield  {journal} {\bibinfo
   {journal} {Phys. Rev. D}\ }\textbf {\bibinfo {volume} {101}},\ \bibinfo
  {pages} {071101} (\bibinfo {year} {2020})},\ \Eprint
  {https://arxiv.org/abs/1912.11389} {arXiv:1912.11389 [hep-ex]} \BibitemShut
  {NoStop}%
\bibitem [{\citenamefont {Riordan}\ \emph {et~al.}(1987)\citenamefont {Riordan}
  \emph {et~al.}}]{Riordan:1987aw}%
  \BibitemOpen
  \bibfield  {author} {\bibinfo {author} {\bibfnamefont {E.~M.}\ \bibnamefont
  {Riordan}} \emph {et~al.},\ }\bibfield  {title} {\bibinfo {title} {{A Search
  for Short Lived Axions in an Electron Beam Dump Experiment}},\ }\href
  {https://doi.org/10.1103/PhysRevLett.59.755} {\bibfield  {journal} {\bibinfo
  {journal} {Phys. Rev. Lett.}\ }\textbf {\bibinfo {volume} {59}},\ \bibinfo
  {pages} {755} (\bibinfo {year} {1987})}\BibitemShut {NoStop}%
\bibitem [{\citenamefont {Bross}\ \emph {et~al.}(1991)\citenamefont {Bross},
  \citenamefont {Crisler}, \citenamefont {Pordes}, \citenamefont {Volk},
  \citenamefont {Errede},\ and\ \citenamefont {Wrbanek}}]{Bross:1989mp}%
  \BibitemOpen
  \bibfield  {author} {\bibinfo {author} {\bibfnamefont {A.}~\bibnamefont
  {Bross}}, \bibinfo {author} {\bibfnamefont {M.}~\bibnamefont {Crisler}},
  \bibinfo {author} {\bibfnamefont {S.~H.}\ \bibnamefont {Pordes}}, \bibinfo
  {author} {\bibfnamefont {J.}~\bibnamefont {Volk}}, \bibinfo {author}
  {\bibfnamefont {S.}~\bibnamefont {Errede}},\ and\ \bibinfo {author}
  {\bibfnamefont {J.}~\bibnamefont {Wrbanek}},\ }\bibfield  {title} {\bibinfo
  {title} {{A Search for Shortlived Particles Produced in an Electron Beam
  Dump}},\ }\href {https://doi.org/10.1103/PhysRevLett.67.2942} {\bibfield
  {journal} {\bibinfo  {journal} {Phys. Rev. Lett.}\ }\textbf {\bibinfo
  {volume} {67}},\ \bibinfo {pages} {2942} (\bibinfo {year}
  {1991})}\BibitemShut {NoStop}%
\bibitem [{\citenamefont {Cudell}\ \emph {et~al.}(2002)\citenamefont {Cudell},
  \citenamefont {Ezhela}, \citenamefont {Gauron}, \citenamefont {Kang},
  \citenamefont {Kuyanov}, \citenamefont {Lugovsky}, \citenamefont
  {Nicolescu},\ and\ \citenamefont {Tkachenko}}]{Cudell:2001pn}%
  \BibitemOpen
  \bibfield  {author} {\bibinfo {author} {\bibfnamefont {J.~R.}\ \bibnamefont
  {Cudell}}, \bibinfo {author} {\bibfnamefont {V.}~\bibnamefont {Ezhela}},
  \bibinfo {author} {\bibfnamefont {P.}~\bibnamefont {Gauron}}, \bibinfo
  {author} {\bibfnamefont {K.}~\bibnamefont {Kang}}, \bibinfo {author}
  {\bibfnamefont {Y.~V.}\ \bibnamefont {Kuyanov}}, \bibinfo {author}
  {\bibfnamefont {S.}~\bibnamefont {Lugovsky}}, \bibinfo {author}
  {\bibfnamefont {B.}~\bibnamefont {Nicolescu}},\ and\ \bibinfo {author}
  {\bibfnamefont {N.}~\bibnamefont {Tkachenko}},\ }\bibfield  {title} {\bibinfo
  {title} {{Hadronic scattering amplitudes: Medium-energy constraints on
  asymptotic behavior}},\ }\href {https://doi.org/10.1103/PhysRevD.65.074024}
  {\bibfield  {journal} {\bibinfo  {journal} {Phys. Rev. D}\ }\textbf {\bibinfo
  {volume} {65}},\ \bibinfo {pages} {074024} (\bibinfo {year} {2002})},\
  \Eprint {https://arxiv.org/abs/hep-ph/0107219} {arXiv:hep-ph/0107219}
  \BibitemShut {NoStop}%
\bibitem [{\citenamefont {Blumlein}\ and\ \citenamefont
  {Brunner}(2011)}]{Blumlein:2011mv}%
  \BibitemOpen
  \bibfield  {author} {\bibinfo {author} {\bibfnamefont {J.}~\bibnamefont
  {Blumlein}}\ and\ \bibinfo {author} {\bibfnamefont {J.}~\bibnamefont
  {Brunner}},\ }\bibfield  {title} {\bibinfo {title} {{New Exclusion Limits for
  Dark Gauge Forces from Beam-Dump Data}},\ }\href
  {https://doi.org/10.1016/j.physletb.2011.05.046} {\bibfield  {journal}
  {\bibinfo  {journal} {Phys. Lett. B}\ }\textbf {\bibinfo {volume} {701}},\
  \bibinfo {pages} {155} (\bibinfo {year} {2011})},\ \Eprint
  {https://arxiv.org/abs/1104.2747} {arXiv:1104.2747 [hep-ex]} \BibitemShut
  {NoStop}%
\bibitem [{\citenamefont {Gninenko}(2012{\natexlab{a}})}]{Gninenko:2011uv}%
  \BibitemOpen
  \bibfield  {author} {\bibinfo {author} {\bibfnamefont {S.~N.}\ \bibnamefont
  {Gninenko}},\ }\bibfield  {title} {\bibinfo {title} {{Stringent limits on the
  $\pi^0 \to \gamma X, X \to e^+e^-$ decay from neutrino experiments and
  constraints on new light gauge bosons}},\ }\href
  {https://doi.org/10.1103/PhysRevD.85.055027} {\bibfield  {journal} {\bibinfo
  {journal} {Phys. Rev. D}\ }\textbf {\bibinfo {volume} {85}},\ \bibinfo
  {pages} {055027} (\bibinfo {year} {2012}{\natexlab{a}})},\ \Eprint
  {https://arxiv.org/abs/1112.5438} {arXiv:1112.5438 [hep-ph]} \BibitemShut
  {NoStop}%
\bibitem [{\citenamefont {Gninenko}(2012{\natexlab{b}})}]{Gninenko:2012eq}%
  \BibitemOpen
  \bibfield  {author} {\bibinfo {author} {\bibfnamefont {S.~N.}\ \bibnamefont
  {Gninenko}},\ }\bibfield  {title} {\bibinfo {title} {{Constraints on sub-GeV
  hidden sector gauge bosons from a search for heavy neutrino decays}},\ }\href
  {https://doi.org/10.1016/j.physletb.2012.06.002} {\bibfield  {journal}
  {\bibinfo  {journal} {Phys. Lett. B}\ }\textbf {\bibinfo {volume} {713}},\
  \bibinfo {pages} {244} (\bibinfo {year} {2012}{\natexlab{b}})},\ \Eprint
  {https://arxiv.org/abs/1204.3583} {arXiv:1204.3583 [hep-ph]} \BibitemShut
  {NoStop}%
\bibitem [{\citenamefont {Bl\"umlein}\ and\ \citenamefont
  {Brunner}(2014)}]{Blumlein:2013cua}%
  \BibitemOpen
  \bibfield  {author} {\bibinfo {author} {\bibfnamefont {J.}~\bibnamefont
  {Bl\"umlein}}\ and\ \bibinfo {author} {\bibfnamefont {J.}~\bibnamefont
  {Brunner}},\ }\bibfield  {title} {\bibinfo {title} {{New Exclusion Limits on
  Dark Gauge Forces from Proton Bremsstrahlung in Beam-Dump Data}},\ }\href
  {https://doi.org/10.1016/j.physletb.2014.02.029} {\bibfield  {journal}
  {\bibinfo  {journal} {Phys. Lett. B}\ }\textbf {\bibinfo {volume} {731}},\
  \bibinfo {pages} {320} (\bibinfo {year} {2014})},\ \Eprint
  {https://arxiv.org/abs/1311.3870} {arXiv:1311.3870 [hep-ph]} \BibitemShut
  {NoStop}%
\bibitem [{\citenamefont {Astier}\ \emph {et~al.}(2001)\citenamefont {Astier}
  \emph {et~al.}}]{NOMAD:2001eyx}%
  \BibitemOpen
  \bibfield  {author} {\bibinfo {author} {\bibfnamefont {P.}~\bibnamefont
  {Astier}} \emph {et~al.} (\bibinfo {collaboration} {NOMAD}),\ }\bibfield
  {title} {\bibinfo {title} {{Search for heavy neutrinos mixing with tau
  neutrinos}},\ }\href {https://doi.org/10.1016/S0370-2693(01)00362-8}
  {\bibfield  {journal} {\bibinfo  {journal} {Phys. Lett. B}\ }\textbf
  {\bibinfo {volume} {506}},\ \bibinfo {pages} {27} (\bibinfo {year} {2001})},\
  \Eprint {https://arxiv.org/abs/hep-ex/0101041} {arXiv:hep-ex/0101041}
  \BibitemShut {NoStop}%
\bibitem [{\citenamefont {Bernardi}\ \emph {et~al.}(1986)\citenamefont
  {Bernardi} \emph {et~al.}}]{Bernardi:1985ny}%
  \BibitemOpen
  \bibfield  {author} {\bibinfo {author} {\bibfnamefont {G.}~\bibnamefont
  {Bernardi}} \emph {et~al.},\ }\bibfield  {title} {\bibinfo {title} {{Search
  for Neutrino Decay}},\ }\href {https://doi.org/10.1016/0370-2693(86)91602-3}
  {\bibfield  {journal} {\bibinfo  {journal} {Phys. Lett. B}\ }\textbf
  {\bibinfo {volume} {166}},\ \bibinfo {pages} {479} (\bibinfo {year}
  {1986})}\BibitemShut {NoStop}%
\bibitem [{\citenamefont {Tsai}\ \emph {et~al.}(2021)\citenamefont {Tsai},
  \citenamefont {deNiverville},\ and\ \citenamefont {Liu}}]{Tsai:2019buq}%
  \BibitemOpen
  \bibfield  {author} {\bibinfo {author} {\bibfnamefont {Y.-D.}\ \bibnamefont
  {Tsai}}, \bibinfo {author} {\bibfnamefont {P.}~\bibnamefont {deNiverville}},\
  and\ \bibinfo {author} {\bibfnamefont {M.~X.}\ \bibnamefont {Liu}},\
  }\bibfield  {title} {\bibinfo {title} {{Dark Photon and Muon $g-2$ Inspired
  Inelastic Dark Matter Models at the High-Energy Intensity Frontier}},\ }\href
  {https://doi.org/10.1103/PhysRevLett.126.181801} {\bibfield  {journal}
  {\bibinfo  {journal} {Phys. Rev. Lett.}\ }\textbf {\bibinfo {volume} {126}},\
  \bibinfo {pages} {181801} (\bibinfo {year} {2021})},\ \Eprint
  {https://arxiv.org/abs/1908.07525} {arXiv:1908.07525 [hep-ph]} \BibitemShut
  {NoStop}%
\bibitem [{\citenamefont {Abrahamyan}\ \emph {et~al.}(2011)\citenamefont
  {Abrahamyan} \emph {et~al.}}]{APEX:2011dww}%
  \BibitemOpen
  \bibfield  {author} {\bibinfo {author} {\bibfnamefont {S.}~\bibnamefont
  {Abrahamyan}} \emph {et~al.} (\bibinfo {collaboration} {APEX}),\ }\bibfield
  {title} {\bibinfo {title} {{Search for a New Gauge Boson in Electron-Nucleus
  Fixed-Target Scattering by the APEX Experiment}},\ }\href
  {https://doi.org/10.1103/PhysRevLett.107.191804} {\bibfield  {journal}
  {\bibinfo  {journal} {Phys. Rev. Lett.}\ }\textbf {\bibinfo {volume} {107}},\
  \bibinfo {pages} {191804} (\bibinfo {year} {2011})},\ \Eprint
  {https://arxiv.org/abs/1108.2750} {arXiv:1108.2750 [hep-ex]} \BibitemShut
  {NoStop}%
\bibitem [{\citenamefont {Merkel}\ \emph {et~al.}(2011)\citenamefont {Merkel}
  \emph {et~al.}}]{A1:2011yso}%
  \BibitemOpen
  \bibfield  {author} {\bibinfo {author} {\bibfnamefont {H.}~\bibnamefont
  {Merkel}} \emph {et~al.} (\bibinfo {collaboration} {A1}),\ }\bibfield
  {title} {\bibinfo {title} {{Search for Light Gauge Bosons of the Dark Sector
  at the Mainz Microtron}},\ }\href
  {https://doi.org/10.1103/PhysRevLett.106.251802} {\bibfield  {journal}
  {\bibinfo  {journal} {Phys. Rev. Lett.}\ }\textbf {\bibinfo {volume} {106}},\
  \bibinfo {pages} {251802} (\bibinfo {year} {2011})},\ \Eprint
  {https://arxiv.org/abs/1101.4091} {arXiv:1101.4091 [nucl-ex]} \BibitemShut
  {NoStop}%
\bibitem [{\citenamefont {Merkel}\ \emph {et~al.}(2014)\citenamefont {Merkel}
  \emph {et~al.}}]{Merkel:2014avp}%
  \BibitemOpen
  \bibfield  {author} {\bibinfo {author} {\bibfnamefont {H.}~\bibnamefont
  {Merkel}} \emph {et~al.},\ }\bibfield  {title} {\bibinfo {title} {{Search at
  the Mainz Microtron for Light Massive Gauge Bosons Relevant for the Muon g-2
  Anomaly}},\ }\href {https://doi.org/10.1103/PhysRevLett.112.221802}
  {\bibfield  {journal} {\bibinfo  {journal} {Phys. Rev. Lett.}\ }\textbf
  {\bibinfo {volume} {112}},\ \bibinfo {pages} {221802} (\bibinfo {year}
  {2014})},\ \Eprint {https://arxiv.org/abs/1404.5502} {arXiv:1404.5502
  [hep-ex]} \BibitemShut {NoStop}%
\bibitem [{\citenamefont {Battaglieri}\ \emph {et~al.}(2015)\citenamefont
  {Battaglieri} \emph {et~al.}}]{Battaglieri:2014hga}%
  \BibitemOpen
  \bibfield  {author} {\bibinfo {author} {\bibfnamefont {M.}~\bibnamefont
  {Battaglieri}} \emph {et~al.},\ }\bibfield  {title} {\bibinfo {title} {{The
  Heavy Photon Search Test Detector}},\ }\href
  {https://doi.org/10.1016/j.nima.2014.12.017} {\bibfield  {journal} {\bibinfo
  {journal} {Nucl. Instrum. Meth. A}\ }\textbf {\bibinfo {volume} {777}},\
  \bibinfo {pages} {91} (\bibinfo {year} {2015})},\ \Eprint
  {https://arxiv.org/abs/1406.6115} {arXiv:1406.6115 [physics.ins-det]}
  \BibitemShut {NoStop}%
\bibitem [{\citenamefont {Adrian}\ \emph {et~al.}(2023)\citenamefont {Adrian}
  \emph {et~al.}}]{Adrian:2022nkt}%
  \BibitemOpen
  \bibfield  {author} {\bibinfo {author} {\bibfnamefont {P.~H.}\ \bibnamefont
  {Adrian}} \emph {et~al.},\ }\bibfield  {title} {\bibinfo {title} {{Searching
  for prompt and long-lived dark photons in electroproduced e+e- pairs with the
  heavy photon search experiment at JLab}},\ }\href
  {https://doi.org/10.1103/PhysRevD.108.012015} {\bibfield  {journal} {\bibinfo
   {journal} {Phys. Rev. D}\ }\textbf {\bibinfo {volume} {108}},\ \bibinfo
  {pages} {012015} (\bibinfo {year} {2023})},\ \Eprint
  {https://arxiv.org/abs/2212.10629} {arXiv:2212.10629 [hep-ex]} \BibitemShut
  {NoStop}%
\bibitem [{\citenamefont {Batley}\ \emph {et~al.}(2015)\citenamefont {Batley}
  \emph {et~al.}}]{NA482:2015wmo}%
  \BibitemOpen
  \bibfield  {author} {\bibinfo {author} {\bibfnamefont {J.~R.}\ \bibnamefont
  {Batley}} \emph {et~al.} (\bibinfo {collaboration} {NA48/2}),\ }\bibfield
  {title} {\bibinfo {title} {{Search for the dark photon in $\pi^0$ decays}},\
  }\href {https://doi.org/10.1016/j.physletb.2015.04.068} {\bibfield  {journal}
  {\bibinfo  {journal} {Phys. Lett. B}\ }\textbf {\bibinfo {volume} {746}},\
  \bibinfo {pages} {178} (\bibinfo {year} {2015})},\ \Eprint
  {https://arxiv.org/abs/1504.00607} {arXiv:1504.00607 [hep-ex]} \BibitemShut
  {NoStop}%
\bibitem [{\citenamefont {Dev}\ \emph {et~al.}(2024)\citenamefont {Dev},
  \citenamefont {Dutta}, \citenamefont {Han}, \citenamefont {Karthikeyan},
  \citenamefont {Kim},\ and\ \citenamefont {Kim}}]{Dev:2023zts}%
  \BibitemOpen
  \bibfield  {author} {\bibinfo {author} {\bibfnamefont {P.~S.~B.}\
  \bibnamefont {Dev}}, \bibinfo {author} {\bibfnamefont {B.}~\bibnamefont
  {Dutta}}, \bibinfo {author} {\bibfnamefont {T.}~\bibnamefont {Han}}, \bibinfo
  {author} {\bibfnamefont {A.}~\bibnamefont {Karthikeyan}}, \bibinfo {author}
  {\bibfnamefont {D.}~\bibnamefont {Kim}},\ and\ \bibinfo {author}
  {\bibfnamefont {H.}~\bibnamefont {Kim}},\ }\bibfield  {title} {\bibinfo
  {title} {{New physics at a neutron beam dump}},\ }\href
  {https://doi.org/10.1103/PhysRevD.110.L051703} {\bibfield  {journal}
  {\bibinfo  {journal} {Phys. Rev. D}\ }\textbf {\bibinfo {volume} {110}},\
  \bibinfo {pages} {L051703} (\bibinfo {year} {2024})},\ \Eprint
  {https://arxiv.org/abs/2311.10078} {arXiv:2311.10078 [hep-ph]} \BibitemShut
  {NoStop}%
\bibitem [{\citenamefont {Aaij}\ \emph {et~al.}(2018)\citenamefont {Aaij} \emph
  {et~al.}}]{LHCb:2017trq}%
  \BibitemOpen
  \bibfield  {author} {\bibinfo {author} {\bibfnamefont {R.}~\bibnamefont
  {Aaij}} \emph {et~al.} (\bibinfo {collaboration} {LHCb}),\ }\bibfield
  {title} {\bibinfo {title} {{Search for Dark Photons Produced in 13 TeV $pp$
  Collisions}},\ }\href {https://doi.org/10.1103/PhysRevLett.120.061801}
  {\bibfield  {journal} {\bibinfo  {journal} {Phys. Rev. Lett.}\ }\textbf
  {\bibinfo {volume} {120}},\ \bibinfo {pages} {061801} (\bibinfo {year}
  {2018})},\ \Eprint {https://arxiv.org/abs/1710.02867} {arXiv:1710.02867
  [hep-ex]} \BibitemShut {NoStop}%
\bibitem [{\citenamefont {Babusci}\ \emph {et~al.}(2013)\citenamefont {Babusci}
  \emph {et~al.}}]{KLOE-2:2012lii}%
  \BibitemOpen
  \bibfield  {author} {\bibinfo {author} {\bibfnamefont {D.}~\bibnamefont
  {Babusci}} \emph {et~al.} (\bibinfo {collaboration} {KLOE-2}),\ }\bibfield
  {title} {\bibinfo {title} {{Limit on the production of a light vector gauge
  boson in phi meson decays with the KLOE detector}},\ }\href
  {https://doi.org/10.1016/j.physletb.2013.01.067} {\bibfield  {journal}
  {\bibinfo  {journal} {Phys. Lett. B}\ }\textbf {\bibinfo {volume} {720}},\
  \bibinfo {pages} {111} (\bibinfo {year} {2013})},\ \Eprint
  {https://arxiv.org/abs/1210.3927} {arXiv:1210.3927 [hep-ex]} \BibitemShut
  {NoStop}%
\bibitem [{\citenamefont {Anastasi}\ \emph {et~al.}(2015)\citenamefont
  {Anastasi} \emph {et~al.}}]{Anastasi:2015qla}%
  \BibitemOpen
  \bibfield  {author} {\bibinfo {author} {\bibfnamefont {A.}~\bibnamefont
  {Anastasi}} \emph {et~al.},\ }\bibfield  {title} {\bibinfo {title} {{Limit on
  the production of a low-mass vector boson in $\mathrm{e}^{+}\mathrm{e}^{-}
  \to \mathrm{U}\gamma$, $\mathrm{U} \to \mathrm{e}^{+}\mathrm{e}^{-}$ with the
  KLOE experiment}},\ }\href {https://doi.org/10.1016/j.physletb.2015.10.003}
  {\bibfield  {journal} {\bibinfo  {journal} {Phys. Lett. B}\ }\textbf
  {\bibinfo {volume} {750}},\ \bibinfo {pages} {633} (\bibinfo {year}
  {2015})},\ \Eprint {https://arxiv.org/abs/1509.00740} {arXiv:1509.00740
  [hep-ex]} \BibitemShut {NoStop}%
\bibitem [{\citenamefont {Anastasi}\ \emph {et~al.}(2018)\citenamefont
  {Anastasi} \emph {et~al.}}]{KLOE-2:2018kqf}%
  \BibitemOpen
  \bibfield  {author} {\bibinfo {author} {\bibfnamefont {A.}~\bibnamefont
  {Anastasi}} \emph {et~al.} (\bibinfo {collaboration} {KLOE-2}),\ }\bibfield
  {title} {\bibinfo {title} {{Combined limit on the production of a light gauge
  boson decaying into $\mu^+\mu^-$ and $\pi^+\pi^-$}},\ }\href
  {https://doi.org/10.1016/j.physletb.2018.08.012} {\bibfield  {journal}
  {\bibinfo  {journal} {Phys. Lett. B}\ }\textbf {\bibinfo {volume} {784}},\
  \bibinfo {pages} {336} (\bibinfo {year} {2018})},\ \Eprint
  {https://arxiv.org/abs/1807.02691} {arXiv:1807.02691 [hep-ex]} \BibitemShut
  {NoStop}%
\bibitem [{\citenamefont {Ablikim}\ \emph {et~al.}(2017)\citenamefont {Ablikim}
  \emph {et~al.}}]{BESIII:2017fwv}%
  \BibitemOpen
  \bibfield  {author} {\bibinfo {author} {\bibfnamefont {M.}~\bibnamefont
  {Ablikim}} \emph {et~al.} (\bibinfo {collaboration} {BESIII}),\ }\bibfield
  {title} {\bibinfo {title} {{Dark Photon Search in the Mass Range Between 1.5
  and 3.4 GeV/$c^2$}},\ }\href {https://doi.org/10.1016/j.physletb.2017.09.067}
  {\bibfield  {journal} {\bibinfo  {journal} {Phys. Lett. B}\ }\textbf
  {\bibinfo {volume} {774}},\ \bibinfo {pages} {252} (\bibinfo {year}
  {2017})},\ \Eprint {https://arxiv.org/abs/1705.04265} {arXiv:1705.04265
  [hep-ex]} \BibitemShut {NoStop}%
\bibitem [{\citenamefont {Lees}\ \emph {et~al.}(2014)\citenamefont {Lees} \emph
  {et~al.}}]{BaBar:2014zli}%
  \BibitemOpen
  \bibfield  {author} {\bibinfo {author} {\bibfnamefont {J.~P.}\ \bibnamefont
  {Lees}} \emph {et~al.} (\bibinfo {collaboration} {BaBar}),\ }\bibfield
  {title} {\bibinfo {title} {{Search for a Dark Photon in $e^+e^-$ Collisions
  at BaBar}},\ }\href {https://doi.org/10.1103/PhysRevLett.113.201801}
  {\bibfield  {journal} {\bibinfo  {journal} {Phys. Rev. Lett.}\ }\textbf
  {\bibinfo {volume} {113}},\ \bibinfo {pages} {201801} (\bibinfo {year}
  {2014})},\ \Eprint {https://arxiv.org/abs/1406.2980} {arXiv:1406.2980
  [hep-ex]} \BibitemShut {NoStop}%
\bibitem [{\citenamefont {Graham}\ \emph {et~al.}(2021)\citenamefont {Graham},
  \citenamefont {Hearty},\ and\ \citenamefont {Williams}}]{Graham:2021ggy}%
  \BibitemOpen
  \bibfield  {author} {\bibinfo {author} {\bibfnamefont {M.}~\bibnamefont
  {Graham}}, \bibinfo {author} {\bibfnamefont {C.}~\bibnamefont {Hearty}},\
  and\ \bibinfo {author} {\bibfnamefont {M.}~\bibnamefont {Williams}},\
  }\bibfield  {title} {\bibinfo {title} {{Searches for Dark Photons at
  Accelerators}},\ }\href {https://doi.org/10.1146/annurev-nucl-110320-051823}
  {\bibfield  {journal} {\bibinfo  {journal} {Ann. Rev. Nucl. Part. Sci.}\
  }\textbf {\bibinfo {volume} {71}},\ \bibinfo {pages} {37} (\bibinfo {year}
  {2021})},\ \Eprint {https://arxiv.org/abs/2104.10280} {arXiv:2104.10280
  [hep-ph]} \BibitemShut {NoStop}%
\bibitem [{\citenamefont {Craik}\ \emph {et~al.}(2022)\citenamefont {Craik},
  \citenamefont {Ilten}, \citenamefont {Johnson},\ and\ \citenamefont
  {Williams}}]{Craik:2022riw}%
  \BibitemOpen
  \bibfield  {author} {\bibinfo {author} {\bibfnamefont {D.}~\bibnamefont
  {Craik}}, \bibinfo {author} {\bibfnamefont {P.}~\bibnamefont {Ilten}},
  \bibinfo {author} {\bibfnamefont {D.}~\bibnamefont {Johnson}},\ and\ \bibinfo
  {author} {\bibfnamefont {M.}~\bibnamefont {Williams}},\ }\bibfield  {title}
  {\bibinfo {title} {{LHCb future dark-sector sensitivity projections for
  Snowmass 2021}},\ }in\ \href@noop {} {\emph {\bibinfo {booktitle} {{Snowmass
  2021}}}}\ (\bibinfo {year} {2022})\ \Eprint
  {https://arxiv.org/abs/2203.07048} {arXiv:2203.07048 [hep-ph]} \BibitemShut
  {NoStop}%
\bibitem [{\citenamefont {Seto}\ \emph {et~al.}(2025)\citenamefont {Seto},
  \citenamefont {Shimomura},\ and\ \citenamefont {Yoshida}}]{Seto:2025mte}%
  \BibitemOpen
  \bibfield  {author} {\bibinfo {author} {\bibfnamefont {O.}~\bibnamefont
  {Seto}}, \bibinfo {author} {\bibfnamefont {T.}~\bibnamefont {Shimomura}},\
  and\ \bibinfo {author} {\bibfnamefont {S.}~\bibnamefont {Yoshida}},\
  }\bibfield  {title} {\bibinfo {title} {{New rare meson decay constraints on a
  light vector in U(1)$_{B-L}$, U(1)$_{R}$ and the dark photon}},\ }\href
  {https://doi.org/10.1007/JHEP09(2025)211} {\bibfield  {journal} {\bibinfo
  {journal} {JHEP}\ }\textbf {\bibinfo {volume} {09}},\ \bibinfo {pages}
  {211}},\ \Eprint {https://arxiv.org/abs/2504.15896} {arXiv:2504.15896
  [hep-ph]} \BibitemShut {NoStop}%
\bibitem [{\citenamefont {Elam}\ \emph {et~al.}(2022)\citenamefont {Elam} \emph
  {et~al.}}]{REDTOP:2022slw}%
  \BibitemOpen
  \bibfield  {author} {\bibinfo {author} {\bibfnamefont {J.}~\bibnamefont
  {Elam}} \emph {et~al.} (\bibinfo {collaboration} {REDTOP}),\ }\bibfield
  {title} {\bibinfo {title} {{The REDTOP experiment: Rare $\eta/\eta^{\prime}$
  Decays To Probe New Physics}},\ }\href@noop {} {\  (\bibinfo {year}
  {2022})},\ \Eprint {https://arxiv.org/abs/2203.07651} {arXiv:2203.07651
  [hep-ex]} \BibitemShut {NoStop}%
\bibitem [{\citenamefont {Pospelov}(2009)}]{Pospelov:2008zw}%
  \BibitemOpen
  \bibfield  {author} {\bibinfo {author} {\bibfnamefont {M.}~\bibnamefont
  {Pospelov}},\ }\bibfield  {title} {\bibinfo {title} {{Secluded U(1) below the
  weak scale}},\ }\href {https://doi.org/10.1103/PhysRevD.80.095002} {\bibfield
   {journal} {\bibinfo  {journal} {Phys. Rev. D}\ }\textbf {\bibinfo {volume}
  {80}},\ \bibinfo {pages} {095002} (\bibinfo {year} {2009})},\ \Eprint
  {https://arxiv.org/abs/0811.1030} {arXiv:0811.1030 [hep-ph]} \BibitemShut
  {NoStop}%
\bibitem [{\citenamefont {Parker}\ \emph {et~al.}(2018)\citenamefont {Parker},
  \citenamefont {Yu}, \citenamefont {Zhong}, \citenamefont {Estey},\ and\
  \citenamefont {M{\"u}ller}}]{Parker:2018vye}%
  \BibitemOpen
  \bibfield  {author} {\bibinfo {author} {\bibfnamefont {R.~H.}\ \bibnamefont
  {Parker}}, \bibinfo {author} {\bibfnamefont {C.}~\bibnamefont {Yu}}, \bibinfo
  {author} {\bibfnamefont {W.}~\bibnamefont {Zhong}}, \bibinfo {author}
  {\bibfnamefont {B.}~\bibnamefont {Estey}},\ and\ \bibinfo {author}
  {\bibfnamefont {H.}~\bibnamefont {M{\"u}ller}},\ }\bibfield  {title}
  {\bibinfo {title} {{Measurement of the fine-structure constant as a test of
  the Standard Model}},\ }\href {https://doi.org/10.1126/science.aap7706}
  {\bibfield  {journal} {\bibinfo  {journal} {Science}\ }\textbf {\bibinfo
  {volume} {360}},\ \bibinfo {pages} {191} (\bibinfo {year} {2018})},\ \Eprint
  {https://arxiv.org/abs/1812.04130} {arXiv:1812.04130 [physics.atom-ph]}
  \BibitemShut {NoStop}%
\bibitem [{\citenamefont {Morel}\ \emph {et~al.}(2020)\citenamefont {Morel},
  \citenamefont {Yao}, \citenamefont {Clad{\'e}},\ and\ \citenamefont
  {Guellati-Kh{\'e}lifa}}]{Morel:2020dww}%
  \BibitemOpen
  \bibfield  {author} {\bibinfo {author} {\bibfnamefont {L.}~\bibnamefont
  {Morel}}, \bibinfo {author} {\bibfnamefont {Z.}~\bibnamefont {Yao}}, \bibinfo
  {author} {\bibfnamefont {P.}~\bibnamefont {Clad{\'e}}},\ and\ \bibinfo
  {author} {\bibfnamefont {S.}~\bibnamefont {Guellati-Kh{\'e}lifa}},\
  }\bibfield  {title} {\bibinfo {title} {{Determination of the fine-structure
  constant with an accuracy of 81 parts per trillion}},\ }\href
  {https://doi.org/10.1038/s41586-020-2964-7} {\bibfield  {journal} {\bibinfo
  {journal} {Nature}\ }\textbf {\bibinfo {volume} {588}},\ \bibinfo {pages}
  {61} (\bibinfo {year} {2020})}\BibitemShut {NoStop}%
\bibitem [{\citenamefont {Fan}\ \emph {et~al.}(2023)\citenamefont {Fan},
  \citenamefont {Myers}, \citenamefont {Sukra},\ and\ \citenamefont
  {Gabrielse}}]{Fan:2022eto}%
  \BibitemOpen
  \bibfield  {author} {\bibinfo {author} {\bibfnamefont {X.}~\bibnamefont
  {Fan}}, \bibinfo {author} {\bibfnamefont {T.~G.}\ \bibnamefont {Myers}},
  \bibinfo {author} {\bibfnamefont {B.~A.~D.}\ \bibnamefont {Sukra}},\ and\
  \bibinfo {author} {\bibfnamefont {G.}~\bibnamefont {Gabrielse}},\ }\bibfield
  {title} {\bibinfo {title} {{Measurement of the Electron Magnetic Moment}},\
  }\href {https://doi.org/10.1103/PhysRevLett.130.071801} {\bibfield  {journal}
  {\bibinfo  {journal} {Phys. Rev. Lett.}\ }\textbf {\bibinfo {volume} {130}},\
  \bibinfo {pages} {071801} (\bibinfo {year} {2023})},\ \Eprint
  {https://arxiv.org/abs/2209.13084} {arXiv:2209.13084 [physics.atom-ph]}
  \BibitemShut {NoStop}%
\bibitem [{\citenamefont {Clade}\ \emph {et~al.}(2006)\citenamefont {Clade},
  \citenamefont {de~Mirandes}, \citenamefont {Cadoret}, \citenamefont
  {Guellati-Khelifa}, \citenamefont {Schwob}, \citenamefont {Nez},
  \citenamefont {Julien},\ and\ \citenamefont {Biraben}}]{Clade:2006zz}%
  \BibitemOpen
  \bibfield  {author} {\bibinfo {author} {\bibfnamefont {P.}~\bibnamefont
  {Clade}}, \bibinfo {author} {\bibfnamefont {E.}~\bibnamefont {de~Mirandes}},
  \bibinfo {author} {\bibfnamefont {M.}~\bibnamefont {Cadoret}}, \bibinfo
  {author} {\bibfnamefont {S.}~\bibnamefont {Guellati-Khelifa}}, \bibinfo
  {author} {\bibfnamefont {C.}~\bibnamefont {Schwob}}, \bibinfo {author}
  {\bibfnamefont {F.}~\bibnamefont {Nez}}, \bibinfo {author} {\bibfnamefont
  {L.}~\bibnamefont {Julien}},\ and\ \bibinfo {author} {\bibfnamefont
  {F.}~\bibnamefont {Biraben}},\ }\bibfield  {title} {\bibinfo {title}
  {{Determination of the Fine Structure Constant Based on Bloch Oscillations of
  Ultracold Atoms in a Vertical Optical Lattice}},\ }\href
  {https://doi.org/10.1103/PhysRevLett.96.033001} {\bibfield  {journal}
  {\bibinfo  {journal} {Phys. Rev. Lett.}\ }\textbf {\bibinfo {volume} {96}},\
  \bibinfo {pages} {033001} (\bibinfo {year} {2006})},\ \Eprint
  {https://arxiv.org/abs/physics/0510101} {arXiv:physics/0510101} \BibitemShut
  {NoStop}%
\bibitem [{\citenamefont {Aliberti}\ \emph {et~al.}(2025)\citenamefont
  {Aliberti} \emph {et~al.}}]{Aliberti:2025beg}%
  \BibitemOpen
  \bibfield  {author} {\bibinfo {author} {\bibfnamefont {R.}~\bibnamefont
  {Aliberti}} \emph {et~al.},\ }\bibfield  {title} {\bibinfo {title} {{The
  anomalous magnetic moment of the muon in the Standard Model: an update}},\
  }\href {https://doi.org/10.1016/j.physrep.2025.08.002} {\bibfield  {journal}
  {\bibinfo  {journal} {Phys. Rept.}\ }\textbf {\bibinfo {volume} {1143}},\
  \bibinfo {pages} {1} (\bibinfo {year} {2025})},\ \Eprint
  {https://arxiv.org/abs/2505.21476} {arXiv:2505.21476 [hep-ph]} \BibitemShut
  {NoStop}%
\bibitem [{\citenamefont {Hook}\ \emph {et~al.}(2011)\citenamefont {Hook},
  \citenamefont {Izaguirre},\ and\ \citenamefont {Wacker}}]{Hook:2010tw}%
  \BibitemOpen
  \bibfield  {author} {\bibinfo {author} {\bibfnamefont {A.}~\bibnamefont
  {Hook}}, \bibinfo {author} {\bibfnamefont {E.}~\bibnamefont {Izaguirre}},\
  and\ \bibinfo {author} {\bibfnamefont {J.~G.}\ \bibnamefont {Wacker}},\
  }\bibfield  {title} {\bibinfo {title} {{Model Independent Bounds on Kinetic
  Mixing}},\ }\href {https://doi.org/10.1155/2011/859762} {\bibfield  {journal}
  {\bibinfo  {journal} {Adv. High Energy Phys.}\ }\textbf {\bibinfo {volume}
  {2011}},\ \bibinfo {pages} {859762} (\bibinfo {year} {2011})},\ \Eprint
  {https://arxiv.org/abs/1006.0973} {arXiv:1006.0973 [hep-ph]} \BibitemShut
  {NoStop}%
\bibitem [{\citenamefont {Curtin}\ \emph {et~al.}(2015)\citenamefont {Curtin},
  \citenamefont {Essig}, \citenamefont {Gori},\ and\ \citenamefont
  {Shelton}}]{Curtin:2014cca}%
  \BibitemOpen
  \bibfield  {author} {\bibinfo {author} {\bibfnamefont {D.}~\bibnamefont
  {Curtin}}, \bibinfo {author} {\bibfnamefont {R.}~\bibnamefont {Essig}},
  \bibinfo {author} {\bibfnamefont {S.}~\bibnamefont {Gori}},\ and\ \bibinfo
  {author} {\bibfnamefont {J.}~\bibnamefont {Shelton}},\ }\bibfield  {title}
  {\bibinfo {title} {{Illuminating Dark Photons with High-Energy Colliders}},\
  }\href {https://doi.org/10.1007/JHEP02(2015)157} {\bibfield  {journal}
  {\bibinfo  {journal} {JHEP}\ }\textbf {\bibinfo {volume} {02}},\ \bibinfo
  {pages} {157}},\ \Eprint {https://arxiv.org/abs/1412.0018} {arXiv:1412.0018
  [hep-ph]} \BibitemShut {NoStop}%
\bibitem [{\citenamefont {Loizos}\ \emph {et~al.}(2024)\citenamefont {Loizos},
  \citenamefont {Wang}, \citenamefont {Thomas}, \citenamefont {White},\ and\
  \citenamefont {Williams}}]{Loizos:2023xbj}%
  \BibitemOpen
  \bibfield  {author} {\bibinfo {author} {\bibfnamefont {B.~M.}\ \bibnamefont
  {Loizos}}, \bibinfo {author} {\bibfnamefont {X.~G.}\ \bibnamefont {Wang}},
  \bibinfo {author} {\bibfnamefont {A.~W.}\ \bibnamefont {Thomas}}, \bibinfo
  {author} {\bibfnamefont {M.~J.}\ \bibnamefont {White}},\ and\ \bibinfo
  {author} {\bibfnamefont {A.~G.}\ \bibnamefont {Williams}},\ }\bibfield
  {title} {\bibinfo {title} {{Constraints on the dark sector from electroweak
  precision observables}},\ }\href {https://doi.org/10.1088/1361-6471/ad4efd}
  {\bibfield  {journal} {\bibinfo  {journal} {J. Phys. G}\ }\textbf {\bibinfo
  {volume} {51}},\ \bibinfo {pages} {075002} (\bibinfo {year} {2024})},\
  \Eprint {https://arxiv.org/abs/2306.13408} {arXiv:2306.13408 [hep-ph]}
  \BibitemShut {NoStop}%
\bibitem [{\citenamefont {Zieli{\'n}ski}\ and\ \citenamefont
  {Gatto}(2025)}]{Zielinski:2025wfa}%
  \BibitemOpen
  \bibfield  {author} {\bibinfo {author} {\bibfnamefont {M.}~\bibnamefont
  {Zieli{\'n}ski}}\ and\ \bibinfo {author} {\bibfnamefont {C.}~\bibnamefont
  {Gatto}},\ }\bibfield  {title} {\bibinfo {title} {{The REDTOP Experiment: an
  \(\eta /\eta ^{\prime }\) Factory to Explore Dark Matter and Physics Beyond
  the Standard Model}},\ }\href {https://doi.org/10.5506/APhysPolBSupp.18.4-A5}
  {\bibfield  {journal} {\bibinfo  {journal} {Acta Phys. Polon. Supp.}\
  }\textbf {\bibinfo {volume} {18}},\ \bibinfo {pages} {4} (\bibinfo {year}
  {2025})},\ \Eprint {https://arxiv.org/abs/2509.26552} {arXiv:2509.26552
  [hep-ex]} \BibitemShut {NoStop}%
\bibitem [{\citenamefont {Gan}\ \emph {et~al.}(2022)\citenamefont {Gan},
  \citenamefont {Kubis}, \citenamefont {Passemar},\ and\ \citenamefont
  {Tulin}}]{Gan:2020aco}%
  \BibitemOpen
  \bibfield  {author} {\bibinfo {author} {\bibfnamefont {L.}~\bibnamefont
  {Gan}}, \bibinfo {author} {\bibfnamefont {B.}~\bibnamefont {Kubis}}, \bibinfo
  {author} {\bibfnamefont {E.}~\bibnamefont {Passemar}},\ and\ \bibinfo
  {author} {\bibfnamefont {S.}~\bibnamefont {Tulin}},\ }\bibfield  {title}
  {\bibinfo {title} {{Precision tests of fundamental physics with
  {\ensuremath{\eta}} and {\ensuremath{\eta}}' mesons}},\ }\href
  {https://doi.org/10.1016/j.physrep.2021.11.001} {\bibfield  {journal}
  {\bibinfo  {journal} {Phys. Rept.}\ }\textbf {\bibinfo {volume} {945}},\
  \bibinfo {pages} {1} (\bibinfo {year} {2022})},\ \Eprint
  {https://arxiv.org/abs/2007.00664} {arXiv:2007.00664 [hep-ph]} \BibitemShut
  {NoStop}%
\bibitem [{\citenamefont {Gherghetta}\ \emph {et~al.}(2019)\citenamefont
  {Gherghetta}, \citenamefont {Kersten}, \citenamefont {Olive},\ and\
  \citenamefont {Pospelov}}]{Gherghetta:2019coi}%
  \BibitemOpen
  \bibfield  {author} {\bibinfo {author} {\bibfnamefont {T.}~\bibnamefont
  {Gherghetta}}, \bibinfo {author} {\bibfnamefont {J.}~\bibnamefont {Kersten}},
  \bibinfo {author} {\bibfnamefont {K.}~\bibnamefont {Olive}},\ and\ \bibinfo
  {author} {\bibfnamefont {M.}~\bibnamefont {Pospelov}},\ }\bibfield  {title}
  {\bibinfo {title} {{Evaluating the price of tiny kinetic mixing}},\ }\href
  {https://doi.org/10.1103/PhysRevD.100.095001} {\bibfield  {journal} {\bibinfo
   {journal} {Phys. Rev. D}\ }\textbf {\bibinfo {volume} {100}},\ \bibinfo
  {pages} {095001} (\bibinfo {year} {2019})},\ \Eprint
  {https://arxiv.org/abs/1909.00696} {arXiv:1909.00696 [hep-ph]} \BibitemShut
  {NoStop}%
\bibitem [{\citenamefont {del Aguila}\ \emph {et~al.}(1988)\citenamefont {del
  Aguila}, \citenamefont {Coughlan},\ and\ \citenamefont
  {Quiros}}]{delAguila:1988jz}%
  \BibitemOpen
  \bibfield  {author} {\bibinfo {author} {\bibfnamefont {F.}~\bibnamefont {del
  Aguila}}, \bibinfo {author} {\bibfnamefont {G.~D.}\ \bibnamefont
  {Coughlan}},\ and\ \bibinfo {author} {\bibfnamefont {M.}~\bibnamefont
  {Quiros}},\ }\bibfield  {title} {\bibinfo {title} {{Gauge Coupling
  Renormalization With Several U(1) Factors}},\ }\href
  {https://doi.org/10.1016/0550-3213(88)90266-0} {\bibfield  {journal}
  {\bibinfo  {journal} {Nucl. Phys. B}\ }\textbf {\bibinfo {volume} {307}},\
  \bibinfo {pages} {633} (\bibinfo {year} {1988})},\ \bibinfo {note} {[Erratum:
  Nucl.Phys.B 312, 751 (1989)]}\BibitemShut {NoStop}%
\bibitem [{\citenamefont {del Aguila}\ \emph {et~al.}(1995)\citenamefont {del
  Aguila}, \citenamefont {Masip},\ and\ \citenamefont
  {Perez-Victoria}}]{delAguila:1995rb}%
  \BibitemOpen
  \bibfield  {author} {\bibinfo {author} {\bibfnamefont {F.}~\bibnamefont {del
  Aguila}}, \bibinfo {author} {\bibfnamefont {M.}~\bibnamefont {Masip}},\ and\
  \bibinfo {author} {\bibfnamefont {M.}~\bibnamefont {Perez-Victoria}},\
  }\bibfield  {title} {\bibinfo {title} {{Physical parameters and
  renormalization of $U(1)_a \times U(1)_b$ models}},\ }\href
  {https://doi.org/10.1016/0550-3213(95)00511-6} {\bibfield  {journal}
  {\bibinfo  {journal} {Nucl. Phys. B}\ }\textbf {\bibinfo {volume} {456}},\
  \bibinfo {pages} {531} (\bibinfo {year} {1995})},\ \Eprint
  {https://arxiv.org/abs/hep-ph/9507455} {arXiv:hep-ph/9507455} \BibitemShut
  {NoStop}%
\bibitem [{\citenamefont {Babu}\ \emph {et~al.}(1996)\citenamefont {Babu},
  \citenamefont {Kolda},\ and\ \citenamefont {March-Russell}}]{Babu:1996vt}%
  \BibitemOpen
  \bibfield  {author} {\bibinfo {author} {\bibfnamefont {K.~S.}\ \bibnamefont
  {Babu}}, \bibinfo {author} {\bibfnamefont {C.~F.}\ \bibnamefont {Kolda}},\
  and\ \bibinfo {author} {\bibfnamefont {J.}~\bibnamefont {March-Russell}},\
  }\bibfield  {title} {\bibinfo {title} {{Leptophobic U(1) $s$ and the R($b$) -
  R($c$) crisis}},\ }\href {https://doi.org/10.1103/PhysRevD.54.4635}
  {\bibfield  {journal} {\bibinfo  {journal} {Phys. Rev. D}\ }\textbf {\bibinfo
  {volume} {54}},\ \bibinfo {pages} {4635} (\bibinfo {year} {1996})},\ \Eprint
  {https://arxiv.org/abs/hep-ph/9603212} {arXiv:hep-ph/9603212} \BibitemShut
  {NoStop}%
\bibitem [{\citenamefont {Dienes}\ \emph {et~al.}(1997)\citenamefont {Dienes},
  \citenamefont {Kolda},\ and\ \citenamefont {March-Russell}}]{Dienes:1996zr}%
  \BibitemOpen
  \bibfield  {author} {\bibinfo {author} {\bibfnamefont {K.~R.}\ \bibnamefont
  {Dienes}}, \bibinfo {author} {\bibfnamefont {C.~F.}\ \bibnamefont {Kolda}},\
  and\ \bibinfo {author} {\bibfnamefont {J.}~\bibnamefont {March-Russell}},\
  }\bibfield  {title} {\bibinfo {title} {{Kinetic mixing and the supersymmetric
  gauge hierarchy}},\ }\href {https://doi.org/10.1016/S0550-3213(97)00173-9}
  {\bibfield  {journal} {\bibinfo  {journal} {Nucl. Phys. B}\ }\textbf
  {\bibinfo {volume} {492}},\ \bibinfo {pages} {104} (\bibinfo {year}
  {1997})},\ \Eprint {https://arxiv.org/abs/hep-ph/9610479}
  {arXiv:hep-ph/9610479} \BibitemShut {NoStop}%
\bibitem [{\citenamefont {Eidelman}\ and\ \citenamefont
  {Jegerlehner}(1995)}]{Eidelman:1995ny}%
  \BibitemOpen
  \bibfield  {author} {\bibinfo {author} {\bibfnamefont {S.}~\bibnamefont
  {Eidelman}}\ and\ \bibinfo {author} {\bibfnamefont {F.}~\bibnamefont
  {Jegerlehner}},\ }\bibfield  {title} {\bibinfo {title} {{Hadronic
  contributions to $g-2$ of the leptons and to the effective fine structure
  constant $\alpha (M_Z^2)$}},\ }\href {https://doi.org/10.1007/BF01553984}
  {\bibfield  {journal} {\bibinfo  {journal} {Z. Phys. C}\ }\textbf {\bibinfo
  {volume} {67}},\ \bibinfo {pages} {585} (\bibinfo {year} {1995})},\ \Eprint
  {https://arxiv.org/abs/hep-ph/9502298} {arXiv:hep-ph/9502298} \BibitemShut
  {NoStop}%
\bibitem [{\citenamefont {Jegerlehner}(2006)}]{Jegerlehner:2006ju}%
  \BibitemOpen
  \bibfield  {author} {\bibinfo {author} {\bibfnamefont {F.}~\bibnamefont
  {Jegerlehner}},\ }\bibfield  {title} {\bibinfo {title} {{Precision
  measurements of sigma(hadronic) for alpha(eff)(E) at ILC energies and
  (g-2)(mu)}},\ }\href {https://doi.org/10.1016/j.nuclphysbps.2006.09.060}
  {\bibfield  {journal} {\bibinfo  {journal} {Nucl. Phys. B Proc. Suppl.}\
  }\textbf {\bibinfo {volume} {162}},\ \bibinfo {pages} {22} (\bibinfo {year}
  {2006})},\ \Eprint {https://arxiv.org/abs/hep-ph/0608329}
  {arXiv:hep-ph/0608329} \BibitemShut {NoStop}%
\end{thebibliography}%

\end{document}